\documentclass[11pt, a4paper, twoside]{Thesis} % Paper size, default font size and one-sided paper

\graphicspath{{./Figures/}} % Specifies the directory where pictures are stored

\usepackage[square, numbers, comma, sort&compress]{natbib}

\hypersetup{linktoc=all,
  colorlinks=true,
  urlcolor=blue
} % Colors hyperlinks in blue - change to black if annoying

\usepackage[T1]{fontenc}
\usepackage[english, polish]{babel}
\usepackage[utf8]{inputenc}
\usepackage{amssymb,amsmath}
\usepackage{multirow}
\usepackage{makecell}
\usepackage{tikz}
\usetikzlibrary{arrows,shapes,trees,backgrounds,snakes,calc,arrows.meta} 
\usepackage{enumitem}
\usepackage{array}
\usepackage{physics}
\usepackage{booktabs}
\usepackage{feynmp}
\usepackage{feynmp-auto}
\usepackage{siunitx}
\usepackage[noend, algochapter]{algorithm2e} %linesnumbered
\usepackage{lipsum}
\usepackage[toc]{appendix}
\usepackage{etoolbox}

\title{\ttitle} % Defines the thesis title - don't touch this

\SetAlCapFnt{\sc \small}

\begin{document}
\selectlanguage{english}

\frontmatter % Use roman page numbering style (i, ii, iii, iv...) for the pre-content pages
% \item
%\setstretch{1.0} % Actual single spacing
\setstretch{1.2} % Line spacing of 1.3
%\setstretch{1.6} % Double line spacing for review

% Define the page headers using the FancyHdr package and set up for one-sided printing
\fancyhead{} % Clears all page headers and footers
\fancyhead[LE,RO]{\thepage} % Sets the right side header to show the page number
\fancyhead[LO,RE]{} % Clears the left side page header

\pagestyle{fancy} % Finally, use the "fancy" page style to implement the FancyHdr headers

\newcommand{\HRule}{\rule{\linewidth}{0.5mm}} % New command to make the lines in the title page

% Apply my LaTeX hacks
\renewcommand{\bar}{\overline}
\newcommand{\kaon}{\mathrm{K}^0}
\newcommand{\akaon}{\bar{\mathrm{K}}^0}
\newcommand{\Ks}{\mathrm{K_S}}
\newcommand{\Kl}{\mathrm{K_L}}
\newcommand{\Ksl}{\mathrm{K_{S,L}}}
\newcommand{\Kp}{\mathrm{K_{+}}}
\newcommand{\Km}{\mathrm{K_{-}}}
\newcommand{\vts}[1]{$\upsilon \, t_{#1} - \mathrm{S}_{#1}$}
\newcommand{\Ts}{T}
\newcommand{\CPs}{CP}
\newcommand{\CPTs}{CPT}

\def\ops/{\mbox{o-Ps}}
\def\jpet/{\mbox{J-PET}}

\newcolumntype{x}[1]{>{\centering\arraybackslash\hspace{0pt}}p{#1}}

\definecolor{darkgreen}{rgb}{0, 0.5, 0}
\definecolor{green}{rgb}{0, 0.5, 0}
\definecolor{darkblue}{rgb}{0, 0, 0.5}
\definecolor{blue}{rgb}{0, 0, 0.5}
\definecolor{black}{rgb}{0.2, 0.2, 0.2}
\tikzset{snake it/.style={decorate, decoration={snake, segment length=2mm, amplitude=0.5mm}}}

\newcommand{\sidecaption}[2]% #1 = label name
{\raisebox{\abovecaptionskip}{\begin{subfigure}[t]{0.45\textwidth}
  \caption[singlelinecheck=off]{#1}% do not center
  \label{#2}
\end{subfigure}}\ignorespaces}

% remove ugly boldface from appendices in TOC
\makeatletter
\g@addto@macro\appendices{%
  \addtocontents{toc}{\protect\patchcmd{\protect\l@chapter}{\bfseries}{}{}{}}
}
\makeatother

% hyphenation for some words
\hyphenation{strange-ness}
\hyphenation{tri-la-tera-tion}

% style settings
\widowpenalty=10000
\clubpenalty=10000

% PDF meta-data
\hypersetup{pdftitle={\ttitle}}
\hypersetup{pdfsubject=\subjectname}
\hypersetup{pdfauthor=\authornames}
\hypersetup{pdfkeywords=\keywordnames}

%----------------------------------------------------------------------------------------
%	TITLE PAGE
%-------------------------------------------------------------------------------
\maketitle

\clearpage % Start a new page
%---------------------------------------------------------------------------------------
%	UGLY BUT REQUIRED STATEMENT
%---------------------------------------------------------------------------------------
\cleardoublepage % Start a new page
\thispagestyle{plain}
\begingroup
\selectlanguage{polish}
\setstretch{1.1}

Wydział Fizyki, Astronomii i Informatyki Stosowanej\\
Uniwersytet Jagielloński\\

\vspace{2cm}

\begin{center}
  \LARGE Oświadczenie
\end{center}
\par
\vspace{3ex}
{\setlength{\parindent}{4ex}
Ja niżej podpisany Aleksander Gajos (nr indeksu: 1040627), doktorant Wydziału Fizyki Astronomii i Informatyki Stosowanej Uniwersytetu Jagiellońskiego, oświadczam, że przedłożona przeze mnie rozprawa doktorska pt.\ ,,Investigations of fundamental symmetries with the electron-positron systems'' jest oryginalna i przedstawia wyniki badań wykonanych przeze mnie osobiście, pod kierunkiem prof. dr. hab. Pawła Moskala oraz dr. Eryka Czerwińskiego. Pracę napisałem samodzielnie.

Oświadczam, że moja rozprawa doktorska została opracowana zgodnie z Ustawą o prawie autorskim i prawach pokrewnych z dnia 4~lutego 1994 r.\ (Dziennik Ustaw 1994 nr~24 poz.~83 wraz z późniejszymi zmianami).

Jestem świadom, że niezgodność niniejszego oświadczenia z prawdą ujawniona w dowolnym czasie, niezależnie od skutków prawnych wynikających z ww.~ustawy, może spowodować unieważnienie stopnia nabytego na podstawie tej rozprawy.
}
\vspace{8ex}

Kraków, dnia ....................... \hfill .............................................

\par
\endgroup
\selectlanguage{english}
\vfill

%---------------------------------------------------------------------------------------
%	FANCY QUOTE
%---------------------------------------------------------------------------------------
\cleardoublepage % Start a new page
\thispagestyle{plain}
\ \\[0.3\textheight]
\begingroup
\setstretch{1.1}
\leftskip4em
\textit{\phantom{aaaa} Poznawczy wysi\l{}ek cz\l{}owieka to ci\k{a}g z granic\k{a} w niesko\'nczoności, a filozofia to pr\'oba osi\k{a}gnięcia tej granicy za jednym zamachem, w kr\'otkim spięciu, daj\k{a}cym pewność wiedzy doskona\l{}ej i niewzruszonej. Nauka posuwa się tymczasem swoim drobnym krokiem, podobnym niekiedy do pe\l{}zania, a nawet okresami do dreptania w miejscu, lecz dociera w ko\'ncu do rozmaitych sza\'nc\'ow ostatecznych, wyrytych przez myśl filozoficzn\k{a}, i nie bacz\k{a}c wcale na to, że tam w\l{}aśnie mia\l{}a przebiegać ultymatywna granica rozumu, idzie dalej.}
\begin{flushright}
  Stanis\l{}aw Lem, \textit{,,G\l{}os Pana''}  
\end{flushright}

\vfill

\textit{Man's quest for knowledge is an expanding series whose limit is infinity, but philosophy seeks to attain that limit at one blow, by a short circuit providing the certainty of complete and inalterable truth. Science meanwhile advances at its gradual pace, often slowing to a crawl, and for periods it even walks in place, but eventually it reaches the various ultimate trenches dug by philosophical thought, and, quite heedless of the fact that it is not supposed to be able to cross those final barriers to the intellect, goes right on.}
\begin{flushright}
  Stanis\l{}aw Lem, \textit{``His Master's Voice''}\\
  Translation by Michael Kandel
\end{flushright}
%\setstretch{1.6} % Double line spacing for review
\par
\endgroup
\vfill

%----------------------------------------------------------------------------------------
%	ABSTRACT PAGE
%----------------------------------------------------------------------------------------
\cleardoublepage % Start a new page

\addtotoc{Abstract} % Add the "Abstract" page entry to the Contents 

 \abstract{\addtocontents{toc}{\vspace{1em}} % Add a gap in the Contents, for aesthetics
   This work concerned two experimental searches for the violation of fundamental discrete symmetries in physical systems originating from electron-positron interactions.

The first study was a direct test of the symmetry under reversal in time in transitions of neutral K mesons, performed with quantum-entangled neutral kaon pairs produced in the $e^+e^-\to\phi\to\Ks\Kl$ process. Data collected by the KLOE experiment operating at the DA$\Phi$NE collider in 2004--2005 were studied to select events of the $\Ks\Kl\to\pi e \nu\;3\pi^0$ and $\Ks\Kl\to \pi^+\pi^-\;\pi e\nu$ processes and compare their rates. For the $\Kl\to 3\pi^0$ decay involving only neutral particles, a dedicated reconstruction technique based on trilateration was devised. Rates of each process identified by two time-ordered neutral kaon decays, determined as a function of a difference between kaon decays, were used to measure the asymptotic level of two \Ts-violation sensitive ratios of double kaon decay rates, yielding the values of $R_2 = 1.020 \pm 0.017_{stat} \pm 0.035_{syst}$ and $R_4 = 0.990 \pm 0.017_{stat} \pm 0.039_{syst}$.
In agreement with expectation based on the size of the the dataset used, these results do not reach the sensitivity needed to probe T violation. However, this measurement proves that the required reconstruction and analysis of the data is feasible and prospects exist for a statistically significant test of the T symmetry with a larger dataset collected by the KLOE-2 experiment is certain systematic effects are eliminated.

The second part of this work comprised a demonstration of the feasibility of using the J-PET detector to search for non-vanishing angular correlations in the decays of ortho-positronium atoms, the lightest purely leptonic systems decaying into photons. The trilateration based reconstruction method prepared for $\Kl\to 3\pi^0$ decay at KLOE was adapted to the ortho-positronium annihilations into three photons. Its performance was validated using Monte Carlo simulations proving it may be applied to determination of spin direction of positrons forming the positronium atoms, thus allowing for control of their polarization in the experiment. Moreover, the feasibility of identification of $3\gamma$ events as well as reconstruction of their origin points was demonstrated using a test measurement performed with the J-PET detector.

%%% Local Variables:
%%% TeX-master: "main"
%%% End: 
 }

 \cleardoublepage
 
 \abstractpol{
   \begin{otherlanguage}{polish}
{
  \setlength{\parindent}{4em}
  \setlength{\parskip}{0em}
  \hspace{4em} Niniejsza praca dotyczyła dwóch eksperymentów poszukujących łamania podstawowych symetrii dyskretnych w układach powstających w oddziaływaniach elektron-pozyton.

Pierwszym z rozważanych eksperymentów był bezpośredni test symetrii względem odwrócenia w czasie w przejściach w układzie neutralnych mezonów K, przy pomocy kwantowo splątanych par neutralnych kaonów wytworzonych w procesie $e^+e^-\to\phi\to\Ks\Kl$. Dane zebrane przez eksperyment KLOE prowadzący pomiary na zderzaczu DA$\Phi$NE w latach 2004--2005 zostały przeanalizowane w celu identyfikacji zdarzeń procesów $\Ks\Kl\to\pi e \nu\;3\pi^0$ i $\Ks\Kl\to \pi^+\pi^-\;\pi e\nu$ oraz w celu porównania ich krotności. Na potrzeby identyfikacji zawierającego wyłącznie neutralne cząstki rozpadu $\Kl\to 3\pi^0$ została opracowana dedykowana technika rekonstrukcji oparta na trilateracji. Krotności obserwacji każdego z badanych procesów, wyrażone w funkcji różnicy czasów własnych rozpadów obydwu kaonów, zostały wykorzystane do wyznaczenia asymptotycznego poziomu dwóch stosunków krotności podwójnych rozpadów kaonów neutralnych, wrażliwych na efekty łamania symetrii T. Otrzymane wartości $R_2 = 1.020 \pm 0.017_{stat} \pm 0.035_{syst}$ oraz $R_4 = 0.990 \pm 0.017_{stat} \pm 0.039_{syst}$, zgodnie z przewidywaniami na podstawie rozmiaru użytej próbki danych, nie osiągają dokładności wymaganej do pomiaru stopnia łamania symetrii T. Przeprowadzony pomiar dowodzi jednak, że niezbędna rekonstrukcja oraz analiza danych jest wykonalna i istnieje możliwość przeprowadzenia znaczącego statystycznie testu symetrii T przy użyciu większego zbioru danych zebranego przez eksperyment KLOE-2, pod warunkiem usunięcia niektórych źródeł niepewności systematycznej.

Druga część niniejszej pracy polegała na wykazaniu możliwości użycia detektora J-PET do poszukiwania niezerowych korelacji kątowych w rozpadach atomów orto-pozytonium, najlżejszych całkowicie leptonowych układów ulegających rozpadowi na fotony. Oparta na trilateracji metoda rekonstrukcji przygotowana dla rozpadów $\Kl\to 3\pi^0$ została zaadaptowana do przypadku anihilacji atomów orto-pozytonium na trzy fotony. Weryfikacja jej skuteczności przy pomocy symulacji Monte Carlo wykazała, że może ona zostać wykorzystana do wyznaczenia kierunku spinu pozytonów tworzących orto-pozytonium, tym samym pozwalając na kontrolę ich polaryzacji w eksperymencie. Możliwość identyfikacji anihilacji na trzy fotony oraz rekonstrukcji punktów anihilacji została ponadto wykazana przy pomocy próbnego pomiaru przeprowadzonego przy pomocy detektora J-PET.

%%% Local Variables:
%%% TeX-master: "main"
%%% End:        
}
   \end{otherlanguage}
 }
 
\cleardoublepage % Start a new page

%----------------------------------------------------------------------------------------
%	LIST OF CONTENTS/FIGURES/TABLES PAGES
%----------------------------------------------------------------------------------------
\pagestyle{fancy} % The page style headers have been "empty" all this time, now use the "fancy" headers as defined before to bring them back

\makeatletter
  \def\@pnumwidth{4em}
  \def\@tocrmarg {3.5em} % Also advisable
\makeatother

\renewcommand{\chaptermark}[1]{%
\markboth{#1}{}}

\fancyhead[LO,RE]{\emph{Contents}} % Set the left side page header to "Contents"

%\setstretch{1.0}
\singlespacing
{
\hypersetup{
  linkcolor=black
} % Colors hyperlinks in blue - change to black if annoying
  
\tableofcontents % Write out the Table of Contents
}

\fancyhead[LO,RE]{\emph{\leftmark}} % Set the left side page header to chapter title

\mainmatter % Begin numeric (1,2,3...) page numbering

%\setstretch{1.0} % Actual single spacing
\setstretch{1.2} % Line spacing of 1.3
%\setstretch{1.6} % Double line spacing for review
%\singlespacing

\pagestyle{fancy} % Return the page headers back to the "fancy" style

% Include the chapters of the thesis as separate files from the Chapters folder
% Uncomment the lines as you write the chapters

\chapter*{Introduction}
\markboth{Introduction}{}
\addtotoc{Introduction}

The concept of symmetries in physics is based on the supposed invariance of physical systems under certain transformations. Whereas symmetries under continuous transformations such as translation in space are of great consequences as they give origin to conservation laws, another extremely important class of symmetry-related transformations is constituted by operations commonly referred to as inversions, which yield the original system when applied to the system twice. The three most relevant inversions are parity, charge conjugation and reversal in time%
\footnote{%
  While the term \textit{time reversal} is commonly used in literature, it has been argued that since the reversal of time as such is clearly unphysical, it is more appropriate to speak of \textit{reversal in time} (the term proposed by John S. Bell) or of \textit{motion reversal} (as originally used by E.~Wigner)~\cite{bernabeu_colloquium, sozzi} In this work, the names \textit{reversal in time} or just \textit{T} will be adopted for the operation and its corresponding symmetry.
}.

Parity operation inverts spatial coordinates with respect to the origin, charge conjugation is an exchange of particles with their antiparticles which makes all charges change sign, and reversal in time applied to a system transforms it to a system with the opposite sense of time, where the spatial coordinates remain unchanged whereas vectors of velocity, momentum and spin change sign. For each of these inversions, a corresponding symmetry of Nature may be considered, i.e.\ parity~($\mathcal{P}$) symmetry, $\mathcal{C}$~symmetry and symmetry under time reversal~(\Ts~symmetry).

Since the introduction of the concept of these fundamental discrete symmetries in microscopic systems described by quantum mechanics by E.~Wigner in 1931~\cite{wigner1931}, large efforts have been made to test these symmetries with various systems and interactions. In fact, all of these symmetries have been found to be violated by the weak interactions but while first deviations from $\mathcal{P}$~and $\mathcal{C}$ symmetries were observed already in 1956 and 1958, respectively~\cite{parity_violation, c_violation}, more than half of a century had passed before direct evidence for nonconservation of the $\mathcal{T}$~symmetry could be found in a measurement with the neutral B meson system performed by the BaBar experiment in 2012~\cite{t_violation_babar}.

After the surprising discovery of violation of the combined \CPs~symmetry by Christenson~\textit{et~al.} in 1964~\cite{cp_violation}, the interest in searches for \Ts~noninvariance had increased as time reversal symmetry-violating effects would be expected from the \CPTs~symmetry. However, even though to date multiple experiments have confirmed \CPs~violation and \CPTs~still appears as a good symmetry of Nature (as tested by numerous experiments~\cite{pdg2016}), the evidence for violation of symmetry under reversal in time is still scarce and tests of this symmetry remain a challenging field of elementary particle physics. On the other hand, despite a multitude of tests of \CPs~(including measurements of the violation level) and test of \CPTs~conducted to date, physical systems and interactions exist for which the fundamental discrete symmetries have been hardly investigated. An example of the latter is constituted by purely leptonic systems and their electromagnetic interactions, for which the violation of both \CPs~and \CPTs~was only recently excluded at the precision level of $10^{-3}$~\cite{cpt_positronium, cp_positronium}. The symmetry under reversal in time has never been studied in this sector.

The aim of this Thesis was to prove the feasibility of two new tests of discrete symmetries. The first one is a direct test of the symmetry under reversal in time in transitions of neutral mesons, following a recently proposed concept~\cite{theory-babar,theory:bernabeu-t}. Such a test, feasible in the systems of flavoured neutral mesons with quantum entanglement, is to date the only experimental technique which provided direct evidence on \Ts~violation through a measurement by the BaBar Collaboration using $B^0$ mesons~\cite{t_violation_babar}. To date, the only possible extension of this test to other physical systems can be performed by the KLOE-2 experiment where quantum-entangled neutral K meson pairs are produced in the decay of $\phi$ mesons created in electron-positron collisions. Hence, a large part of the work presented herein was concentrated on providing the tools required to conduct a direct test of symmetry under reversal in time with neutral kaons at the KLOE-2 experimental setup.
As the latter is in the course of collecting data at the time of writing of this Thesis, an analysis was performed with a dataset collected by the KLOE experiment in 2004--2005. Although sensitivity of \Ts~violation measurement results obtained with this data is limited due to statistics, the goal of this work was to devise steps needed to extract certain transitions of neutral kaons between their flavour and CP-definite states from the data taken by the general-purpose KLOE detector and to demonstrate the feasibility of the \Ts~test with a view to its realization with larger amount of data collected by the KLOE-2 experiment. To this end, a complete analysis of the KLOE 2004--2005 data was prepared including determination of the two \Ts-asymmetric observables of the test.

The second experimental search for discrete symmetry violation elaborated on in this work concerned the lightest purely leptonic system constituted by the bound state of an electron and a positron, i.e.\ a positronium atom. A search for non-vanishing expectation values of operators odd under certain symmetries in angular correlations between photons momenta in the decays of ortho-positronium (\ops/) atoms is one of the objectives of the J-PET (Jagiellonian Positron Emission Tomograph) experiment, capable of improving the present $\order{10^{-3}}$ limits on \CPs~and \CPTs~violation in the purely leptonic systems. A~similar approach may be used to perform the first test of the \Ts~symmetry with positronium atoms~\cite{moskal_potential}. Therefore, a part of this Thesis is devoted to a first attempt to identify \ops/$\to 3\gamma$ events in the data collected with the J-PET system. Reconstruction of such decays and the necessary introductory steps to test the discrete symmetries with orto-positronium decay into photons with J-PET are also described.

Although the two experiments concerned in this Thesis are seemingly different as concerning various physical systems, interactions and energy scales as well as employing different strategies of searching for discrete symmetries' violation, their common traits are more than originating from an electron-positron system. Both experimental cases involve decays of neutral particles into final states comprising several photons, reconstructed solely on the basis of the photons' interactions in a calorimetric detector. Such processes, namely $\Kl\to 3\pi^0\to 6\gamma$ in KLOE and \ops/$\to 3\gamma$ in J-PET, are reconstructed using a trilateration-based approach.

%
% TODO: update if structure changes!
%
This Thesis is divided into eight chapters. Chapter~1 comprises the most essential information about the operation of reversal in time, present state of the \Ts~violation searches as well as properties of the system of neutral K mesons used for the direct \Ts~test, whose theoretical principle is described in Chapter~2. Chapter~3, in turn, discusses the scheme of discrete symmetry tests with ortho-positronium decays at J-PET\@. These discussions of the concept of each experiment are followed by descriptions of the KLOE and J-PET experimental systems, given in Chapter~4. The trilateration-based reconstruction of neutral particle decays into photons, constituting a common point of both presented data analyses, is introduced in Chapter~5. Details of the analysis of KLOE data in view of the \Ts~symmetry test, along with obtained results and a discussion of perspectives for a measurement with the KLOE-2 detector are  discussed in Chapter~6. Subsequently, Chapter~7 contains the results of feasibility tests of reconstruction of $e^+e^-$ annihilations into three photons at J-PET in the context of planned searches for non-vanishing angular correlations of photon momenta in these decays. A summary of the results of both studies and an outlook for the capabilities of future measurements based on the presented work follow in the last Chapter. Finally, the text is closed by a set of Appendices comprising details and derivations of several numerical and data analysis methods used in this work.

%%% Local Variables:
%%% TeX-master: "../main"
%%% End: 

\chapter{Symmetry under reversal in time and its violation}\label{chapter:symmetries}

\section{Properties of reversal in time}
A transformation of reversal in time in the Quantum Mechanics is understood as a bijection of the Hilbert space $T: \mathcal{H} \to \mathcal{H}$ which transforms the state vectors and observables of a system in such a way that the resulting system has an inverse sense of time, i.e.\ (in the Schr\"odinger and Heisenberg picture respectively):
\begin{eqnarray*}
  \ket{\psi} \to& \ket{\psi_T} &= T\ket{\psi},\\
  O \to& O_T &= TOT^{\dagger},
\end{eqnarray*}
where $\ket{\psi}$ and $O$ denote a given state and operator, respectively, the $T$ subscript indicates their counterparts in the $T$-transformed spaces, and $T^{\dagger}$ denotes a Hermitian conjugate of the $T$ operator. Correspondence with reversal in time in the classical case requires that
\begin{equation*}
  T\mathbf{X}T^{\dagger} = \mathbf{X}, \qquad T\mathbf{P}T^{\dagger} = -\mathbf{P}, \qquad   T\boldsymbol{\sigma}T^{\dagger} = -\boldsymbol{\sigma},
\end{equation*}
where $\mathbf{X}$ and $\mathbf{P}$ stand for the operators of position and momentum, respectively, and $\boldsymbol{\sigma}$ is the spin operator.

A symmetry operator U which preserves transition probabilities must satisfy the condition:
\begin{equation}
  \forall_{\phi,\psi\in\mathcal{H}}: \left| \braket{\phi}{\psi}\right|^2 = \left| \braket{\phi'}{\psi'}\right|^2 = \left| \braket{U\phi}{U\psi}\right|^2,
\end{equation}
from which a conclusion known as Wigner's theorem follows that a symmetry operation must be represented by an operator which is either unitary or anti-unitary. The operators of parity and charge conjugation are indeed unitary. In case of reversal in time, however, it can be shown that its corresponding operator T may not be unitary, e.g~by considering the time evolution of a state $\psi$ for an infinitesimal time interval $(t,t+\delta t)$ in a system reversed in time:
\begin{equation}
  \ket{\psi(t)}\to\ket{\psi(t+\delta t)} \xrightarrow{T} \ket{\psi_T(t+\delta t)}\to\ket{\psi_T(t)}.
\end{equation}
An evolution of a reversed state $\psi_T(t+\delta t)$  must give the reversed state $\psi_T(t)$ in time $t$~\cite{sozzi}:
\begin{align}
  (1-iH\delta t)\ket{\psi_T(t+\delta t)} & = (1-iH\delta t)T\ket{\psi(t+\delta t)} \nonumber \\
 & = (1-iH\delta t)T(1-iH\delta t)\ket{\psi(t)} = \ket{\psi_T(t)}\\
  T(-i)H & = iHT.
\end{align}
From the above relation it follows that $TH=-HT$ if T is a unitary operator or $TH=HT$ is T is anti-unitary.
% \begin{equation}
%   TH=-HT.
%   \label{eq:unitary_relation}
% \end{equation}
If unitarity of T is assumed, however, an eigenstate of a Hamiltonian~$\ket{n}$:
\begin{equation}
  H\ket{n} = E_n\ket{n},
\end{equation}
transformed by T would result in the eigenstate reversed in time $T\ket{n}$ having a negative energy $-E_n<0$. This observation along with Wigner's theorem lead to a conclusion that the operator of reversal in time must be anti-unitary. A more comprehensive discussion of the anti-unitarity of the T~operator can be found in references~\cite{sozzi, sachs}.

The anti-unitarity differentiates the reversal in time from other fundamental discrete symmetry transformations and has grave consequences for the properties of the \Ts~symmetry as well as for experimental capabilities of putting it to test. 

It can be shown that for a particle with zero spin $T^2=1$ and for spin of $\frac{1}{2}$, $T^2=-1$~\cite{sachs}. The action of the time reversal operator in a state $\psi$ is therefore
\begin{equation}
  T\ket{\psi(t)} = \eta_T\ket{\psi(-t)}^*,
\end{equation}
where $|\eta_T|^2=1$. The $\eta_T$ phase factor can, however, always be replaced with unity with an appropriate choice of the phase. As a consequence, there are no states with physical eigenvalues of the T operator and thus the invariance under reversal in time does not imply existence of any conserved quantity which could be used for testing the symmetry e.g.\ through experimental verification of selection rules~\cite{sozzi}.
%Moreover, no null experiment using a single observable can prove the violation of \Ts symmetry~\cite{sozzi}. 

Although a test of the symmetry under reversal in time can be performed through a comparison between time-inverted processes, experimentally preparation of such a pair of states presents serious difficulties. Use of decay as the process is practically unfeasible due to impossibility to obtain the initial state from the decay products in the same conditions. However, other phenomena can be used as the processes under comparison, i.e. oscillations of flavoured neutral mesons as well as their transitions between \CPs~and flavour eigenstates. Both of these cases will be discussed in the next sections.

Despite aforementioned complications in experimental testing of the \Ts~symmetry, anti-unitarity of the T operator also opens the possibility to design other tests. One of the consequences of T anti-unitarity is the following relation, valid for any operator $O$:
\begin{equation}
  \begin{split}
    \mel{\phi}{O}{\psi} & = \mel{\phi}{T^{\dagger}TOT^{\dagger}T}{\psi} \\ & = \mel{\phi_T}{TOT^{\dagger}}{\psi_T}^* = \mel{\psi_T}{TO^{\dagger}T^{\dagger}}{\phi_T}.
  \end{split}
\label{eq:rel_any_operator}
\end{equation}
If the operator O is even or odd with respect to reversal in time so that
\begin{equation}
  O_T = TOT^{\dagger} = \pm O,
\end{equation}
relation~(\ref{eq:rel_any_operator}) applied to that operator yields:
\begin{equation}
  \mel{\phi}{O}{\psi} = \mel{\phi_T}{O_T}{\psi_T}^*,
\end{equation}
which for identical states $\phi=\psi$ and for a Hermitian operator $O^{\dagger}=O$ reduces to
\begin{equation}
  \expval{O}_T = \pm\expval{O},
  \label{eq:expectation_value_any_operator}
\end{equation}
where the sign refers to the parity (+) or oddity (-) of the operator O w.r.t.\ reversal in time.
As a consequence, the expectation value of a Hermitian operator odd under reversal in time must vanish if the system is time reversal invariant~\cite{sozzi}. This conclusion opens the possibility to search for \Ts~symmetry violation by means of measurements of expectation values of operators specially constructed from observables in the system so that the operators are T-odd.

\section{Experimental searches for \Ts~violation}
The peculiar properties of reversal in time as compared to other discrete transformations result in a very specific range of feasible experimental approaches to testing the \Ts~noninvariance. In fact, available experimental approaches can be divided into four categories following a comprehensive summary by L.~Wolfenstein~\cite{wolfenstein_summary}.

The first class of searches for \Ts~violation is constituted by attempts to measure a non-zero value of a T-odd operator in an elementary system, such as the electric dipole moment of particles. Since the electric field $\mathbf{E}$ is even under reversal in time and spin is odd, a non-vanishing term of the form $\mathrm{d}\boldsymbol{\sigma}\cdot\mathbf{E}$ in the Hamiltonian would break the \Ts~symmetry. A notable example of such an experiment is the search for neutron electric dipole moment, presently reaching a sensitivity at the level of $10^{-26}\:\mathrm{e}\cdot\mathrm{cm}$ in a direct measurement~\cite{nedm}. While in case of the electron, its electric dipole moment can only be probed indirectly by the dipole moments of paramagnetic atoms, the measurements to date (with Thorium atoms) reach a sensitivity of $10^{-28}\:\mathrm{e}\cdot\mathrm{cm}$~\cite{eedm}. Despite excellent precision, no evidence for \Ts~violation was provided by such experiments to date.

Another possible way to test the time reversal symmetry is based on the property of the T transformation derived in Equations~(\ref{eq:rel_any_operator})--(\ref{eq:expectation_value_any_operator}). If observables in a system allow to construct a T-odd operator and if the Hamilitonian of this system $H$ satisfies the following condition for the initial and final states $\ket{i},\;\ket{f}$ of the process under investigation:
\begin{equation}
  \mel{f}{H}{i} = \mel{-f}{H}{-i},
\end{equation}
then the measured non-zero expectation value of a T-odd scalar would be a signal of violation of the symmetry under time reversal~\cite{wolfenstein_summary}. One of such measurements, performed at the KEK-E246 experiment, studied the muon polarization transverse to the decay plane in the \mbox{$K^+\to\pi^0\mu^+\nu$} weak decay. Transverse polarization was defined as
\begin{equation*}
  \mathcal{P}_T = \mathcal{P}_{K}\cdot (\mathbf{p}_{\pi}\times\mathbf{p}_{\mu}) /|\mathbf{p}_{\pi}\times\mathbf{p}_{\mu}|,
\end{equation*}
where $\mathcal{P}_{K}$ denotes initial kaon polarization and $\mathbf{p}_{\pi}$ and $\mathbf{p}_{\mu}$ are the momentum vectors of pion and muon in the final state. The experiment yielded a value of
\begin{equation*}
  \mathcal{P}_T = (-1.7\pm 2.3\pm 1.1)\times 10^{-3},
\end{equation*}
showing no evidence for \Ts~violation~\cite{operatory_kek}. A similar principle was exploited in several measurements of transverse polarization of electrons from the $\beta$ decay performed at Paul Scherrer Institute, which studied decays of polarized $^8$Li nuclei~\cite{bodek_li} and of free neutrons~\cite{bodek_free_n}. In both experiments, the transverse polarization of electrons was defined as a T-odd operator whose non-vanishing expectation value was sought through measurement of angular correlations in the final state. These measurements also yielded results consistent with \Ts~invariance, amounting to $R=(0.9\pm2.2)\times 10^{-3}$ with lithium nuclei~\cite{bodek_li} and $R=0.004\pm 0.012\pm 0.005$ with free neutrons~\cite{bodek_free_n}.

%
% TODO: dopisac, ze trzba dobrz wziac pod uwage pseudo trv
%

It is worth noting that the measurements exploiting operators odd under reversal in time have been conducted using processes governed by the weak interactions only. In principle, however, this technique is not limited to a specific class of processes. In fact, non-zero expectation values of odd operators have been employed in searches of \CPs~and \CPTs~violation in a purely leptonic system constituted by ortho-positronium atoms, i.e. triplet bound states of electron and positron, undergoing a decay into three photons~\cite{cpt_positronium, cp_positronium}. Notably, no searches for signals of time reversal symmetry violation in such purely leptonic systems have been reported to date.

Besides the aforementioned searches for non-zero expectation values of T-odd operators, a range of tests of the \Ts~symmetry is possible with the systems of neutral mesons, in particular the $\kaon$ mesons. The oscillation between flavour eigenstates of neutral mesons allows to obtain time-conjugated pairs of particle transitions in a process different from particle decay.
While a direct comparison of rates of the transitions back and forth in neutral meson oscillations presents certain difficulties in terms of preparing a direct \Ts~test independent of \CPs~effects, another class of tests has been devised which utilizes neutral meson transitions between flavour and CP eigenstates and allows for a genuine test of the symmetry under reversal in time. Preparing tools for data analysis necessary to perform such a test with the neutral K meson system was one of the objectives of this Thesis.

Next sections will discuss properties of the system of neutral K mesons and the test of the symmetry under reversal in time in their oscillations, together with a discussion of the interpretation controversy arising from its result. Subsequently, principle of the novel concept of the \Ts~symmetry test designed to avoid these issues, which was employed experimentally in the work presented in this Thesis, will be covered in detail in Chapter~\ref{chapter:test_kloe}.

\section{The system of neutral K mesons}
\label{sec:kaons}
K mesons, commonly referred to as kaons, are the lightest particles containing the strange~(s) quark. Thanks to that fact, not only did they give rise to a new field of flavour physics being the first strange particles to be discovered, but also they were one of the most studied particle systems due to their great availability for medium energy range accelerator-based experiments such as NA31~\cite{Kleinknecht:1994ns}, NA48~\cite{Winhart:2012bv}, KTeV~\cite{Wanke:2003vp} or KLOE~\cite{kloe_results}.

The four K mesons constitute two isospin doublets: ($\mathrm{K}^{+}$, $\mathrm{K}^{0}$) with S=1 and ($\mathrm{K}^{-}$, $\mathrm{\overline{K}}^{0}$) with S=-1. The quark content of each of the K mesons is presented in~\tref{tab:quarks}.

%%%%%%%%%%%%%%%%%%%%%%%%%%%% table:quarks %%%%%%%%%%%%%%%%%%%%%%%%%%%%
\begin{table}[h!]
\centering
\caption{Quark content, strangeness and isospin of the K mesons.}\label{tab:quarks}
\begin{tabular}{crrrr}
  \toprule
  & $\mathrm{K}^{+}$ & $\mathrm{K}^{-}$ & $\kaon$ & $\akaon$ \\
  \midrule
  quarks & u\={s} & \={u}s & d\={s} & \={d}s  \\
  strangeness & 1 & -1 & 1 & -1 \\
  isospin & $+\frac{1}{2}$ & $-\frac{1}{2}$ & $+\frac{1}{2}$ & $-\frac{1}{2}$\\
  \bottomrule
\end{tabular}
\end{table}
%%%%%%%%%%%%%%%%%%%%%%%%%% end table:quarks %%%%%%%%%%%%%%%%%%%%%%%%%%

The K mesons are produced in strong interactions in which strangeness is conserved and thus their production in conveniently described in the basis of states with definite strangeness,
i.e.\ $\{\kaon,\akaon\}$ for neutral
and $\left\{\mathrm{K}^{+},\mathrm{K}^{-}\right\}$ for charged kaons.
Moreover, strangeness conservation demands that production of kaons in decays of non-strange particles results in a pair of kaons or a kaon accompanied by another strange particle like $\Lambda^0$~($\Lambda^0=$uds). The possibility to infer the strangeness of a kaon based on the flavour of the associated particle at its production is a very useful feature of the K meson system, extensively used in kaon experiments.

The remainder of this Section will focus only on the properties of the system of neutral K mesons due to its great relevance for discrete symmetry tests. As the neutral kaons decay through weak interactions which do not conserve strangeness, it is useful to express their states not only in the aforementioned flavour basis but also in the basis of eigenstates of the CP~operator. Even though $\kaon$ and $\akaon$, being antiparticles of each other, clearly are not CP eigenstates
\[ \mathcal{CP} \ket{ \kaon } = \ket{ \akaon }, \quad \mathcal{CP} \ket{ \akaon} = \ket{ \kaon }, \]
a proper basis of CP eigenstates $\left\{\Kp,\Km\right\}$%
\footnote{%
  This work will use a convention where the CP eigenstates are denoted by a subscript indicating CP parity of the neutral kaon state. States of charged kaons are indicated by the charge marked in superscript.
}
may be constructed with linear combinations of these flavour-definite kaon states:
\begin{equation}
  \label{eq:cp_basis}
  \begin{split}
    \ket{\Kp} & =\frac{1}{\sqrt{2}}\left(\ket{\kaon}+\ket{\akaon}\right)\qquad \mathcal{CP}\ket{\Kp}=+1\ket{\Kp},  \\
    \ket{\Km} & =\frac{1}{\sqrt{2}}\left(\ket{\kaon}-\ket{\akaon}\right)\qquad \mathcal{CP}\ket{\Km}=-1\ket{\Km}.
  \end{split}
\end{equation}

Conservation of the \CPs~symmetry requires that in case of neutral kaon decays into hadronic final states with pions, the $\Kp$ state may only decay into two pions, while for $\Km$ only a three-pion final state is allowed. Therefore, if \CPs~violation in kaon decays is neglected%
\footnote{%
In fact, CP is known to be violated in kaon decays. Influence of this violation on further considerations contained in this Thesis is, however, negligible~\cite{theory:bernabeu-t} as will be discussed in Chapter~\ref{chapter:test_kloe}.
},
observation of a decay of a neutral K meson into a certain final state with pions can be used to identify the decaying state in the $\left\{\Kp,\Km\right\}$ basis as summarized in \tref{tab:tagging}. In case of the anti-symmetric CP eigenstate, determination of the CP value for a $\pi^+\pi^-\pi^0$ final state requires consideration of the orbital momentum, but the decay into three neutral pions can be used as a clear indication of the decaying $\Km$ state.

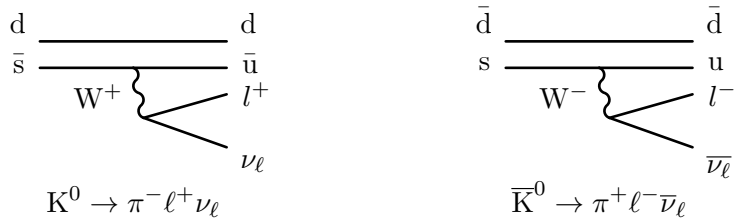
\begin{figure}[h!]
  \centering
  \begin{fmffile}{fgraphs}
    \vspace{1 em}
    \begin{minipage}{.4\textwidth}
          \centering
            \begin{fmfgraph*}(70,40)
              % bottom and top verticies
              \fmfstraight
              \fmfleft{i2,a,i3,i4,i5}
              \fmfright{o2,a,o3,o4,o5}
              % incoming proton to gluon vertices
              \fmf{plain}{i5,o5}
              \fmflabel{d}{i5}
              \fmflabel{d}{o5}
              \fmf{plain}{i4,v2}
              \fmflabel{\={s}}{i4}
              \fmf{plain}{v2,o4}
              \fmflabel{\={u}}{o4}
              \fmffreeze
              \fmf{photon, tension=1,label=$\mathrm{W}^{+}$}{v2,v3}
              \fmf{plain}{v3,o3}
              \fmflabel{$l^{+}$}{o3}
              \fmf{plain}{v3,o2}
              \fmflabel{$\nu_{\ell}$}{o2}
              % phantom centres the W->cs vertex
              \fmf{phantom,tension=1.5}{i2,v3}
            \end{fmfgraph*}
 %           \vspace{1 em}
          \[ \kaon \to \pi^- \ell^+ {\nu}_{\ell} \]
          \end{minipage}
            %
%            \hspace{0.1\textwidth}
            %
    \begin{minipage}{.4\textwidth}
          \centering
            \begin{fmfgraph*}(70,40)
              % bottom and top verticies
              \fmfstraight
              \fmfleft{i2,a,i3,i4,i5}
              \fmfright{o2,a,o3,o4,o5}
              % incoming proton to gluon vertices
              \fmf{plain}{i5,o5}
              \fmflabel{\={d}}{i5}
              \fmflabel{\={d}}{o5}
              \fmf{plain}{i4,v2}
              \fmflabel{s}{i4}
              \fmf{plain}{v2,o4}
              \fmflabel{u}{o4}
              \fmffreeze
              \fmf{photon, tension=1,label=$\mathrm{W}^{-}$}{v2,v3}
              \fmf{plain}{v3,o3}
              \fmflabel{$l^{-}$}{o3}
              \fmf{plain}{v3,o2}
              \fmflabel{$\overline{\nu_{\ell}}$}{o2}
              % phantom centres the W->cs vertex
              \fmf{phantom,tension=1.5}{i2,v3}
            \end{fmfgraph*}
%           \vspace{1 em}
            \[ \akaon \to \pi^+ \ell^- \bar{\nu}_{\ell} \]
          \end{minipage}
     \end{fmffile}
  \caption{Feynman diagrams of the semileptonic decays of neutral K mesons. The $\Delta S=\Delta Q$ rule allows the neutral kaon states with +1 and -1 strangeness to decay into a semileptonic final sate only with a positively or negatively charged lepton, respectively. }
  \label{fig:dsdq}
\end{figure}

Similarly to the identification (later on referred to as \textit{tagging}) of the CP eigenstates with hadronic decays, semileptonic final states into a pion, lepton (generally denoted by $\ell$ in the following considerations) and neutrino may be utilized to identify flavour of the decaying kaon state. This is based on the so-called $\Delta S=\Delta Q$ selection rule
%stating that a change in the strangeness in a weak decay must be matched by the same change in the 
related to CPT conservation whose violation has not been observed to date~\cite{pdg2016}. As a consequence of $\Delta S=\Delta Q$, the $\kaon$ state can only decay semileptonically into a state with a positively charged lepton while $\akaon$ may only produce a negative lepton as illustrated by the Feynman diagrams in Figure~\ref{fig:dsdq}.
The right-hand-side of~\tref{tab:tagging} summarizes the possibilities of tagging the flavour states of neutral kaons with semileptonic decays.

\begin{table}[h!]
  \centering
  \caption{Possibilities of identification of the neutral kaon states in the CP and flavour bases by observation of kaon decays into specific final states.}\label{tab:tagging}
  \begin{tabular}[center]{cp{5ex}ccp{5ex}cc}
    \toprule
    kaon state & & $\Kp$ & $\Km$ & & $\kaon$ & $\akaon$ \\
               & & CP=+1 & CP=-1 & & S=+1    & S=-1     \\
    \midrule
    identifying & & $\pi^+\pi^-$ & $\pi^0\pi^0\pi^0$ & & $\pi^-\ell^+\bar{\nu}$ & $\pi^+\ell^-\nu$ \\
    decay      & & $\pi^0\pi^0$ &                   & &\\
    \bottomrule
  \end{tabular}
\end{table}

The decays of neutral K mesons involve weak interactions violating the \CPs~symmetry. In fact, it was the kaon system in which the \CPs~violation was discovered in the famous experiment of Christenson, Cronin, Fitch and Turlay~\cite{cp_violation} where the state supposed as pure $\Km$ was observed to decay into a two-pion final state. As a result of \CPs~noninvariance, the ``physical'' states of neutral kaons, i.e.\ eigenstates of the full system Hamiltonian including weak interactions, are not exactly CP-definite states $\Kp$ and $\Km$. Instead, the former must be corrected to allow for state impurities whose size is measured by small complex parameters $\epsilon_S$ and $\epsilon_L$ (compare with Equations~\ref{eq:cp_basis}):
 \begin{equation}
   \label{eq:kskl}
   \begin{split}
     \ket{\Ks} & = \frac{1}{\sqrt{2(1+|\epsilon_S|^2)}}\left((1+\epsilon_S)\ket{\kaon}+(1-\epsilon_S)\ket{\akaon}\right),  \\
     \ket{\Kl} & = \frac{1}{\sqrt{2(1+|\epsilon_L|^2)}}\left((1+\epsilon_L)\ket{\kaon}-(1-\epsilon_L)\ket{\akaon}\right). 
   \end{split}
 \end{equation}
The above states properly describe the decaying kaons. The indices S and L attributed to these physical states correspond respectively to \textit{short-lived} and \textit{long-lived} neutral kaons as the lifetimes of $\Ks$ and $\Kl$ differ almost by three orders of magnitude (see~\tref{tab:kaon_properties}). This effect originates from smaller phase space for the decays of $\Kl$ which predominantly decays into three pions as opposed to $\Ks$ for which two-pion decays dominate as shown in the list of major neutral kaon decays in~\tref{tab:kaon_properties}.

%%%%%%%%%%%%%%%%%%%%%%%%%%% kaon_properties %%%%%%%%%%%%%%%%%%%%%%%%%%%
\begin{table}[h!]
  \small
  \centering
  \caption{Selected properties and major decay modes of neutral kaons \cite{pdg2016}.}\label{tab:kaon_properties}
  \begin{tabular}{ccrcr}
  \toprule
  {}                &         \multicolumn{2}{c}{  $\mathrm{K_S}$} & \multicolumn{2}{c}{ $\mathrm{K_L}$}   \\
  \midrule
  mean life time    & \multicolumn{2}{c}{(89.54 $\pm$ 0.04) ps} & \multicolumn{2}{c}{(51.16 $\pm$ 0.21) ns} \\
  \midrule
                  &	     $\pi^+\pi^-$ & (6.920 $\pm$ 0.005)$\times 10^{-1}$  &    $\pi^\pm e^{\mp} \nu_e$ & (4.055 $\pm$ 0.011)$\times 10^{-1}$ \\      	      
major decay &	     $\pi^0\pi^0$ & (3.069 $\pm$ 0.005)$\times 10^{-1}$	&    $\pi^\pm \mu^{\mp} \nu_{\mu}$ & (2.704 $\pm$ 0.007)$\times 10^{-1}$ \\      	   
modes  &	     $\pi^+\pi^-\gamma$ & (1.79 $\pm$ 0.05) $\times 10^{-3}$		 &   	    3$\pi^0$ & (1.952 $\pm$ 0.012)$\times 10^{-1}$   \\
 (branching ratio)  &	     $\pi^{\pm} e^{\mp} \nu_e$ & (7.04 $\pm$ 0.08) $\times  10^{-4}$	   	 &	    $\pi^+\pi^-\pi^0$ & (1.254 $\pm$ 0.005)$\times 10^{-1}$  \\
  &	          $\pi^{\pm} \mu^{\mp} \nu_{\mu}$ & (4.69 $\pm$ 0.05) $\times  10^{-4}$          &	$\pi^{\pm}e^{\mp}\nu_e\gamma$ & (3.79 $\pm$ 0.06) $\times 10^{-3}$ \\
&  \multicolumn{2}{c}{} &  $\pi^+\pi^-$ & (1.966 $\pm$ 0.010) $\times 10^{-3}$	 \\
                    & \multicolumn{2}{c}{}  &  $\pi^0\pi^0$ & (8.64 $\pm$ 0.06) $\times 10^{-4}$       \\
    \midrule
mass              &	     \multicolumn{4}{c}{ (497.611 $\pm$ 0.013) MeV/$\mathrm{c^2}$ } \\ 
  mass difference   &          \multicolumn{4}{c}{ (3.484 $\pm$ 0.006)$\times 10^{-12}$ MeV/$\mathrm{c^2}$ } \\
  \bottomrule
\end{tabular}
\end{table}
%%%%%%%%%%%%%%%%%%%%%%%%% end kaon_properties %%%%%%%%%%%%%%%%%%%%%%%%%

The full Hamiltonian of the neutral K meson system including strong, electromagnetic and weak interactions is a 2$\times$2 non-hermitian matrix which may be decomposed into hermitian and anti-hermitian parts~\cite{interf_handbook}
\begin{equation}
  \mathbf{H} =  \mathbf{M} - \frac{i}{2} \mathbf{\Gamma},
\end{equation}
where the hermitian \textbf{M} and $\mathbf{\Gamma}$ matrices are commonly referred to as mass and decay matrices. The eigenvalues of $\mathbf{H}$ which correspond to eigenstates of $\Ks$ and $\Kl$ take the following form:
\begin{equation}
  \begin{split} 
    \label{eq:eigenvalues}
    \lambda_S & = m_S - i \frac{\Gamma_S}{2}, \\
    \lambda_L & = m_L - i \frac{\Gamma_L}{2},
  \end{split}
\end{equation}
where $m_{S(L)}$ denotes mass of $\mathrm{K_{S(L)}}$ state and $\Gamma_{\mathrm{S(L)}}$ denotes its decay width. The time evolution of the physical states of neutral kaons is thus described by pure exponentials
\begin{eqnarray}
  \ket{\Ks(t)} & = e^{-i\lambda_S t}\ket{\Ks}, \\
  \ket{\Kl(t)} & = e^{-i\lambda_L t}\ket{\Kl}.
\end{eqnarray}

Properties of the system of neutral K mesons, expressed by its Hamiltonian, are inherently connected to fundamental discrete symmetries. The kaons' relation to \Ts, \CPs~and \CPTs~symmetries may be expressed in an elegant form of constraints on particular elements of $\mathbf{H}$. It is therefore useful to expand the Hamiltonian as well as mass and decay matrices and consider particular elements:
\begin{equation}
  \label{eq:hamiltonian}
  \left(
    \begin{array}{cc}
      \mathrm{H_{11}}  &        \mathrm{H_{12}} \\
      \mathrm{H_{21}}  &        \mathrm{H_{22}}
    \end{array}
    \right)
    =
    % = \mathbf{M} - \frac{i}{2} \mathbf{\Gamma} = 
\left(
  \begin{array}{cc}
    \mathrm{M_{11}}  &        \mathrm{M_{12}} \\
    \mathrm{M^*_{12}}  &        \mathrm{M_{22}}
  \end{array}
\right)
- \frac{i}{2}
\left(
  \begin{array}{cc}
    {\Gamma_{11}}  &        {\Gamma_{12}} \\
    {\Gamma^*_{12}}  &        {\Gamma_{22}}
  \end{array}
\right).
\end{equation}

The following constraints must be satisfied if the neutral kaon system is invariant under certain symmetry transformations~\cite{book_cp_violation}:
\begin{eqnarray}
  \label{eq:discrete_constraints}
  \mathcal{CPT}\text{ invariance:}\quad & \mathrm{H}_{11} = \mathrm{H}_{22}, \label{eq:constr_cpt} \\
  \mathcal{T}\text{ invariance:}\quad & |\mathrm{H}_{12}| = |\mathrm{H}_{21}|, \label{eq:constr_t} \\
  \mathcal{CP}\text{ invariance:}\quad & \mathrm{H}_{11} = \mathrm{H}_{22} \;\land\; |\mathrm{H}_{12}| = |\mathrm{H}_{21}|. \label{eq:constr_cp}
\end{eqnarray}

A convention commonly used in the description of discrete symmetries' violation in the $\Ks\Kl$ system  redefines the \CPs-violating parameters $\epsilon_S$ and $\epsilon_L$ from Eqs.~\ref{eq:kskl} in the following manner:
\begin{eqnarray}
  \label{eq:epsilon_delta}
  {\epsilon} & \equiv \frac{\epsilon_S+\epsilon_L}{2}, \\
  \delta & \equiv \frac{\epsilon_S-\epsilon_L}{2}.
\end{eqnarray}
Thus defined $\delta$ and ${\epsilon}$ parameters are related to Hamiltonian elements up to leading order terms as below~\cite{interf_handbook}:
\begin{eqnarray}
  \label{eq:epsilon_delta_H}
  {\epsilon} & = \frac{\mathrm{H}_{12}-\mathrm{H}_{21}}{2(\lambda_S-\lambda_L)} & =
                   \frac{-i\Im(\mathrm{M}_{12})-\frac{1}{2}\Im(\Gamma_{12})}{\Delta m +\frac{i}{2}\Delta \Gamma}, \\
  \delta & = \frac{\mathrm{H}_{11}-\mathrm{H}_{22}}{2(\lambda_S-\lambda_L)} & =
                   \frac{\frac{1}{2}\left(\mathrm{M}_{22} - \mathrm{M}_{11}-\frac{i}{2}(\Gamma_{22}-\Gamma_{11})\right)}{\Delta m + \frac{i}{2}\Delta \Gamma}.
\end{eqnarray}

It can be shown that with a certain choice of an arbitrary phase in ${\epsilon}$~\cite{fidecaro_pedagogical}
\begin{equation}
  \label{eq:kabir_epsilon}
  \frac{|\mathrm{H}_{12}|^2-|\mathrm{H}_{21}|^2}{|\mathrm{H}_{12}|^2+|\mathrm{H}_{21}|^2} \simeq 4\Re({\epsilon}).
\end{equation}
In accordance with constraints~(\ref{eq:constr_t}) and (\ref{eq:constr_cp}), the above term violates the \CPs~and \Ts~symmetries and is known as the Kabir asymmetry~\cite{Kabir1970}. More generally, non-zero value of the real part of $\epsilon$ would imply that symmetry under reversal in time is violated, whereas $\Re(\delta)$ is a parameter sensitive to \CPTs~violation in the neutral kaon system~\cite{interf_handbook}.

\section{Test of \Ts~symmetry in oscillations of neutral kaons}
% Due to zero spin of neutral mesons such as kaons, their flavour eigenstates are invariant under the operation of reversal in time.
The states of particles of zero spin are invariant under the operation of reversal in time. Therefore, a direct test of the \Ts~symmetry for such particles may be defined in a way very close to the commonsense understanding of ``time reversal'', i.e.\ through a comparison of a process $\ket{i}\to \ket{f}$ and the same process with the initial and final states exchanged $\ket{f}\to \ket{i}$. A natural property to compare experimentally are the probabilities of both processes measurable with the rates of observed events.

Such a T-conjugated pair of processes is, however, not easily available in experiment. No decay phenomena may be used due to the practical impossibility to obtain a reverse process with exactly the same parameters of the system. A unique case in which a pair of processes mutually reverse in time can be observed, is the oscillation of neutral mesons such as kaons.
The latter oscillate between the states with $+1$ and $-1$ strangeness so that the transitions $\kaon\to\akaon$ and $\akaon\to\kaon$ satisfy the requirements for a pair of time-reversal conjugated processes and can be used for a \Ts~symmetry test by measuring the probability asymmetry between them.

Such an experiment was performed in 1998 with the CPLEAR detector, where $\kaon$ and $\akaon$ were produced in strong interactions $p\bar{p}\to \mathrm{K}^-\pi^+\kaon$ and $p\bar{p}\to \mathrm{K}^+\pi^-\akaon$ and the other products were used to identify initial ($t=0$) strangeness of the kaon. Subsequently, state of the K meson at the time $\tau$ of its decay was tagged by certain semileptonic final states of the decay (see~\tref{tab:tagging}). The observable of the test was the following asymmetry of transition probabilities:
\begin{equation}
  \label{eq:cplear_prob}
  A = \frac{\text{P}(\akaon\to\kaon) - \text{P}(\kaon\to\akaon)}{\text{P}(\akaon\to\kaon) + \text{P}(\kaon\to\akaon)},
\end{equation}
probed experimentally through the average decay rate asymmetry~\cite{cplear}:
\begin{equation}
  \label{eq:cplear_exp}
  A_{CPLEAR} = \left< \frac{R(\akaon(t=0)\to e^+\pi^-\nu(t=\tau))-R(\kaon(t=0)\to e^-\pi^+\bar{\nu}(t=\tau))}{R(\akaon(t=0)\to e^+\pi^-\nu(t=\tau))+R(\kaon(t=0)\to e^-\pi^+\bar{\nu}(t=\tau)} \right>,
\end{equation}
Measurement of the above quantity with CPLEAR has yielded a significant non-zero asymmetry:
\begin{equation}
  \label{eq:cplear_result}
  A_{CPLEAR} = (6.6\pm 1.3_{stat}\pm 1.0_{syst})\times 10^{-3}.
\end{equation}

This result was interpreted by the CPLEAR collaboration as  as a measurement of the Kabir asymmetry~\cite{Kabir1970}, which can be related to the neutral kaon system Hamiltonian (see Equation~(\ref{eq:hamiltonian})) elements as shown in Eq.~(\ref{eq:kabir_epsilon}).

Since non-zero value of the Kabir asymmetry would clearly imply violation of the symmetry under reversal in time, the result of CPLEAR was quoted as the first direct observation of time-reversal noninvariance in the neutral kaon system~\cite{cplear}. However, it has been pointed out by L.~Wolfenstein that the measured value may not be identical with the Kabir asymmetry due to an essential role of decay in the phenomenon studied by CPLEAR~\cite{wolfenstein_other_paper, wolfenstein_summary}.
Due to the fact that this measurement inevitably uses the decay as initial state interaction, the asymmetry between $\kaon\to\akaon$ and $\akaon\to\kaon$ transitions results from an interference between two effects. Besides the dispersive component of neutral kaon mixing, which includes short-distance box diagrams and long-range interactions mediated by particles off the mass shell, there is also a contribution from initial state interaction. As the latter involves decays into intermediate particles on the mass shell, this contribution is at a leading order proportional to the decay width difference $\Delta \Gamma$ between $\kaon$ and $\akaon$~\cite{bernabeu_colloquium}.

In fact, Wolfenstein has shown that the dependence of the asymmetry measured by CPLEAR on the decay width difference takes the form
\begin{equation}
  A_{CPLEAR} = \frac{2\Delta \Gamma (\Im(\mathrm{M}_{12})+\frac{1}{2}i \Im(\Gamma_{12}))}{\left(\frac{1}{2}\Delta \Gamma\right)^2+(\Delta m)^2} \xrightarrow{\Delta \Gamma \to 0} 0,
\end{equation}
so that in the limit of small $\Delta \Gamma$ the asymmetry would diverge from the Kabir asymmetry and its value would vanish even if the symmetry under reversal in time was substantially violated~\cite{wolfenstein_summary}. This claim is in agreement with a measurement of the equivalent of this asymmetry in the oscillations of neutral B mesons for which the decay width difference is negligible, which was performed by the BaBar collaboration and yielded a result consistent with zero~\cite{babar_zero_result}. The requirement that an observable for a \Ts-violation test must not vanish in the $\Delta \Gamma \to 0$ limit is recognized as the Wolfenstein criterion.

While several other authors disputed the critique of the interpretation of the CPLEAR result and others argued that the decay is not relevant in the case of neutral K meson oscillations~\cite{Ellis:1999xh,Gerber:2004hc},
another way to perform a direct test of the symmetry under reversal in time which would be free of the aforementioned issues was clearly desirable.

\section[Tests of \Ts~symmetry in neutral meson transitions between\newline flavour and CP eigenstates]{Tests of \Ts~symmetry in neutral meson transitions between flavour and CP eigenstates}
 A solution to the interpretation problems of the \Ts-violation measurement in neutral meson oscillations was first proposed for the $\mathrm{B}^0\overline{\mathrm{B}}^0$
system soon after the discussion of CPLEAR result~\cite{wolfenstein_summary, banuls_first_bmesons} and was later devised in detail by Bernabeu~\textit{et al.}~\cite{babar_theory}. A proposition of a similar experiment using the neutral kaon system followed in 2013 with a view to its application in the only existing facility capable of performing such experiment, the KLOE detector at the DA$\Phi$NE $\phi$-factory~\cite{theory:bernabeu-t}.

\begin{figure}[h!]
  \centering
      \begin{tikzpicture}[scale=0.9]
      \node[text width=3cm, text centered] (a1) at (0,0) {\small $\kaon$};
      \node[text width=3cm, text centered] (a2) at (2,0) {\small $\Kp$};
      \node[text width=3cm, text centered] (a3) at (0,-1) {\small $\akaon$};
      \node[text width=3cm, text centered] (a4) at (2,-1) {\small $\Km$};
      \draw[thick, <->] ($(a1)+(0.5,0)$) -- ($(a2)-(0.5,0)$);
      \draw[thick, <->] ($(a3)+(0.5,0)$) -- ($(a4)-(0.5,0)$);
      \draw[thick, <->] ($(a1)+(0.5,-0.2)$) -- ($(a4)-(0.5,-0.2)$);
      \draw[thick, <->] ($(a3)+(0.5,0.2)$) -- ($(a2)-(0.5,0.2)$);
    \end{tikzpicture} 
  \caption{Possible choice of transitions between flavour and CP definite states for a \Ts~symmetry test with the neutral kaon system. For each transition, its time-reversal conjugate can be observed experimentally in an entangled neutral kaon system.}
  \label{fig:transitions}
\end{figure}
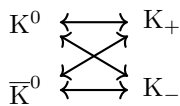

Similarly as in the case of neutral K or B meson oscillations, this concept follows the idea of a direct \Ts~symmetry test based on a comparison of probabilities for a certain transition $\ket{i}\to\ket{f}$ and its T-conjugate where initial and final states are inversed, $\ket{f}\to\ket{i}$. The $\ket{i}$ and $\ket{f}$ states, however, can be chosen as eigenstates of the CP operator in addition to the flavour-definite states of the mesons. Figure~\ref{fig:transitions} shows possible choices of the transitions between eigenstates in the flavour and CP bases for the system of neutral kaons. A $\kaon\to\Kp$ process, for example, may be compared to $\Kp\to\kaon$ leading to determination of the probability ratio $\mathcal{P}(\kaon\to\Kp)/\mathcal{P}(\Kp\to\kaon)$ as a \Ts~violation observable.

It is worth noting that preparation and observation of such a pair of processes mutually reverse in time requires that the initial state of each of the two transitions must be identified in the respective basis before the particle decays. In the scheme proposed in References~\cite{babar_theory, theory:bernabeu-t} this problem is overcome by using quantum-entangled pairs of neutral mesons produced in certain strong decays. As a result, the decay does not play a crucial role in the processes under study and observation of transitions between eigenstates in the flavour and CP bases  allows for a definition of measurable asymmetries sensitive to violation of the time reversal symmetry, which are independent of \CPs~violation and do not vanish in the $\Delta \Gamma \to 0$ limit, thus satisfying the Wolfenstein criterion. 

The requirement of quantum entanglement in a system of two neutral mesons, however, limits the possible realizations of such symmetry tests to B and $\phi$ factory facilities. A first direct test of the symmetry under reversal in time performed in this manner came from the BaBar setup and used the $\mathrm{B}^0$ meson system to determine \Ts-violating parameters in its time evolution~\cite{t_violation_babar}. The obtained values amount to:
\begin{eqnarray}
  \label{eq:babar_results}
  \Delta S^+_T & = 1.37 \pm 0.14_{stat} \pm 0.06_{syst}, \\
  \Delta S^-_T & = 1.17 \pm 0.18_{stat} \pm 0.11_{syst},
\end{eqnarray}
showing a significant offset from \Ts~noninvariance at the level of 14 $\sigma$. To date, this result constitutes the only existing direct evidence for violation of the symmetry under reversal in time obtained by an exchange of initial and final states.

It is therefore of great interest to perform such tests of the \Ts~symmetry in systems other than $\mathrm{B}^0\overline{\mathrm{B}}^0$. A natural candidate is the system of neutral K mesons renowned for symmetry-related discoveries it enabled through the last 70 years. The entangled pairs of neutral kaons are presently uniquely available to the KLOE detector operating on the DA$\Phi$NE collider in Laboratori Nazionali di Frascati, INFN, Italy. The work presented in this Thesis was largely devoted to devising analysis tools required to perform the direct test of \Ts~symmetry with the KLOE experiment and to conducting a preliminary test with the data already collected by KLOE. Details of the \Ts~test with entangled neutral kaons employed in this Thesis are presented in Chapter~\ref{chapter:test_kloe} whereas Chapter~\ref{chapter:detectors} contains a description of the experimental setup of KLOE and DA$\Phi$NE.

%%% Local Variables:
%%% TeX-master: "../main"
%%% End:  
\chapter{A direct test of time reversal symmetry with neutral K mesons}
\label{chapter:test_kloe}

%
% TODO: dopisac, ze niezalezny od CP
% 

%% introductory words
Reversal of a process in time, in its most intuitive sense, means an inversion of the arrow of time so that a process of transition from state A to state B becomes a passage from B to A. In a more detailed consideration, one can note that also the states A and B may not be symmetric under the \Ts~transformation, so that the transition reversed in time is in fact $\mathcal{T}(\mathrm{A}\to\mathrm{B}) = \mathcal{T}(\mathrm{B})\to\mathcal{T}(\mathrm{A})$. In the case of spinless particles such as neutral K mesons, however, the possible particle states are in fact \Ts-symmetric so that the notion of reversing a transition between states in time remains very close to the natural understanding of \textit{``time reversal''}. The beauty of the concept of a direct test of the \Ts~symmetry therefore lies in its proximity to the commonsense intuition behind testing whether Nature is symmetric under reversal in time by comparing whether a process A$\to$B is equally likely as B$\to$A. This Chapter presents the principles of such \Ts~symmetry test using transitions of neutral kaons between strangeness and \CPs~eigenstates along with prospects for its experimental realization with the KLOE setup. A more detailed description of this measurement concept can be found in Reference~\cite{theory:bernabeu-t}.

\section{Definition of the compared processes}
As explained in Section~\ref{sec:kaons}, the state of a neutral K meson can be described either in the flavour basis, i.e.\ the basis of $\{\kaon,\akaon\}$ states of definite strangeness or in the basis of eigenstates of the \CPs~operator $\left\{\Kp,\Km\right\}$. In order to prepare a direct test of the symmetry under reversal in time, transitions between certain kaon states must be observed experimentally as well as their \Ts-conjugates obtained by an exchange of initial and final states. To this end, transitions between pure flavour and \CPs-definite neutral kaon states can be used, allowing to define a total of eight transitions as shown in \fref{fig:transitions}. As half of these processes are time-reversal conjugates of the others,
four of them are chosen as reference transitions and compared to their T-inverted counterparts.

The direct symmetry test is based on a comparison of probabilities among such process pairs, therefore four observables $R_{i}$ ($i=1,\ldots 4$) are defined as ratios of probability of each reference and its \Ts-conjugated process:
\begin{eqnarray}
  R_1(\Delta t)  = \frac{P[\kaon(0) \to \Kp(\Delta t)]}{P[\Kp(0) \to \kaon(\Delta t)]}, \label{eq:r1t}\\
  {R_2(\Delta t)}  = {\frac{P[\kaon(0) \to \Km(\Delta t)]}{P[\Km(0) \to \kaon(\Delta t)]} },\label{eq:r2t}\\
  R_3(\Delta t)  = \frac{P[\akaon(0) \to \Kp(\Delta t)]}{P[\Kp(0) \to \akaon(\Delta t)]},\label{eq:r3t} \\
  {R_4(\Delta t)}  = {\frac{P[\akaon(0) \to \Km(\Delta t)]}{P[\Km(0) \to \akaon(\Delta t)]} }.\label{eq:r4t}
\end{eqnarray}
It can be shown that such probability asymmetries must be time-dependent~\cite{theory:bernabeu-t}, therefore the above ratios are functions the time $\Delta t$ of the evolution of the kaon between given initial and final states. If invariance under reversal in time holds, all these ratios are expected to be equal to unity for any value of $\Delta t$.

\section{Principle of observation of transitions}
In an experiment, the probabilities used to define the ratios from Equations~\ref{eq:r1t}--\ref{eq:r4t} are measurable through rates of events where a given transition is observed to occur in a certain time interval $\Delta t$. Registration of such events requires identification of kaon state in two moments of its evolution, $t=0$ and $t=\Delta t$. In the latter case, final state of a transition is easily recognized by observing the kaon decay into a certain final state, where semileptonic decays are only possible for $\kaon$ and $\akaon$ and final states with two or three pions indicate that the decaying kaon was in a pure $\Kp$ or $\Km$ state, as summarized in~\tref{tab:tagging}.

\begin{figure}[h!]
  \centering
  \begin{tikzpicture}[scale=1.0] %, show background rectangle]                   
    \node (first) at (-2,0){};
    \node (same) at (2,0){};
    \node (second) at (4.5,0){};
    \node[circle, inner sep=0.06cm, fill=gray,draw] at (0,0) {} node[above] {$\phi$};
    \draw[thick, dashed, ->] (0.1,0) -- (2,0);
    \draw[thick, dashed, ->] (-0.1,0) -- (-2,0);
    \node at (-1,0.3) {$\mathrm{K}?$};
    \node at (1,0.3)  {$\mathrm{K}?$};
      \node (t12) at (2,-0.5) {$t_1$};
      \node (t11) at (-2,-0.5) {$t_1$};
      \draw[black, ->] (first) -- (-2.8,0) node[] {$\pi^-\;\;\;\;$};
      \draw[black, ->] (first) -- (-2.8,0.4) node[] {$\ell^+\;\;\;\;$};
      \draw[black, ->] (first) -- (-2.8,-0.4) node[] {$\nu\:\;\;\;\;$};
      \node at (-1.7,0.7) {$\ket{\kaon}$};
      \node (decay1) at (-2,0.5) {};
      \node[above of=decay1, text width=8em, text centered] {(a)};
%      \draw[thick, red, ->] (decay1) -- (first);
      \node (decay2) at (2,0.5) {};
      \node[above of=decay2, text width=11em, text centered] {(b)};
%      \draw[thick, red, ->] (decay2) -- (same);
      \node at (2,0.7) {$\ket{\akaon}$};
      \draw[ultra thick, black, ->] (2,0) -- (4.5,0) node[midway,below] {$\Delta t$};
      % \node[star,star points=10, inner sep=0.06cm, draw, thick, fill=orange!100] at (second) {};
      \node (t21) at (4.5,-0.5) {$t_2$};
      \draw[black, ->] (second) -- (5.3,0) node[] {$\;\;\;\;\pi^0$};
      \draw[black, ->] (second) -- (5.3,0.4) node[] {$\;\;\;\;\pi^0$};
      \draw[black, ->] (second) -- (5.3,-0.4) node[] {$\;\;\;\;\pi^0$};
      \node at (4.3,0.7) {$\ket{\Km}$};
      \node (decay3) at (4.5,0.5) {};
      \node[above of=decay3, text width=8em, text centered] {(c)};
%      \draw[thick, red, ->] (decay3) -- (second);
  \end{tikzpicture}
  \caption{\label{fig:principle}
    Scheme of experimental observation of a $\akaon\to\Km$ transition using quantum  entanglement of $\kaon\akaon$ pairs. Dashed arrows denote initial propagation of the entangled kaons produced in a $\phi$ decay when neither of the states is known. (a) At time $t_1$, first of the kaons undergoes a $\mathrm{K}\to \pi^- \ell^+ \nu$ decay and is identified as $\kaon$ (see~\tref{tab:tagging}). (b)~At~the same time, its entangled partner is known to be in an $\akaon$ state. (c) After further propagation, its decay into $3\pi^0$ at time $t_2$ identifies the kaon as the $\Km$ state. Thick arrow denotes thus observed $\akaon\to\Km$ process happening in time $\Delta t=t_2-t_1$.
  }
\end{figure}
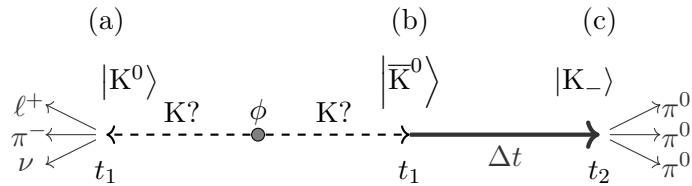

Identification of an initial state for a transition must be performed in a non-destructive way in order to observe further evolution of the kaon, to which end the peculiar properties of pairs of neutral K mesons produced in certain strong decays can be employed. In the decay of a $\phi$ meson with $J^{PC}=1^{--}$ into two neutral kaons, conservation of quantum numbers by the strong interaction results in the products created in a quantum-entangled antisymmetric state which may be expressed in the flavour basis as
\begin{equation}
  \label{eq:entangled1}
  \ket{i} = \frac{1}{\sqrt{2}}\left( \ket{\kaon(+\vec{p})}\ket{\akaon(-\vec{p})} - \ket{\akaon(+\vec{p})}\ket{\kaon(-\vec{p})} \right),
\end{equation}
or equivalently with the \CPs~eigenstates as
\begin{equation}
  \label{eq:entangled2}
  \ket{i} = \frac{1}{\sqrt{2}}\left( \ket{\Kp(+\vec{p})}\ket{\Km(-\vec{p})} - \ket{\Km(+\vec{p})}\ket{\Kp(-\vec{p})} \right),
\end{equation}
where the two kaons are distinguished by the sense of their momentum in the center-of-mass reference frame. Such neutral kaon pairs, exhibiting entanglement in the genuine sense first described by Einstein, Podolsky and Rosen~\cite{Einstein:1935rr}, allow for determination of the state of a kaon without its decay by observation of an earlier decay of its entangled partner. Entanglement requires that at the moment of decay of the first of two kaons, when observed products allow to tag its state in one of the bases, the second kaon which did not decay yet must be in an orthogonal state. This state, inferred from the other kaon decay, is used as initial state of the transitions investigated in the direct \Ts~symmetry test. This principle is depicted schematically for the case of a $\akaon\to\Km$ transition in Figure~\ref{fig:principle}. As a consequence of this measurement scheme, each of the kaon transitions entering the ratios shown in Equations~\ref{eq:r1t}--\ref{eq:r4t} is identified experimentally by observation of a time-ordered pair of certain neutral kaon decays. \tref{tab:processes} summarizes the compared transitions and their experimental signatures.

It is worth noting that in the above considerations the kaon states identified by their decays into two and three pions are supposed to be pure \CPs~eigenstates $\Kp$ and $\Km$ and to be exactly orthogonal, which is crucial for the use of quantum entanglement. However, as a result of direct \CPs~violation in the neutral kaon system, the decaying states $\Ks$ and $\Kl$ as defined in Equations~\ref{eq:kskl}, are not perfectly orthogonal, with $\braket{\Ks}{\Kl}\approx \epsilon_L + \epsilon_S^*$. Therefore, the considerations described herein explicitly neglect \CPs~violation in neutral kaon decays so that:
\begin{eqnarray}
  \label{eq:cp_neglect}
  \ket{\Kp} \simeq \ket{\Ks},\\
  \ket{\Km} \simeq \ket{\Kl}.
\end{eqnarray}
It has been shown in Reference~\cite{theory:bernabeu-t} that \CPs~violation indeed does not have a substantial effect on the measurement of the asymmetric ratios defined in Equations~\ref{eq:r1t}--\ref{eq:r4t}.

\begin{table}[h]
  \centering
  \caption{\label{tab:processes}Four transitions between flavour and \CPs~eigenstates available in the neutral kaon system and their time-reversal conjugates with initial and final states exchanged. Each of the listed processes is experimentally identified by a time-ordered pair of kaon decays into final states indicated in parentheses. }
    \begin{tabular}{lllll}
      \toprule
      No. &  \multicolumn{2}{c}{Transition}   & \multicolumn{2}{c}{\Ts-conjugate}  \\
      \midrule
      1 &  $\kaon \to \Kp$ & $(\pi^+\ell^-\bar{\nu},\pi\pi)$ & $\Kp \to \kaon$ & $(3\pi^0, \pi^-\ell^+\nu)$ \\
      2 &  $\kaon \to \Km$ & $(\pi^+\ell^-\bar{\nu}, 3\pi^0)$ & $\Km \to \kaon$ & $(\pi\pi,\pi^-\ell^+\nu)$  \\
      3 &  $\akaon \to \Kp$ & $(\pi^-\ell^+\nu,\pi\pi)$ & $\Kp \to \akaon$ & $(3\pi^0, \pi^+\ell^-\bar{\nu})$ \\
      4 &  $\akaon \to \Km$ & $(\pi^-\ell^+\nu,3\pi^0)$ & $\Km \to \akaon$ & $(\pi\pi, \pi^+\ell^-\bar{\nu})$ \\
      \bottomrule
    \end{tabular}
\end{table}

%TODO: sprawdzic czy mathrm w dsdq bylo w poprzednim rodziale
A second assumption which is taken in the direct test of the symmetry under time reversal, is the validity of the $\Delta S=\Delta Q$ rule in semileptonic kaon decays, which is employed in identification of flavour states of neutral kaons. This rule, however, has been thoroughly tested to date and no evidence exists for its violation~\cite{pdg2016}.

\section{Observables of the test}\label{sec:observables}
The principle of observation of neutral kaon transitions between flavour and \CPs~states using quantum entanglement requires that two kaon decays are observed in a single \mbox{$\phi\to\kaon\akaon$} event in order to identify a certain transition, and that their proper decay times are determined precisely to measure $\Delta t$. Each of the transitions listed in~\tref{tab:processes} is therefore identified by registration of a time-ordered pair of specific neutral kaon decays, indicated next to the transitions in the table. It should be emphasized that the first decay in each pair corresponds to an entangled partner of the kaon whose transition if observed and thus tags a state opposite to the initial one in the transition of interest.

Directly observable values comprise distributions of rates of events $I(f_1,f_2;\Delta t)$ characterized by two kaon decays into final states $f_1$ and $f_2$ as a function of the time interval $\Delta t$ between both decays measured in the kaons' proper reference frames. For the case of neutral K meson pairs produced in $\phi$ decays, the theoretical \Ts-asymmetric ratios defined in Equations~\ref{eq:r1t}--\ref{eq:r4t} are linked with the experimantally-measurable ratios of decay rates $R^{exp}$ defined in the following way:
  \begin{eqnarray}
    R_1^{exp}(\Delta t) & = \frac{\mathrm{I}(\pi^+\ell^-\bar{\nu},\pi\pi;\Delta t)}{\mathrm{I}(3\pi^0,\pi^-\ell^+\nu;\Delta t)} {  = R_1(\Delta t)\times  \frac{C(\pi^+\ell^-\bar{\nu},\pi\pi)}{C(3\pi^0,\pi^-\ell^+\nu)}},  \label{eq:r1e}\\
    {R_2^{exp}(\Delta t)} & = \frac{\mathrm{I}(\pi^+\ell^-\bar{\nu},3\pi^0;\Delta t)}{\mathrm{I}(\pi\pi,\pi^-\ell^+\nu;\Delta t)} {  = R_2(\Delta t) \times \frac{C(\pi^+\ell^-\bar{\nu},3\pi^0)}{C(\pi\pi,\pi^-\ell^+\nu)}}, \label{eq:r2e} \label{eq:r2def}\\
    R_3^{exp}(\Delta t) & = \frac{\mathrm{I}(\pi^-\ell^+\nu,\pi\pi;\Delta t)}{\mathrm{I}(3\pi^0,\pi^+\ell^-\bar{\nu};\Delta t)} {  = R_3(\Delta t) \times \frac{C(\pi^-\ell^+\nu,\pi\pi)}{C(3\pi^0,\pi^+\ell^-\bar{\nu})}}, \label{eq:r3e}\\
    {R_4^{exp}(\Delta t)} & = \frac{\mathrm{I}(\pi^-\ell^+\nu,3\pi^0;\Delta t)}{\mathrm{I}(\pi\pi,\pi^+\ell^-\bar{\nu};\Delta t)} {  = R_4(\Delta t) \times \frac{C(\pi^-\ell^+\nu,3\pi^0)}{C(\pi\pi,\pi^+\ell^-\bar{\nu})}} \label{eq:r4e} \label{eq:r4def},
  \end{eqnarray}
  where $C(f_1,f_2)$ denote coefficients dependent on transition amplitudes of neutral kaons to particular final states $f_1$ and $f_2$~\cite{theory:bernabeu-t}. It can be shown that these coefficients are symmetric with respect to an interchange of final states of the first and second decay, i.e.:
  \begin{equation}
    \label{eq:c_sym}
    C(f_1,f_2) = C(f_2,f_1).
  \end{equation}

Although all of the experimental ratios $R^{exp}_{1}$--$R^{exp}_{4}$ use independent physical processes, one can note that they are pairwise equivalent to their inverses if the order of observed decays is reversed. Therefore, using the symmetry of the $C$ coefficients from \eref{eq:c_sym}, one can express possibilities by only two \Ts-asymmetric ratios if the time difference between kaon decays is allowed to be negative:
\begin{eqnarray}
  R_2^{exp}({-\Delta t}) & = \left({R_3^{exp}(\Delta t)}\right)^{-1}, \\
  R_4^{exp}({-\Delta t}) & = \left({R_1^{exp}(\Delta t)}\right)^{-1}.
\end{eqnarray}

The above observation makes it sufficient to consider only two out of four ratios if any order of the kaon decays is allowed so that $\Delta t \in (-\infty,+\infty)$. In the further considerations only the $R_2$ and $R_4$ will thus be included, following the convention used in Ref.~\cite{theory:bernabeu-t}. 

\begin{figure}[htb]
  \centering
  \includegraphics[width=0.6\textwidth]{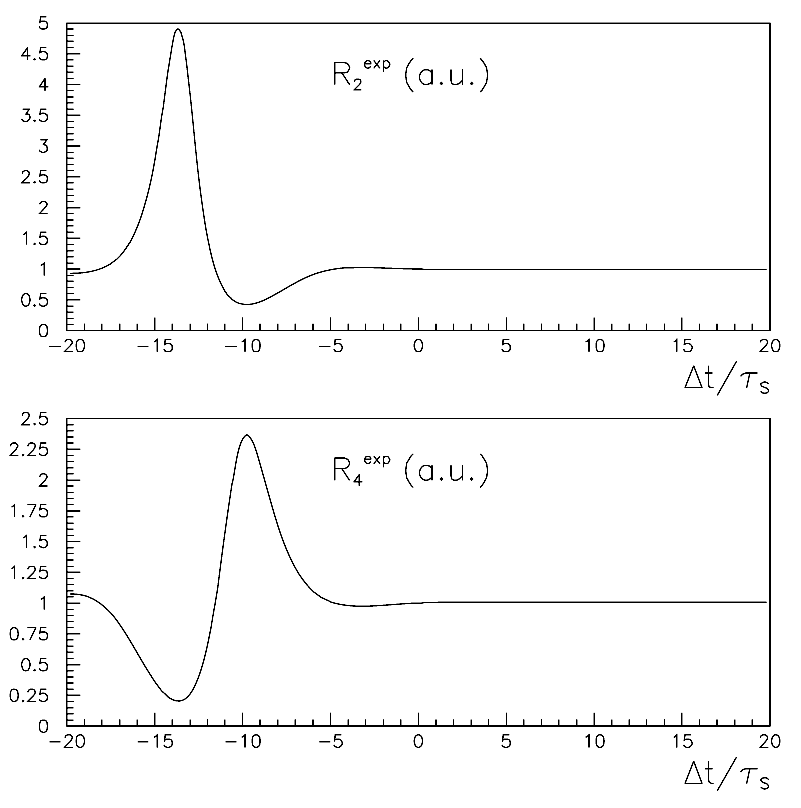}
  \caption{\label{fig:strategy}
    Expected dependence of the observable asymmetric ratios  $R^{exp}_2(\Delta t)$ and $R^{exp}_4(\Delta t)$ on the time difference between kaon decays in presence of \Ts~violation.
    Symmetry violation effects can be sought either as significant $\Delta t$ dependence for small and negative values of $\Delta t$ or as deviation of the asymptotic value $R(\Delta t \gg \tau_{S})$ from unity.
    Figure adapted from~\cite{theory:bernabeu-t}.
  }
\end{figure}

\section{Possible measurement strategies}\label{sec:strategies}
The observable double decay rate ratios defined in Equations~\ref{eq:r2e} and~\ref{eq:r4e} in case of violation of the symmetry under reversal in time are expected to depend on the kaons' decay time difference $\Delta t$ as shown on the plots in~\fref{fig:strategy}. For small time differences and especially for $\Delta t < 0$ one can search for effects of \Ts~violation manifested in the form of structures in the $\Delta t$ dependence of the $R^{exp}_2(\Delta t)$ and $R^{exp}_4(\Delta t)$ ratios. Authors of Reference~\cite{theory:bernabeu-t} have shown that significant deviations from constant value are expected in case of broken symmetry (see \fref{fig:strategy}). On the other hand, in the region of $\Delta t \gg \tau_{S}$ (where $\tau_S$ denotes $\Ks$ lifetime) the ratios are expected to reach a constant asymptotic level. Therefore, a measurement of the asymptotic level of $R^{exp}_2(\Delta t)$ and $R^{exp}_4(\Delta t)$ constitutes another available strategy to search for breaking of he \Ts~symmetry. In agreement with the commonsense expectation, any deviation of the theoretical ratios $R_2$ and $R_4$ (extracted from $R^{exp}_2$ and $R^{exp}_4$ using the respective $C(f_1,f_2)$ factors as in Equations~\ref{eq:r2e} and~\ref{eq:r4e}) from unity would be an indication of violated \Ts~symmetry. It can be shown that such deviations are proportional to the real part of the \Ts-violating $\epsilon$ parameter (see Equations~\ref{eq:epsilon_delta} and~\ref{eq:kabir_epsilon})~\cite{theory:bernabeu-t}:
\begin{equation}
  \begin{aligned}
    \label{eq:theor_deviations}
    & R_2(\Delta t \gg \tau_{S}) \simeq 1-4\Re\epsilon, \\
    & R_4(\Delta t \gg \tau_{S}) \simeq 1+4\Re\epsilon.
  \end{aligned}
\end{equation}

The first of the aforementioned measurement strategies requires collection of a large number of double kaon decay events within a small range of decay time differences, where additionally the negative $\Delta t$ region would be populated by events where first kaon decayed into three neutral pions and the second one into the $\pi\ell\nu$ final state.

However, the probability of the long-lived neutral kaon decaying before its short-lived partner is small (at the level below \SI{1}{\percent}) and a decay of $\Ks$ into three neutral pions is a \CPs-violating decay whose branching ratio is below 2.6$\times 10^{-8}$ by the searches performed so far, which found no candidate events~\cite{Babusci:2013tr}.
%Practically, due to large difference of lifetimes between $\Ks$ and $\Kl$ this would have to result mostly from $\Ks\to 3\pi^0$ and
As a consequence, observation of the $R_{2(4)}^{exp}$ region for small and negative $\Delta t$ would require very high statistics and is thus not feasible with the KLOE detector at the DA$\Phi$NE collider.
Moreover, this experimental approach would require reaching a resolution of kaon decay times difference sufficient to observe the structures in $\Delta t$ dependence of $R_{2(4)}^{exp}$.

Conversely, the test exploiting the asymptotic region of large time differences can be performed with KLOE and DA$\Phi$NE as large numbers of events are expected for $\Delta t \gg \tau_{S}$ and as such measurement would comprise averaging over a broad range of time differences. This is an advantage of KLOE, whose size allows it to capture double decay  events with $\Delta t$ up to almost 400 $\tau_{S}$. Moreover, kaons' decay time reconstruction is not required at a level as precise as in the case of searching for structures in the time dependence. Thus in case of this measurement, resolution of the $\Kl\to 3\pi^0$ decay is not a limiting factor despite challenges related to its reconstruction which will be discussed in Chapter~\ref{chapter:gps}.

Application of the test of symmetry under reversal in time by determination of the asymptotic values of the $R^{exp}_2(\Delta t)$ and $R^{exp}_4(\Delta t)$ ratios is one of the goals of the KLOE-2 experiment and is expected to yield a statistically significant result using an integrated luminosity of the order of 5 fb$^{-1}$~\cite{theory:bernabeu-t}. In the time when this data is being collected, data reconstruction and analysis methods required for this test were devised using a dataset of the KLOE experiment. Their preparation and application to KLOE data constitute a major part of the work reported on in this Thesis. 

\section{Determination of the constant coefficient in asymmetric ratios}\label{sec:d_determination}
As shown in Equations~\ref{eq:r1e}--\ref{eq:r4e}, the experimentally-observable ratios $R^{exp}_{1-4}$ are linked to the theoretical ratios of certain transitions' probability with proportionality constants dependent on particular kaon decays identifying the transitions used to define the ratios. 

\begin{eqnarray}
 \frac{C(\pi^+\ell^-\bar{\nu},3\pi^0)}{C(\pi\pi,\pi^-\ell^+\nu)} = \left|\frac{\expval{\pi^+\ell^-\bar{\nu}|\mathbb{T}|\akaon}\expval{3\pi^0|\mathbb{T}|\Km}}{\expval{\pi^-\ell^+\nu|\mathbb{T}|\kaon}\expval{\pi\pi|\mathbb{T}|\Kp}}\right|^2,\\  
  \frac{C(\pi^-\ell^+\nu,3\pi^0)}{C(\pi\pi,\pi^+\ell^-\bar{\nu})} = \left|\frac{\expval{\pi^-\ell^+\nu|\mathbb{T}|\kaon}\expval{3\pi^0|\mathbb{T}|\Km}}{\expval{\pi^+\ell^-\bar{\nu}|\mathbb{T}|\akaon}\expval{\pi\pi|\mathbb{T}|\Kp}}\right|^2,
\end{eqnarray}
where $\mathbb{T}$ denotes the transition matrix of the neutral kaon system.

If \CPTs~violating effects in semileptonic decays of neutral kaons are not allowed (which indeed have not been observed to date), the amplitudes $\expval{\pi^+\ell^-\bar{\nu}|T|\akaon}$ and $\expval{\pi^-\ell^+\nu|T|\kaon}$ are equal and both of the above coefficients reduce to the same value which will be referred to as the $D$ coefficient in further considerations:
\begin{equation}
 D \equiv \frac{C(\pi^+\ell^-\bar{\nu},3\pi^0)}{C(\pi\pi,\pi^-\ell^+\nu)} = \frac{C(\pi^-\ell^+\nu,3\pi^0)}{C(\pi\pi,\pi^+\ell^-\bar{\nu})} =  \left|\frac{\expval{3\pi^0|\mathbb{T}|\Km}}{\expval{\pi\pi|\mathbb{T}|\Kp}}\right|^2.
  \label{eq:c_coeff_1}
\end{equation}
The decay probabilities which define the above can be expressed with branching fractions of these decays and the respective kaon lifetimes $\tau_S$ and $\tau_L$ of the short and long-lived kaons, respectively. If additionally the \CPs~violation in neutral kaon decays is neglected as in \eref{eq:cp_neglect}, the coefficient from~\ref{eq:c_coeff_1} can be completely expressed in terms of experimentally-measured quantities:
\begin{equation}
  \label{eq:D}
  D \equiv \left|\frac{\expval{3\pi^0|\mathbb{T}|\Km}}{\expval{\pi\pi|\mathbb{T}|\Kp}}\right|^2 \simeq \frac{\left|\expval{3\pi^0|\mathbb{T}|\Kl}\right|^2}{\left|\expval{\pi\pi|\mathbb{T}|\Ks}\right|^2} = \frac{\text{BR}(\Kl\to 3\pi^0)\Gamma_L}{\text{BR}(\Ks\to\pi\pi)\Gamma_S} = \frac{\text{BR}(\Kl\to 3\pi^0)\tau_S}{\text{BR}(\Ks\to\pi\pi)\tau_L}.
\end{equation}

Determination of the $D$ parameter value is required in order to extract the ratios of kaon transition probabilities $R_2(\Delta t)$ and $R_4(\Delta t)$ from the observable ratios of decay rates defined in equations~\ref{eq:r2e} and~\ref{eq:r4e}:
\begin{eqnarray}
  R_2(\Delta t) = R^{exp}_2(\Delta t) / D, \\
  R_4(\Delta t) = R^{exp}_4(\Delta t) / D.
\end{eqnarray}

As the four values entering the calculation of $D$ (see \eref{eq:D}), one may choose either averages of available measurements, e.g.\ the ones extracted by the Particle Data Group (PDG)~\cite{pdg2016} or results of measurements performed by the same experimental system used for the T symmetry test. In case of the KLOE experiment, all the quantities of interest have been already measured with the same setup, thus allowing for a self-contained determination of the $R_2$ and $R_4$ asymmetries. It should also be noted that the results of KLOE measurements are in most cases (except the lifetime of $\Ks$) the most precise results entering the PDG averages. The value of $\Kl$ branching ratio for $\Kl\to 3\pi^0$ was determined in a comprehensive study of dominant $\Kl$ decays performed by the KLOE experiment in 2006~\cite{kloe_kl3pi0_br}. Although this study involved a fit which yielded the $\Kl$ lifetime, for the following calculations a result of a direct measurement, also conducted by KLOE~\cite{kloe_kl_lifetime}, was used. Similarly, for the values of $\Ks\to\pi^+\pi^-$ branching ratio and lifetime, results of high precision KLOE measurements~\cite{kloe_kspipi_br,kloe_ks_lifetime} were taken.

\tref{tab:D_inputs} presents a comparison of the values of the relevant branching ratios and lifetimes as determined by the KLOE experiment and as the PDG averages, as well as the D parameter values resulting from calculations based on these inputs.

\begin{table}[h!]
  \centering
  \caption{Quantities entering the calculation of the D parameter determined by KLOE and by PDG\@. Last row contains the values of D evaluated with each set.}\label{tab:D_inputs}
  \begin{tabular}[h!]{cS[table-format=2.12]S[table-format=2.12]}
    \toprule 
    & {KLOE measurement~\cite{kloe_kl3pi0_br,kloe_kspipi_br,kloe_kl_lifetime,kloe_ks_lifetime}} & {PDG average~\cite{pdg2016}} \\
    \midrule
    BR($\Kl\to 3\pi^0$) & 0.1997(20) & 0.1969(26) \\
    BR($\Ks\to \pi\pi$) & 0.6919(51) & 0.6920(5) \\
    $\tau_L$ [ns] & 50.92(30) & 50.99(21) \\
    $\tau_S$ [ps] & 89.562(52)  & 89.54(4) \\
    \midrule
    D & {0.5076(59)$\times 10^{-3}$} & {0.4998(69)$\times 10^{-3}$} \\
    \bottomrule
  \end{tabular}
\end{table}

As the values of the D constant parameter calculated using KLOE measurements and world averages from PDG presented in~\tref{tab:D_inputs} are consistent within their uncertainties, in the following considerations the D value obtained solely with KLOE results
\begin{equation}
  \label{eq:D_kloe}
  D = \frac{\text{BR}(\Kl\to 3\pi^0)\tau_S}{\text{BR}(\Ks\to\pi\pi)\tau_L} = 0.5076(59)\times 10^{-3},
\end{equation}
will be used in order allow for a self-contained KLOE determination of the \Ts-asymmetric observables.

%%%Local Variables:
%%% TeX-master: "../main"
%%% End:
 
\chapter[Discrete symmetry tests with odd operators in ortho-positronium decays]{Discrete symmetry tests with odd operators in ortho-positronium decays}\label{chapter:test_jpet}

Although the \Ts,~\CPs~and~\CPTs~symmetries have been investigated thoroughly in a multitude of physical systems during the past half of a century, experimental results on their tests in purely leptonic systems have been scarce~\cite{Arbic:1988pv,Skalsey:1991vt,cp_positronium,cpt_positronium}. Among the latter, the positronium exotic atoms constitute a likely candidate for precise searches of discrete symmetries' violations~\cite{Bernreuther:1988tt}. The most recent experimental attempts, testing the \CPs~and \CPTs~symmetries in the ortho-positronium decays, both obtained null results with a precision at the level of $10^{-3}$~\cite{cp_positronium,cpt_positronium}. Possible searches for symmetries' violation in the positronium decays are only limited by photon-photon final state interactions which may cause false asymmetries~\cite{Bernreuther:1988tt,Arbic:1988pv} and by weak interactions~\cite{sozzi}. These effects are, however, expected at the levels of $10^{-9}$ and $10^{-14}$, respectively, which leaves a vast space still to be explored. Pushing the presently obtained precision limits in the discrete symmetry studies with o-Ps atoms is one of the aims of the \jpet/ experiment. This Chapter discusses the planned measurements with a special emphasis on their dependency on the trilateration-based reconstruction technique for o-Ps$\to 3\gamma$ decays, presented in~\cref{chapter:gps} of this Thesis.

\section{Properties of the ortho-positronium atoms}
Positronium is a bound state of an electron and a positron, which may be formed prior to their annihilation. Having a reduced mass only twice smaller than a hydrogen atom, it has a similar energy level structure and is thus often referred to as an exotic atom. While being an eigenstate of the parity operator like atoms, positronium is also symmetric under charge conjugation operation as a particle-antiparticle system. As the positronium spin is a linear combination of electron and positron spins, four possible states may be formed depending on their orientation~\cite{Harpen:2003zz}:
\begin{eqnarray*}
  \ket{S=1,S_Z=1} &=& \ket{\uparrow}\ket{\uparrow},\\
  \ket{S=1,S_Z=0} &=& \frac{1}{\sqrt{2}}\left[\ket{\uparrow}\ket{\downarrow}+\ket{\downarrow}\ket{\uparrow}\right],\\
  \ket{S=1,S_Z\!=\!-1} &=& \ket{\downarrow}\ket{\downarrow},\\
  \ket{S=0,S_Z=0} &=& \frac{1}{\sqrt{2}}\left[\ket{\uparrow}\ket{\downarrow}-\ket{\downarrow}\ket{\uparrow}\right],
\end{eqnarray*}
where $\ket{\uparrow}$ and $\ket{\downarrow}$ denote $S_z=+\frac{1}{2}$ and $S_z=-\frac{1}{2}$ for a single electron or positron. The three states with $S=1$ constitute a triplet known as ortho-positronium~(\ops/) whereas the $S=0$ singlet state is referred to as para-positronium~(p-Ps). In the ground state with $l=0$, the charge conjugation parity of such a system depends on the spin as $({-1})^S$ and is therefore different for para- and ortho-positronium. As the electromagnetic interactions conserve charge parity, the allowed final states of para-positronium annihilation must contain an even number of photons ($C=(-1)^n$ for a system of $n$ photons) whereas o-Ps can only annihilate into an odd number of photons. In practice, final states with larger photon numbers are suppressed by few orders of magnitude and the positronium annihilations are dominated by p-Ps$\to 2\gamma$ and o-Ps$\to 3\gamma$~\cite{Harpen:2003zz}.

The difference of allowed decays is also a reason for much larger mean lifetime of ortho-positronium atoms due to a smaller available phase space and smallness of the fine structure constant ($\approx$1/137). In vacuum, the lifetimes of ortho and para states amount to~\cite{PhysRevLett.72.1632,PhysRevLett.90.203402,JINNOUCHI2003117}:
\begin{eqnarray*}
  \tau_{\text{p-Ps}} &\approx& 0.125\ \text{ns}, \\
  \tau_{\text{o-Ps}} &\approx& 142\ \text{ns}.
\end{eqnarray*}

Such a lifetime discrepancy makes it possible to distinguish the p-Ps and \ops/ states experimentally if the positronium creation and decay time can be measured. The former may be estimated e.g.\ when positronia are created with $e^+$ from a source in which $\beta^+$ emission results is creation of a nucleus in an excited state, where the prompt de-excitation photon provides a time signal approximately corresponding to the positronium creation~\cite{daria_epjc}.

\section{Symmetry tests with angular correlations in o-Ps$\to 3\gamma$}
\label{sec:ops_operators}
\begin{figure}[h!]
  \centering
  \begin{tikzpicture}[scale=0.45]
    \filldraw[fill=gray!20] (-5,4) -- (2,4) --  (6,8) -- (-1,8) -- (-5,4);
    \draw[black,thick,->] (0,6.5) node[black,yshift=-5,xshift=8]{\small o-Ps} -- (1,9) node[anchor=west] {$\vec{S}$};
    \draw[black, thick,->] (0,6.5) -- (4,7)node[midway, below,xshift=15,yshift=3] {$\vec{k_1}$};
    \draw[black, thick,->] (0,6.5) -- (-0.8,7.5)node[yshift=-7.0,xshift=-7.0] {$\vec{k_3}$};
    \draw[black, thick,->] (0,6.5) -- (-2.5,4.5)node[midway, below] {$\vec{k_2}$};
  \end{tikzpicture}
  \caption{Vectors in an o-Ps$\to 3 \gamma$ event (in the \ops/ frame of reference) used to construct the operators listed in~\tref{tab:jpet_operators}. The photons' momentum vectors are ordered by magnitude so that $|\vec{k_1}| > |\vec{k_2}| > |\vec{k_3}|$. $\vec{S}$ denotes the spin of the decaying positronium.}\label{fig:ops_decay_scheme}
\end{figure}
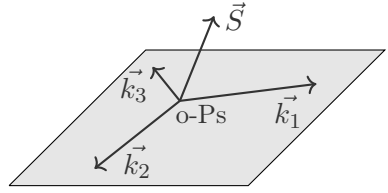

Ortho-positronium annihilations into three photons allow for experimental searches of discrete symmetry violations which could be manifested by specific angular correlations in the events' topology~\cite{Bernreuther:1988tt}. To measure such asymmetries, several operators can be constructed using four vectors measurable in a single annihilation event: momenta of the three photons $\vec{k_i},\ i=1,2,3$ and spin $\vec{S}$ of the decaying ortho-positronium atom. As depicted in~\fref{fig:ops_decay_scheme}, the photons' momentum vectors are constrained by momentum conservation to lie in a single plane, and are not correlated with the \ops/ spin. In order to distinguish the three photons, they can be ordered by decreasing energy so that $|\vec{k_1}| > |\vec{k_2}| > |\vec{k_3}|$.

\begin{table}
  \centering
  \caption{Operators constructed with photons' momenta and ortho-positronium spin $\vec{S}$ which can be used for discrete symmetry tests in the o-Ps$\to 3\gamma$ decays. For each listed symmetry, $+$ and --- indicate that the operator is even or odd, respectively.\label{tab:jpet_operators}}
  \begin{tabular}[h!]{cp{3em}p{3em}p{3em}p{3em}p{3em}}
    \toprule
    operator & C & P & T & CP & CPT \\
    \midrule
    ${\vec{S} \cdot \vec{k_1}}$ & + & --- & + & --- & ---\\      
    ${\vec{S} \cdot (\vec{k_1}\times\vec{k_2})}$ & + & + & --- & + & ---\\
    $({\vec{S} \cdot \vec{k_1})(\vec{S} \cdot (\vec{k_1}\times\vec{k_2})})$ & + & --- & --- & --- & +\\
    \bottomrule
  \end{tabular}
\end{table}

\tref{tab:jpet_operators} lists three operators which can be built with the aforementioned vectors, along with their parity with respect to chosen symmetry operations~\cite{moskal_potential}.
%
% @TODO: sprawdzić czy to jest prawda w ogóle
%
The considerations for the expectation value of an operator odd under time reversal from Equations~\ref{eq:rel_any_operator}--\ref{eq:expectation_value_any_operator} can be applied to other symmetry transformations (in case of \CPs, which is unitary, the hermicity of the operator O is not required). As a result, any of the operators from~\tref{tab:jpet_operators} which is odd under a certain symmetry transformation can be used to construct a test of this symmetry by searching for the operator's non-vanishing expectation value.

As the aforementioned operators correspond to particular angular correlations in the \ops/$\to 3\gamma$ events, their non-zero mean value can be experimentally sought in the following asymmetry:
\begin{equation}
  A = \frac{N_+ - N_-}{N_+ + N_-},
\end{equation}
where $N_+$ and $N_-$ respectively are the numbers of observed events with a positive or negative value of a certain correlations, e.g.\ for the T-odd operator ${\vec{S} \cdot (\vec{k_1}\times\vec{k_2})}$, $N_-$ will be the count of events with the decay plane normal vector $\vec{k}_1\times\vec{k}_2$ anti-parallel to the \ops/ spin.

%%% Local Variables:
%%% TeX-master: "../main"
%%% End: 
\chapter{The KLOE and J-PET detectors}
\label{chapter:detectors}

This Thesis covers two approaches to searching for violations of the symmetry under reversal in time: a comparison of probabilities for A$\to$B and B$\to$A transitions in the system of quantum-entangled neutral K mesons and a determination of the expected value of a \Ts-odd operator constructed using angular correlations between photons from an ortho-positronium decay. The former is performed using the KLOE experimental setup, constituted by a large general-purpose particle detector recording the decays of $\phi$ mesons into kaons, taking place in electron-positron collisions at 1 GeV energy scale. The latter will be realized with the J-PET detector, a device dedicated to recording photons from electron-positron annihilation and positronium atom decays, at low energies of positron thermalization in matter.

Although these two measurements are based on substantially different principles and are conducted in very different conditions, the physical processed involved in these experiments and the setups used to record them are not without similarities. Not only do they both use an electron-positron system to produce a desired physical state, but also both these measurements involve a decay of a neutral state into several photons, where the decay location must be reconstructed in order to calculate event properties essential for the experiment. In each case, reconstruction is performed using the same technique. Application of the same reconstruction scheme in the KLOE and J-PET experimental setups alike was possible due to several properties shared by them in terms of recording neutral particle decays into photons.

This Chapter presents details of the KLOE and J-PET detectors used in the studies covered by this Thesis, with a special emphasis on the links between them which made it possible to employ the same reconstruction algorithm (described in detail in Chapter~\ref{chapter:gps}) in each of these detectors.

\section{The KLOE detector and the DA$\Phi$NE $\phi$-factory}
% The KLOE experimental setup was designed with a primary goal to probe direct \CPs~violation in the decays of neutral K mesons by measurement of the $\Re\left(\frac{\epsilon'}{\epsilon}\right)$ parameter through the following observable double ratio:
% \[ R = \frac{\Gamma(\Ks\to\pi^0\pi^0)\Gamma(\Kl\to\pi^+\pi^-) }{\Gamma(\Ks\to \pi^+\pi^-)\Gamma(\Kl\to\pi^0\pi^0)}, \]
% at the same time allowing for a broad range of other studies in the field of kaon physics~\cite{Campana:1997ai}.
With a broad range of studies of the physics of neutral K meson system as one of its major goals, the properties of the KLOE (K LOng Experiment) detector have been largely dictated by requirements to record the decays of both the short and long-lived neutral kaons with high efficiency. The same reasons influenced the location of KLOE to be at the DA$\Phi$NE $e^+e^-$ collider which provides it with copious pairs of both neutral and charged kaons, produced in the decays of $\phi$ mesons. Next Section presents details of this collider relevant for the studies presented in this work, whereas in the further Sections the features of the KLOE detector are discussed.

%%%%%%%%%%%%%%%%%%%%%%%%%%%%%%%%%%%%%%%%%%%%%%%%%%%%%%%%%%%%%%%%%%%%%%%%%%
% DAFNE                                                                  %
%%%%%%%%%%%%%%%%%%%%%%%%%%%%%%%%%%%%%%%%%%%%%%%%%%%%%%%%%%%%%%%%%%%%%%%%%%
\subsection{The DA$\Phi$NE collider}
\label{sec:dafne}

\begin{figure}[h!]
  \centering
  \includegraphics[width=0.7\textwidth]{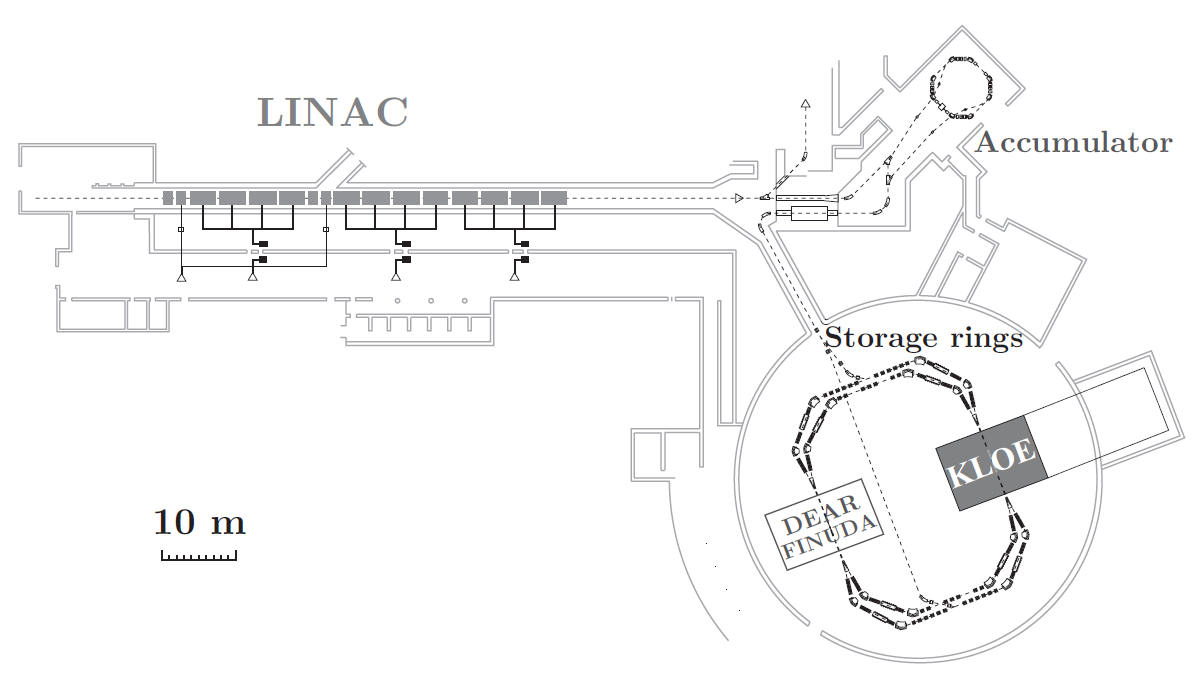}
  \caption{Scheme of the acelerator complex of National Laboratories of Frascati which includes the two storage rings of the DA$\Phi$NE collider. Figure adapted from~\cite{kloe_results}.\label{fig:dafne}}
\end{figure}

DA$\Phi$NE, whose name is an acronym for Double Annular $\Phi$-factory for Nice Experiments, is an electron-positron collider operating since 1999 in the National Laboratories of Frascati (LNF) located near Rome, Italy. The whole accelerator complex of LNF is schematically presented in Figure~\ref{fig:dafne}. A linear accelerator (LINAC) provides electron and positrons (where the latter are obtained with conversion of 250 MeV electrons on a metallic target) in 10~ns pulses with a frequency of 50 Hz~\cite{Vignola:1996mt}. Subsequently, both electrons and positrons are transferred to an accumulator ring where they are cooled in order to damp the emittances of the beams before their injection into DA$\Phi$NE. Finally, electrons and positrons are injected in turns into the storage rings of DA$\Phi$NE as a single bunch at a time.

In order to reduce beam-beam interactions, the DA$\Phi$NE collider consists of separate storage rings for $e^-$ and $e^+$, which intersect at an angle of 25 mrad in two interaction regions. One of them is occupied in turns by the KLOE and SIDDHARTA~\cite{CurceanuPetrascu:2007zz} detectors, while the other served the DEAR~\cite{Lucherini:2003fk} and FINUDA~\cite{Agnello:2007rk} experiments. KLOE, which is of interest in this work, surrounds its interaction region and is oriented so that its $x$ axis is horizontal and perpendicular to the beam direction, $y$ axis is vertical whereas the $z$ axis of the detector frame of reference is parallel to bisector of an angle between the beams.

\begin{table}[h!]
  \centering
  \caption{Selected parameters of the DA$\Phi$NE collider~\cite{Vignola:1996mt}.\label{tab:dafne}}
  \begin{tabular}{cc r}
    \toprule
    \multicolumn{2}{l}{DA$\Phi$NE property} & value \\
    \midrule
    \multicolumn{2}{l}{beam energy} & 510 MeV \\
    \multicolumn{2}{l}{RF frequency} & 368.25 MHz \\
    \multicolumn{2}{l}{inter-bunch-crossing period} & 2.72 ns \\
    \multicolumn{2}{l}{bunches per beam} & 120 \\
    \multicolumn{2}{l}{particles per bunch} & 8.9$\times 10^{10}$ \\
    \multirow{3}{*}{bunch spread}           & $\sigma_x$ & 0.02 cm \\ 
    \multirow{3}{*}{} & $\sigma_y$ & 0.002 cm \\ 
    \multirow{3}{*}{}           & $\sigma_z$ & 3 cm \\
    \multicolumn{2}{l}{bunch crossing angle} & 25 mrad \\
    \bottomrule
  \end{tabular}
\end{table}

The storage rings of the collider contain a total of 120 particle bunches and operate at a RF frequency of 368.25 MHz, which results in an interval of 2.72 ns between subsequent bunch crossings in an interaction region. A single bunch, consisting of about $9\times10^{10}$ electrons or positrons, has a very small spatial spread in the $y$ (vertical) direction perpendicular to the storage ring plane, and is largest in the longitudinal direction ($z$), where its spread reaches 3 cm ($\sigma$). A summary of the DA$\Phi$NE parameters is given in~\tref{tab:dafne}.

The energies of electron and positron beams circulating in DA$\Phi$NE are symmetrical and chosen so as to obtain a center-of-mass energy of the collisions equal to the mass of the $\phi$ meson, i.e.~$\sqrt{s}\approx m_{\phi}=1019.46\pm0.016$ MeV. Thus, in the $e^+e^-$ collisions, $\phi$ mesons are predominantly produced with a cross section of about 3.1 $\mu$b, due to which fact the name of a $\phi$-factory is often attributed to DA$\Phi$NE\@. In fact, during the operation of the KLOE experiment at this collider, a total of about $10^{10}$ of $\phi$ mesons were produced and their decays recorded by KLOE\@.

Due to the small crossing angle of the colliding beams, the produced $\phi$ mesons are almost at rest ($\beta_{\phi}\approx 0.015$) with only a small momentum component in the $x$ direction:
\begin{equation}
  \label{eq:phi_px}
  p_x^{\phi} = 13.1 \ \text{MeV/c}.
\end{equation}

Strong interactions governing the $\phi$ decays make them almost immediately ($\tau_{\phi} = 1.5500\pm0.0058\times 10^{-22}$ s~\cite{pdg2016}) produce a pair of either neutral or charged K mesons or hadronic states such as $\rho\pi^0$, with branching fractions presented in~\tref{tab:phi_brs}.

\begin{table}[h!]
  \centering
  \caption{Branching fractions of major decays of the $\phi$ meson~\cite{pdg2016}.}\label{tab:phi_brs}
  \begin{tabular}{c c}
    \toprule
    Decay & Branching fraction \\
    \midrule
    $\phi\to\mathrm{K}^+\mathrm{K}^-$ & 48.9$\pm$0.5\% \\
    $\phi\to\kaon\akaon$ & 34.2$\pm$0.4\% \\
    $\phi\to\rho\pi^0/\pi^+\pi^-\pi^0$ & 15.32$\pm$0.32\% \\
    \bottomrule
  \end{tabular}
\end{table}

%%%%%%%%%%%%%%%%%%%%%%%%%%%%%%%%%%%%%%%%%%%%%%%%%%%%%%%%%%%%%%%%%%%%%%%%%%
% KLOE                                                                   %
%%%%%%%%%%%%%%%%%%%%%%%%%%%%%%%%%%%%%%%%%%%%%%%%%%%%%%%%%%%%%%%%%%%%%%%%%%
\subsection{General properties of the KLOE detector}
\label{sec:kloe}

The design of KLOE follows a general scheme often used in general purpose particle detectors, with a cylindrical tracking device symmetrically surrounding the interaction region enclosed by a calorimetric detector. However, the properties of these components are specifically driven by the aim to capture a large part of decays of the $\Kl$ meson (whose mean free path in the KLOE conditions is about 3.4 m~\cite{LeeFranzini:2007hj}), minimize kaon regeneration and register photons from hadronic decays into neutral pions.

\begin{figure}[h!]
  \centering
  \includegraphics[width=0.6\textwidth]{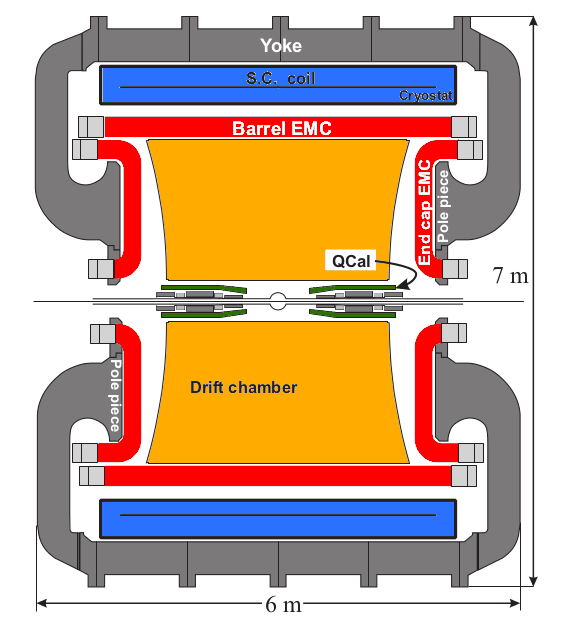}
  \caption{Longitudinal cross section of the KLOE detector. Its main components comprise a drift chamber~(orange), electromagnetic calorimeter~(red), two calorimeters surrounding quadrupole beam-focusing magnets (green) and a superconducting solenoid~(blue). Figure adapted from~\cite{LeeFranzini:2007hj}.\label{fig:kloe}}
\end{figure}

Figure~\ref{fig:kloe} shows a longitudinal section of the detector. Around the interaction region of DA$\Phi$NE, located in the geometrical center of KLOE, the beam pipe is specially shaped as a sphere of 10 cm radius (which corresponds to about 17$\lambda_{K_S}$ where $\lambda_{K_S}=\beta_{K_S}c\tau_S$) and made of a light Al-Be alloy of 1 mm thickness in order to reduce the probability of neutral kaon regeneration due to nuclear interactions in the beam pipe material. Further from the interaction region, two final focusing quadrupole magnets of the collider are surrounded with tile calorimeters (QCal, marked green in~\fref{fig:kloe}) which decrease the loss of registration of photons from the $\Kl\to 3\pi^0\to 6\gamma$ decays going into the volume of the magnets~\cite{Adinolfi:2002me}.

Most of the KLOE detector volume is filled by a cylindrical drift chamber (DC, marked orange in~\fref{fig:kloe}) whose tracking and particle identification capabilities are aided by the presence of 0.52~T magnetic field directed along the $z$ axis and provided by a superconducting coil (marked blue in~\fref{fig:kloe}). The drift chamber is surrounded by an electromagnetic calorimeter (EMC, red in~\fref{fig:kloe}) which records interactions of photons and other particles. Good hermeticity of the EMC is ensured by its composition of a barrel-shaped part and a set of C-shaped endcaps. As the drift chamber and electromagnetic calorimeter of KLOE are essential for the work presented in this Thesis, they will be discussed in detail in the following Sections.

\subsection{The KLOE drift chamber}
\label{sec:dc}
The tracking of charged particles in KLOE is provided by a large drift chamber shaped as a cylinder with a 2~m outer radius and length of 3.3~m. Such a large sensitive volume allows it to record about \SI{40}{\percent} of decays of the long-lived neutral K mesons before they leave the detector~\cite{LeeFranzini:2007hj}. The region close to the axis of the chamber is hollow up to an inner radius of 25~cm to allow space for the spherical beam pipe, beam-focusing magnets and QCal calorimeters. As the design of the KLOE DC was largely driven by the need to reduce material budget and thus minimize chances of regeneration for neutral kaons and multiple scattering for the charges ones, the DC walls are fabricated with light carbon fiber composite and thin (0.1~mm) aluminum foil. Moreover, a low-Z gas mixture of isobutane~(\SI{10}{\percent}) and helium~(\SI{90}{\percent}), is used, which makes the chamber transparent to photons of energy as low as 20~MeV in addition to minimizing kaon interactions.

\begin{figure}[h!]
  \centering
  \includegraphics[height=31ex]{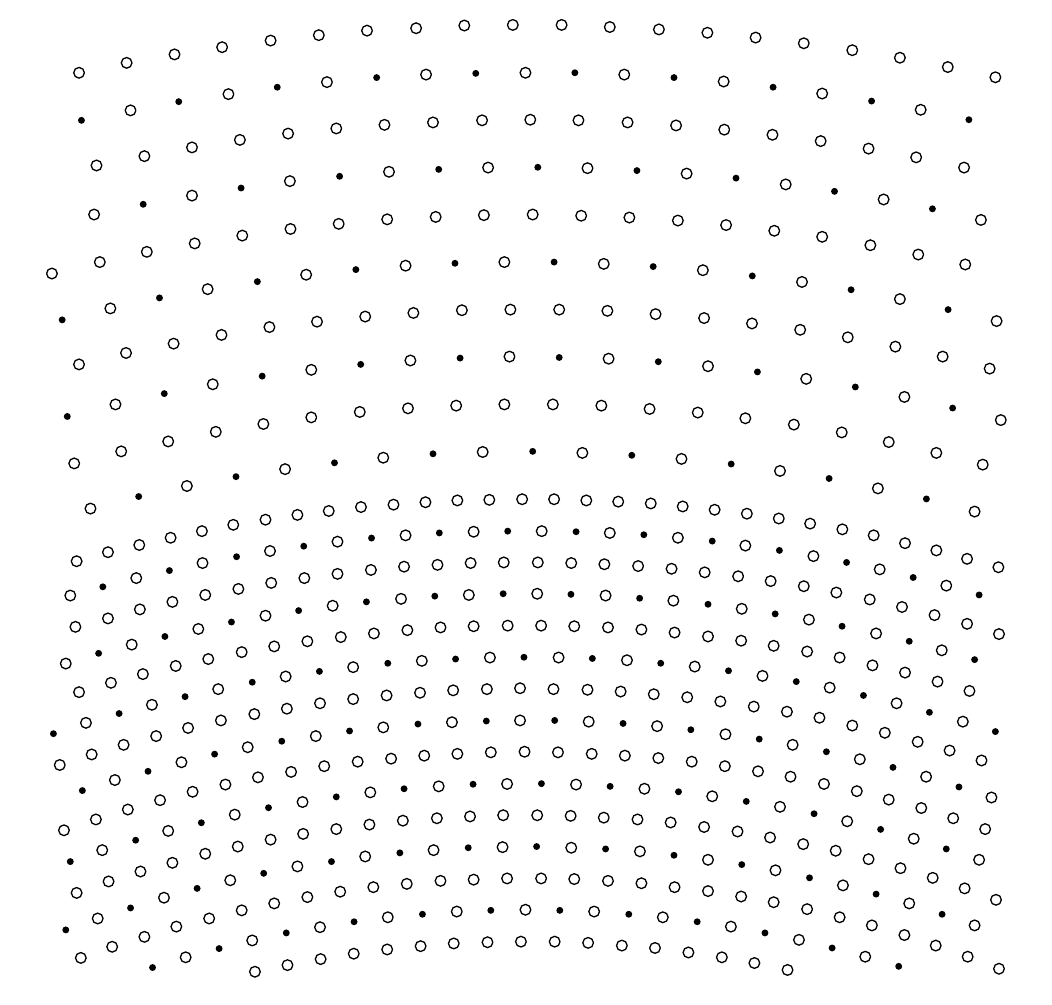}
  \hspace{1em}
  \includegraphics[height=31ex]{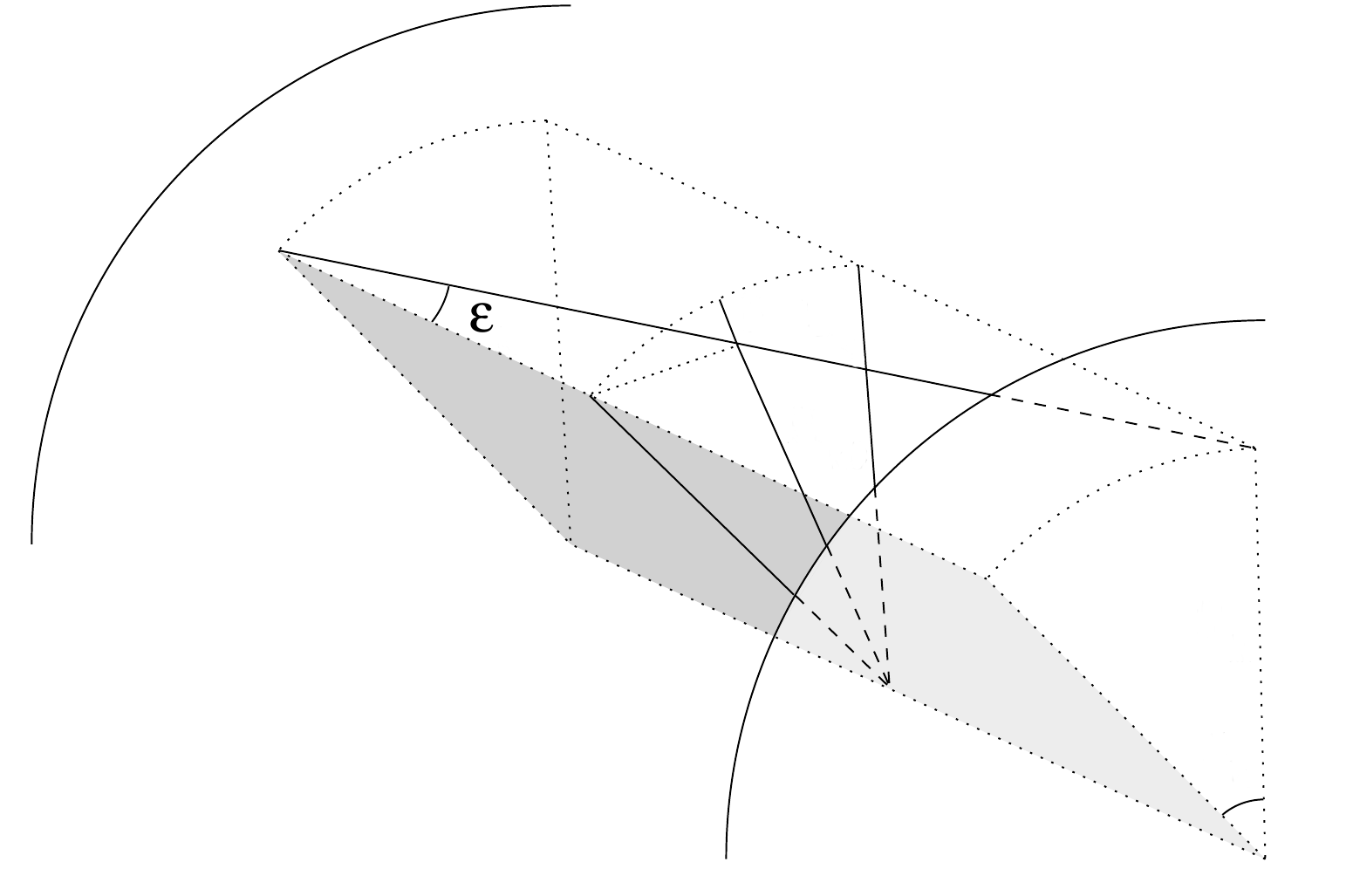}
  \caption{Left: structure of wires and cells in a polar slice of the drift chamber. Field wires are denoted with open circles and dots indicate sense wires. Right: geometry of a wire in the KLOE Drift Chamber, mounted so as to deviate from the direction parallel to the detector $z$ axis by a small angle $\varepsilon$ in order to allow for determination of ionizing particle interaction location on the $z$ axis. Both figures were adapted from~\cite{Adinolfi:2002uk}.\label{fig:dc}}
\end{figure}

A total of almost 40000 field wires are stretched between end plates of the chamber to form electric field in the drift cells, each of which is defined  along a sense wire. The wire and cell structure of the DC shown in~\fref{fig:dc} (left) was dictated by the fact that due to small momenta of $\phi$ decay products, the track density at KLOE is higher close to the interaction point. Therefore the cells are organized radially in 58 layers, with the 12 innermost layers constituted by small 2$\times$2~cm$^2$ cells and the remainder by larger 3$\times$3~cm$^2$ cells~\cite{Adinolfi:2002uk}.
While position of gas ionization by a charged particle in the transverse ($xy$) plane is determined using spatial location of a certain sense wire and the space-time relation for particular drift cell, localization of the interaction along the $z$ detector axis is obtained with a stereo geometry of the wires, presented schematically in the right panel of~\fref{fig:dc}. In this geometry, the angle $\varepsilon$ between a wire and the chamber axis ranges between $\pm$60~mrad and $\pm$150~mrad whereas its sign alternates between subsequent layers, which ensures high and uniform coverage of the detector sensitive volume~\cite{Adinolfi:2002uk}.

Reconstruction of charged particle tracks uses information on wire geometry, external magnetic field and space-time relations for drift cells and comprises three steps: pattern recognition, track fitting and vertex fitting. The resulting spatial resolution of the KLOE drift chamber is below 200~$\mu$m in the transverse plane and 2~mm along the $z$ axis. Decay vertices, determined on the basis of closest approach of two tracks, are resolved up to 1~mm~\cite{Adinolfi:2002uk}. Finally, the momentum is attributed to a registered track using its curvature caused by the external magnetic field, yielding a relative momentum resolution at a level of \SI{0.4}{\percent}~\cite{kloe_results}. A comprehensive description of the KLOE DC reconstruction procedures can be found in Reference~\cite{data_handling}.

\subsection{The electromagnetic calorimeter of KLOE}
\label{sec:emc}
In order to cover almost the full (\SI{98}{\percent}) solid angle around the interaction point and record interactions of both photons and mesons in the energy range between 20 and 510~MeV, the KLOE electromagnetic calorimeter was built in a sampling technology using lead foils alternated with scintillating fibers of 1~mm diameter. The latter are shaped so as to enable light propagation in a single mode, thus allowing for excellent timing resolution. A single module of the EMC is composed of 200 layers of lead foil and 200 layers of fibers glued with epoxy in a lead:fiber:epoxy ratio of 42:48:10. Such composition ensures a good energy resolution due to high scintillator content. The radiation length in the calorimeter material is about 1.5~cm and the total thickness of the calorimeter amounts to about 15~radiation lengths~\cite{Adinolfi:2002zx}.

\begin{figure}[h!]
  \centering
  \includegraphics[height=27ex]{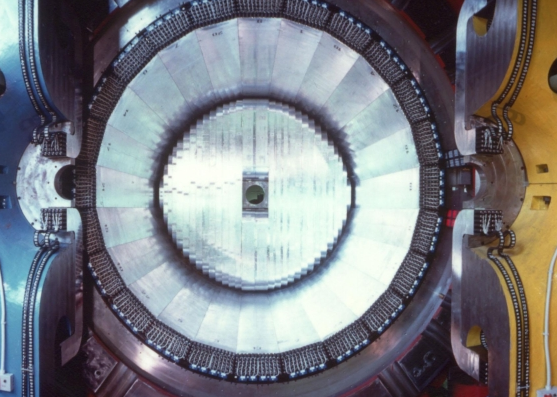}
  \hspace{1em}
  \includegraphics[height=17ex]{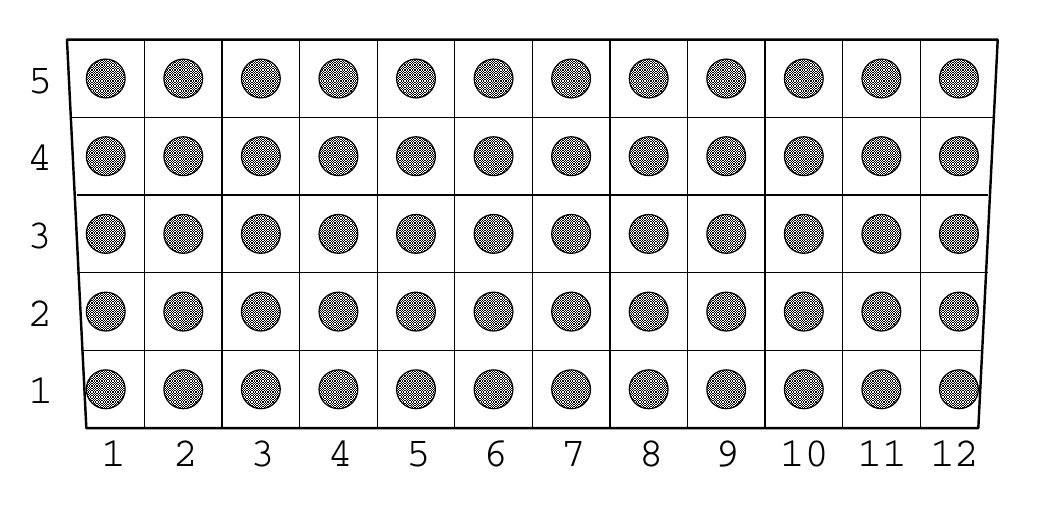}
  \caption{Left: view inside the KLOE calorimeter through one of the endcaps (the other endcap is already closed). Structure of the 24 modules of the EMC barrel is visible. Figure adapted from~\cite{kloe_web}. Right: structure of EMC cells in a single barrel module with 12 columns and 5 layers. Black circles represent photomultiplier locations, each of which determined a single cell. Figure adapted from~\cite{zdebik_mgr}.}
  \label{fig:emc}
\end{figure}

The barrel part of the calorimeter is composed of 24 modules of trapezoidal cross section, located at different azimuthal angles and spanning the whole length of the barrel, as visible in the photograph in the left panel of~\fref{fig:emc}. Each of these modules is read out at both ends by photomultipliers attached through light guides and organized in a grid presented schematically in~\fref{fig:emc} (right). Each volume of a single module read out by a separate photomuliplier constitutes a distinct cell of the calorimeter, i.e.\ the smallest entity in which and energy deposition in the EMC can be localized in the transverse plane. Each cell in a module is identified by its coordinates: layer (1--5) and column number (1--12). Such segmentation of the readout results in a $xy$ resolution of energy deposition point of about 1.3~cm~\cite{Adinolfi:2002zx}. Each of the EMC endcaps consists of 32 similarly structured modules, although these modules are bent at the edges in order to position the photomultipliers parallel to the magnetic field~(see~\fref{fig:kloe}) and oriented vertically. Segmentation of the readout similar as in the barrel modules allows for position determination in the $xz$ plane in case of the endcaps.

The position of energy deposit along the calorimeter modules $s$ (w.r.t.\ center of the module) and time of interaction $t$ are estimated using the times of light registration at both ends of the modules, $t^A$ and $t^B$ (neglecting known time offsets originating from hardware):
\begin{eqnarray}
  \label{eq:emc_ts}
  s & = & \frac{\upsilon(t^A-t^B)}{2}, \label{eq:emc_s}\\
  t & = & \frac{t^A+t^B}{2} - \frac{L}{2\upsilon}, \label{eq:emc_t}
\end{eqnarray}
where $L$ denotes the length of the module and $\upsilon$ is the velocity of light propagation in the scintillating fibers~\cite{Adinolfi:2002zx}. The deposited energy is obtained from a mean of energies recorded at both ends of the calorimeter cell, measured through photomultiplier signal amplitudes corrected for light attenuation in the scintillating fibers.

As a result of the above scheme, the resolutions of energy, time and $z$ coordinate of interaction exhibit a dependence on the deposited energy, which can be approximated by the following relations~\cite{Adinolfi:2002zx}:
\begin{eqnarray}
  \label{eq:emc_res}
  \sigma(E) & = & \frac{0.057\:E}{\sqrt{E(GeV)}}, \\
  \sigma(t) & = & \frac{54\:ps}{\sqrt{E(GeV)}} \oplus 140\:ps, \\
  \sigma(z) & = & \frac{1.2\:cm}{\sqrt{E(GeV)}}.
\end{eqnarray}

After energy depositions in single cells are recorded, an algorithm searching for groups of spatially-adjacent cells with deposited energy is applied in order to identify sets of cells corresponding to a single particle interaction~\cite{data_handling, zdebik_mgr}, commonly referred to as \textit{calorimeter clusters}. The time and location of the physical interaction corresponding to a cell cluster is calculated as an energy-weighted average of member cell properties.

The excellent time resolution of the KLOE EMC for registration of photons from the $\Kl\to 3\pi^0$ process is one of the key factors allowing for a trilateration-based reconstruction of this decay described in this work. Moreover, even though the calorimeter was designed to record energy depositions starting from 20~MeV, it has proven capable of detecting photons of energies as low as 7~MeV~\cite{difalco_phd}. 

\subsection{The trigger system}\label{sec:trigger}
The trigger of the KLOE data readout is designed to accept a broad range of processes while discriminating mostly the machine background and the Bhabha $e^+e^-$ scattering. It operates on local energy deposits in the electromagnetic calorimeter and multiplicity of signals from drift chamber wires. In order to assure minimal readout delay, the trigger is divided into two levels with level 1 requiring at least one of the following:
\begin{itemize}
\item energy deposit larger that 50~MeV in the EMC barrel,
\item energy deposit larger that 150~MeV in the EMC endcaps,
\item 15 hits registered by the DC within a time window of 250~ns from a bunch crossing.
\end{itemize}
Moreover, at this level, the $e^+e^-\to e^+e^-$ events are identified by presence of clusters in the EMC with~E$>$350~MeV and are mostly rejected with only a pre-defined fraction preserved for detector calibration and accelerator luminosity monitoring. The level~1 trigger initiates early conversion of signals by the front-end electronics. However, in order to account for the full response of the drift chamber, the readout of its front-end electronics is extended to about 2~$\mu$s which corresponds to maximum drift time of electrons in the DC cells. Only then does the second trigger level process the DC-based event information along with more advanced topological requirements on the energy deposits in the calorimeter in order to validate the level 1 trigger. At level 2, a cosmic ray veto is also applied which rejects events with two $>$30~MeV energy deposits in the outer layer of the EMC\@.

As the $\Kl$ mesons at KLOE propagate in the detector with average $\beta\approx0.22$, their decays can occur tens of nanoseconds after their creation. Therefore, the trigger cannot be directly synchronized with each bunch crossing of DA$\Phi$NE\@. Instead, the readout is phase-locked to a demultiplied RF clock of the collider with a period of 4 bunch crossing ($t_{sync}=10.85$ ns)~\cite{data_handling}. The association of a recorded event to a particular bunch crossing number $N_{BC}$, is thus performed in the offline analysis in order to determine the event start time $T_0=N_{BC}\cdot T_{RF}$ ($T_{RF}=2.72$ ns), which must then be subtracted from all the times measured in the event. A first approximation of $T_0$ is obtained using the assumption that the first recorder calorimeter cluster was created by a prompt photon originating at the primary interaction point. The following relation is used to extract $T_0$ from the measured cluster time $T_{clu}$ and path length $R_{clu}$ travelled by the supposed photon to this cluster (electronics-related delays are neglected here for clarity)~\cite{data_handling}:
\begin{equation}
  T_0 = \text{nint}\footnote{nint(x) denotes the nearest integer for a real value x.} \left(\frac{T_{clu} -R_{clu} / c}{T_{RF}} \right) T_{RF}.
  \label{eq:t0_prompt}
\end{equation}
%%% @TODO: replace 'one of the further chapters' by proper subsection reference 
For events where no prompt photons are present, start time is estimated using other techniques considering flight time of heavy particles. A case of $T_0$ estimation for $K_{S}\to\pi e \nu$ events relevant for this work will be described in detail in~\sref{sec:t1_tof}.

\subsection{Background filters and data streaming}\label{sec:streaming}
Before tracking algorithms are applied to the recorded DC information, an additional background filter FILFO\footnote{\textit{FILtro di FOndo}, Italian for \textit{Background Filter}} is used for early reduction of the amount of present background to minimize the CPU time needed for event tracking. FILFO is dedicated to remove machine background relying solely on information from the electromagnetic calorimeter~\cite{data_handling}.

After complete reconstruction, a single event is assigned to one or more categories referred to as \textit{event streams} based on topological criteria. Streams constitute sets of event candidates for particular physical analyses. The following streams are used at KLOE:
\begin{itemize}
\item charged kaons,
\item neutral kaons,
\item $\phi\to\pi^+\pi^-\pi^0$ events,
\item radiative $\phi$ decays,
\item Bhabha scattering events,
\item unclassified events.
\end{itemize}
The last category contains a fraction (1/20) of all events before classification and is therefore useful for evaluation of the efficiency of event assignment to particular streams. Details of criteria on event topology required for each of the streams can be found in Reference~\cite{memo_225}.  

% @TODO: move this to analysis section
% Although all of the background rejection mechanisms described in this and the previous section are optimized to impose only minimal reduction of the physics processes of interest in KLOE, an analysis of the data collected by this setup requires accounting for finite selection efficiencies of the following filtering stages:
% \begin{itemize}
% \item cosmic ray veto,
% \item FILFO background filter,
% \item event assignment to the stream used in the analysis.
% \end{itemize}

\subsection{Detector upgrade for the KLOE-2 experiment}\label{sec:kloe2}
The KLOE detector operated in two periods, in the years 2000--2001 and 2004--2005, collecting datasets of $\phi$ decays amounting to about 450 nb$^{-1}$ and 1.7 fb$^{-1}$, respectively. Although the data analysis reported on in this Thesis was conveyed using the 2004--2005 data, an important objective of this work was to prepare tools for selection and reconstruction of relevant events, which can be applied in a future analysis of a larger dataset to be collected by the KLOE-2 experiment. KLOE-2 is constituted by the KLOE detector which has been upgraded with new components, comprising a novel inner tracking device created in the cylindrical-GEM (Gas Electron Multiplier) technology~\cite{kloe_inner_tracker} and filling the volume between the spherical beam pipe and inner wall of the drift chamber, as well as new crystal calorimeters covering in the final focusing regions of the beam~\cite{Cordelli:2013mka}, and upgraded QCAL calorimeters~\cite{Cordelli:2009xb}.

In addition to detector improvements, in particular enhanced tracking resolution due to the new inner tracker, the KLOE-2 experiment is expected to collect a larger dataset thanks to a new interaction scheme of the DA$\Phi$NE collider allowing for a higher instantaneous luminosity~\cite{Zobov:2007xw}. At the moment of writing of this Thesis, the KLOE2 experiment was in the course of its datataking with a goal to collect an integrated luminosity of at least 5 fb$^{-1}$. Reference~\cite{kloe2_general} contains a comprehensive description of the KLOE-2 physics programme.

\section{The J-PET detector}\label{sec:jpet}
The J-PET (Jagiellonian Position Emission Tomograph) is a photon detector constructed entirely at the Jagiellonian University in Krak\'ow and designed for registration of photons from electron-positron annihilations as well as decays of positronium exotic atoms. J-PET was built with a two-fold purpose in mind: as a prototype of a novel design of a Positron Emission Tomography (PET) scanner for medical imaging~\cite{moskal_patent} and for basic research, including tests of discrete symmetries in the decays of ortho-positronium states~\cite{moskal_potential}.

\begin{figure}[h!]
  \centering
  \includegraphics[width=0.4\textwidth]{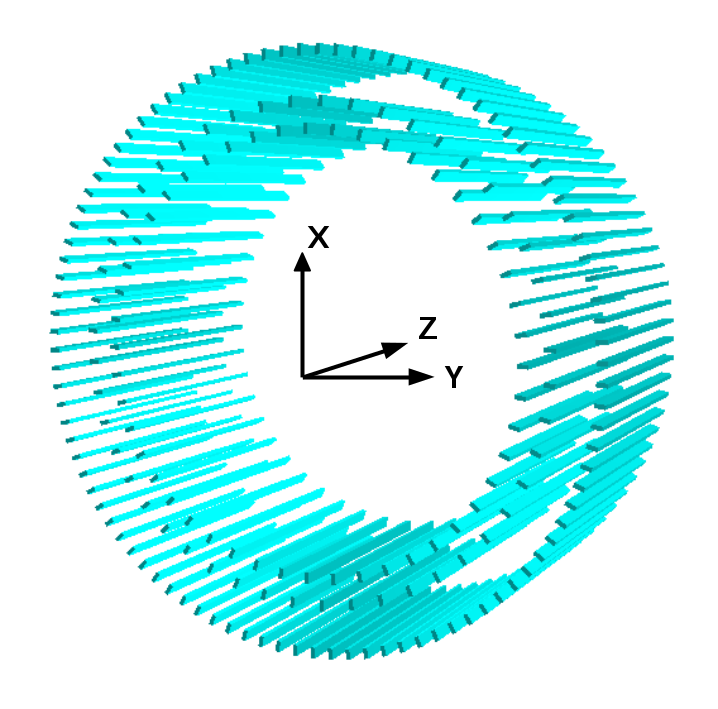}
  \hspace{1em}     
  \includegraphics[width=0.5\textwidth]{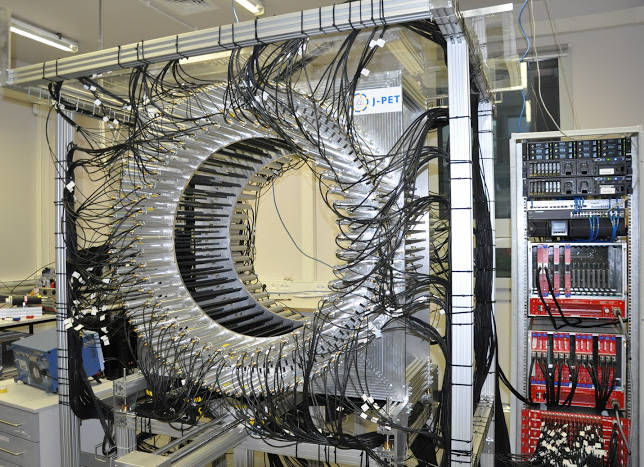} 
  \caption{Left: arrangement of the scintillator strips constituting the J-PET detector barrel in three concentric layers and the J-PET coordinate system. There are 48 strips in each of the two innermost layers and 96 strips in the outer layer. The strips are oriented along the Z axis whereas the X direction is vertical. Right: photograph of the J-PET detector and DAQ setup. The scintillator strips (covered with black foil) are mounted between two metal plates while the photomultipliers attached to strips' ends are located outside the plates.}\label{fig:jpet_detector}
\end{figure}

The detector is constituted by a total of 192 strips of plastic scintillator (EJ-230) arranged in three concentric layers which form a barrel geometry as shown in the left panel of~\fref{fig:jpet_detector}. Each scintillator strip has a length of 50~cm and rectangular cross-section with dimensions of 7$\times$19~mm$^2$ and is oriented along the barrel axis. Both ends of each scintillator strip are optically coupled with separate photomultiplier tubes (PMTs).

Plastic scintillators used in J-PET offer fast signals with a decay time of about 1.5~ns thus allowing for a high time resolution and for use of high-activity sources~\cite{gajos_gps}. However, due to their composition mostly of elements with a low atomic number, probability of photon interaction through the photoelectric effect is negligibly small for this material. Therefore, $\gamma$ detection in J-PET is based on Compton scattering instead. A photon scattered in a plastic scintillator strip transfers a fraction of its energy to the scattered electron whose further interactions in the material lead to production of photons from the visible light spectrum~\cite{anna_scints}. The sides along a scintillator strip are covered with reflective foil to ensure total internal reflection of these photons, which leads to their propagation to the two ends of the strip with losses only due to light attenuation in the scintillator. The ends of the strip are optically connected to the windows of \mbox{PMT-s} which collect photons produced in the $\gamma$ interaction and transform them into electric signals.

\begin{figure}[h!]
  \centering
  \begin{tikzpicture}[
  scale=0.4
  ]
  \definecolor{scint}{rgb}{0.12, 0.56, 1.0}
  \definecolor{metal}{rgb}{0.75, 0.75, 0.75}
  \coordinate (interaction) at (-2,0);
  % scint
  \draw[fill, scint] (-7,-0.7) -- (-7,0.7) -- (7,0.7) -- (7,-0.7);
  % pmts
  \draw[fill, metal] (7,-1.2) -- (7,1.2) -- (11,1.2) -- (11,-1.2) node[black] at (9,0) {\large PMT B};
  \draw[fill, metal] (-7,-1.2) -- (-7,1.2) -- (-11,1.2) -- (-11,-1.2) node[black] at (-9,0) {\large PMT A};;
  % z = 0
  \draw[black, line width=1, dashed, dash pattern=on 7pt off 4pt] (0,1.5) -- (0,-2) node[right] {\Large $z=0$};
  % interaction
  \draw[black, line width=1, dashed] (-2,0) -- (-2,3);
  \node[black, above] at (-2,3) {\large $t_{\gamma}$};
  \node[star,star points=10, inner sep=0.1cm, draw, thick, fill=red!100] at (interaction) {};
  % time arrows
  \draw[line width=1.5, black, <->] (-7,2) -- (-2,2) node[midway, above] {\large $s_{A}$};
  \draw[line width=1.5, black, <->] (-2,2) -- (7,2) node[midway, above] {\large $s_{B}$};
  \draw[snake it, red, thick, ->] (0,4) -- (-1.75,0.4);
  \end{tikzpicture}
  \caption{Determination of the $z$ coordinate of a $\gamma$ interaction point (red star) in a J-PET scintillator strip (blue). Position along the scintillator is calculated as a difference of the $s_A$ and $s_B$ paths travelled by scintillation light along the strip from the interaction point to the photomultiplier tubes, obtained from the PMT signal arrival times and effective velocity of light in the scintillator.} \label{fig:jpet_strip}
\end{figure}
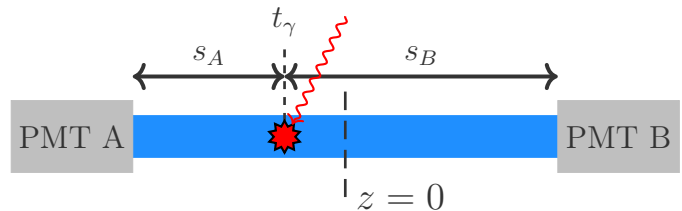

In terms of reconstruction of the time and place of particle interactions, the J-PET detector shares several similarities with the electromagnetic calorimeter of KLOE described in~\sref{sec:emc}. While position of a $\gamma$ interaction point in the XY plane of the detector is determined up to the location of a single scintillator (which corresponds to an azimuthal angle resolution at the level of 1 degree~\cite{gajos_gps}), the Z coordinate of the interaction is inferred from the difference between times of light propagation to both ends of the scintillator strip as shown in~\fref{fig:jpet_strip}. While exact time needed by the optical photons to travel from the interaction point to a strip end depends on a particular sequence of internal reflections, the average light propagation along the strip can be described by an effective velocity $\upsilon_{e}$, determined as a calibration constant of the detector~\cite{jpet_single_module}. Assuming the $\gamma$ was scattered in the scintillator at time $t_{\gamma}$, the times of light arrival to the two ends of the strip (labeled A and~B) $t_A$ and $t_B$ can be expressed as:
\begin{eqnarray}
  \label{eq:jpet_zpos}
  t_A &=& t_{\gamma} + \frac{s_A}{\upsilon_{e}},\\
  t_B &=& t_{\gamma} + \frac{s_B}{\upsilon_{e}}.
\end{eqnarray}
The interaction position along the $z$ axis, measured with respect to the strip center, is proportional to the difference of the light registration times at PMT-s A and B and is thus obtained using the same formula as in case of the KLOE electromagnetic calorimeter (compare~\eref{eq:emc_s}):
\begin{equation}
  \label{eq:jpet_zpos2}
  z_{\gamma} = \frac{s_A - s_B}{2} = \frac{\upsilon_{e}\left( t_A - t_B \right)}{2}.
\end{equation}
Similarly, the time of the interaction $t_{\gamma}$ can be obtained from a sum of the light registration times at sides A and B of the scintillator, again using the same recipe in the KLOE EMC (compare~\eref{eq:emc_t}):
\begin{equation}
  \label{eq:jpet_hittime}
  t_{\gamma} = \frac{t_A + t_B}{2}  - \frac{L}{2\upsilon_{e}},
\end{equation}
where $L=s_A+s_B$ is the length of the scintillator strip.

The majority of $\gamma$ interactions in the plastic scintillators are through Compton scattering, where amount of energy deposited by the incident $\gamma$ depends on the scattering angle with a differential cross-section for scattering in a solid angle element d$\Omega$ given by the Klein-Nishina formula~\cite{Griffiths}:
\begin{equation}
  \label{eq:klein_nishina}
  \frac{\text{d}\sigma}{\text{d}\Omega} = \frac{r_0^2}{2}\left(\frac{E'}{E}\right)^2\left[ \frac{E}{E'} + \frac{E'}{E} - sin^2\vartheta \right],
\end{equation}
where $r_0$ is the classical electron radius, $E$ and $E'$ denote energies of the primary and scattered photon and $\vartheta$ is the planar angle between their momenta.

\begin{figure}[h!]
  \centering
  \includegraphics[width=0.5\textwidth]{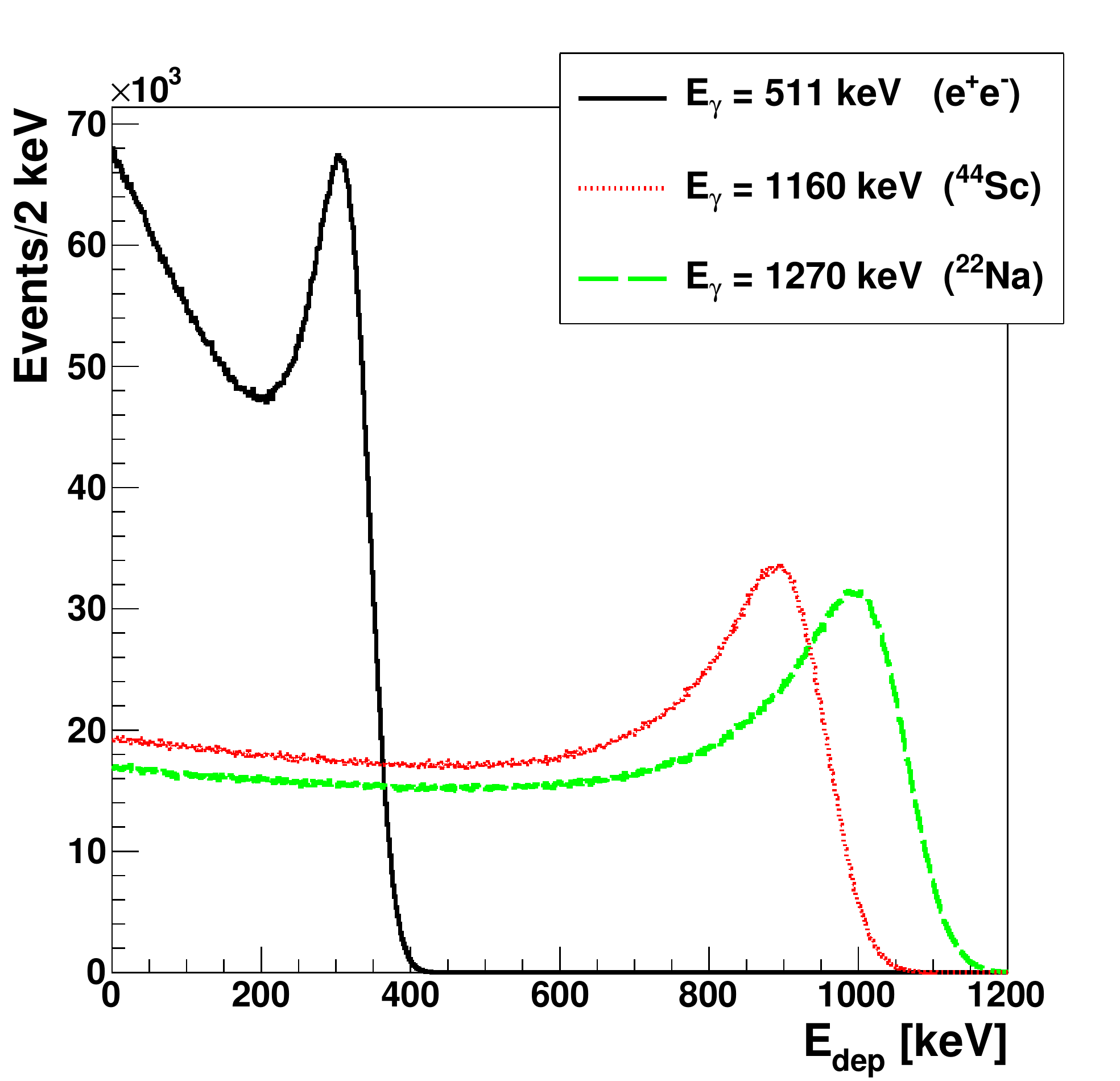}
  \caption{Simulated energy deposition spectra for Compton scattering of $\gamma$ with various energies in the J-PET plastic scintillators. Figure was adapted from~\cite{daria_epjc}.}
  \label{fig:compton}
\end{figure}

What follows is that a scattered photon can deposit any energy from a continuous spectrum between zero and a certain boundary value smaller than its total energy. This maximal transferred energy, corresponding to $\gamma$ back-scattering, smeared by experimental effects constitutes a characteristic cut-off in the spectrum, referred to as the Compton edge. \fref{fig:compton} presents exemplary simulated energy deposition spectra for three different energies of incident $\gamma$~quanta, where different positions of Compton edges are visible. Consequently, although it is impossible to measure the photon energy directly through registration of a single interaction, $\gamma$ quanta of energies significantly larger than annihilation products can be distinguished by the amount of deposited energy above the Compton spectrum edge for 511~keV photons.

\begin{figure}[h!]
  \centering
  \includegraphics[width=0.6\textwidth]{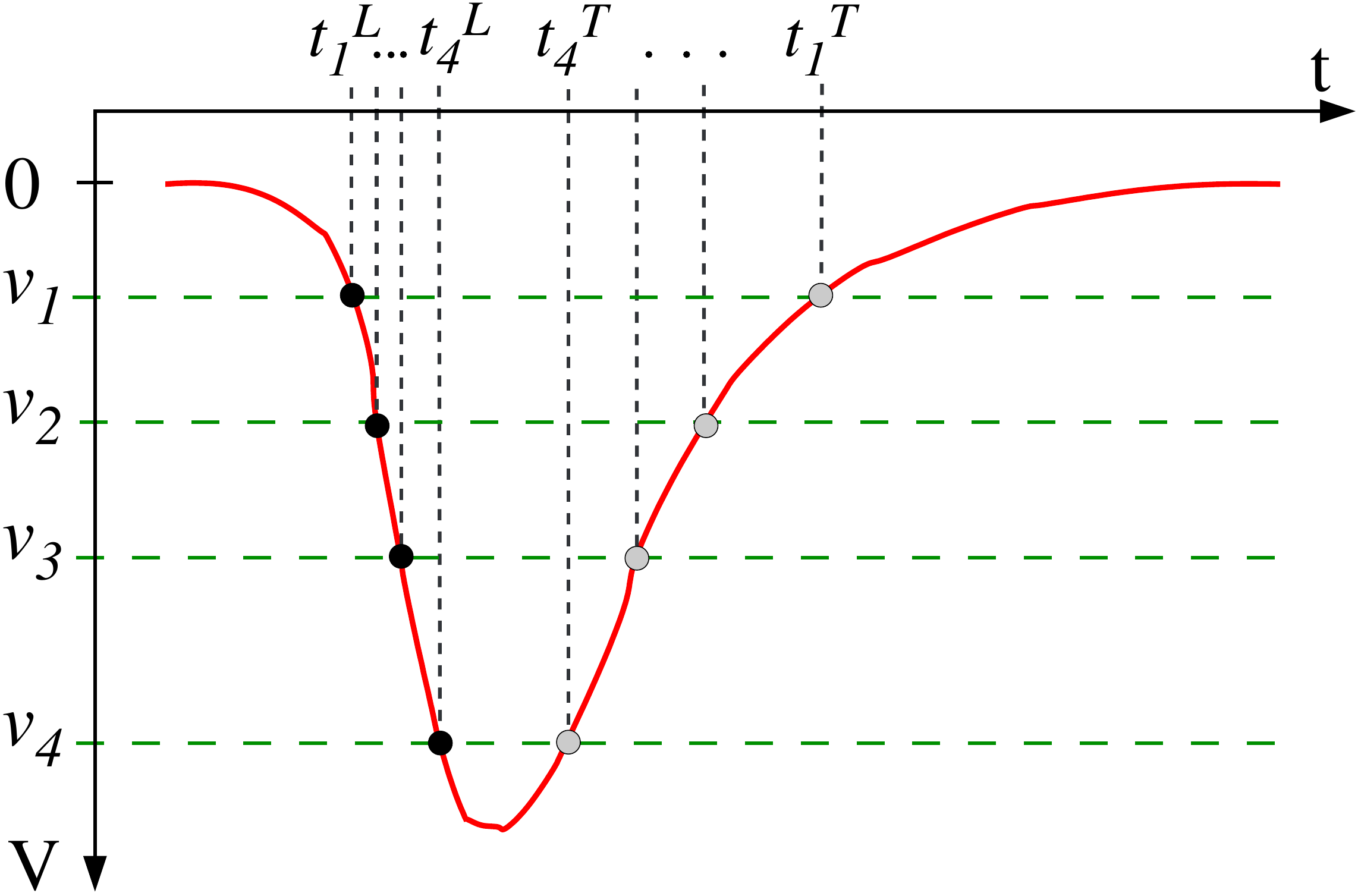}
  \caption{Scheme of digitalization of a photomultiplier signal by the J-PET front-end electronics. An electric signal (red) is sampled at four preset voltage threshold levels
(green) $\nu_{i}, \ i=1,\ldots,4$. Times of the signal voltage crossing the thresholds are recorded at both leading ($t_i^{L}$) and trailing edge ($t_i^{T}$) of the signal (black and gray dots respectively).}\label{fig:jpet_signal}
\end{figure}

J-PET therefore attempts to measure the deposited energy indirectly through estimation of an integral of electric signals produced by the PMT-s in response to the recorded scintillation light. This estimation is based on the time-over-threshold (TOT) measurement enabled by dedicated front-end electronics (FEE)~\cite{marek_fee}. The J-PET front-end electronics are connected to the photomultiplier tubes which produce negative electric signals (understood as voltage changes in time). The FEE probe electric signals in the time domain at four preset threshold values of voltage $\upsilon_i, \: i=1,\ldots,4$ as depicted in~\fref{fig:jpet_signal}. For each of the threshold voltages, the FEE output a logical signal at the moment when the signal voltage crosses the threshold, separately for the leading and trailing edge of the signal~\cite{marek_fee}. These logical signals are passed to the Time-to-Digital Converters (TDC-s) of the data acquisition (DAQ) system~\cite{greg_daq} which produces up to 8 timestamps per a single PMT signal. Times recorded for both the leading and trailing signal edge at the same threshold $i$ may be used to calculate the time-over-threshold value:
\begin{equation}
  \label{eq:tot_i}
  TOT_i = t^T_i-t^L_i.
\end{equation}
A set of $TOT_i$ values at certain thresholds constitute a measure of the signal size (understood as its the integral, compare~\fref{fig:jpet_signal}), the sum of TOT-s at all thresholds crossed by the signals on both A and B sides of a scintillator allow for an estimation of the energy deposited in the strip by a $\gamma$ quantum. In further considerations, such a sum of TOT values will be referred to as TOT for a certain $\gamma$ interaction recorded in J-PET:
\begin{equation}
  \label{eq:tot_def}
  TOT_{\gamma} = \sum_{i=1}^4 TOT_i^{A} + \sum_{i=1}^4 TOT_i^{B}.
\end{equation}

In addition to providing a measure of deposited energy, this signal digitalisation scheme allows for a partial retrieval of the original signal shape based on the recorded points with techniques such as compressive sensing~\cite{lech_compressive}. Moreover, it provides additional information on the signal leading edge useful for precise reconstruction of its arrival time, needed in turn for determination of the interaction $z$ position using relations~\ref{eq:jpet_zpos2} and~\ref{eq:jpet_hittime}. A more advanced approach to the logitudinal interaction position reconstruction has also been studied with a simplified J-PET prototype~\cite{neha_synchronized}.

Since any non-trivial reconstruction preformed in the J-PET detection 
is based on measurements of times, 
calibration of the recorded time values is crucial for the obtained accuracy. Time calibration is performed with a point-like source of annhilation photon pairs coupled with small auxiliary reference detector approaching subsequently every scintillator strip and positioned precisely at its center in order to provide reference events for which $t_A=t_B$. A detailed description of the time calibration procedure can be found in~\cite{jpet_time_calibration}.
Although the exact resolution of the time and $z$ position determination of $\gamma$ scattering in J-PET is only under study at the moment of writing of this Thesis, studies involving a two-dimensional prototype of the same design have shown that the achievable resolution can be as low as $\sigma(t_{\gamma}) = 80$ ps and  $\sigma(z_{\gamma}) = 0.93$ cm~\cite{jpet_single_module}.

\subsection{Polarization control}\label{sec:ops_polarization}
\begin{figure}[h!]
  \centering
  \includegraphics[width=0.6\textwidth]{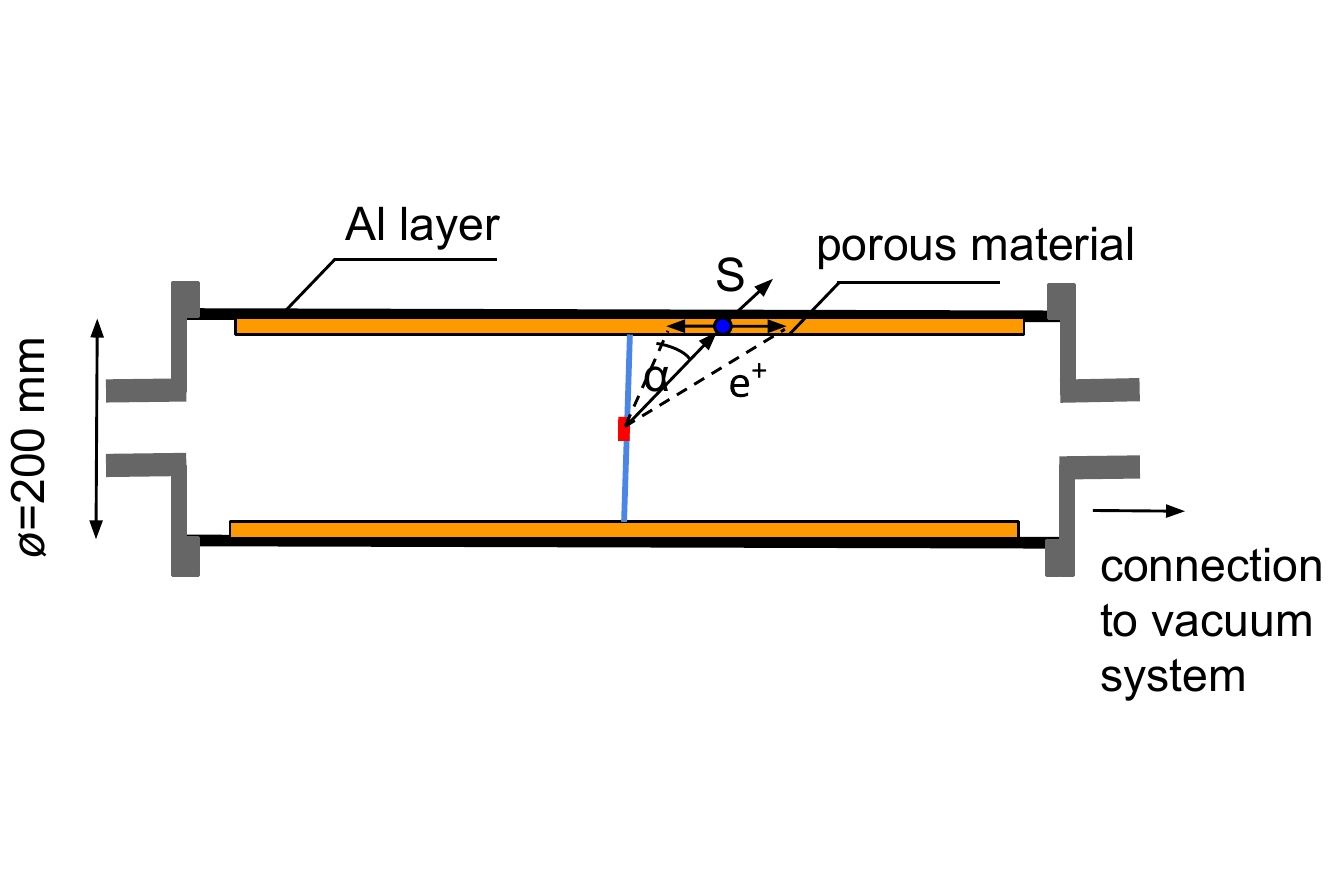}
  \hspace{1em}
  \includegraphics[width=0.35\textwidth]{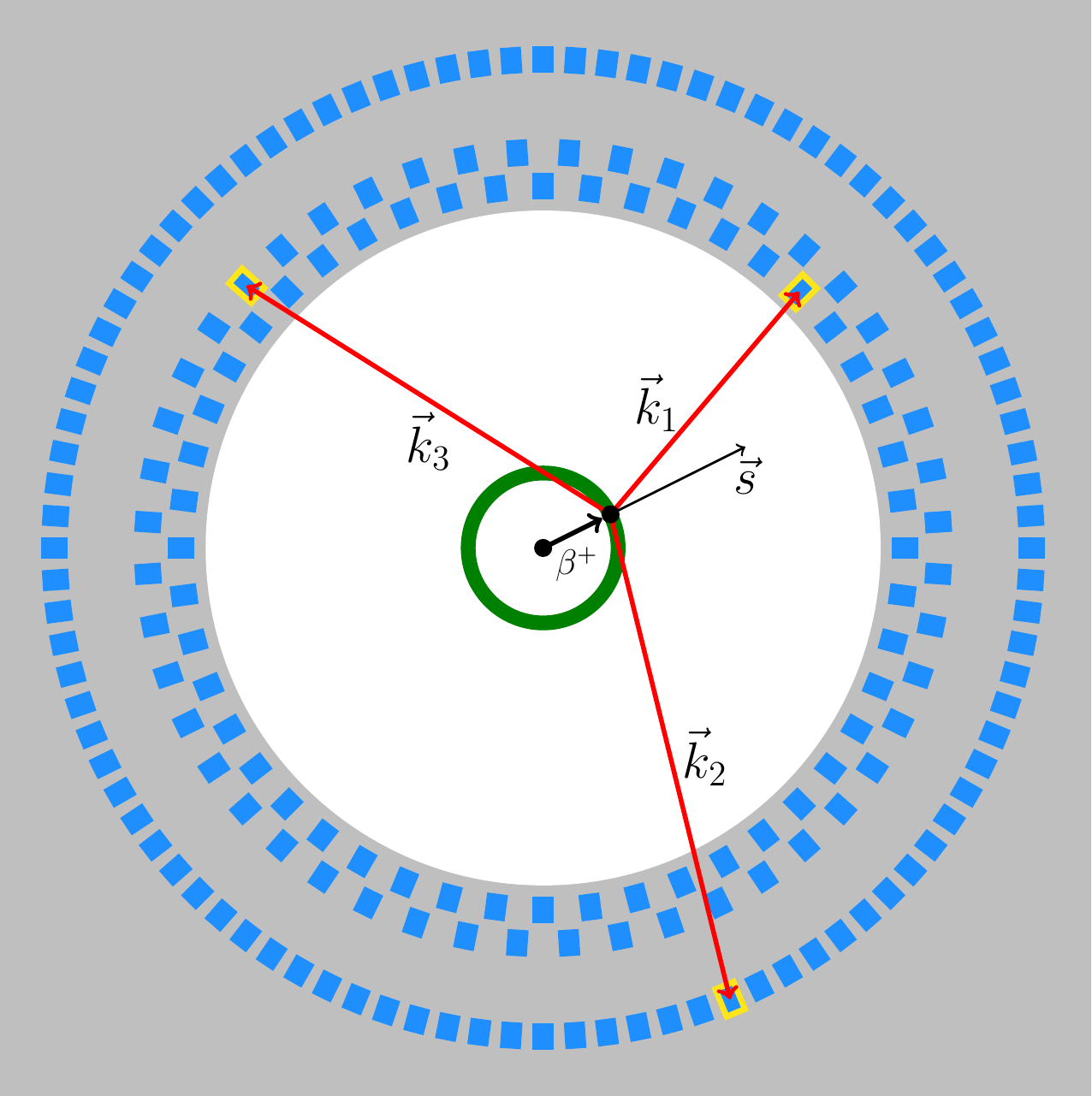}
  \caption{Left: longitudinal section of an o-Ps annihilation chamber allowing for determination of ortho-positronium spin direction throught spin of the positron which creates it. Right: scheme of the o-Ps spin reconstruction in the J-PET detector presented in its transverse plane.}
  \label{fig:ops_spin_determination}
\end{figure}

Calculation of the angular correlation operators described in~\cref{chapter:test_jpet} relies on determination of the spin of decaying ortho-positronia. With an appropriate experimental setup presented schematically in~\fref{fig:ops_spin_determination}, this can be performed on an event-by-event basis to an extent limited by several statistical factors. In the planned setup, a $\beta^+$ radioactive point-like source will be placed in the center of a cylindrical vacuum chamber whose walls will be coated on the inner side with a porous medium for \ops/ production. Certain materials such as silica aerogels can be used as the medium to maximize the ratio of \ops/$\to 3\gamma$ to total annihilation rate and to other $3\gamma$ annihilations not originating from an ortho-positronium state~\cite{Jasinska:2016qsf,moskal_potential}. Positrons from the source, after reaching the medium on the chamber walls, undergo thermalization and may form ortho-positronium atoms. As the possible distance travelled by the \ops/ atom before its decay is negligible, reconstruction of the decay point, along with the fixed $\beta^+$ source location, allows for an estimation of the original positron flight direction (see~\fref{fig:ops_spin_determination}). Positrons from a $\beta^+$ decay are spin-polarized along their velocity vector $\vec{\upsilon}_e$ with an average polarization $\vec{P}=\frac{\vec{\upsilon}_e}{c}$ due to parity violation, therefore knowledge of the $\vec{\upsilon}_e$ direction defines the positron spin direction with an average uncertainty dependent on the mean positron energy and thus on the chosen $\beta^+$ emitter~\cite{moskal_potential}.

The average uncertainty on the determined \ops/ spin further depends on the polarization loss during positron thermalization~\cite{PhysRevLett.43.1281} and the fact that only 2/3 of the created positronia retain the spin of the positron~\cite{Arbic:1988pv}. Moreover, an indeterminacy of the positron direction within a cone with an $\alpha$ opening angle (indicated in the left panel of~\fref{fig:ops_spin_determination}) diminishes the polarization by an additional factor of $\frac{1}{2}(1+cos(\alpha))$~\cite{Coleman}. Since the impacts of thermalization and \ops/ formation on the final spin estimation are inevitable, it is crucial to minimize the effect from positron flight direction uncertainty. This can be achieved with a precise reconstruction of the ortho-positronium annihilation point, providing which was one of the objectives of this Thesis. Next Chapter contains a description of a suitable \ops/$\to 3\gamma$ reconstruction technique whereas the results of its preliminary performance studies are presented in~\cref{chapter:analysis_jpet}.

%%%
% do zacytowania:
%%%
% rekonstrukcja \cite{neha_synchronized} \cite{lech_compressive}
% kalibracja \cite{jpet_time_calibration}
% jak zrekonstruować energie na podstawie kątów \cite{daria_epjc}
% 80ps i efektywna predkosc \cite{jpet_single_module}
% pierwsze studia odrzucania tła, suma vs roznica kątów etc. \cite{jpet_commissioning}

%%% Local Variables:
%%% TeX-master: "../main"
%%% End:
\chapter{Trilateration-based reconstruction of  decays into photons}\label{chapter:gps}

Reconstruction of a point of a particle decay, commonly performed for charged particles using tracking devices such as drift chambers or solid state pixel detectors, becomes a much more challenging task in case of processes which only contain neutral states. The two cases discussed in this work, the \ops/$\to 3\gamma$ annihilation and the long-lived neutral K meson decay $\Kl\to 3\pi^0\to 6\gamma$, although occurring at vastly different energy scales and governed by different mechanisms, possess the same experimental difficulties when it comes to their reconstruction using only the resulting photons recorded by a detector. This Chapter presents a technique based on trilateration which allows for determination of the decay point and time for both of the aforementioned processes.

\section{Application of trilateration to reconstruction of particle decays}
\label{sec:general_trilateration}
Trilateration is well-established method of localizing points which are known to lie simultaneously on surfaces of several spheres or circles whose centers correspond to certain reference points. In the classical trilateration problem, both reference points and spheres' radii are well defined and knowledge of three non-identical and mutually intersecting spheres is sufficient to calculate the position of the sought common point. In fact, trilateration using three references can yield up to two solutions (in general, the intersection of two spheres is a circle which can have 0, 1 (in case of tangency) or 2 intersection points with a third sphere) and additional criteria must be applied to disambiguate the point being localized.

Practical applications of trilateration commonly measure the spheres' radii through the time of propagation of a certain signal between the localized point and the known references. This is the case in e.g. the Global Positioning System (GPS)~\cite{Langley2005TheMO, Carter1999PrinciplesOG} where references constituted by satellite locations exchange radio signals with the receiver whose position is sought. In such an approach, however, both the references and the localized object must measure the signal emission and recording time with respect to a single origin, which is often impossible. To resolve this issue, radii of the spheres must be parametrized with an additional unknown (besides the three spatial coordinates) being either the start or stop time of the exchanged signal. As the reconstruction problem with three references and variable spheres' radii is underdetermined, additional information must be included in the system, e.g.\ by using a fourth reference point. On the other hand, a useful consequence in this case is that the initially unknown time is obtained from the reconstruction as a by-product additional to the spatial coordinates, e.g.\ in the GPS system the localized receiver determines the current time precisely besides its position.

Such extended trilateration approach similar to GPS positioning has been adapted to reconstruction of neutral states into several photons. In this case, the place of decay is the localized object, final state photons serve as the exchanged signal whereas the recording points and times of their interactions in the detector provide the references. Details of the general reconstruction problem as well as its solution are discussed in the next Section on the example of $\Kl\to 3\pi^0\to 6\gamma$ decay while~\sref{sec:gps_jpet} presents a slightly different case of \ops/$\to 3\gamma$, where the decay point is obtained with three reference points only.

\section{Reconstruction of $\Kl\to 3\pi^0 \to 6 \gamma$ in KLOE}\label{sec:gps_kloe}
As discussed in Sections~\ref{sec:kaons} and~\ref{sec:kloe}, the lifetime of the $\Kl$ meson in the KLOE detector frame of reference is sufficiently large for the meson to decay anywhere in the detector volume. This fact calls for reconstruction procedures capable for localizing a $\Kl$ decay based on its products with comparable performance for any distance of the decay from the detector center. In case of decays involving charged products, this is catered for by the drift chamber filling almost the entire detector. In case of the $\Kl\to 3 \pi^0$ process%
\footnote{As the mean $\pi^0$ lifetime, $(8.52\pm0.18)\times 10^{-17}$~s, is unregisterably small for KLOE and the branching fraction for the $\pi^0\to\gamma\gamma$ decay is $98.823\pm0.034$~\%, further in this work the process $\Kl\to 3\pi^0\to 6\gamma$ will always be considered.}%
, the most prominent among all-neutral decays of the long-lived neutral K meson, multiple studies performed in KLOE have augmented its reconstruction with information coming from the decay of a second kaon ($\Ks$) originating from the same $\phi$ decay~\cite{difalco_phd,kloe_kl3pi0_br,kloe_kl_lifetime}. If the accompanying kaon decay is chosen e.g.~as $\Ks\to\pi^+\pi^-$ for which DC-based reconstruction of the two pion tracks yields a very precise determination of the $\Ks$ momentum vector, the kinematics of the $\phi\to\Ks\Kl$ decay along with known average $\phi$ boost provide an estimate of the $\Kl$ momentum direction as shown in~\sref{sec:ks_from_kl}. Consequently, the $\Kl\to 3 \pi^0\to 6 \gamma$ decay vertex can be found along the known $\Kl$ line of flight by considering the sum of times of flight of the kaon and and even a single photon~\cite{data_handling}. This approach, however, is not applicable in case of the \Ts~symmetry test described in~\cref{chapter:test_kloe} where the process under study is defined by final states of both kaon decays and $\Kl\to 3\pi^0$ must be accompanied by a semi-leptonic decay of $\Ks$ (see~\tref{tab:processes}). Then, the unregistered neutrino momentum prevents a good reconstruction of $\vec{p}_{\Ks}$ and thus also the $\Kl$ line of flight. To avoid these difficulties, the trilateration technique has been employed in the localization of the $\Kl\to 3\pi^0$ decay without relying on any information except up to 6 photons recorded in the electromagnetic calorimeter of KLOE.

\begin{figure}[h!]
  \centering
  % PIC 1
  \captionsetup[sub]{margin=1ex}
  \begin{subfigure}{0.45\textwidth}
    \begin{tikzpicture}[scale=0.45,photon/.style={->, ultra thick, black},]
    \coordinate (clu1) at (-4.58,2);
    \coordinate (clu2) at (1,4.90);
    \coordinate (clu3) at (4,3);
    \coordinate (clu4) at (4.90,-1);
    \coordinate (clu5) at (-2,4.58);
    \coordinate (clu6) at (4.90, 1.0);
    \coordinate (vertex) at (1.0,2.0);

    % draw a fake arc and text to keep the same size of the picture as the next ones
    \draw[thick, white, rotate=85, dashed] (clu1)+(-5.58,0) arc (180:320:5.58);
    \node[white,anchor=south] at (clu1) {\scriptsize$(\!X_1\!,\!Y_1\!,\!Z_1\!,\!T_1\!)$};
    
    \draw[gray, line width=4pt] (0,0) circle (5);
    \draw[red,fill] (clu1) circle (0.2);
    \draw[darkgreen,fill] (clu2) circle (0.2);
    \draw[blue,fill] (clu3) circle (0.2);
    \draw[orange,fill] (clu4) circle (0.2);
    \draw[yellow,fill] (clu5) circle (0.2);
    \draw[violet,fill] (clu6) circle (0.2);

    \draw[black,dashed,->] (0,0) -- (vertex) node[midway, right,yshift=-2mm] {\footnotesize $\Kl$};
    \draw[darkgray,fill] (0,0) circle (0.1) node[left]{$\phi$};
    \draw[darkgray,fill] (vertex) circle (0.1);
    
    \draw[snake it,->] (vertex) -- (clu1) node[midway, above]{$\gamma$};
    \draw[snake it,->] (vertex) -- (clu2) node[midway, left]{$\gamma$};
    \draw[snake it,->] (vertex) -- (clu3) node[midway, above]{$\gamma$};
    \draw[snake it,->] (vertex) -- (clu4) node[midway, above]{$\gamma$};
    \draw[snake it,->] (vertex) -- (clu5) node[midway, above]{$\gamma$};
    \draw[snake it,->] (vertex) -- (clu6) node[midway, above]{$\gamma$};
    
  \end{tikzpicture}
  \caption{Schematic presentation of a \mbox{$\Kl\to 3\pi^0\to 6 \gamma$} decay in the transverse view of KLOE EMC (gray band) which records at maximum 6 photon interaction points (labeled with dots of different colors).}\label{fig:trilateration_kloe_a}.
\end{subfigure}
% PIC 2
  \begin{subfigure}{0.45\textwidth}
  \begin{tikzpicture}[scale=0.45,photon/.style={->, ultra thick, black},]
    \coordinate (clu1) at (-4.58,2);
    \coordinate (clu2) at (1,4.90);
    \coordinate (clu3) at (4,3);
    \coordinate (clu4) at (4.90,-1);
    \coordinate (clu5) at (-2,4.58);
    \coordinate (clu6) at (4.90, 1.0);
    \coordinate (vertex) at (1.0,2.0);
    
    \draw[gray, line width=4pt] (0,0) circle (5);    \draw[red,fill] (clu1) circle (0.2);
    \draw[darkgreen,fill] (clu2) circle (0.2);
    \draw[blue,fill] (clu3) circle (0.2);
    \draw[orange,fill] (clu4) circle (0.2);
    \draw[yellow,fill] (clu5) circle (0.2);
    \draw[violet,fill] (clu6) circle (0.2);

    \node[black,anchor=south] at (clu1) {\scriptsize$(\!X_1\!,\!Y_1\!,\!Z_1\!,\!T_1\!)$};
    \draw[thick, red,rotate=85, dashed] (clu1)+(-5.58,0) arc (180:320:5.58);
    
    \draw[red,->] (0.63, 0) -- (-4.3,1.9) node[midway, above] {$\gamma$} node[midway, below, black,xshift=-3mm,yshift=-2mm] {\footnotesize $R^\gamma_1\!=\!c(T_1\!-\!t_d)$};
  \end{tikzpicture}
  \caption{For each of the points of $\gamma$ interaction in the detector, a set of possible creation points of this $\gamma$ quantum is a sphere (dashed line) centered at the known location of the EMC cluster and with a radius parametrized by an unknown time $t_d$ of $\Kl$~decay.}\label{fig:trilateration_kloe_b}
\end{subfigure}
  % PIC 3
  \begin{subfigure}{0.45\textwidth}
  \begin{tikzpicture}[scale=0.45,photon/.style={->, ultra thick, black},]
    \coordinate (clu1) at (-4.58,2);
    \coordinate (clu2) at (1,4.90);
    \coordinate (clu3) at (4,3);
    \coordinate (clu4) at (4.90,-1);
    \coordinate (clu5) at (-2,4.58);
    \coordinate (clu6) at (4.90, 1.0);
    \coordinate (vertex) at (1.0,2.0);

    % draw a fake text to keep the same size of the picture as the next ones
    \node[white,anchor=south] at (clu1) {\scriptsize$(\!X_1\!,\!Y_1\!,\!Z_1\!,\!T_1\!)$};
    
    \draw[gray, line width=4pt] (0,0) circle (5);
    \draw[red,fill] (clu1) circle (0.2);
    \draw[darkgreen,fill] (clu2) circle (0.2);
    \draw[blue,fill] (clu3) circle (0.2);
    \draw[orange,fill] (clu4) circle (0.2);
    \draw[yellow,fill] (clu5) circle (0.2);
    \draw[violet,fill] (clu6) circle (0.2);

    \draw[thick, red,rotate=85, dashed] (clu1)+(-5.58,0) arc (180:320:5.58);
    \draw[thick, darkgreen,rotate=-10,dashed] (clu2)+(-2.90,0) arc (180:360:2.90);
    \draw[thick, blue,rotate=-55,dashed] (clu3)+(-3.16,0) arc (180:360:3.16);
    \draw[thick, orange,rotate=-90, dashed] (clu4)+(-4.92,0) arc (180:330:4.92);

    % \draw[black,fill] (vertex) circle (0.15) node[below, yshift=1mm, xshift=-8mm]{\footnotesize $(x\!,y\!,z\!,t)$};
    \draw[black,fill] (vertex) circle (0.15) node[below, yshift=2mm, xshift=-11mm]{\footnotesize $(x_d,y_d,z_d,t_d)$};        
  \end{tikzpicture} 
  \caption{The $\Kl\to 3\pi^0 \to 6 \gamma$ decay vertex is found as an analytically-calculated intersection of the possible origin spheres of at least 4 recorded photons.}\label{fig:trilateration_kloe_c}
\end{subfigure}
% PIC 4
\begin{subfigure}{0.45\textwidth}
    \begin{tikzpicture}[scale=0.45,photon/.style={->, ultra thick, black},]
    \coordinate (clu1) at (-4.58,2);
    \coordinate (clu2) at (1,4.90);
    \coordinate (clu3) at (4,3);
    \coordinate (clu4) at (4.90,-1);
    \coordinate (clu5) at (-2,4.58);
    \coordinate (clu6) at (4.90, 1.0);
    \coordinate (vertex) at (1.0,2.0);

    % draw a fake text to keep the same size of the picture as the next ones
    \node[white,anchor=south] at (clu1) {\scriptsize$(\!X_1\!,\!Y_1\!,\!Z_1\!,\!T_1\!)$};
    
    \draw[gray, line width=4pt] (0,0) circle (5);
    \draw[red,fill] (clu1) circle (0.2);
    \draw[darkgreen,fill] (clu2) circle (0.2);
    \draw[blue,fill] (clu3) circle (0.2);
    \draw[orange,fill] (clu4) circle (0.2);
    \draw[yellow,fill] (clu5) circle (0.2);
    \draw[violet,fill] (clu6) circle (0.2);

    \draw[thick, red, rotate=85, dashed] (clu1)+(-5.58,0) arc (180:320:5.58);
    \draw[thick, darkgreen,rotate=-10,dashed] (clu2)+(-2.90,0) arc (180:360:2.90);
    \draw[thick, blue,rotate=-55,dashed] (clu3)+(-3.16,0) arc (180:360:3.16);
    \draw[thick, orange,rotate=-90, dashed] (clu4)+(-4.92,0) arc (180:330:4.92);
    \draw[thick, yellow,rotate=20,dashed] (clu5)+(-3.96,0) arc (180:360:3.96);
    \draw[thick, violet,rotate=-60, dashed] (clu6)+(-4.03,0) arc (180:330:4.03);
    
    \draw[black,fill] (vertex) circle (0.15) node[below, yshift=2mm, xshift=-11mm]{\footnotesize $(x_d,y_d,z_d,t_d)$};        
  \end{tikzpicture} 
  \caption{If 5 or 6 photons were registered, the additional reference points can be used to find the intersection numerically, minimizing effects if uncertainties on determination of EMC cluster locations and times.}\label{fig:trilateration_kloe_d}
\end{subfigure}

\caption{Scheme of reconstruction of $\Kl\to 3\pi^0 \to 6 \gamma$ decays in KLOE.}\label{fig:trilateration_kloe}
\end{figure}
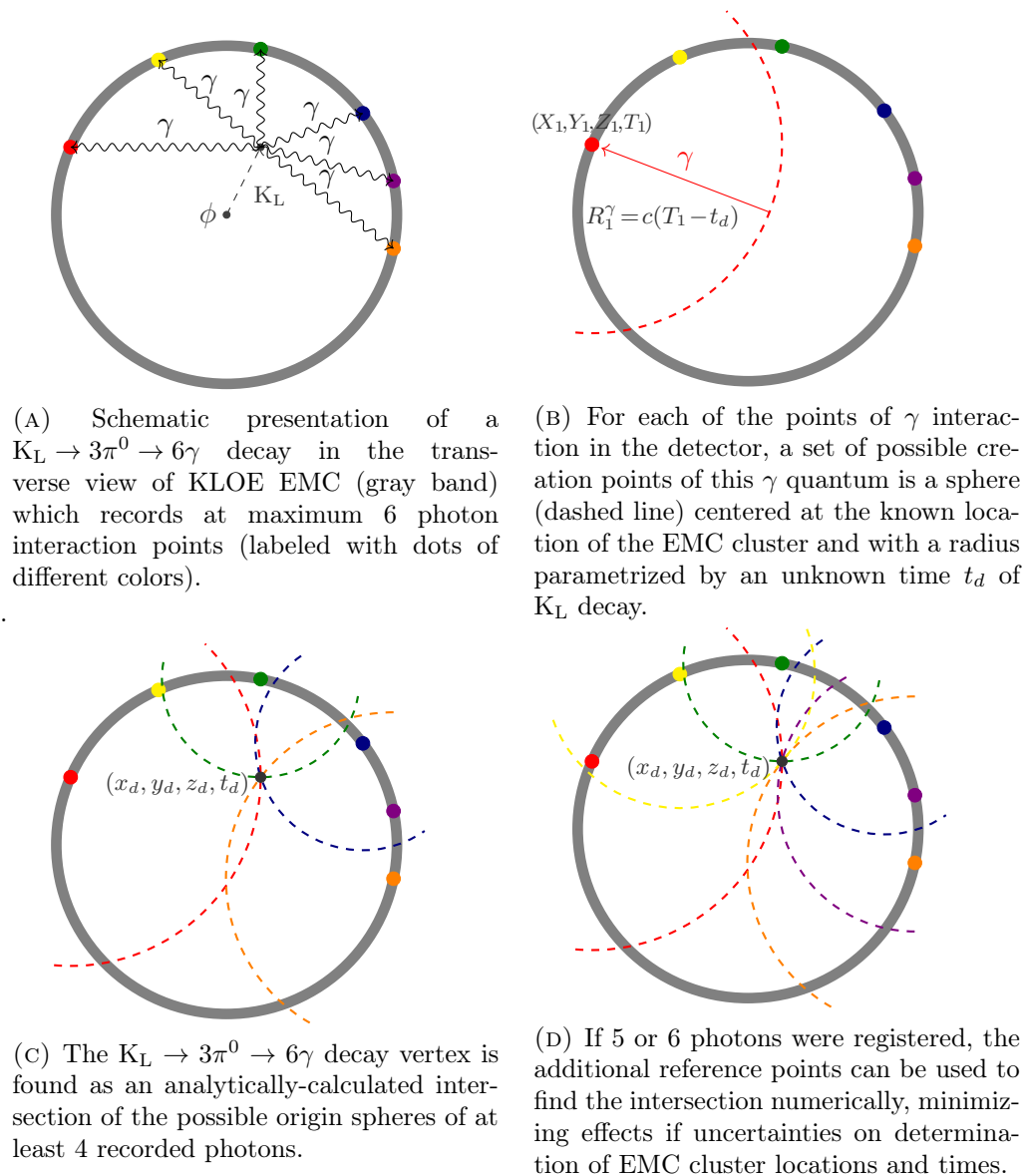

\fref{fig:trilateration_kloe_a} shows a schematic representation of a $\Kl\to 3\pi^0\to 6\gamma$ event inside the KLOE detector. Although the particles' momenta lie in a 3-dimensional space, a projection on the detector transverse plane is shown for simplicity. The electromagnetic calorimeter of KLOE records energy deposition clusters corresponding to up  to 6 photons%
\footnote{I may happen that some photons are not registered due to their insufficient energy deposition or escaping through a region not covered by the calorimeter.}
from this process.
 For each cluster $i$ ($i=1,\ldots,6$), its spatial coordinates ($X_i,Y_i,Z_i$) and recording time $T_i$ are determined using the procedures described in~\sref{sec:emc}.
A single cluster results from an incident photon which arrived from an unknown direction but is known to have originated from a decay of $\Kl$ happening at time $t_{d}$ (also unknown). A set of possible creation points of this photon is a sphere centered at the EMC cluster and with a radius given by the product of the $\gamma$ time of flight (from its creation to interaction in the detector) and velocity of light (see~\fref{fig:trilateration_kloe_b}):
 \begin{equation}
   \label{eq:gps_radius}
   R^{\gamma}_i = c(T_i - t_d).
 \end{equation}
The radii of each of the spheres of possible $\gamma$ origin points are thus parametrized by the same unknown decay time of the $\Kl$ meson, $t_d$.

It is easily observed that the $\Kl\to 3\pi^0\to 6 \gamma$ decay vertex must be a common creation point of all recorded photons and thus its location $(x_d,y_d,z_d)$ is defined by the following system of equations:
\begin{eqnarray}
  \label{eq:gps_6_eqns}
  (X_i-x_d)^2 - (Y_i-y_d)^2 - (Z_i-z_d)^2 = c^2(T_i-t_d)^2, \quad i=1,\ldots,6,
\end{eqnarray}
 whose solution corresponds to finding an intersection of the spheres of potential origins, defined for each of the photons as shown in~\fref{fig:trilateration_kloe_c}.

 In the above system, all quantities denoted with capital letters are known, as is $c$. There are therefore 4 unknowns and, consequently, four reference points are required to obtain a solution. Thus a necessary condition for the decay vertex reconstruction is that at least four out of possible six photons are recorded in the KLOE EMC.

 An analytical solution of the equation system~\ref{eq:gps_6_eqns} for $i=1,\ldots,4$ can be obtained through pairwise subtraction of chosen equations, which removes terms quadratic in unknowns, and solving thus obtained underdetermined linear system for $(x_d,y_d,z_d)$ parametrized by $t_d$. The latter solution practically describes a parametric line on which the solutions of the original problem must be located, which in turn can be found by intersecting this line with any of the initial spheres. As a result, zero, one or two solutions can be found for the decay vertex location and time. In case the procedure yields an ambiguous solution, additional physical criteria must be used to distinguish the real decay point from a mathematical artifact of the reconstruction. As the solution of the equation system was first elaborated for a similar case of reconstructing the $\Kl\to\pi^0\pi^0$ decay at KLOE, a detailed recipe can be found in Reference~\cite{gajos_mgr}. Alternatively, this system can also be solved with the method of Kleusberg known in GPS positioning~\cite{Kleusberg2003}. Although in this work the former procedure was implemented, it was compared with the Kleusberg algorithm  and validated to provide consistent results.

In case more than the 4 photons required for an analytical solution are recorded by the detector, the redundant information on additional 1 or 2 reference points can be utilized to minimize the uncertainty due to errors on determination of clusters' positions and times. Although these uncertainties will usually make the overdetermined system of 5--6 equations inconsistent, a numerical solution which minimizes the residual error of all the equations can enhance the resolution. A procedure for such a solution of the overdetermined system for the case of available information on all 6 photons is proposed in~\aref{appendix:numerical_6_equations}.
 
An important property of the reconstruction technique discussed above is that, unlike typical vertexing methods, it yields not only the spatial vertex position but also an independent estimate of the decay time (with respect to the same starting point as the measured $\gamma$ interaction times $T_i$). This, in turn, allows for imposing additional constraints on the reconstruction problem (e.g\ consistency of reconstructed kaon decay time with its time of flight expected from the travelled distance) in order to improve resolution of the decay properties as shown in~\sref{sec:kinfit}. Studies of the resolution of $\Kl\to 3\pi^0$ reconstruction at KLOE using the described technique are presented in~\sref{sec:kl3pi0}.

\section{Reconstruction of o-Ps$\to 3\gamma$ decay in J-PET}\label{sec:gps_jpet}
Whereas the reconstruction of $\Kl\to 3\pi^0$ decay point in a three-dimensional space presented in the previous Section is very close to the GPS positioning in its mathematical form, the \ops/$\to 3\gamma$ process requires a slightly modified approach. Since no more than three photons can be registered by a detector in this case, it is apparent from the previous considerations that the decay position cannot be obtained without any additional information. However, the lack of a fourth reference point can be compensated by the fact that all three $\gamma$ quanta originate from the same annihilation%
\footnote{It should be noted that is is not the case for $\Kl\to 3\pi^0 \to 6 \gamma$ where three $\gamma\gamma$ pairs are created in independent decays of $\pi^0$.}
and thus their momenta must be constrained to a single plane in the center-of-mass reference frame. As the momentum of the annihilating \ops/ atom is relatively small, it has been verified with MC simulations that the effect of its boost in the laboratory frame is negligible when resolutions at the $\order{1~\text{cm}}$ are considered~\cite{daria_epjc}. The coplanarity can therefore be assumed to hold also in the laboratory frame. \fref{subfig:a}~depicts an exemplary \ops/$\to 3\gamma$ annihilation event recorded by the \jpet/ detector where decay plane is marked with the shaded surface. It is clear that the decay plane is determined by the positions of the three points where $\gamma$ interactions had been recorded in the scintillator strips and that the unknown annihilation point must also belong to that plane. Therefore, the reconstruction problem may be reduced to a 2-dimensional case contained in the decay plane~(\fref{subfig:b}).

\begin{figure}[h!]
  \centering
  \sidecaption{An o-Ps atom annihilates inside the detector into three photons which are recorded by the scintillator strips. Paths travelled by the photons are marked with red, yellow and purple lines and all lie in a single plane denoted by the grey shaded surface.}{subfig:a}
  \raisebox{-0.7\height}{\hspace{1cm}\includegraphics[width=0.4\textwidth]{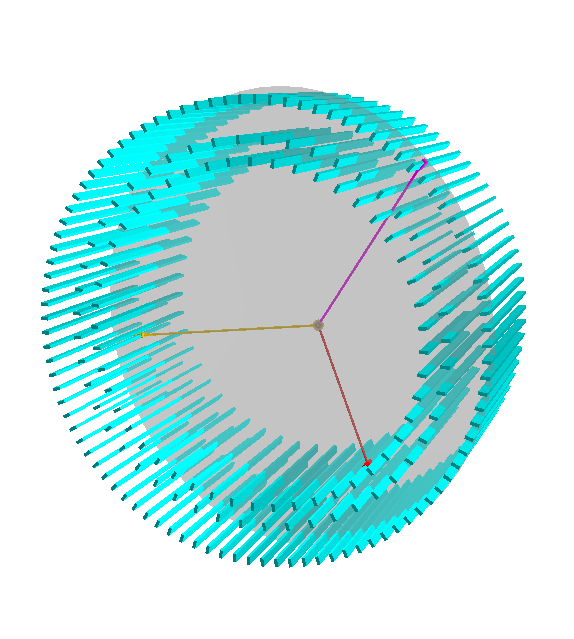}}
  \sidecaption{The reconstruction problem from (A) is transformed to a 2-dimensional problem contained in the decay plane. In this plane, the recording point of each photon is characterized by 2 Cartesian coordinates $X'_i$ and $Y'_i$ (where prime denotes 3D coordinates transformed to the decay plane) and recording time $T_i$}{subfig:b}
  \raisebox{-0.7\height}{\hspace{1cm}\includegraphics[width=0.4\textwidth]{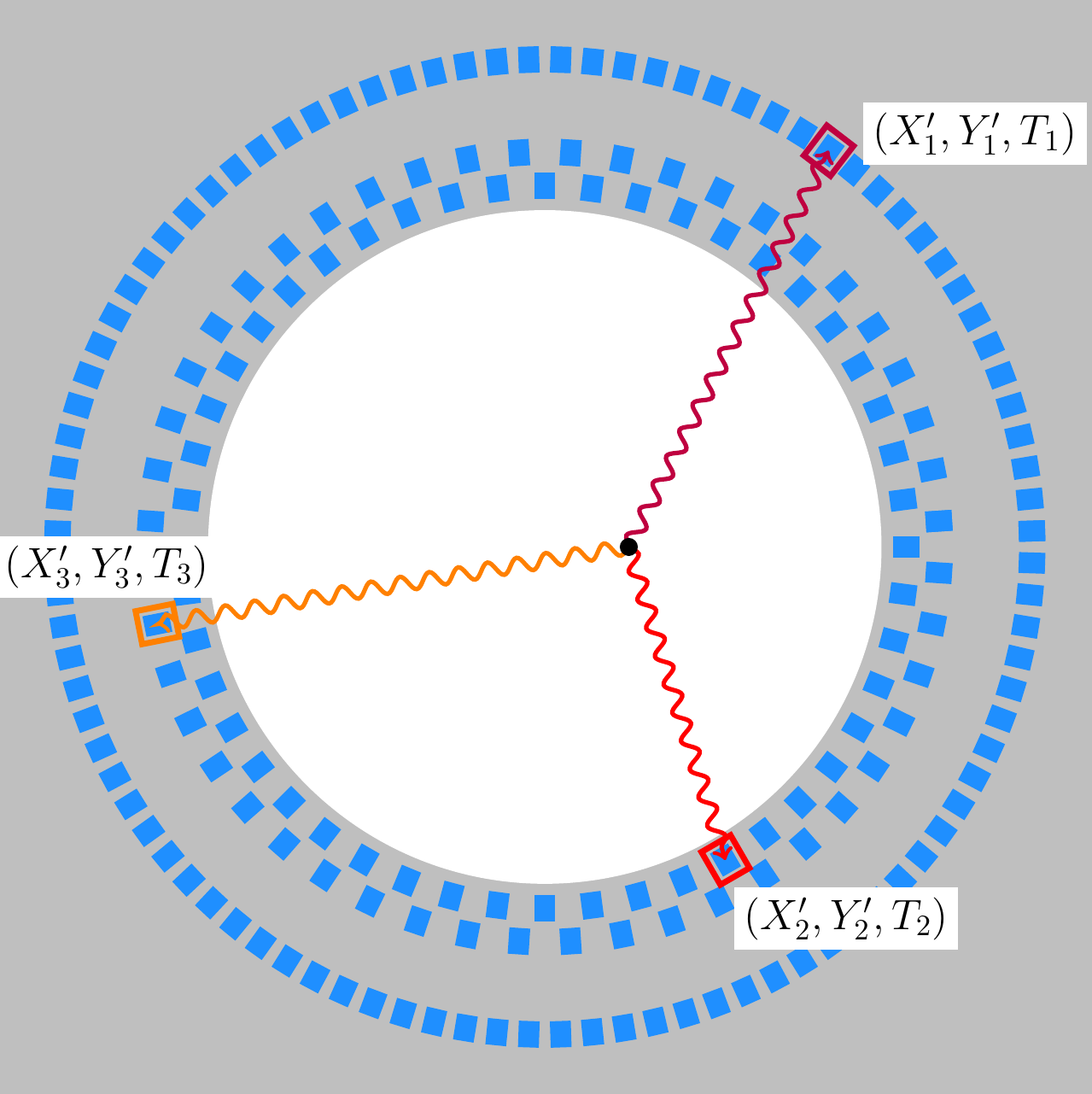}}
  \sidecaption{For each of the three recorded photon interactions, a set of possible photon creation points, which would allow for its interaction in time $T_i$ at the position $X'_i,Y'_i$, is constituted by a circle centered at $X'_i,Y'_i$ with a variable radius given by $R_i(t)=c(T_i-t)$ where $t$ is an unknown time of the o-Ps annihilation. The annihilation point $(x',y')$, being a common origin of the three photons, can be found as an intersection of such three circles. \ops/ annihilation time $t$ is obtained as a by-product of the reconstruction.}{subfig:c}
  \raisebox{-0.9\height}{\hspace{1cm}\includegraphics[width=0.4\textwidth]{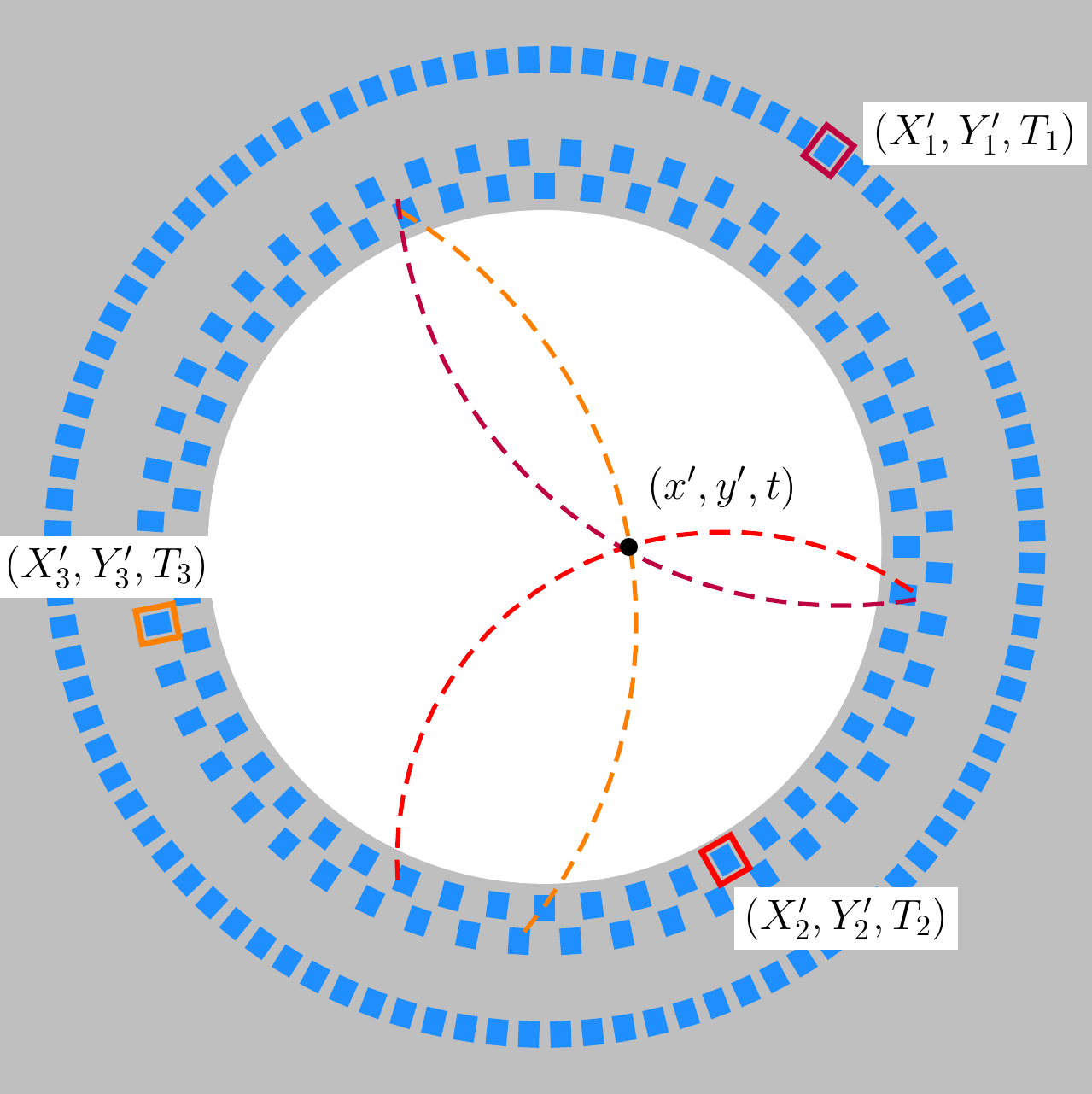}}
  \caption{Scheme of o-Ps$\to 3 \gamma$ decay reconstruction in J-PET.}\label{fig:trilateration_jpet}
\end{figure}

For each of the recorded photon interaction points, its spatial coordinates and time $(\!X_i,\!Y_i,\!Z_i,\!T_i\!)$ are determined as described in~\sref{sec:jpet}. If their locations are expressed with pointing vectors $\vec{r}_1,\vec{r}_2,\vec{r}_3$ with respect to origin of the \jpet/ coordinate system (\fref{fig:jpet_detector}, left), a normal to the decay plane is given by:
\begin{equation}
  \label{eq:jpet_rec_normal}
  \hat{n} = \frac{(\vec{r}_2 - \vec{r}_1) \times (\vec{r}_3 - \vec{r}_1)}{\abs{(\vec{r}_2 - \vec{r}_1) \times (\vec{r}_3 - \vec{r}_1)}}. 
\end{equation}
Before vertex reconstruction, the pointing vectors of $\gamma$ recording points are transformed in the following way:
\begin{equation}
  \label{eq:jpet_gps_transformation}
  \vec{r_i}' \equiv (X'_i,Y'_i,0)  = \mathbf{T}(\vec{v})\mathbf{R}(\angle (\hat{n},\hat{z}),\;\hat{n}\times\hat{z}) \vec{r}_i, \quad i=1,2,3,
\end{equation}
where $\hat{z}$ is the $z$ axis of the detector frame of reference, $\mathbf{R}(\alpha, \hat{k})$ denotes a rotation by an angle $\alpha$ around a $\hat{k}$ axis and $\mathbf{T}(\vec{v})$ is a translation by a vector required to set the decay plane after rotation to $z=0$ in the detector frame. As a result, the problem is reduced to a 2-dimensional space where for each $\gamma$ interaction point in the detector, the set of potential creation points of the photon is constituted by a circle centered in that point and with a radius parametrized by an unknown time of \ops/ decay $t$:
\begin{equation}
  \label{eq:jpet_gps_circles}
  (X'_i-x)^2 + (Y'_i-y)^2 = c^2(T_i-t)^2, \quad i=1,2,3.
\end{equation}

The \ops/$\to 3\gamma$ annihilation point can again be found as a common origin point of all three photons in an event, i.e.\ at an intersection of the variable-radius spheres shown in~\fref{subfig:c}. Similarly as in the 3-dimensional case, to solve such a system one more reference point is needed with respect to the problem of intersecting fixed-radii circles, so that all three photons must be detected to enable event reconstruction. The annihilation point on the decay plane $(x',y',0)$ is found analytically as a solution of the above equation system, which also yields the time $t$ of the decay alike the $\Kl\to 3\pi^0$ reconstruction. A comprehensive recipe for an analytical solution of the system from~\eref{eq:jpet_gps_circles} is presented in~\aref{appendix:jpet_solution}. %As shown therein, the decay point and time may be determined with a two-fold ambiguity. In such cases, additional criteria must be used to distinguish the physical solution from a reconstruction artifact.

The final reconstruction step comprises transformation of the solution from the decay plane back to the 3-dimensional frame of reference of the detector:
\begin{equation}
  \label{eq:jpet_gps_transformation_inverse}
  (x,y,z) = \mathbf{R}(-\angle (\hat{n},\hat{z}),\;\hat{n}\times\hat{z}) \mathbf{T}(-\vec{v}) (x',y',0).
\end{equation}

The reconstruction method presented above has been comprehensively tested with MC simulations in order to assess its resolution and applicability for the studies described in~\cref{chapter:test_jpet}~\cite{gajos_gps}. Moreover, its employment in a novel medical imaging technique is the subject of a patent application~\cite{imaging_patent}. Preliminary tests with the first recorded sample of annihilations into three photons have also been performed and their results are presented in~\cref{chapter:analysis_jpet}.

%%% Local Variables:
%%% TeX-master: "../main"
%%% End: 
\chapter{Realization of the direct T symmetry test at KLOE}\label{chapter:analysis_kloe}

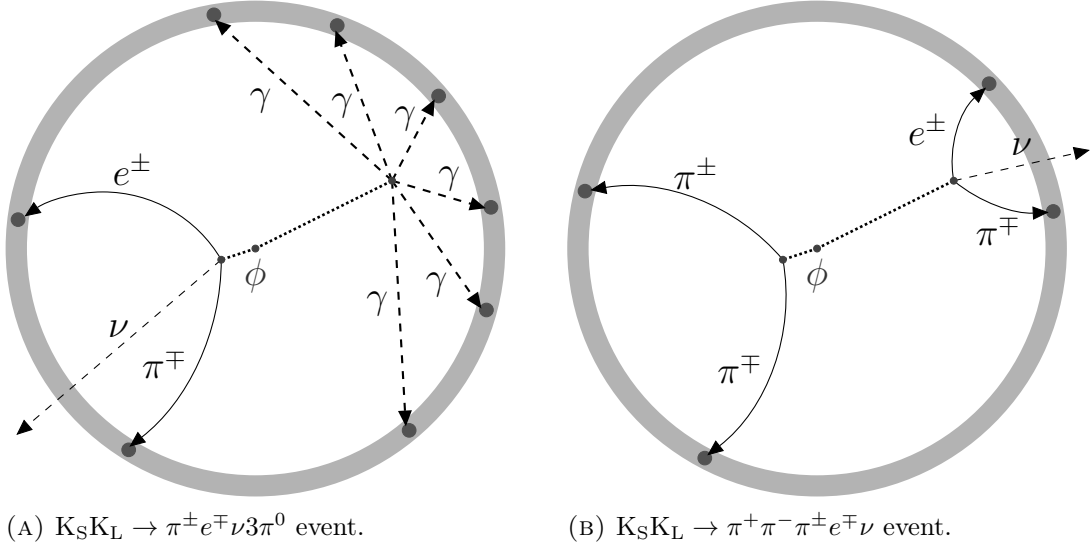
\begin{figure}[h!]
  \centering
  \begin{subfigure}{0.45\textwidth}
  \begin{tikzpicture}[scale=0.45,
  font={\fontsize{14pt}{12}\selectfont}
  ]
  \coordinate (kldec) at (4.0,2);
  \coordinate (ksdec) at (-1,-0.33);
  \draw[gray!60, line width=8] (0,0) circle (7.0);
  
  \draw[line width=1,densely dotted] (0,0) -- (kldec);% node[black,midway,above]{$\kaon$};    
  \draw[line width=1,densely dotted] (0,0) -- (ksdec);% node[black,midway,above]{$\kaon$};    

  % clusters
  \draw[black!85,fill] (100:7) circle (0.2);
  \draw[black!85,fill] (70:7) circle (0.2);
  \draw[black!85,fill] (40:7) circle (0.2);
  \draw[black!85,fill] (10:7) circle (0.2);
  \draw[black!85,fill] (-15:7) circle (0.2);
  \draw[black!85,fill] (-50:7) circle (0.2);

  \draw[black!85,fill] (173:7) circle (0.2);
  \draw[black!85,fill] (-122:7) circle (0.2);
  
  \draw[arrows={-{Triangle}[width=5,length=6]},thick, dashed] (kldec) -- (100:6.85) node[midway, left, xshift=-8] {$\gamma$};
  \draw[arrows={-{Triangle}[width=5,length=6]},thick, dashed] (kldec) -- (70:6.85) node[midway, left] {$\gamma$};
  \draw[arrows={-{Triangle}[width=5,length=6]},thick, dashed] (kldec) -- (40:6.85) node[midway, left, yshift=8,xshift=5] {$\gamma$};
  \draw[arrows={-{Triangle}[width=5,length=6]},thick, dashed] (kldec) -- (10:6.85) node[midway, left, yshift=6,xshift=12] {$\gamma$};
  \draw[arrows={-{Triangle}[width=5,length=6]},thick, dashed] (kldec) -- (-15:6.85) node[midway, left, yshift=-15,xshift=8] {$\gamma$};
  \draw[arrows={-{Triangle}[width=5,length=6]},thick, dashed] (kldec) -- (-50:6.85) node[midway, left] {$\gamma$};

  \draw[arrows={-{Triangle}[width=5,length=6]}] (ksdec) arc (30:125:4) node[midway,above]{$e^\pm$};
  \draw[arrows={-{Triangle}[width=5,length=6]}] (ksdec) arc (360:308:7) node[midway,left]{$\pi^\mp$};
  \draw[dashed, arrows={-{Triangle}[width=5,length=6]}] (ksdec) -- (-7, -5.5) node[midway, above]{$\nu$};

  % decay dots
  \draw[darkgray,fill] (0,0) circle (0.1) node[black,below]{$\phi$};
  \draw[darkgray,fill] (ksdec) circle (0.1);
  \draw[darkgray,fill] (kldec) circle (0.1);
\end{tikzpicture}        
\caption{$\Ks \Kl \to \pi^{\pm}e^{\mp}\nu 3\pi^0$ event.}\label{fig:event_schemes:a}
\end{subfigure}
\hspace{1em}
\begin{subfigure}{0.45\textwidth}
\begin{tikzpicture}[scale=0.45,
  font={\fontsize{14pt}{12}\selectfont}
  ]
  \coordinate (kldec) at (4.0,2);
  \coordinate (ksdec) at (-1,-0.33);
  \draw[gray!60, line width=8] (0,0) circle (7.0);

  \draw[line width=1,densely dotted] (0,0) -- (kldec);% node[black,midway,above]{$\kaon$};    
  \draw[line width=1,densely dotted] (0,0) -- (ksdec);% node[black,midway,above]{$\kaon$};    
  
  % ks decay
  \draw[black!85,fill] (166:7) circle (0.2);
  \draw[arrows={-{Triangle}[width=5,length=6]}] (ksdec) arc (40:100:6) node[midway,above]{$\pi^\pm$};a
  \draw[black!85,fill] (242:7) circle (0.2);
  \draw[arrows={-{Triangle}[width=5,length=6]}] (ksdec) arc (10:-52:6) node[midway,left]{$\pi^\mp$};

  % kl decay
  \draw[black!85,fill] (44:7) circle (0.2);
  \draw[arrows={-{Triangle}[width=5,length=6]}] (kldec) arc (190:132:3) node[midway,left]{$e^\pm$};a
  \draw[black!85,fill] (9:7) circle (0.2);
  \draw[arrows={-{Triangle}[width=5,length=6]}] (kldec) arc (230:273:4) node[midway,below]{$\pi^\mp$};
  \draw[dashed, arrows={-{Triangle}[width=5,length=6]}] (kldec) -- (20:8.5) node[midway, above]{$\nu$};  

  % decay dots
  \draw[darkgray,fill] (0,0) circle (0.1) node[black, below]{$\phi$};
  \draw[darkgray,fill] (ksdec) circle (0.1);
  \draw[darkgray,fill] (kldec) circle (0.1);
\end{tikzpicture}
\caption{$\Ks \Kl \to \pi^{+}\pi^{-} \pi^{\pm}e^{\mp}\nu$ event.}\label{fig:event_schemes:b}
\end{subfigure}
\caption{Schemes of the two processes studied in KLOE for the direct T symmetry test. Dotted lines represent paths travelled by neutral kaons before their decays and dashed lines denote photons and neutrinos, both invisible for the detector DC. Tracks of charged particles recorded by the DC are marked with solid lines. Black circles denote energy deposition clusters in the EMC barrel (grey band).}\label{fig:event_schemes}
\end{figure}

The direct test on the symmetry under reversal in time at KLOE described in~\cref{chapter:test_kloe} requires mutual comparison of time-dependent rates of four classes of processes~(listed in~\tref{tab:processes}), each characterized by specific decays of both neutral K mesons present in a $\phi\to\Ks\Kl$ event. In a practical data analysis at KLOE, it is reasonable to limit the study to using semileptonic decays with electrons and positrons rather than muons and in case of decays into two pions, to choose $\pi^+\pi^-$ rather than $\pi^0\pi^0$ as the two-pion state due to good reconstruction of the former in the drift chamber. Moreover, it can be noted that the processes indicated in~\tref{tab:processes} comprise two general classes, each of which is additionally split into two categories by charge of the lepton in the semileptonic decay. These two classes, presented schematically in~\fref{fig:event_schemes}, are:

\begin{enumerate}
\item events with an earlier kaon decay into a semileptonic final state $\pi^-e^+\nu_e$ or $\pi^+e^-\bar{\nu}_e$, characterized by two tracks recorded in the DC and with an unregistered neutrino, and a later kaon decay into $3\pi^0$ resulting in up to 6 points of $\gamma$ quanta interaction in the EMC~(\fref{fig:event_schemes:a}),
\item events with an earlier neutral kaon decay into two charged pions, characterized by two DC tracks originating close to the $\phi$ decay point, and a later semileptonic decay into $\pi^-e^+\nu_e$ or $\pi^+e^-\bar{\nu}_e$ recorded as two tracks with a common origin anywhere in the drift chamber volume, where the neutrino escapes detection~(\fref{fig:event_schemes:b}).
\end{enumerate}

Although these sequences of final states can in principle originate from any order of $\Ks$ and $\Kl$ decays, due to the suppression of \CPs-violating decay modes $\Ks\to 3\pi^0$%
\footnote{The $\Ks\to 3\pi^0$ decay has, in fact, never been observed~\cite{Babusci:2013tr}.}
and $\Kl\to\pi^+\pi^-$~(see~\tref{tab:kaon_properties}) the earlier decays are expected to come from short-lived kaons in the regime of large time differences between both kaon decays which is under study in this work (see~\sref{sec:strategies}). As a result, the final state $\pi^{\mp}e^{\pm}\nu$ in the first class of processes and $\pi^+\pi^-$ in the second one originate close to the $\phi$ decay point defined by $e^+e^-$ collisions, with a distance corresponding to several times the mean free path of $\Ks$ at KLOE, $\lambda_S\approx 6$~mm. On the other hand, the second decay in each class may originate anywhere in the detector volume.

In the analysis presented in this work, the data collected by the KLOE experiment in the years 2004-5 were explored, with a total integrated luminosity of about~1.7~fb$^{-1}$. Performance of each of the analysis steps was evaluated using a sample of Monte Carlo (MC) simulated events corresponding to the conditions and integrated luminosity of the experimental data and including simulation of the full detector response~\cite{data_handling}. As mentioned in~\sref{sec:streaming}, the data of KLOE is pre-sorted into several streams and in this work the selection of relevant event classes was started from the 
% MEMO TODO: napisac explicite ze KSL
neutral kaon stream. The same streaming selection was applied to the data and MC samples used.

This Chapter presents details of event selection and reconstruction for the two aforementioned classes of processes, respectively in Sections~\ref{sec:t-analysis-1} and~\ref{sec:t-analysis-2}.
The direct \Ts~symmetry test imposes two crucial requirements on its experimental realization. Firstly, the samples of events of interest must be selected from all processes recorded by KLOE with a high purity as any background contamination will have a direct impact on the sensitivity of the test. From the branching ratios contained in~\tref{tab:kaon_properties} is is apparent that achieving the desired purity is the most challenging task in case of the $\Ks\to \pi^{\mp}e^{\pm}\nu$ decay as its probability is only at the level of $10^{-4}$. Therefore its selection comprises, in addition to a set of cuts based on the signal properties~(\sref{sec:ksemil}), also two steps dedicated to removing particular background components described in Sections~\ref{sec:ks2pi0_rejection} and~\ref{sec:pimu_rejection}. The second requirement of this analysis is that the time differences between two kaon decays, and thus also the decay times, must be reconstructed with a precision sufficient to study the double decay rates as functions of $\Delta t$~(see~\sref{sec:observables}). Whereas for all the involved decays which include charged particles, the decay times are easily obtained with an accuracy at the level of~1~$\tau_S$, the $\Kl\to 3\pi^0$ process requires special treatment to this end. With its reconstruction limited by the resolution of the calorimeter (inferior to the DC vertexing), a special kinematic fit is introduced to enhance time resolution for this decay (\sref{sec:kinfit}).

\section{Selection and reconstruction of the $\Ks \Kl \to \pi^{\pm}e^{\mp}\nu 3\pi^0$ process}\label{sec:t-analysis-1}

\subsection{Preselection of events}\label{sec:preselection-t1}
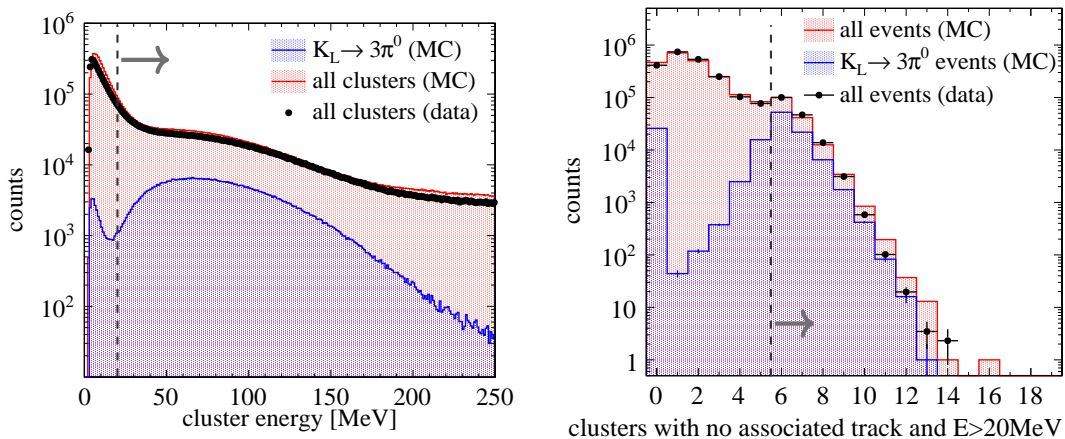
\begin{figure}[h!]
  \centering
  \begin{tikzpicture}
    \node[anchor=south west,inner sep=0] at (0,0) 
    {\includegraphics[width=0.45\textwidth]{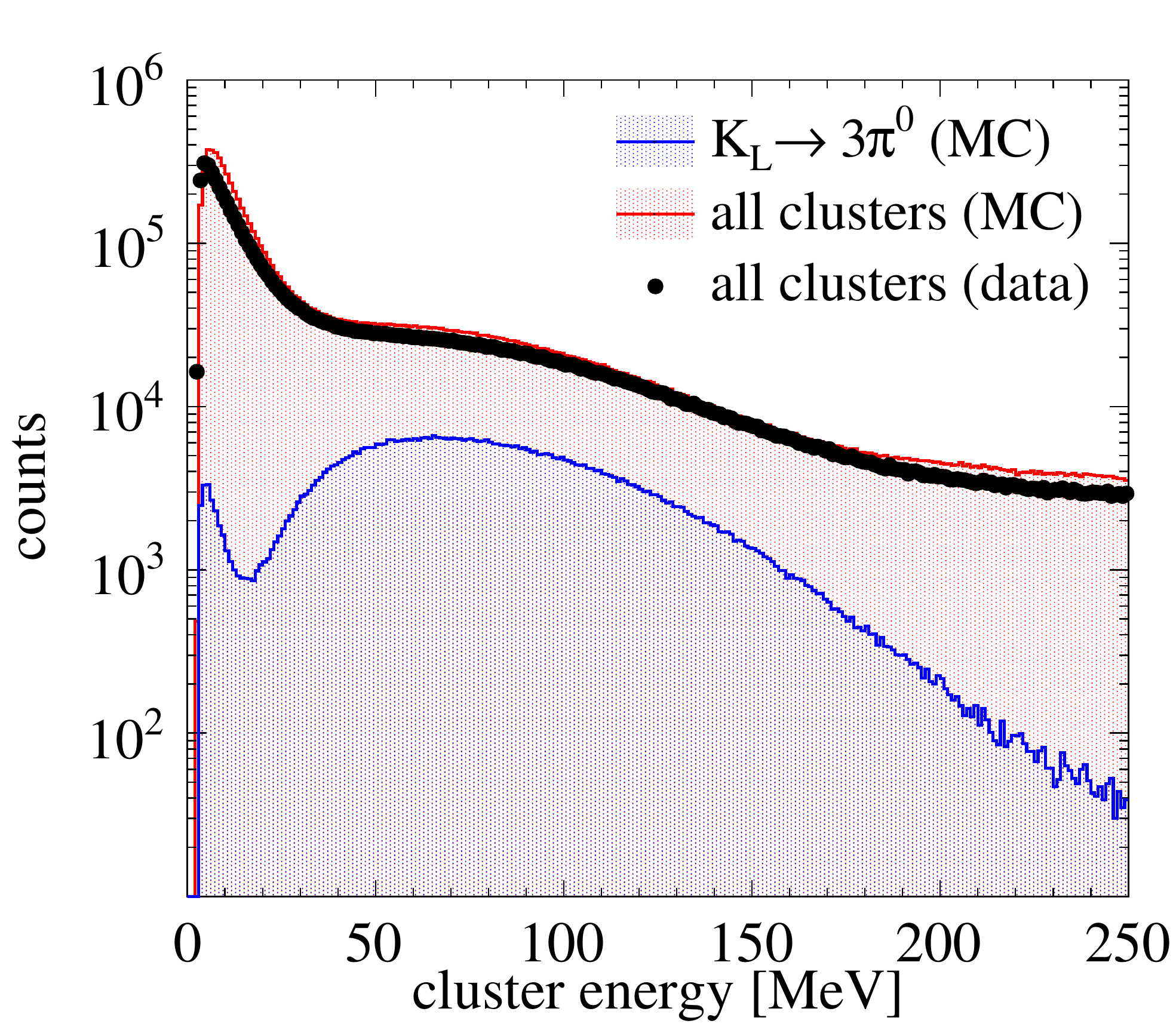}};
    \draw[black, thick, dashed] (1.52,0.8) -- (1.52,5.46);
    \draw[ultra thick, black!70!white, ->] (1.57, 5) -- (2.2, 5);
  \end{tikzpicture}
  \hspace{1em}
  \begin{tikzpicture}
    \node[anchor=south west,inner sep=0] at (0,0) 
    {\includegraphics[width=0.45\textwidth]{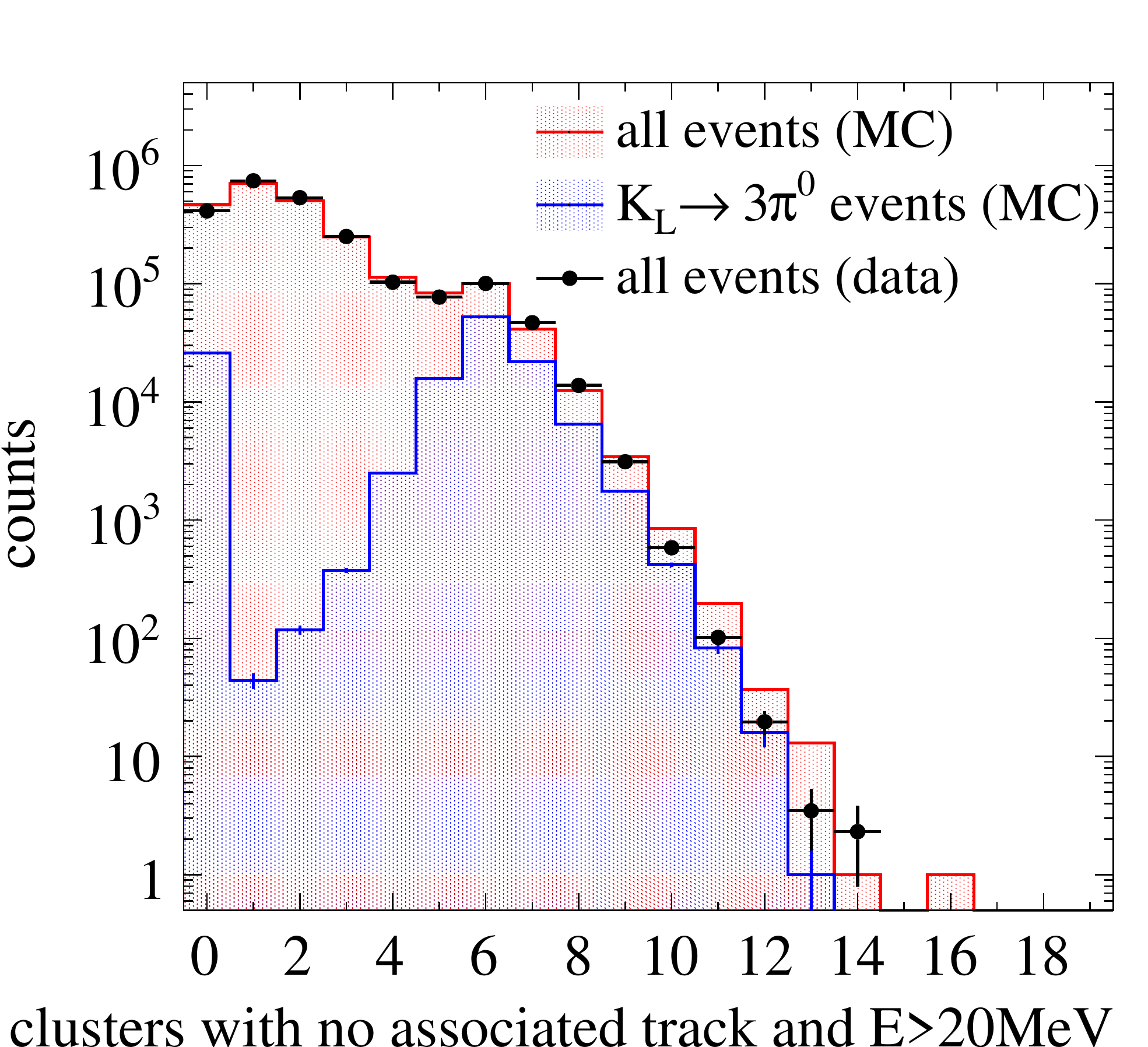}};
    % \draw[very thin,step=0.2] (0,0) grid +(5.5,5.5);
    % \draw[very thin,step=0.2] (0,0) grid +(5.5,5.5);
    \draw[black, thick, dashed] (2.73,0.8) -- (2.73,5.46);
    \draw[ultra thick, black!70!white, ->] (2.78, 1.5) -- (3.3, 1.5);
  \end{tikzpicture}
  \caption{Left: energy distributions of clusters not associated to DC tracks. Dashed line denotes minimum required energy.
    %for all clusters recorded in any events in the neutral kaon stream (red) and for clusters originating from $\Kl\to 3\pi^0 \to 6 \gamma$ (blue) as well as all data (black points).
    Right: numbers of $E>20$~MeV clusters not associated to DC tracks in a single event. Events with at least 6 such clusters are accepted as marked with the dashed line and gray arrow.
    % present in all MC events (red) and MC events containing a $\Kl\to 3\pi^0$ decay (blue) as well as data (black points).
  }
  \label{fig:t1-early-cuts}
\end{figure}

In order to properly reconstruct the decay time $\Delta t$ in a $\Ks \Kl \to \pi^{\pm}e^{\mp}\nu 3\pi^0$ event, decay vertices of both kaons must be available. Therefore, the first requirements imposed on events from the neutral kaon stream comprise (compare~\fref{fig:event_schemes:a}):
\begin{itemize}
\item presence of a vertex constituted by two approaching DC tracks in a cylindrical volume close to the $e^+e^-$ collision point limited by~$r_T=\sqrt{x^2+y^2}<$15~cm and $|z|<$10~cm,
% MEMO TODO: dopisac, ze wymagany qualv
\item presence of at least~6 EMC clusters not associated to DC tracks and with deposited energies $E>20$ MeV.
\end{itemize}
The volume for the 2-track vertex location is broad enough to contain almost all $\Ks$ decays~\cite{daria_memo, daria_article}. For the second requirement, track-to-cluster associations are determined by identifying sets of tracks corresponding to decay chains and extrapolating last track in each chain to the calorimeter inner surface. An EMC cluster is considered as associated to a track if distance between its centroid and the point of track incidence on EMC, measured in a plane perpendicular to the direction of incidence, is closer than 30~cm~\cite{data_handling}. Clusters associated to DC tracks constitute candidates for interaction points of charged particles which reached the EMC (e.g.\ all the clusters in~\fref{fig:event_schemes:b}), therefore only those without track association are considered for the identification of $\Kl\to 3\pi^0\to 6\gamma$. \fref{fig:t1-early-cuts} (left) shows the distribution of energy for all clusters with no associated tracks and for those originating from $\Kl\to 3\pi^0$ (signal) according to MC simulations. The dip below 20~MeV in the signal spectrum corresponds to the EMC threshold for efficient registration of $\gamma$~quanta while the peak for lower energies arises from effects such as cluster splitting in which one particle interaction is recorded by the calorimeter as several clusters.
% TODO: cytowac splitting
% TODO: sprawdzic czy to prawda
A cut on $E>$20~MeV thus allows to discriminate a large part of background clusters as well as reject poorly reconstructed events.

The distribution of the number of clusters with sufficient energy in a single event is presented in the right panel of~\fref{fig:t1-early-cuts}. Although for MC-simulated $K_L\to 3\pi^0$ events the distribution peaks at 6 in accordance with intuition, there is a considerable probability of recording only 5 photons, as well as of presence of additional clusters arising either from background from cluster splitting. The peak at 0 corresponds to cases where $\Kl$ escaped from the EMC without interacting. Even though events with 4 and 5 clusters only could in principle be used in the analysis, their acceptance would entail a significant increase in the amount of background and reduce the possibilities of improving vertex reconstruction with additional reference points described in~\aref{appendix:numerical_6_equations}. On the other hand, inclusion of events with 7 and more clusters considerably increases the statistics without hindering the reconstruction as the proper set of clusters from $\Kl\to 3\pi^0$ can be identified with the procedure shown in the next Section.

\subsection{Selection and reconstruction of $\Kl \to 3\pi^{0}$ decays}\label{sec:kl3pi0}
As there may be more than 6 EMC clusters satisfying the preselection criteria (later on referred to as candidate clusters), identification of $K_L\to 3\pi^0$ (signal) events comprises two tasks, accomplished simultaneously with the event selection process:
\begin{itemize}
\item recognition of signal events and rejection of background,
\item for signal events containing 7 and more candidate clusters, selection of the correct 6-element subset of clusters originating from the 3$\pi^0\to 6\gamma$, which is then used for decay reconstruction.
\end{itemize}

\begin{figure}[ht!]
  \centering
  \captionsetup[sub]{margin=1ex}
  % SUB 1
  \begin{subfigure}{0.45\textwidth}
    \begin{tikzpicture}
      \node[anchor=south west,inner sep=0] at (0,0) 
      {\includegraphics[width=1.0\textwidth]{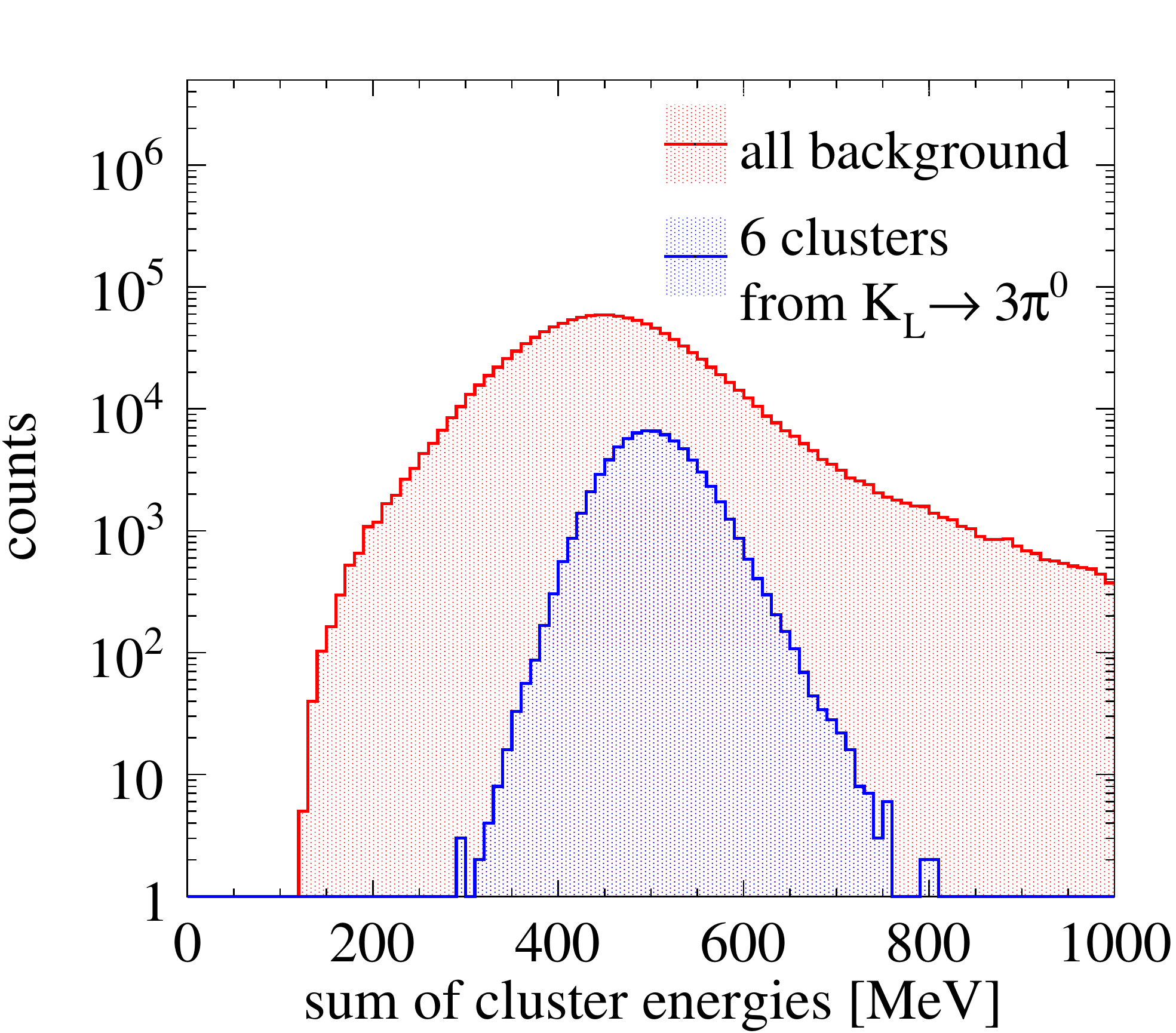}};
      \draw[black, thick, dashed] (2.95,0.8) -- (2.95,5.46);
      \draw[black, thick, dashed] (4.81,0.8) -- (4.81,4.0);
    \draw[ultra thick, black!70!white, <->] (2.95, 1.5) -- (4.81, 1.5);
    \end{tikzpicture}
    \caption{Total energy of 6 clusters for a correct 6-element cluster set in signal events (blue) and for background (red, contains both physical and combinatorial background).}\label{fig:kl3pi0selection:a}
  \end{subfigure}
  % SUB 2
    \begin{subfigure}{0.45\textwidth}
    \begin{tikzpicture}
      \node[anchor=south west,inner sep=0] at (0,0) 
      {\includegraphics[width=1.0\textwidth]{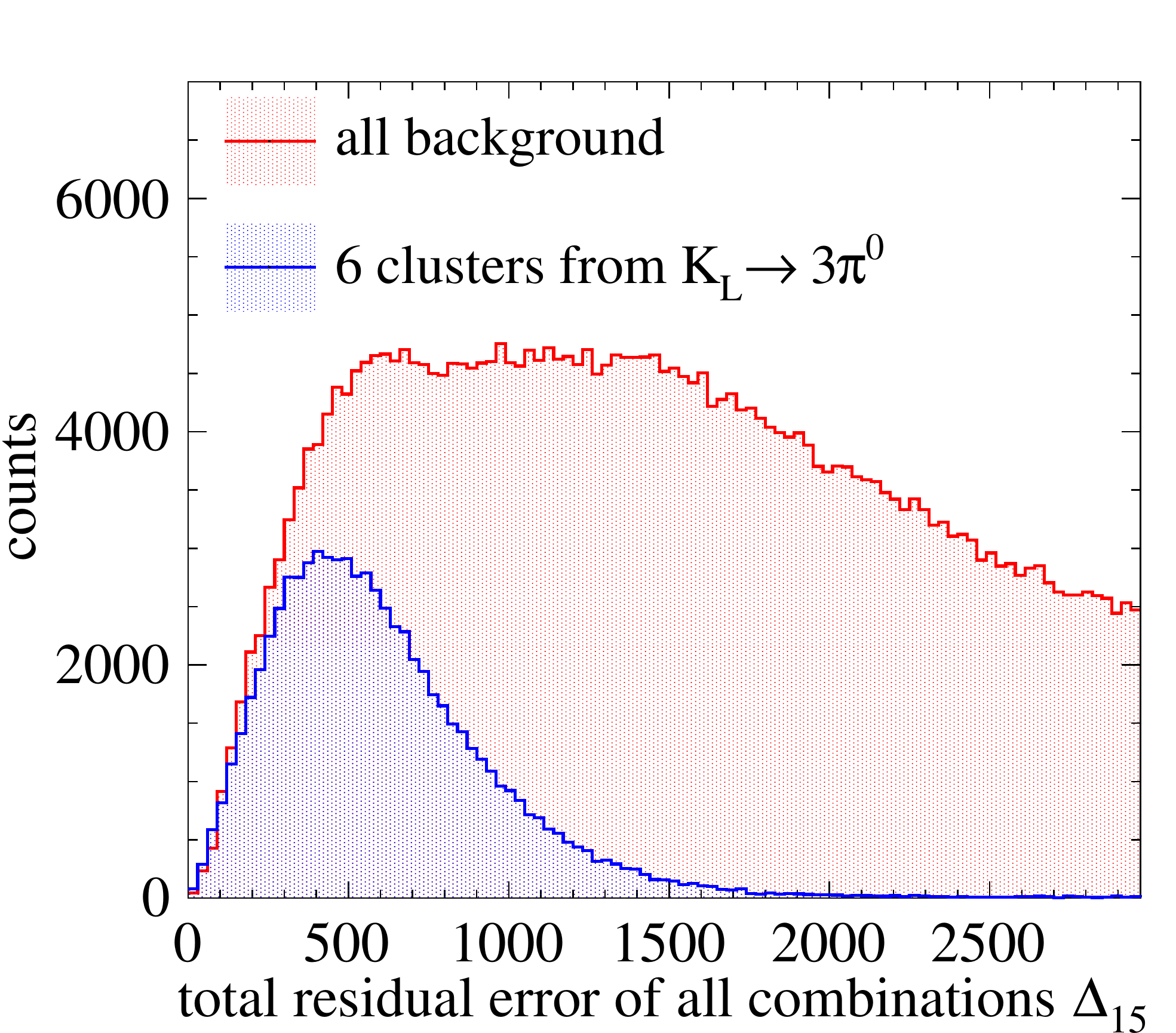}};
      \draw[black, thick, dashed] (4.8,0.8) -- (4.8,4.2);
    \draw[ultra thick, black!70!white, <-] (4.2, 1.5) -- (4.72, 1.5);
    \end{tikzpicture}
    \caption{Distribution of $\Delta_{15}$ (see~\eref{eq:delta_15}), a measure of consistency of a 6-cluster set with the $\Kl \to 3\pi^0\to 6\gamma$ hypothesis. Background plot (red) contains both physical and combinatorial background.}\label{fig:kl3pi0selection:b}
  \end{subfigure}
  % SUB 3
    \begin{subfigure}{0.45\textwidth}
    \begin{tikzpicture}
      \node[anchor=south west,inner sep=0] at (0,0) 
      {\includegraphics[width=1.0\textwidth]{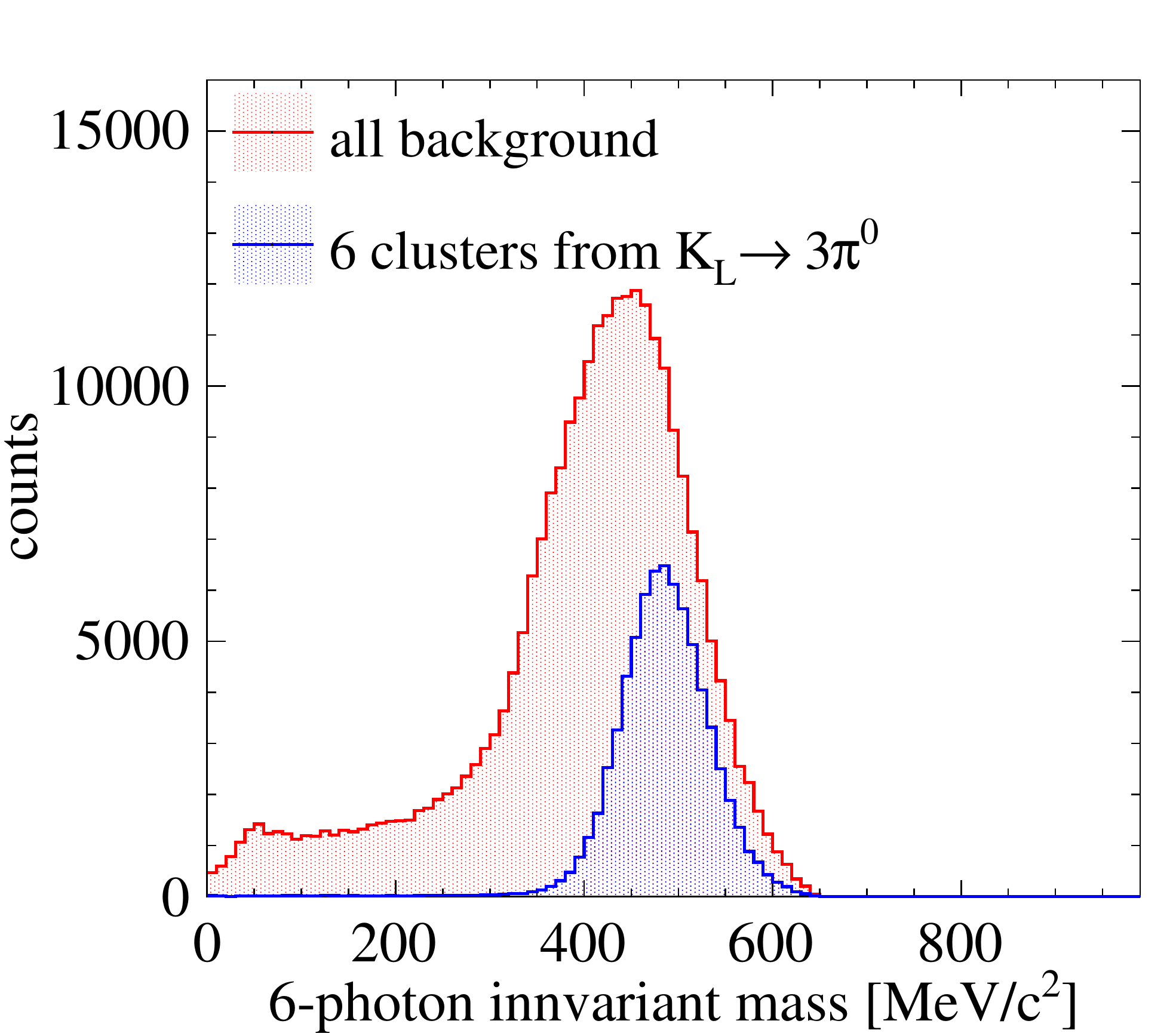}};
      \draw[black, thick, dashed] (3.1,0.8) -- (3.1,4.2);
      \draw[ultra thick, black!70!white, ->] (3.14, 2.2) -- (3.5, 2.2);
    \end{tikzpicture}
    \caption{Invariant mass of original $\Kl$ reconstructed using 6 hypothetical photons associated to the clusters. Background plot (red) contains both physical and combinatorial background.}\label{fig:kl3pi0selection:c}
  \end{subfigure}
  % SUB 4
    \begin{subfigure}{0.45\textwidth}
    \begin{tikzpicture}
      \node[anchor=south west,inner sep=0] at (0,0) 
      {\includegraphics[width=1.0\textwidth]{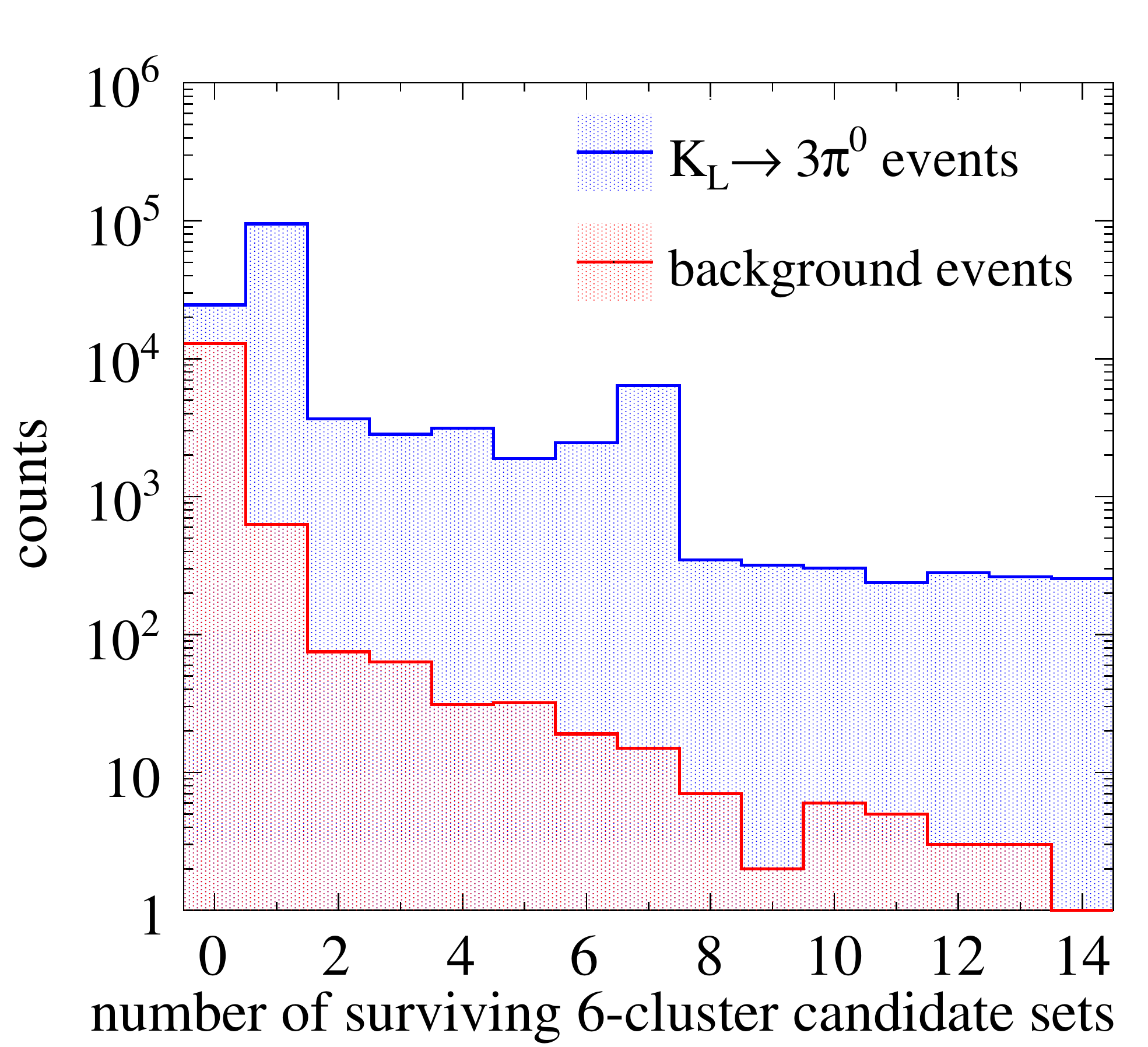}};
      \draw[ultra thick, black!70!white, ->] (1.49, 5.4) -- (1.84, 5.4);
      \draw[black, thick, dashed] (1.45,0.8) -- (1.45,5.46);
    \end{tikzpicture}
    \caption{Number of 6-cluster candidate sets passing the criteria depicted with dashed lines in panels A--C. Events with at least 1 surviving set are retained for further analysis.}\label{fig:kl3pi0selection:d}
  \end{subfigure}
  \caption{MC-based distributions of the variables used for selection of 6-cluster sets among all combinatorial possibilities in an event. Dashed vertical lines denote the cuts used and gray arrows indicate the accepted distribution regins.}
  \label{fig:kl3pi0selection}
\end{figure}
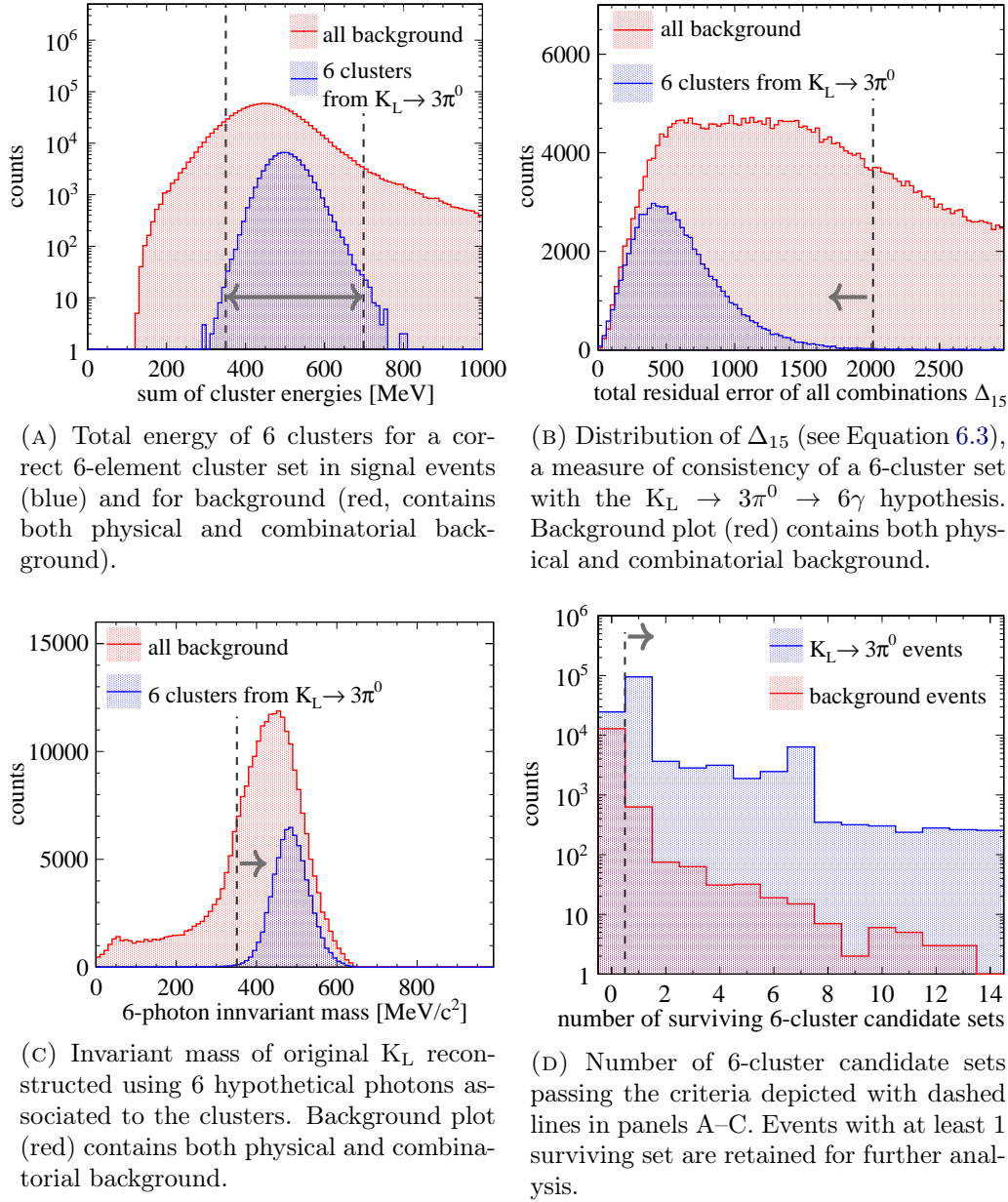

To this end, all combinatorically possible choices of 6 elements from all candidate clusters present in an event are considered and a set of criteria is subsequently applied to each combination. The first criterion is based on a simple sum of energies of 6 clusters:
\begin{equation}
  350\:\text{MeV} <  \sum_{i=1}^6 E_i < 700\:\text{MeV},
\end{equation}
which, as tested with MC simulations, is well confined to a region around the neutral kaon mass in case of 6-cluster sets from $\Kl\to 3\pi^0$ as shown in~\fref{fig:kl3pi0selection:a}.
% It is important to note that in the 
In the Figures~\ref{fig:kl3pi0selection:a}--\ref{fig:kl3pi0selection:c}, the background spectra account for two types of events:
\begin{itemize}
\item physical background, i.e.\ processes with a decay other than $\Kl\to 3\pi^0$ (one count per event),
\item combinatorial background arising from all possible wrong choices of 6-cluster sets in an event (${N\choose 6} - 1$ counts per event with $N$ candidate clusters).
\end{itemize}

The second criterion applied to 6-cluster sets is based on the consistency of decay properties reconstructed using particular 4-cluster subsets. In the trilateration-based method described in~\sref{sec:gps_kloe} finding an analytical solution only requires 4 reference points, therefore all possible ${6\choose 4}=15$ choices of 4-cluster subsets are subsequently used to reconstruct the $\Kl\to 3\pi^0$ decay point and time. Each time, the obtained solution is inserted into the two equations left out (labeled $j$ and $k$) and their residual error is calculated as~(compare~\eref{eq:gps_6_eqns}):
\begin{equation}
  \label{eq:residual_error}
  \begin{split}
  r_{jk} & =  \left((X_j-x_{jk})^2 + (Y_j-y_{jk})^2 + (Z_j-z_{jk})^2 - c^2(T_j-t_{jk})^2\right)\\
  & + \left( (X_k-x_{jk})^2 + (Y_k-y_{jk})^2 + (Z_k-z_{jk})^2 - c^2(T_k-t_{jk})^2   \right).
  \end{split}
\end{equation}
where $(x_{jk},y_{jk},z_{jk},t_{jk})$ is the reconstruction result obtained by leaving out equations $j$ and $k$. For a set of 6 clusters all created by $\gamma$ quanta from the same decay, the residuals of the remaining equations should be small independently of the choice of 4 clusters used to find the decay point and time. However, if one or more clusters is not related to the decay, certain combinations will yield solutions inconsistent with the unused reference points. Therefore, the sum of residual errors for unused equations computed for all 15 possible 4-cluster set choices:
\begin{equation}
  \Delta_{15} = \sum_{j=1}^{6}\sum_{k=j+1}^{6} r_{jk},
  \label{eq:delta_15}
\end{equation}
constitutes a measure of a 6-cluster set consistency with a $\Kl \to 3\pi^0\to 6\gamma$ hypothesis useful for discrimination of wrong combinations as well as non-$3\pi^0$ background as shown in~\fref{fig:kl3pi0selection:b}. Combinations with $\Delta_{15}>2000$ are rejected.

Finally, reconstruction of invariant mass of the original neutral K meson is attempted using momenta of the supposed photons associated with the 6 clusters in the set under consideration. The $\gamma$ momenta are estimated using directions from the reconstructed $\Kl\to 3\pi^0$ decay point to the clusters' locations and the clusters' energies. \fref{fig:kl3pi0selection:c} presents the resulting 6-photon invariant mass distributions, where 6-cluster combinations are rejected if $M_{6\gamma} < 350\:\text{MeV/c}^2$.

The number of 6-cluster combinations which satisfy all of the above criteria is shown in~\fref{fig:kl3pi0selection:d}. Events where no candidate cluster sets survive are rejected. The ``plateau'' region extending between 1 and 7 sets arises from the relatively high probability of total 7 candidate clusters in an event (compare~\fref{fig:t1-early-cuts}, Right) as several consistent 6-cluster sets can be found e.g.\ due to splitting of a cluster originating from a signal event. In order to maximize selection efficiency, events with more than 1 candidate cluster set are retained for the analysis. In such cases, the set is chosen which minimizes the $\Delta_{15}$ consistency measure.

\begin{figure}[h!]
  \centering
  \includegraphics[width=0.5\textwidth]{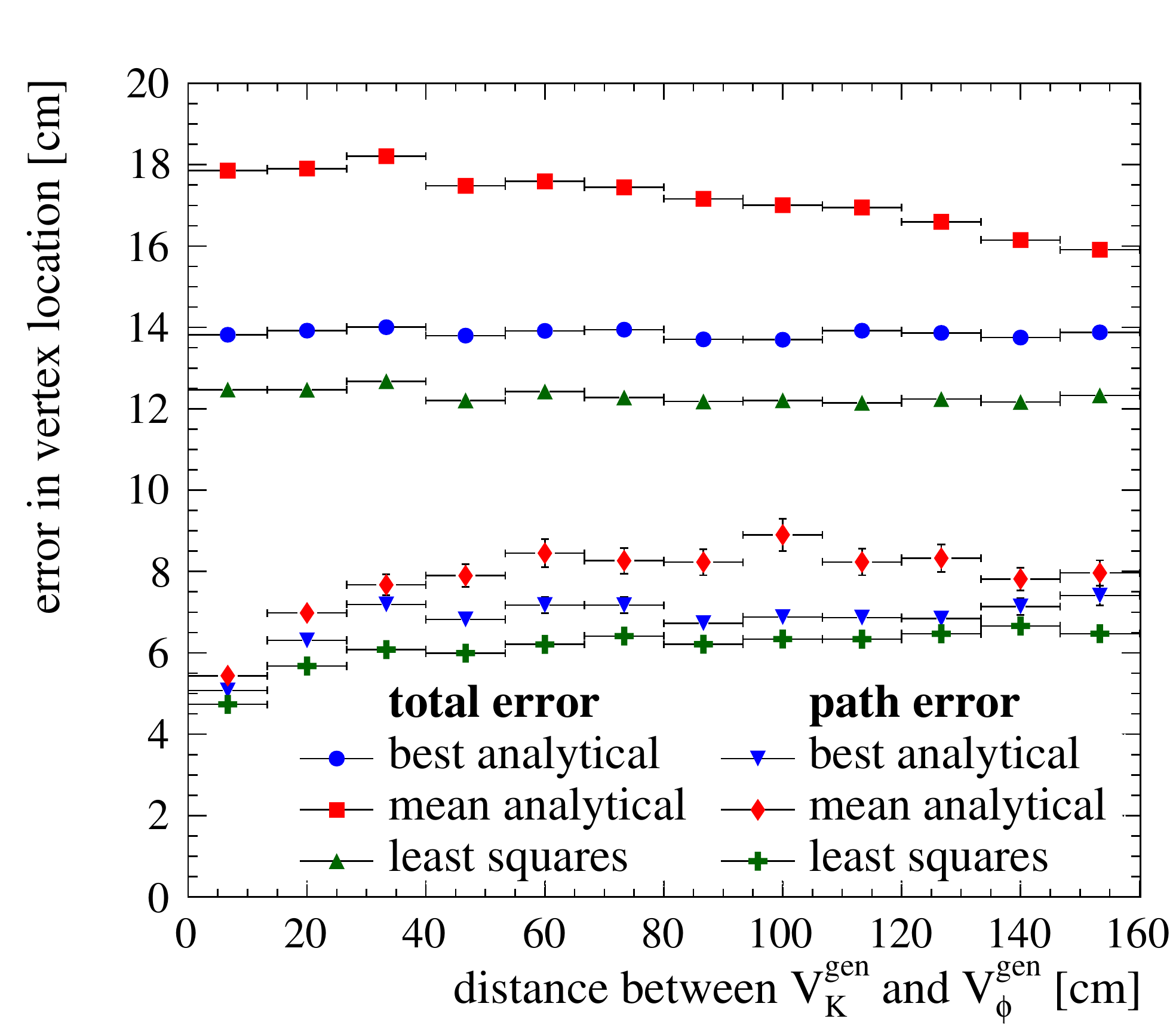}
  \caption{MC-based study of the resolution of $\Kl\to 3\pi^0$ decay vertex reconstruction expressed as total error on reconstructed w.r.t.\ MC-generated location of the decay point, as well as error on determination of the kaon travelled path, for subsequent ranges of decay distance from $\phi$. Three different reconstruction methods are presented.}\label{fig:resolutions_nofit}
\end{figure}

% TODO: dopisac, dlaczego rozdzielczosc w funkcji odleglosci of phi
Once an event is identified as $\Kl\to 3\pi^0$ by the presence of at least one candidate set of 6 clusters, reconstruction of the decay point and time using the trilateration technique can be attempted in several manners:
\begin{itemize}
\item using a mean solution obtained from all ${6\choose 4}$ possible choices of a subset of 4 out of 6 clusters (referred to as \textit{mean analytical} in~\fref{fig:resolutions_nofit}),
\item using the subset of 4 equations for which the residual error (see~\eref{eq:residual_error}) was smallest (\textit{best analytical} in~\fref{fig:resolutions_nofit}); this has the advantage of being able to yield a correct solution even if the chosen 6-cluster set is contaminated,
\item solving the system for all 6 reference points with iterative least squares method to minimize measurement uncertainties.
\end{itemize}

The resolution available with each of these approaches was studied using a sample of MC-simulated $\Kl\to 3 \pi^0$ events for which the choice of 6 clusters used for decay reconstruction was done as described above. As the performance of reconstruction may change with decay vertex location in the detector, resolution was studied as the distance between MC-generated and reconstructed decay point for multiple regions of decay distance from the $\phi$ decay. In addition to the total error in determination of the decay vertex position, the error on path travelled by the kaon was studied as being relevant to the calculation of decay time, needed for the \Ts~symmetry test. \fref{fig:resolutions_nofit} shows results of the resolution studies, from which it is evident that the numerical (least squares) solution using all 6 clusters as reference points allows for the most accurate determination of the $\Kl$ decay point. However, the error on the kaon travelled path at the 6~cm level still corresponds to about 1~ns uncertainty in the $\Kl$ decay time whereas the desired temporal resolution for the \Ts~test should be $\order{1\:\tau_S}\approx\order{0.1\:\text{ns}}$. Therefore, at a later analysis stage the decay time resolution is further enhanced with a dedicated kinematic fit described in~\sref{sec:kinfit}.
%
% TODO: wydajności oraz % identyfikacji dobrego zestawu klastrów
%

\subsection{Estimation of $\Ks$ momentum}\label{sec:ks_from_kl}
The selection of semileptonic decays of $\Ks$ presented in the next Sections relies partially on an estimate of the momentum vector of this kaon. However, momentum of the escaping neutrino prevents a precise reconstruction of $\vec{p_{K_S}}$ using a sum of the recorded products' momenta. Alternatively, the $\Ks$ momentum can be inferred from momentum conservation for the $\phi\to\Ks\Kl$ decay using the associated long-lived neutral kaon. $\phi\to\Ks\Kl$, as a two-body decay into particles of identical mass, produces both kaons with equal energies and opposite momenta in the center-of-mass frame of reference:
\begin{equation}
  \label{eq:momentum_tagging}
  p_{K}^{CM} = \sqrt{\frac{1}{4}m_{\phi}^2-m_{K_0}^2}.
\end{equation}
which moves with respect to the laboratory system with the momentum of initial $\phi$~meson, defined by properties of colliding electron and positron beams (see~\sref{sec:dafne}). The Lorentz transformation required to know the momenta of $\Ks$ and $\Kl$ in the laboratory frame only acts on the their components parallel to $\vec{p_{\phi}}$. Therefore, the complete kinematical configuration of a $\phi\to\Ks\Kl$ event, including moduli and momenta of both neutral kaons, depends only on the angle of their emission with respect to the $\phi$ momentum. Consequently, to determine $\vec{p}_{K_S}$ and $\vec{p}_{K_L}$, it is sufficient to know the direction of momentum of one of the kaons~\cite{daria_memo}. This can be estimated in a number of ways in the $\Ks \Kl \to \pi^{\pm}e^{\mp}\nu\:3\pi^0$ events:
\begin{enumerate}
\item using the vector spanning between the $\phi$ decay point and $\Ks\to\pi e\nu$ vertex obtained from DC tracks as direction of the $\Ks$ momentum,
\item using the vector spanning between the $\phi$ decay point and $\Kl\to 3\pi^{0}$ decay point obtained with the trilaterative reconstruction as direction of the $\Kl$ momentum,
\item using the $\Kl$ momentum obtained from a sum of the 6 photons' momenta calculated from their corresponding $\Kl$ vertex--EMC cluster directions and energies deposited in the clusters.
\end{enumerate}

\begin{figure}[h!]
  \centering
  \includegraphics[width=0.45\textwidth]{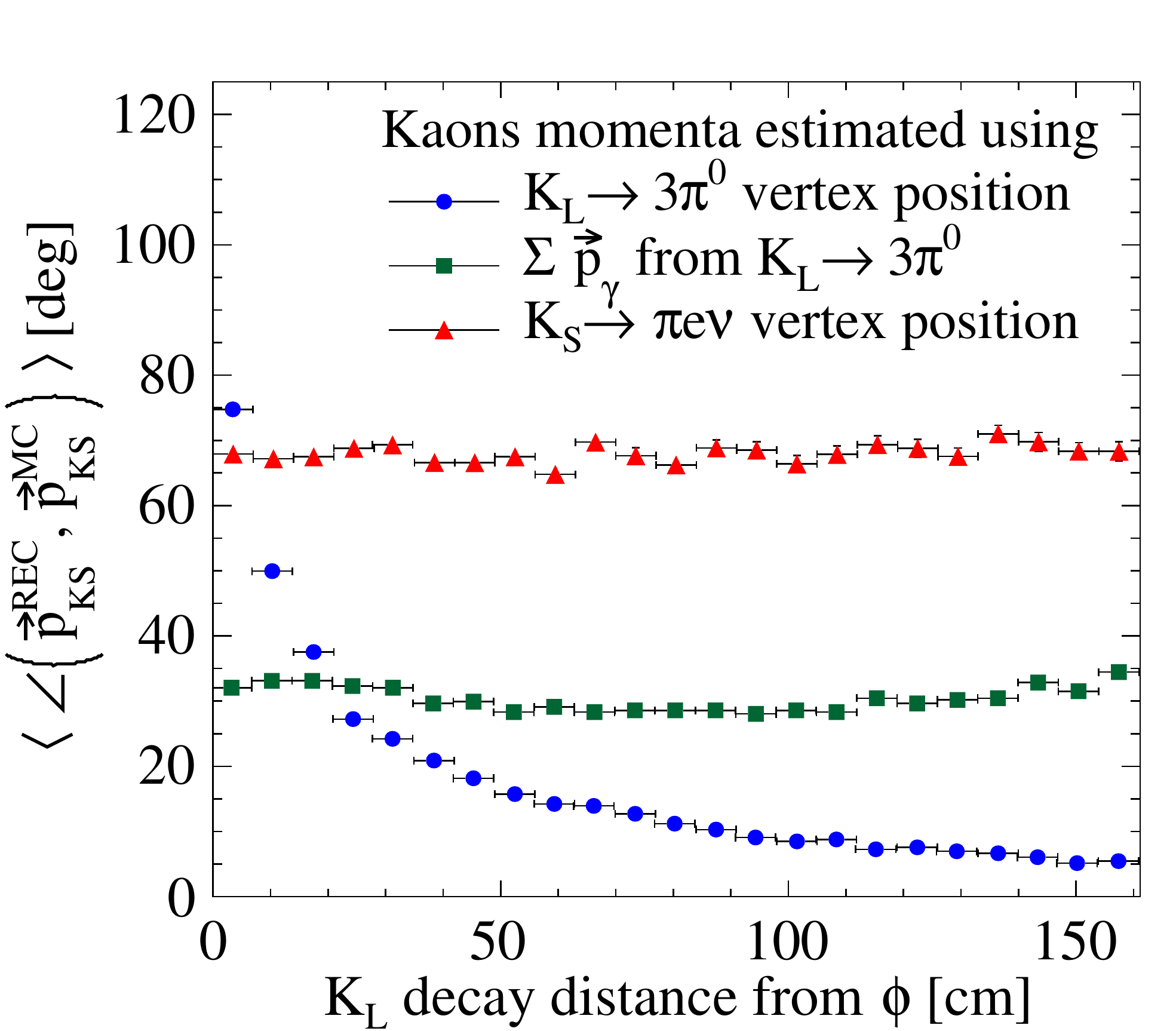}
  \hspace{1em}
  \includegraphics[width=0.45\textwidth]{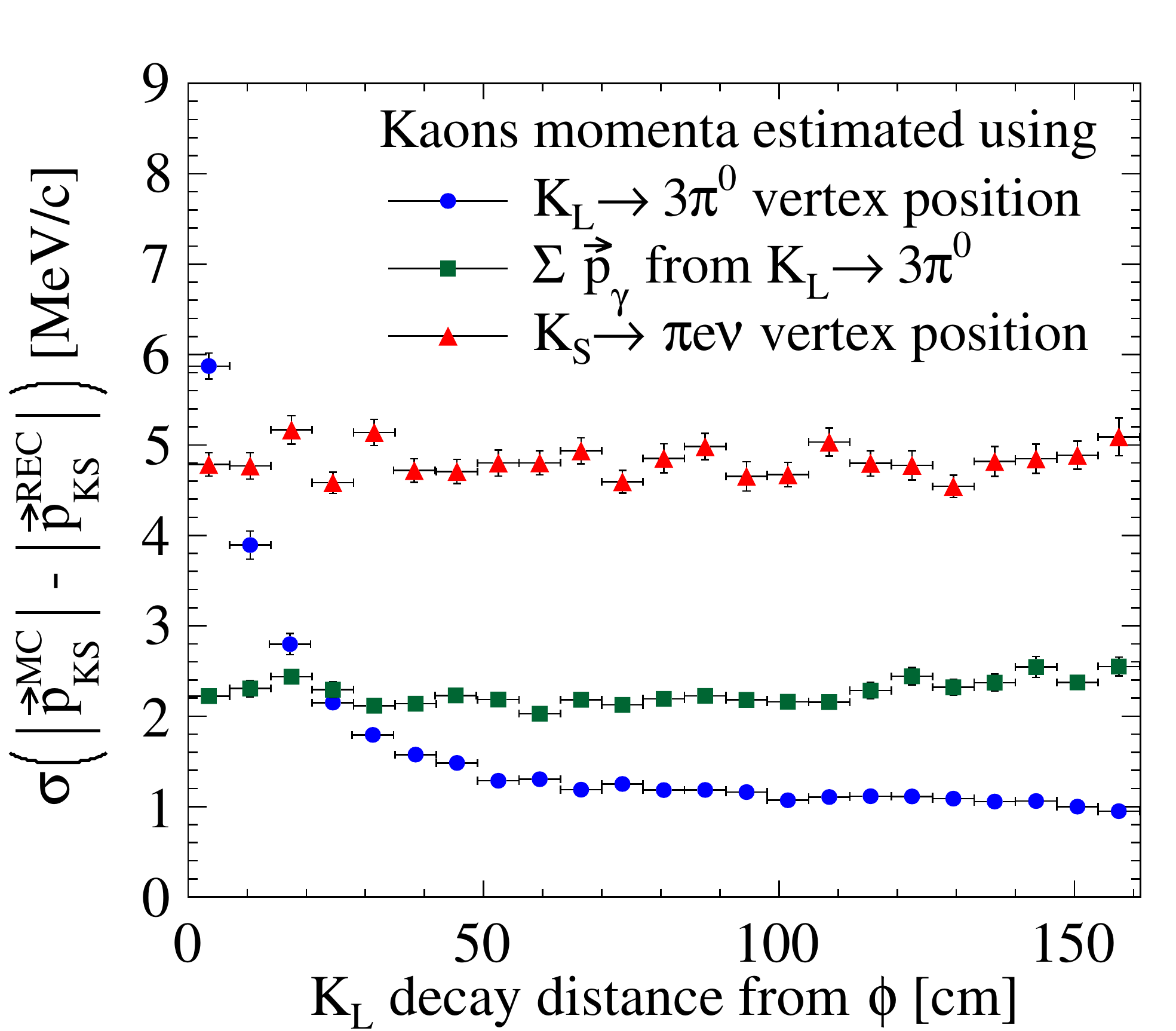}
  \caption{Resolution of direction (left) and modulus (right) of the $\Ks$ momentum estimated using the kinematics of 2-body $\phi\to\Ks\Kl$ decay and three kinds of available information on kaons' momentum directions.}\label{fig:ks_angle_resolution}
\end{figure}

\fref{fig:ks_angle_resolution} presents a comparison of the resolutions of $\Ks$ momentum in terms of its direction (left) and modulus (right), along with their dependence on the distance of $\Kl$ decay from the primary interaction point, obtained with momentum conservation for the $\phi\to \Ks\Kl$ decay and the three approaches to evaluate a kaon momentum direction listed above. Method 1 (red in~\fref{fig:ks_angle_resolution}) does not exhibit any resolution variations with the $\Kl$ decay position being based solely on information related to $\Ks$. Although it uses the precise position of $\Ks\to\pi e\nu$ vertex obtained with the drift chamber, larger uncertainty on the average $\phi$ decay point and the fact that both points are located close to each other result in large errors on the $\Ks$ momentum. While the same uncertainty of average $\phi$ vertex influences the second approach (blue), significantly larger possible distances between the two considered points provide better resolution of $\vec{p}_{K_S}$ which improves with distance between $\Kl$ decay and $\phi$. For early $\Kl$ decays, however, performance of this technique is reduced due to a relatively low resolution of the $\Kl\to 3\pi^{0}$ vertex location. The third estimation method (green) does not directly utilize information on decay vertices, instead relying on photons' momentum information reconstructed using the $\Kl$ decay point and EMC cluster energies. Resolution of this method, constant independently of the $\Kl$ decay distance, is superior to the other approaches for the events with early decays of $\Kl$.

In order to obtain the best estimate of $\vec{p}_{K_S}$ for the use in next analysis steps, the following strategy was adopted:
\begin{itemize}
\item using $\Kl\to 3\pi^{0}$ vertex position (method 2) if $|V_{\Kl}-V_{\phi}| > 20$~cm,
\item using $\sum \vec{p}_{\gamma}$ (method 3) if $|V_{\Kl}-V_{\phi}| \leq 20$~cm.
\end{itemize}

\subsection{Selection of semileptonic $\Ks$ decays}\label{sec:ksemil}
The most abundant background for the semileptonic $\Ks$ decays is constituted by the two-pion final states $\pi^+\pi^-$ and $\pi^0\pi^0$ which together account for~\SI{99.89}{\percent} of all decays of this meson. While the $\pi^0\pi^0$ is easily differentiated by the lack of charged particle tracks recorded in the KLOE drift chamber%
\footnote{In fact, $\Ks\to\pi^0\pi^0$ may be confused with an early $\Kl\to 3\pi^0$ decay when the accompanying kaon produces a semileptonic final state. Identification of such cases is discussed in~\sref{sec:ks2pi0_rejection}.},
$\pi^+\pi^-$ must be distinguished from the signal $\pi e\nu$ events using kinematical properties of the two recorded tracks.

\begin{figure}[h!]
  \centering
  \begin{tikzpicture}
    \node[anchor=south west,inner sep=0] at (0,0) 
    {\includegraphics[width=0.45\textwidth]{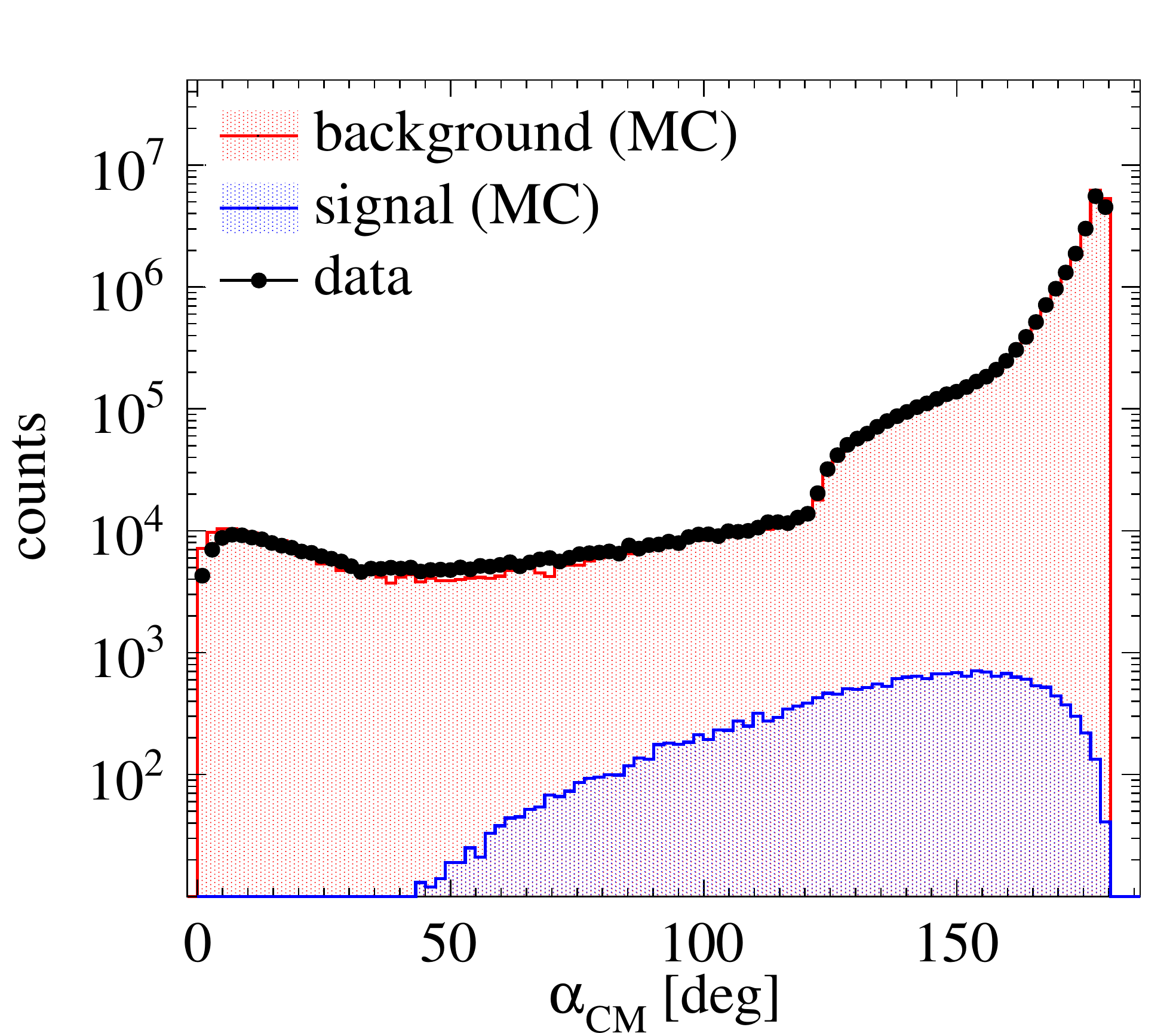}};
    \draw[black, thick, dashed] (2.32,0.8) -- (2.32,4.1);
    \draw[black, thick, dashed] (6.1,0.8) -- (6.1,5.46);
  \end{tikzpicture}
  \hspace{1em}
    \begin{tikzpicture}
    \node[anchor=south west,inner sep=0] at (0,0) 
    {\includegraphics[width=0.45\textwidth]{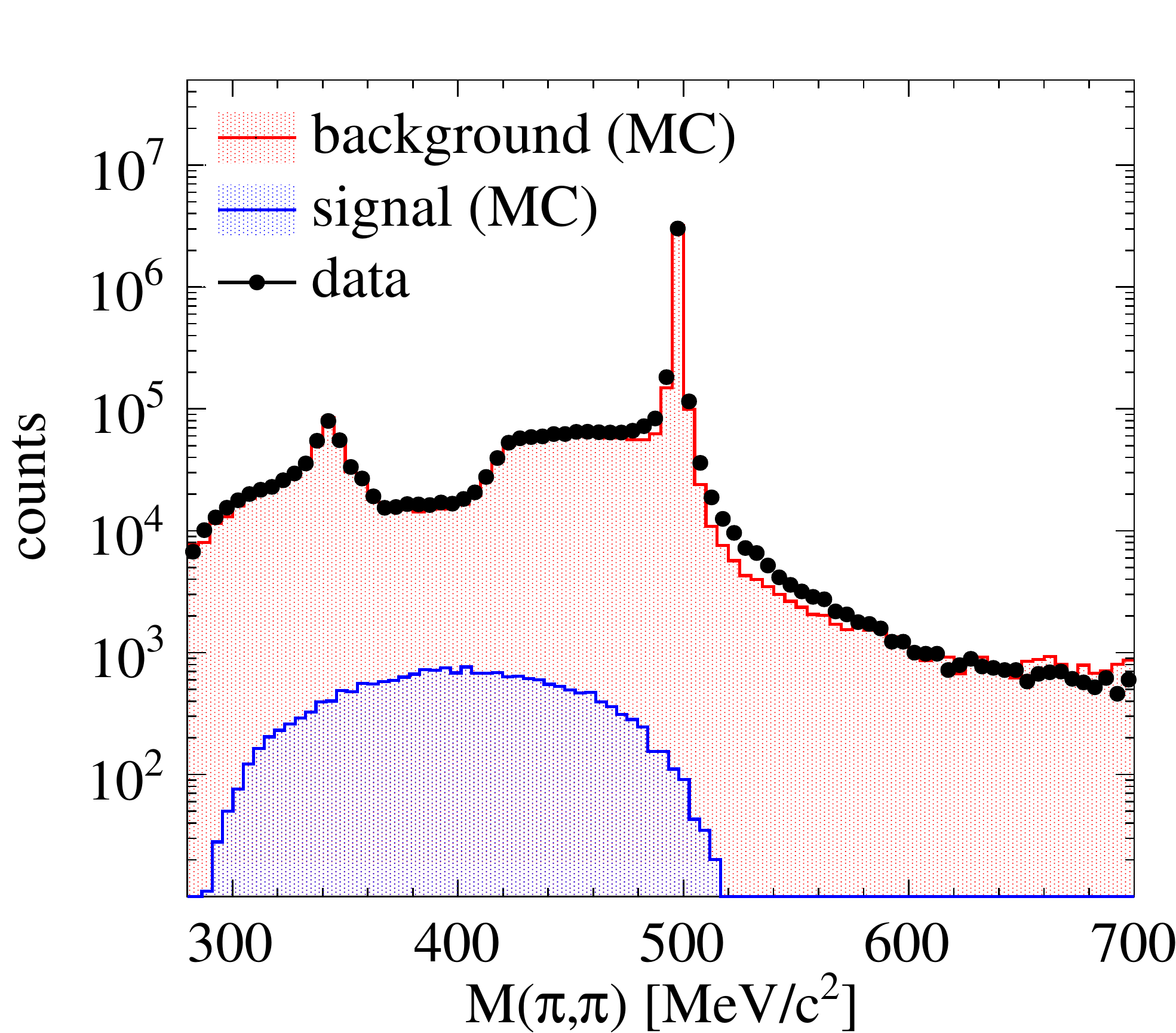}};
    \draw[black, thick, dashed] (3.81,0.8) -- (3.81,5.0);
  \end{tikzpicture}
  \caption{Distributions used for preselection of semileptonic $\Ks$ decays (referred to as \textit{signal}). Dashed lines denote the values of cuts. Left: Angle between momenta of the two decay products whose tracks are recorded by the KLOE DC, expressed in the center-of-mass reference frame. Right: invariant mass of the decaying state calculated with momenta of the two recorded decay products assuming that both were charged pions.}
  \label{fig:ksemil_basic_cuts}
\end{figure}
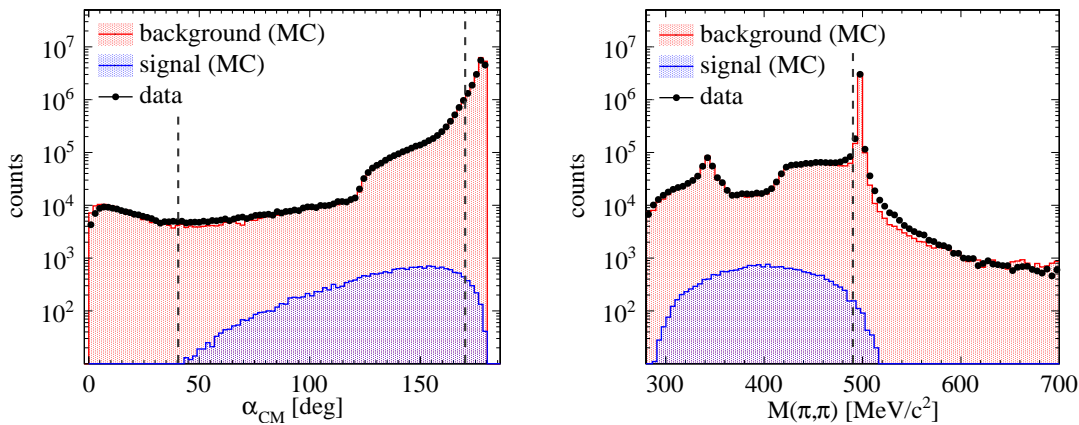

In the center of mass reference frame, the $\Ks\to\pi^+\pi^-$ process is characterized by momenta of both products being exactly opposite whereas the angle between momenta of the two recorded particles in a semileptonic decay come from a broad spectrum with small probability of a back-to-back decay. The first step of selection of $\Ks\to\pi e\nu$ is therefore based on the angle $\alpha_{CM}$ constituted by momenta associated to the two recorded DC tracks at their common vertex, expressed in the center-of-mass frame. To this end, momenta reconstructed in the detector frame were subjected to a Lorentz transformation to the $\Ks$ frame of reference using its momentum estimated as shown in the previous Section. The obtained distributions of the $\alpha_{CM}$ angle are presented in the left panel of~\fref{fig:ksemil_basic_cuts} where signal is defined as events with the $\Ks\to\pi e\nu$ decay and background is dominated by $\Ks\to\pi^+\pi^-$. Events were retained for further analysis if:
\begin{equation*}
   \ang{40} < \alpha_{CM} < \ang{170}.
\end{equation*}

At the event preselection stage, the two recorded tracks with a common vertex are not yet identified as pions, electrons or muons. Due to well-defined kinematics of $\Ks\to\pi^+\pi^-$,
invariant mass of the decaying state, calculated attributing the mass of a charged pion to both tracks, peaks at the neutral kaon mass with any possible deviation caused only by detector and reconstruction resolution. By contrast, the semileptonic decay where a neutrino carries away unknown momentum, may yield an invariant mass from a broad spectrum mostly below $m_{\kaon}$ as shown in~\fref{fig:ksemil_basic_cuts} (right). In order to reject a large part of the $\pi^+\pi^-$ background, only events for which:
\begin{equation*}
  \text{M}(\pi,\pi) < 490\:\text{MeV/c}^{2},
\end{equation*}
are processed further. After the selection of $\Kl\to 3\pi^0$ and the aforementioned $\Ks\to \pi e\nu$ preselection, the signal to background (S/B) ratio is about 0.13 which calls for a refined event selection, achieved with a time of flight (TOF) analysis of the decay products.

\subsection{Time of flight analysis for charged particles}\label{sec:t1_tof}
The analysis of time of flight is performed for particles corresponding to the two DC tracks associated to the $\Ks$ decay vertex candidate chosen at preselection stage.
The general scheme of this procedure is based on the approaches used in previous studies of $\Ks\to \pi e \nu$ at KLOE~\cite{Ambrosino:2006si,daria_article}. However, these studies allowed a certain amount of background events in the final event sample and perform a MC-based background subtraction and event counting. Moreover, they were based on events with a $\Kl$ interaction in the calorimeter which provided a high resolution of $\Kl$ and $\Ks$ momenta (calculated as discussed in~\sref{sec:ks_from_kl}). Conversely, in the analysis described in this work, kaons' momenta estimation with $\Kl\to 3\pi^0$ is not as precise and the event selection is required to yield a highly pure sample of semlileptonic decays of $\Ks$. Therefore, the details of the time of flight analysis have been adjusted to perform a more stringent event selection and additional cuts have been introduced to reject particular types of remaining background.

In addition to background discrimination, the time of flight study serves two other purposes. Firstly, it allows to identify the tracks created by pions and by electrons or positrons by testing two possible hypotheses about masses of the two charged particles in an event. Secondly, TOF information is used to calculate the correct time of the $\phi$ decay, which may be different from the DAQ start time by a multiple of the collider bunch crossing periods as discussed in~\sref{sec:trigger}.

%
% MEMO TODO: opisac i zacytowac datizza i pezzetto itd.
%
In order to calculate the expected time of flight of a particle associated with a DC track, the track is extrapolated to its incidence point on the electromagnetic calorimeter of KLOE and total length $L$ of the particle track up to its interaction in the EMC is calculated. Velocity is calculated from the momentum reconstructed for a track, with a certain assumption on the particle mass $m_x$:
\begin{equation}
  \label{eq:track_beta}
  \beta(m_x) = \frac{p}{\sqrt{p^{2}+m_x^{2}c^{2}}},
\end{equation}
so that the expected particle time of flight reads $L/(c\beta(m_{x}))$. To compare this value with a recorded particle time of flight, the particle must undergo an interaction in the calorimeter whose time $t_{cl}$ is recorded. Although such requirement removes a fraction of the signal events where poor track reconstruction or a $\pi^{\pm}\to\mu^{\pm}\nu$ decay inside the drift chamber may prevent a correct extrapolation of a track to the EMC or association to an energy deposition cluster (efficiency of this step is about \SI{43}{\percent}), it provides information which is powerful in terms of background rejection and particle identification.

For each of the two recorded tracks, a discrepancy between expected and recorded time of flight of a particle with mass $m_{x}$ is calculated as:
\begin{equation}
  \label{eq:dtof_t1}
  \delta t({m_x}) = t_{cl} - \frac{L}{c\beta(m_x)},
\end{equation}
where $x=\pi^{\pm}$ or $x=e^{\pm}$ denotes the adopted particle mass hypothesis. It should be noted that, as mentioned in~\sref{sec:trigger}, at this stage the recorded time of particle interaction in the EMC (and thus also $\delta t(m_{x})$) may be determined with respect to an incorrect event start time ($t_{0}$) and therefore all variables used in the TOF analysis are defined as differences of such values attributed to both tracks in an event. Any possible time offset, which must be common for both particles, is then cancelled out. 

\begin{figure}[h!]
  \centering
    \begin{tikzpicture}
    \node[anchor=south west,inner sep=0] at (0,0) 
    {\includegraphics[width=0.45\textwidth]{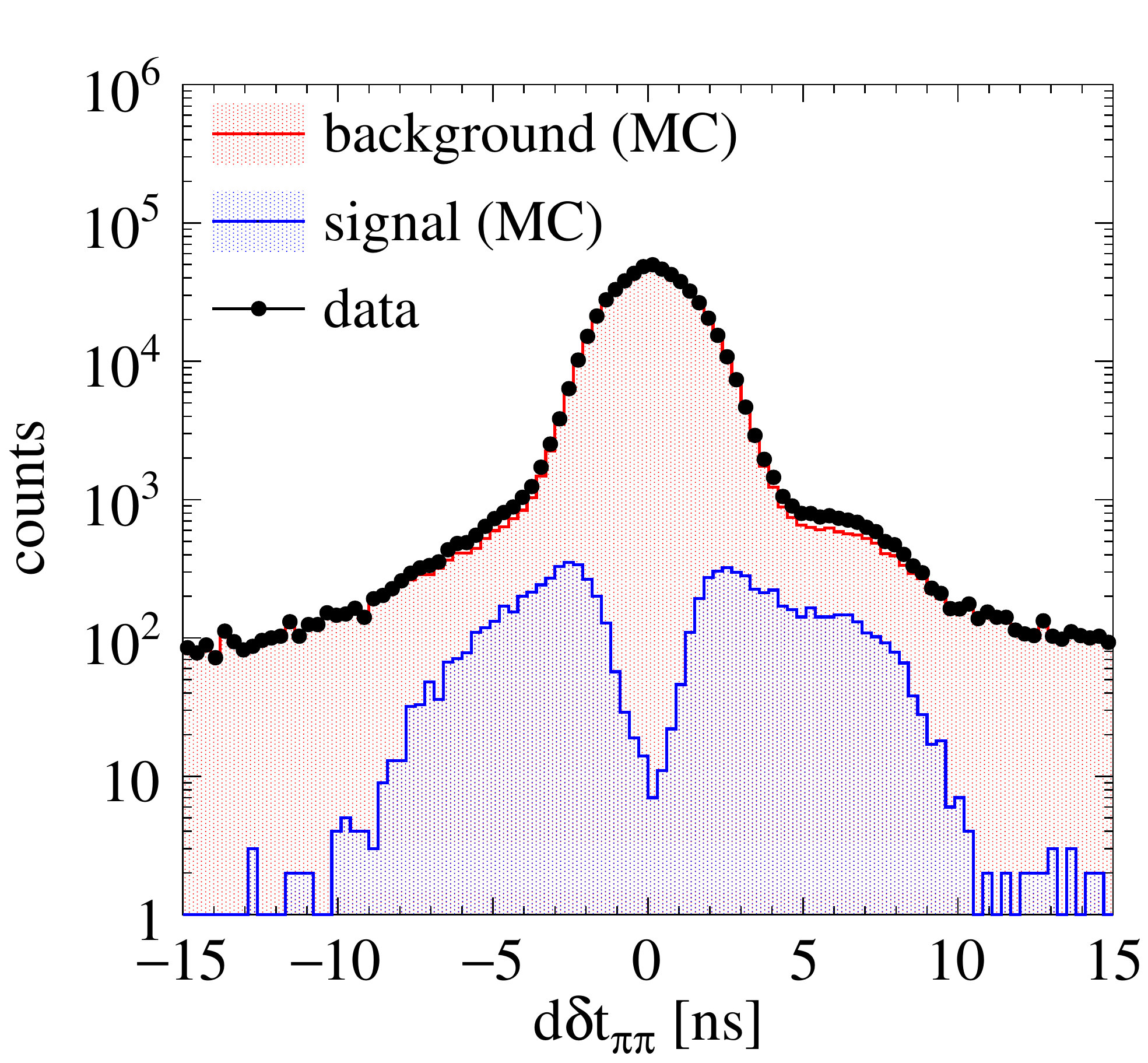}};
    \draw[black, thick, dashed] (2.0,0.8) -- (2.0,4.0);
    \draw[black, thick, dashed] (3.55,0.8) -- (3.55,4.0);
    \draw[ultra thick, black!70!white, <->] (2.05, 1.2) -- (3.5, 1.2);
    \draw[black, thick, dashed] (4.1,0.8) -- (4.1,5.2);
    \draw[black, thick, dashed] (5.65,0.8) -- (5.65,5.46);
    \draw[ultra thick, black!70!white, <->] (4.15, 1.2) -- (5.6, 1.2);
  \end{tikzpicture}
  \caption{TOF discrepancy of two tracks with a $\pi^+\pi^-$ assumption. Dashed lines and gray arrows denote the two accepted regions.}\label{fig:tof_12_t1_1}
\end{figure}
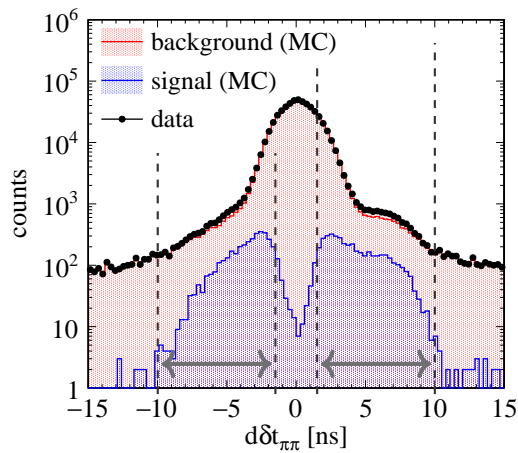

Wrong assignment of particle mass hypotheses $x$ and $y$ can be revealed in the difference between $\delta t$ values for the two tracks (referred to as tracks 1 and 2) in an event:
\begin{equation}
  \label{eq:tof2_t1}
  d\delta t_{x,y} = \delta t_1(m_x) - \delta t_2(m_y),
\end{equation}
expected to vanish if both particles are correctly identified. Therefore, rejection of background is first performed with the distribution of $d\delta t_{\pi,\pi}$ for which both tracks are assumed to come from charged pions. As shown in~\fref{fig:tof_12_t1_1}, this distribution peaks at zero for the $\pi^+\pi^-$-dominated background while being shifted off zero for semileptonic events where one of the pion track assumptions is clearly incorrect. Events are accepted if they satisfy the following requirement:
\begin{equation*}
  |d\delta t_{\pi,\pi}| \in (1.5,\;10)\:\text{ns}.
\end{equation*}

\begin{figure}[h!]
  \centering
  \captionsetup[subfigure]{justification=centering}
  \begin{subfigure}{0.45\textwidth}
  {\includegraphics[width=1.0\textwidth]{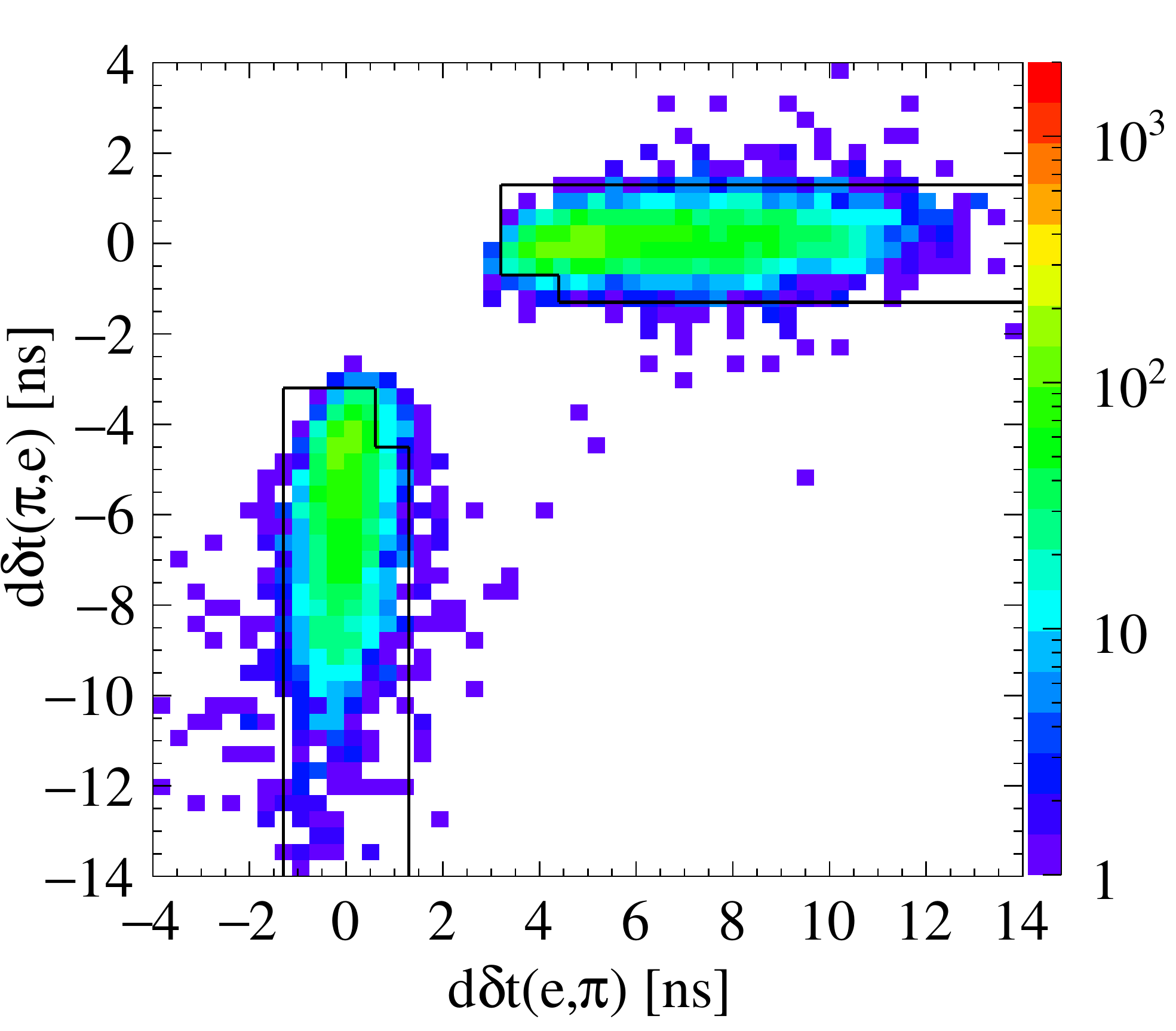}}
  \caption{Signal events (MC).}\label{fig:tof_12_t1_2}
\end{subfigure}
  \hspace{1em}
\begin{subfigure}{0.45\textwidth}
  {\includegraphics[width=1.0\textwidth]{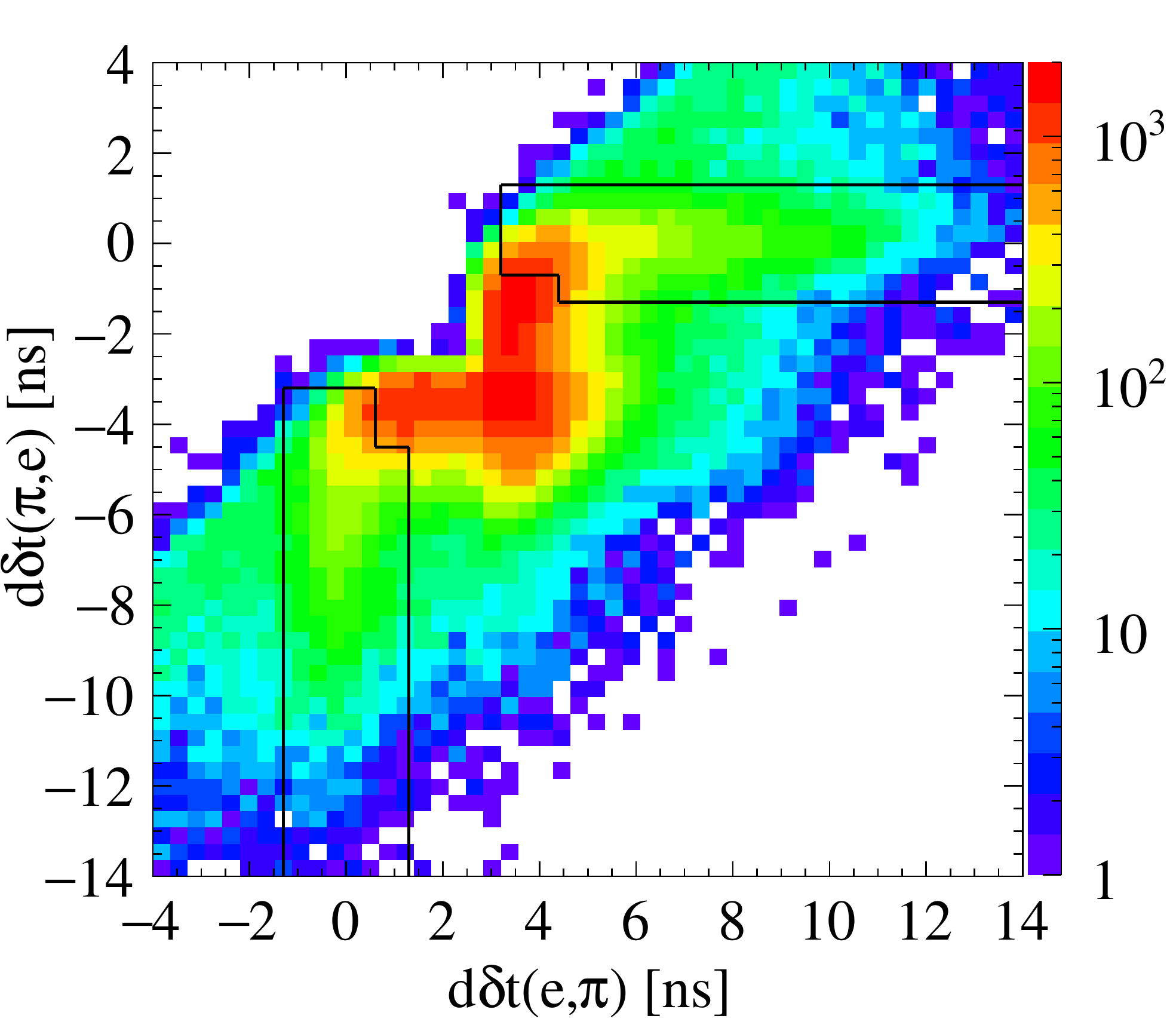}}
  \caption{Background events (MC).}\label{fig:tof_12_t1_3}
  \end{subfigure}
  \begin{subfigure}{0.45\textwidth}
  {\includegraphics[width=1.0\textwidth]{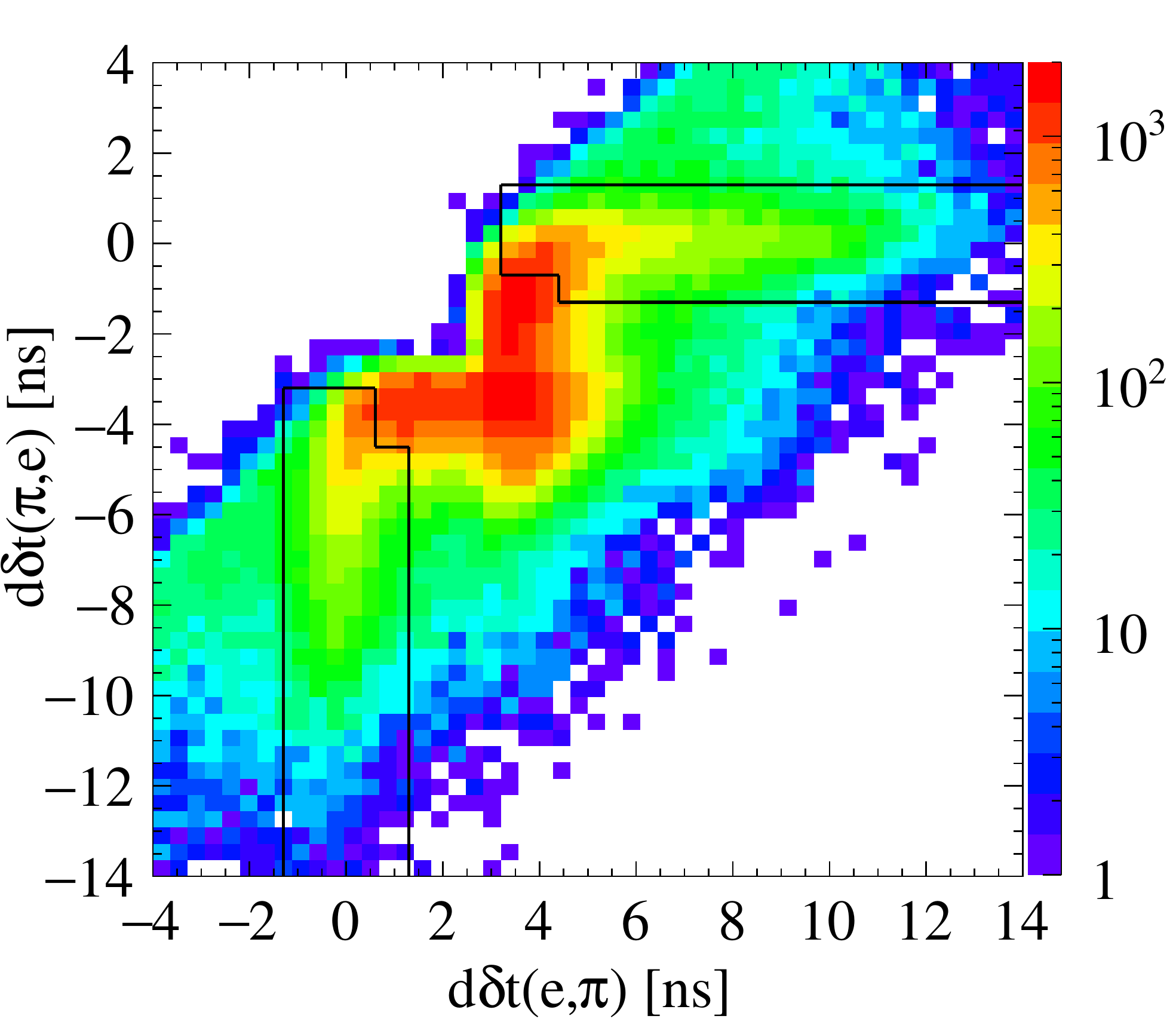}}
  \caption{All data events.}\label{fig:tof_12_t1_4}
  \end{subfigure}
  \caption{Errors of two possible $\pi^{\pm}$ and $e^{\mp}$ mass hypotheses assignments to tracks. Solid lines denote accepted regions, each corresponding to one posible assignment of pion and electron mass hypotheses' assignment to the two DC tracks in an event.}\label{fig:tof_12_t1}
\end{figure}

Subsequently, an assignment of $\pi^{\pm}$ and $e^{\mp}$ mass hypotheses to the two tracks is tried in two possible configurations. The correct assignment is identified using the relative distributions of $d\delta t (\pi,e)$ and $d \delta t(e,\pi)$ presented in Figure~\ref{fig:tof_12_t1}. Signal events belong to one of two regions characterized by a small value of one of the $d\delta(x,y)$ discrepancies and significant value of the other one (marked with solid line in the figures). Affiliation to a particular region indicates the correct attribution of pion and electron masses to each of the tracks. Background events, on the other hand, mostly populate the region closer to diagonal where both $d\delta t(\pi,e)$ and $d\delta t(e,\pi)$ are non-zero. Consequently, events are selected for further analysis if they belong to one of the regions marked with solid line in Figure~\ref{fig:tof_12_t1}.
%
% TODO: wypisac explicite wartosci ciec
%

\subsection{Correction of event start time}\label{sec:t0_corr}
%
% TODO: dopisac, ze standardowe T0 nie dziala bo nie ma prompt fotonow w zdarzeniu
%
\begin{figure}[b]
  \centering
  \begin{subfigure}{0.45\textwidth}
  {\includegraphics[width=1.0\textwidth]{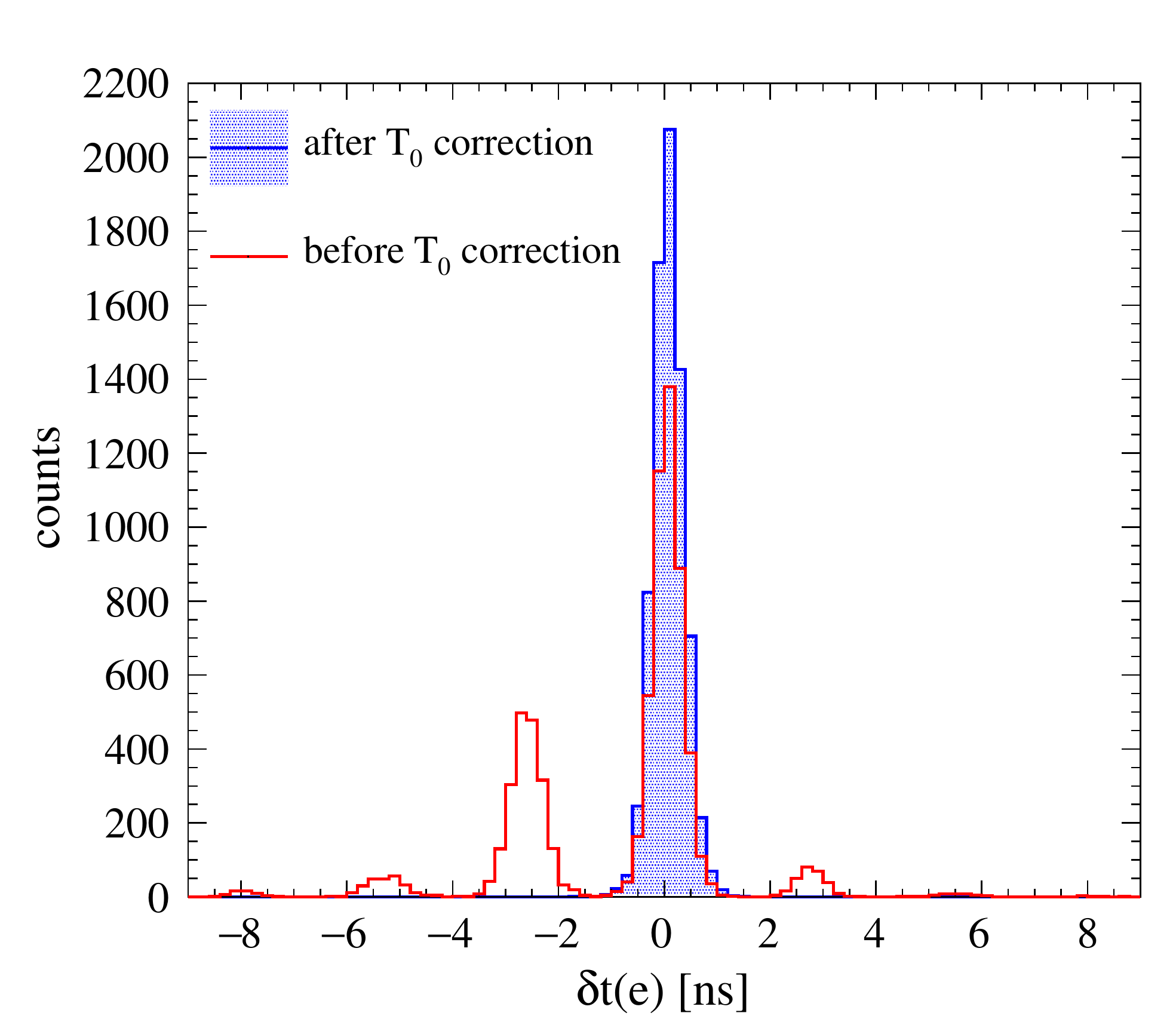}}
\end{subfigure}
\begin{subfigure}{0.45\textwidth}
  {\includegraphics[width=1.0\textwidth]{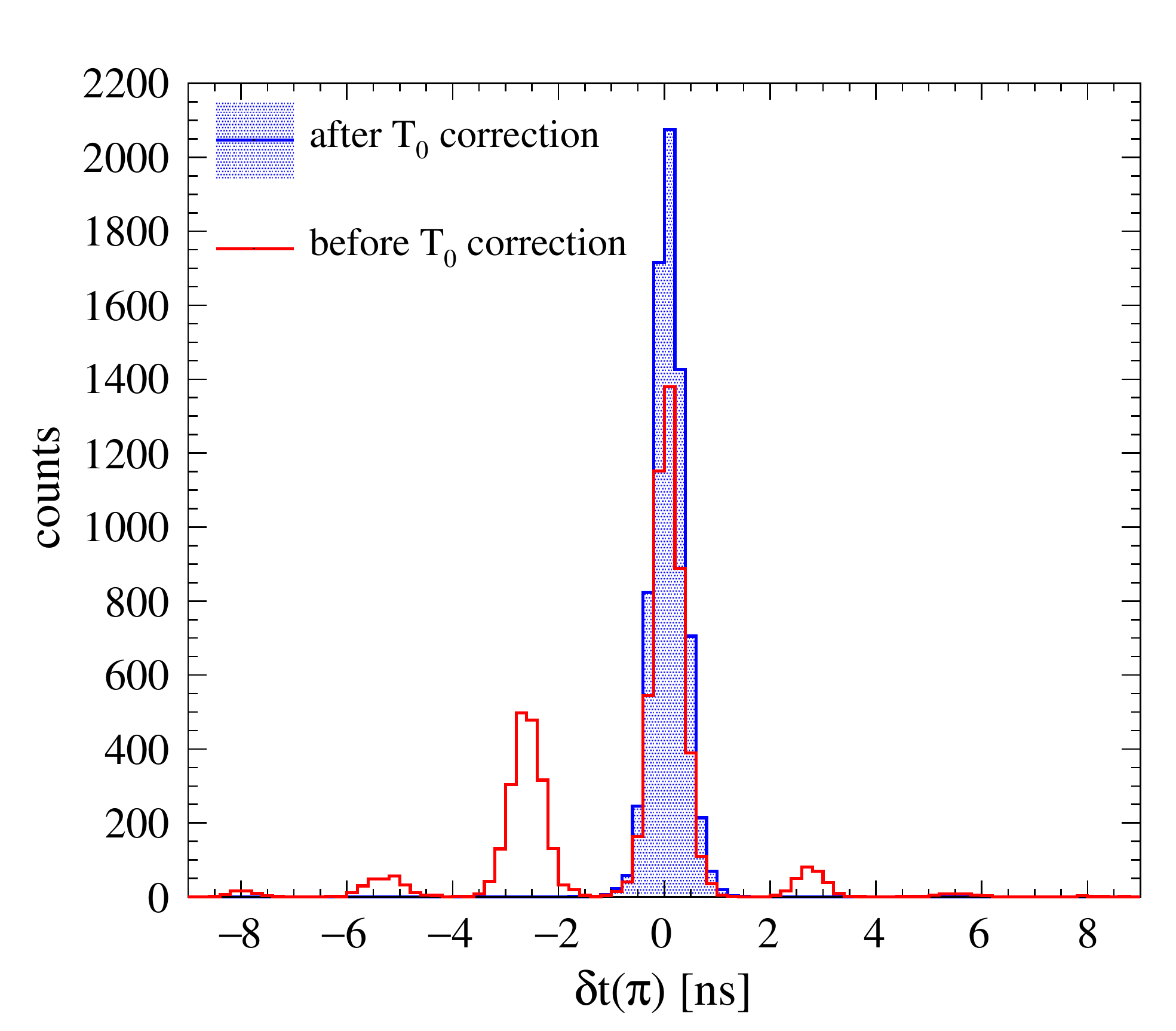}}
\end{subfigure}
\begin{subfigure}{0.45\textwidth}
  {\includegraphics[width=1.0\textwidth]{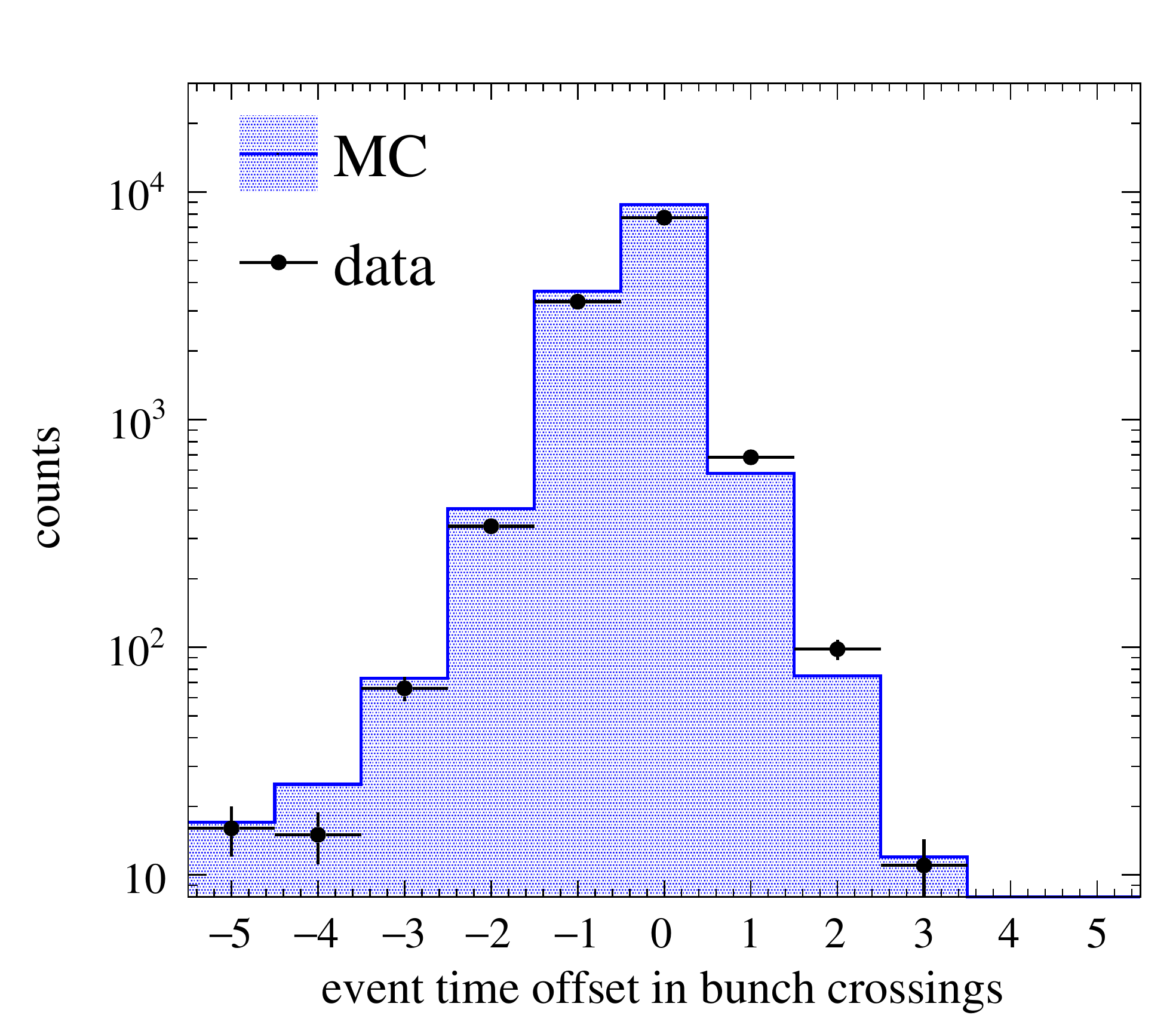}}
\end{subfigure}
  \caption{Top: discrepancies between recorded and expected time of flight for the track identified as originating from $e^{\pm}$ (left) and $\pi^{\mp}$ (right) before and after correction of event $T_0$. Distributions were obtained with data. For clarity, the  presented distributions were obtained after the background rejection steps discussed in~\sref{sec:t1-fine-selection} to avoid contamination from incorrectly identified tracks. Bottom: number of collider bunch crossing periods shifted as a result of event $T_0$ correction, for data and MC simulations.}\label{fig:t0_corr}
\end{figure}

Once the tracks originating from pion and electron/positron are identified, their TOF calculated with correct mass hypotheses can be used to determine a correction to the event start time which is later subtracted from all times recorded in the event. If the discrepancies between recorded and expected time of flight for the pion and lepton (compare~\eref{eq:dtof_t1}) are denoted as $\delta t(\pi)$ and $\delta t(e)$ respectively, the correction to event $T_0$ is obtained as an average of these values for both tracks, rounded to a multiple of DA$\Phi$NE bunch crossing periods $T_{RF}$~(compare~\eref{eq:t0_prompt}):
\begin{equation}
  \label{eq:t0_correction_semil}
  T'_{0} = T_{RF}\cdot nint\left(\frac{\delta t(\pi) + \delta t(e)}{2T_{RF}}\right).
\end{equation}

\fref{fig:t0_corr} shows the distributions of $\delta t(\pi)$ and $\delta t(e)$ for each of the charged particles in an event, before and after application of the $T_0$ correction the recorded times of EMC clusters. The uncorrected cluster times are most commonly offset forward by one bunch crossing, i.e.\ by 2.72~ns resulting in several peaks of the $\delta t$ distributions. The bottom panel of~\fref{fig:t0_corr} presents the total number of bunch crossings which needed to be shifted in an event. The $T_0$ correction is required in about \SI{32}{\percent} of the considered events.

\subsection{Fine selection of $\Ks\to\pi e\nu$ decays}\label{sec:t1-fine-selection}
\begin{figure}[h!]
  \centering
\captionsetup[subfigure]{justification=centering}
  \begin{subfigure}{0.45\textwidth}
    \includegraphics[width=1.0\textwidth]{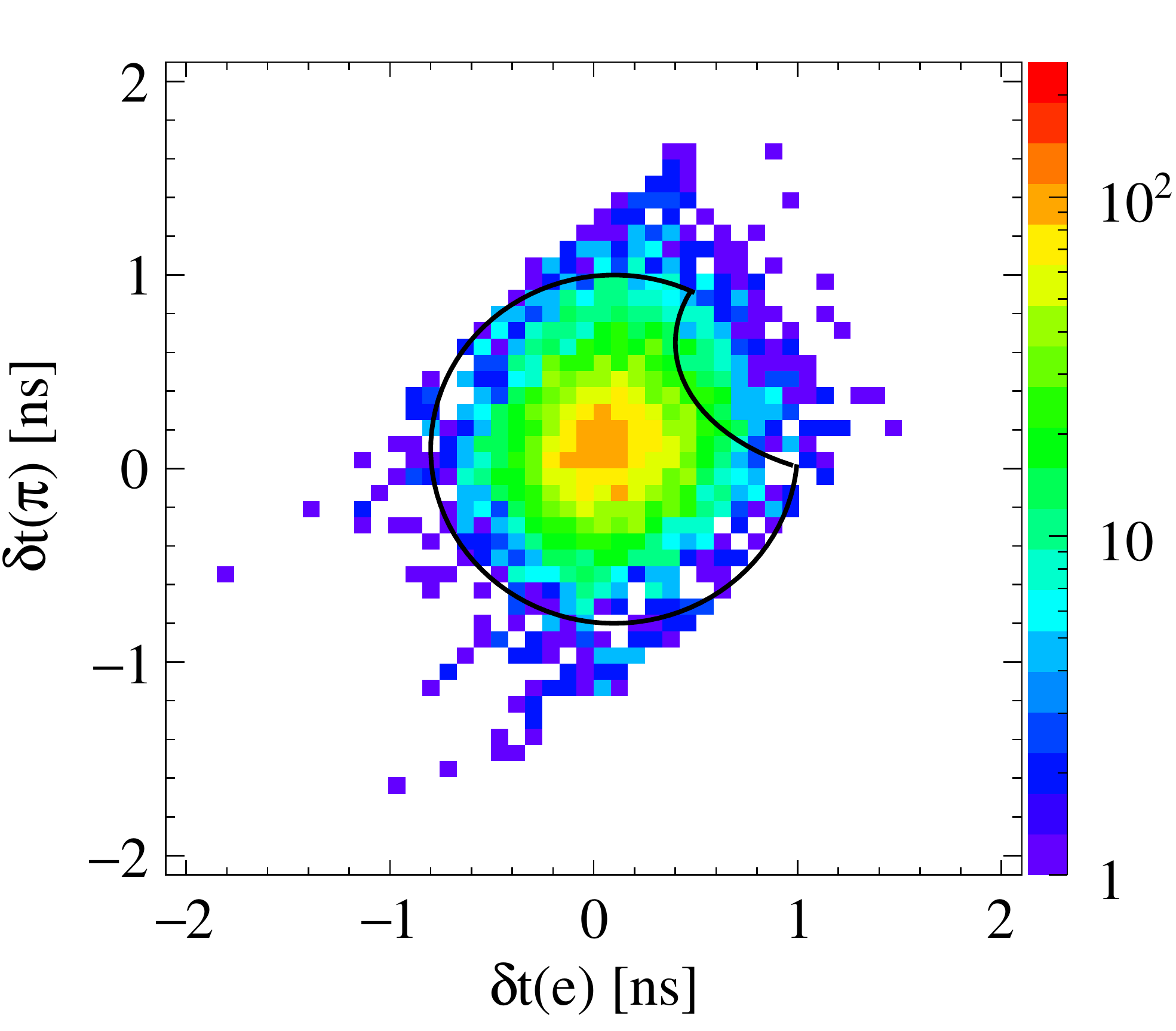}
    \caption{Signal events (MC)}\label{fig:tof3_cut_1}
  \end{subfigure}
  \begin{subfigure}{0.45\textwidth}
    \includegraphics[width=1.0\textwidth]{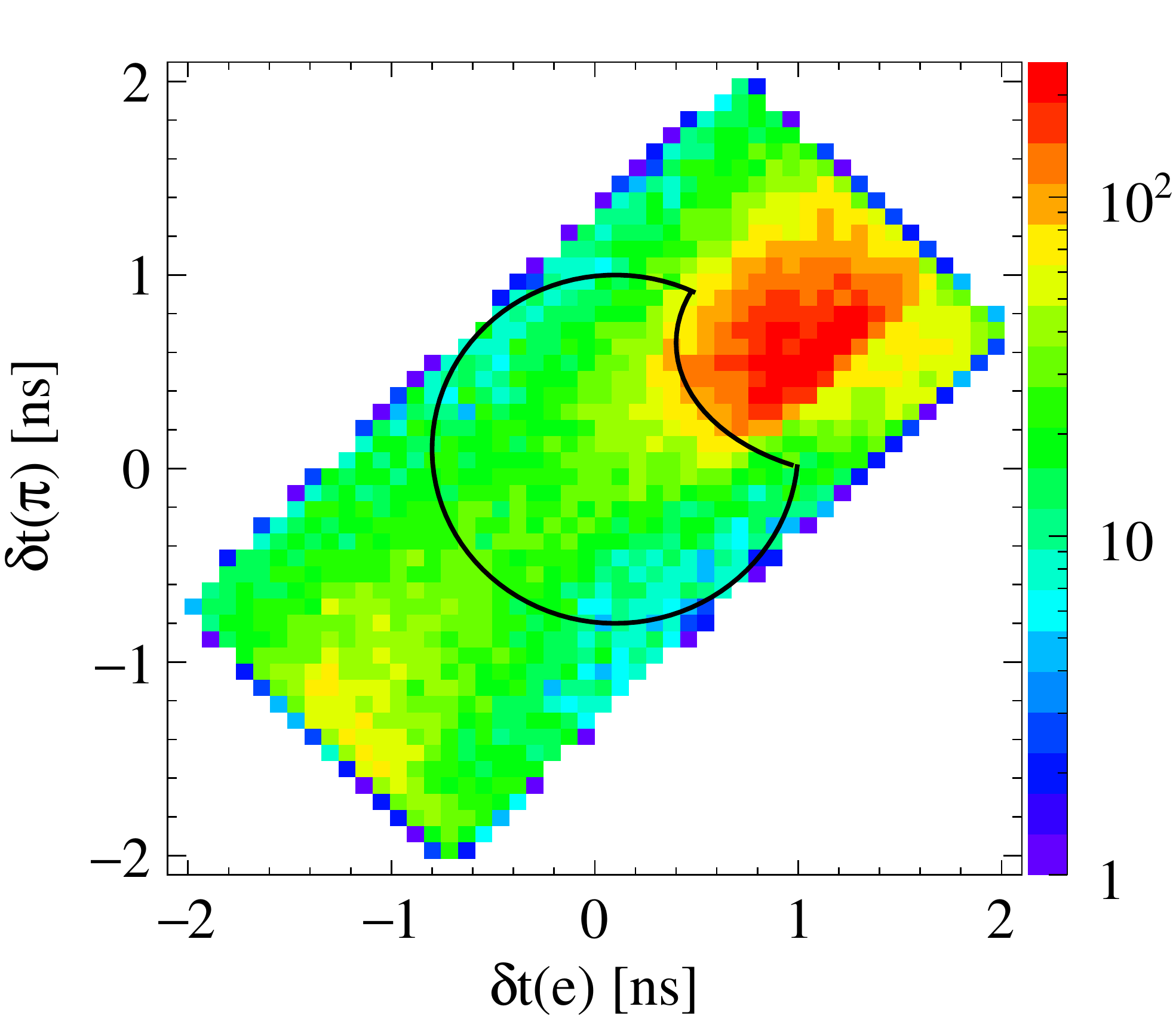}
    \caption{Background events (MC)}
  \end{subfigure}
  \begin{subfigure}{0.45\textwidth}
    \includegraphics[width=1.0\textwidth]{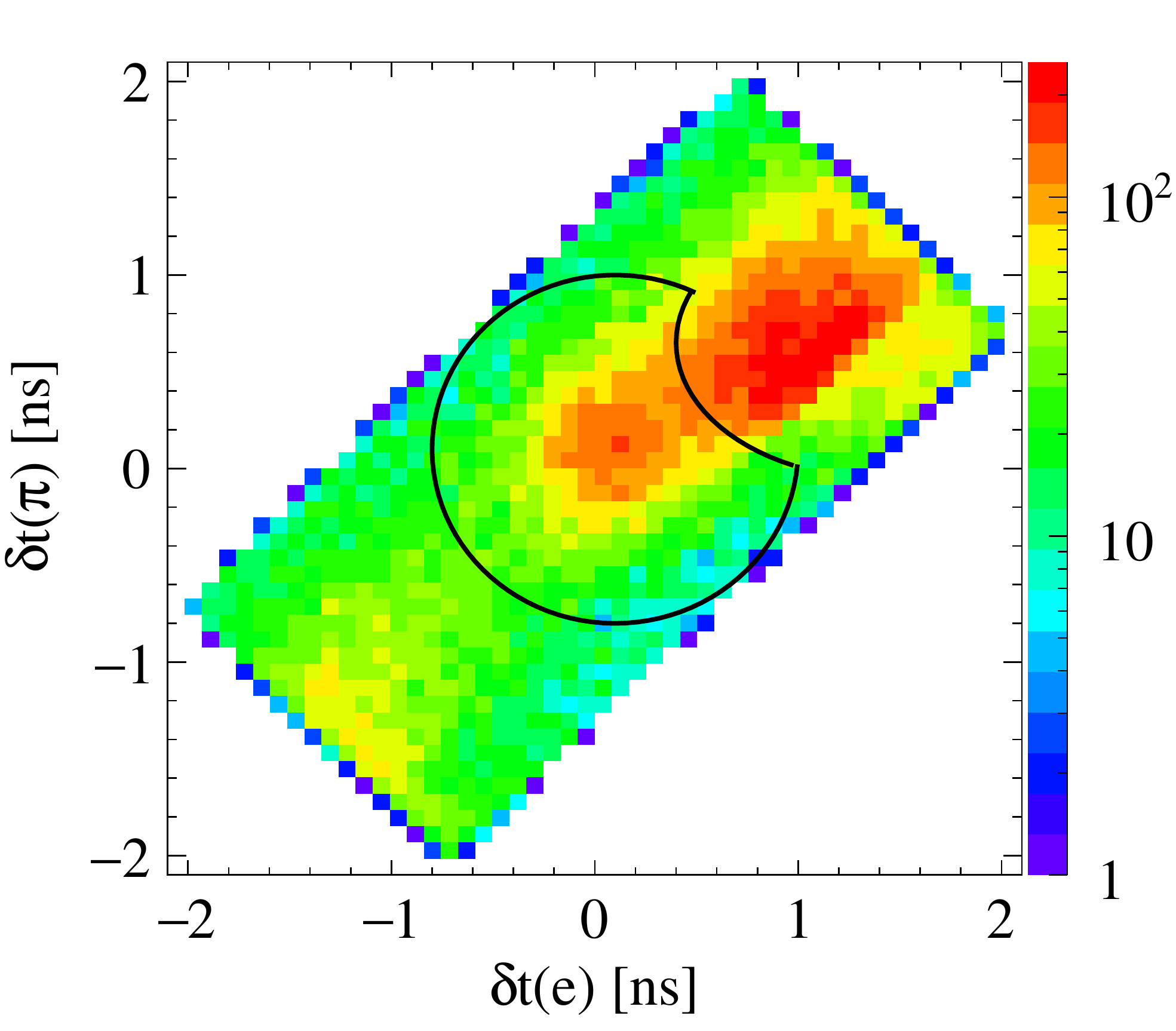}
    \caption{All data events}
  \end{subfigure}
  \caption{Relative distribution of TOF discrepancies for tracks identified as pion and electron/positron. Limits of the populated region result from cuts introduced at a previous stage of the TOF analysis~(see~\fref{fig:tof_12_t1}). Events are retained if they lie inside the region marked with black solid line.}\label{fig:tof3_cut}
\end{figure}

With a correct event start time, the TOF discrepancies $\delta t(\pi)$ and $\delta t(e)$ are concentrated around zero for both the pion and electron/positron tracks~(see~\fref{fig:tof3_cut}\subref{fig:tof3_cut_1}). On the other hand, background events in which at least one of the particles was incorrectly identified must be shifted off the center of the distributions presented in~\fref{fig:tof3_cut}. In fact, the shift is usually exhibited by both values as the error in calculation of one of $\delta t(\pi)$ or $\delta t(e)$ is eventually propagated to both through an incorrect event $T_0$ value subtracted from EMC cluster times.

Therefore, the relative distribution of both TOF discrepancies from~\fref{fig:tof3_cut} is useful for rejection of background form non-semileptonic decays of $\Ks$ which is concentrated in the top upper part of the distributions. To achieve the best purity of $\Ks\to\pi e \nu$ events, only events lying inside a circle in the center of the distribution are kept for further analysis, with exception of those lying in a ellipse surrounding the background region, i.e.:
\begin{eqnarray*}
  \label{eq:tof_cut_3}
  \left(\frac{\delta t(\pi)-0.1\:\text{ns}}{0.9\:\text{ns}}\right)^2 &+& \left(\frac{\delta t(e)-0.1\:\text{ns}}{0.9\:\text{ns}}\right)^2 < 1\\
  &\land&\\
  {\left(\frac{\delta t(\pi)-0.65\:\text{ns}}{0.7\:\text{ns}}\right)}^2 &+& {\left(\frac{\delta t(e)-1.4\:\text{ns}}{1.0\:\text{ns}}\right)}^2 > 1,
\end{eqnarray*}
as marked with the black line in~\fref{fig:tof3_cut}.

\begin{figure}[h!]
  \centering
  \includegraphics[width=0.4\textwidth]{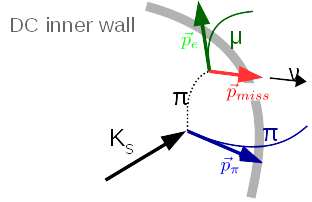}
  \caption{Scheme of a $\Ks\to\pi^+\pi^-\to \pi\mu\nu$ decay occurring before the inner wall of the KLOE drift chamber so that the muon track is misidentified as electron or positron and extrapolated to an artificial vertex common with the pion track.}\label{fig:kspimu}
\end{figure}

After the two-dimensional cut on the $\delta t$ values, a majority of the surviving background comes from three classes:
\begin{itemize}
\item $\Ks\to\pi^+\pi^-$ with imperfectly reconstructed tracks where wrong momentum estimation for one of the pions allows it to pass the selection criteria for an electron/positron track,
\item radiative $\Ks\to\pi^+\pi^-\gamma$ decays where a photon carries away a part of the momentum,
\item $\Ks\to\pi^+\pi^-\to \pi\mu\nu$ decay chains where one of the charged pions decayed into a muon and a neutrino before the inner wall of the KLOE drift chamber so that its decay was not recorded.
\end{itemize}
Since the decay vertices located close to the beamline and thus not inside the drift chamber are reconstructed using extrapolations of tracks from the DC towards the $\phi$ decay point, in the $\Ks\to\pi^+\pi^-\to \pi\mu\nu$ case one of the recorded tracks in fact corresponds to a muon not originating at the same vertex as the pion track as depicted schematically in~\fref{fig:kspimu}.

\begin{figure}[h!]
  \centering
\captionsetup[subfigure]{justification=centering}
  \begin{subfigure}{0.45\textwidth}
    \includegraphics[width=1.0\textwidth]{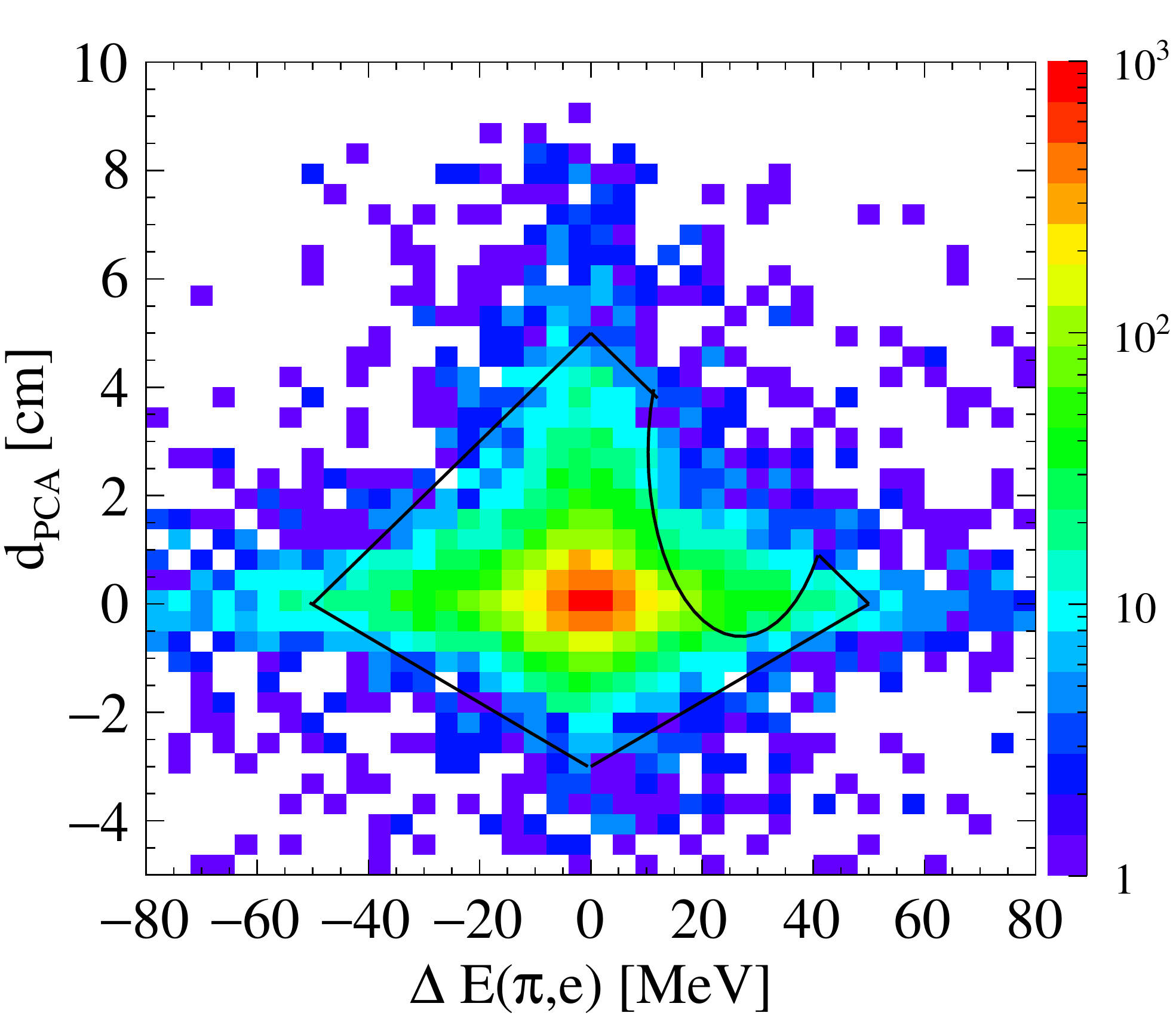}
    \caption{Signal events (MC)}
  \end{subfigure}
  \begin{subfigure}{0.45\textwidth}
    \includegraphics[width=1.0\textwidth]{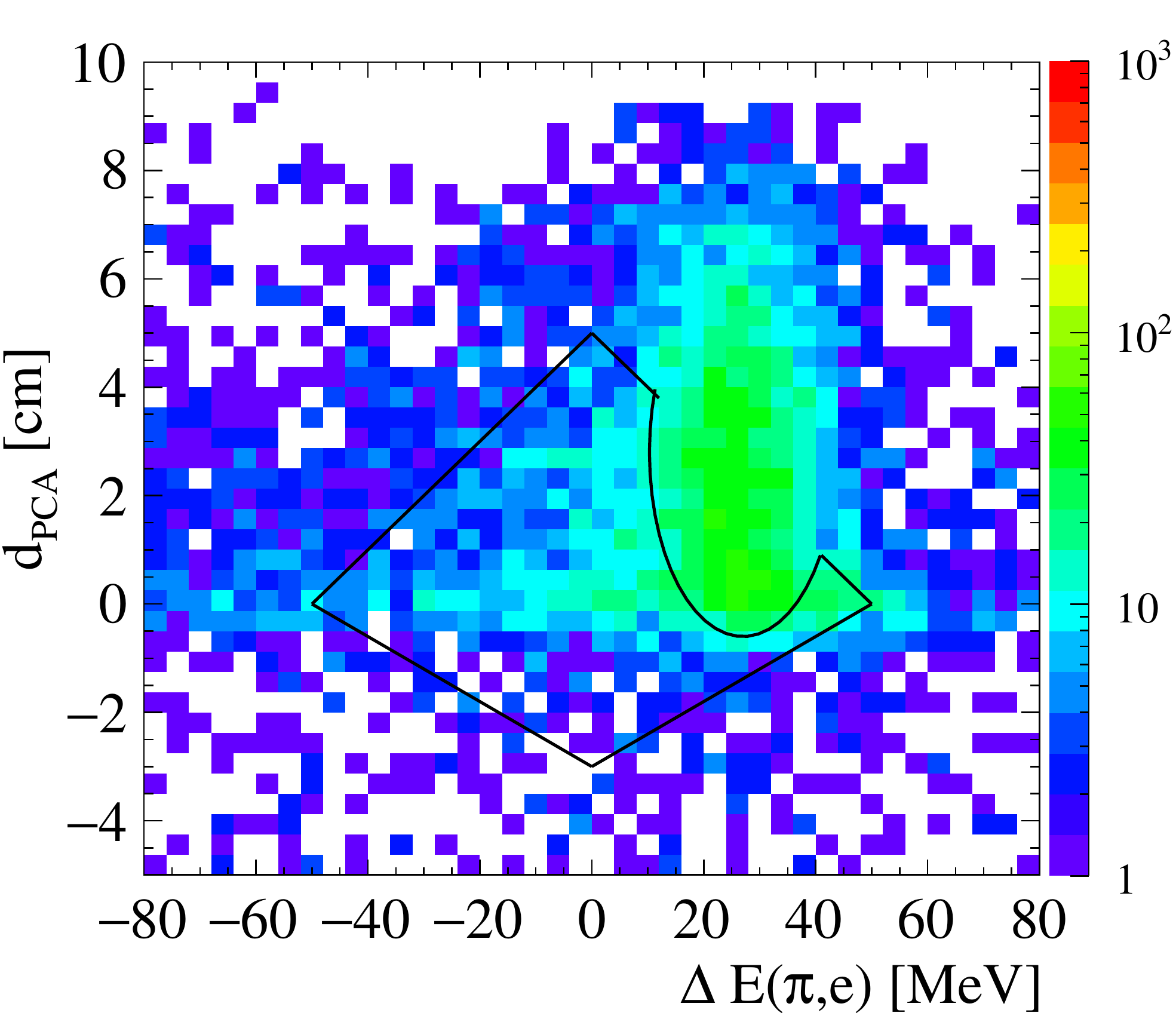}
    \caption{Background events (MC)}
  \end{subfigure}
  \begin{subfigure}{0.45\textwidth}
    \includegraphics[width=1.0\textwidth]{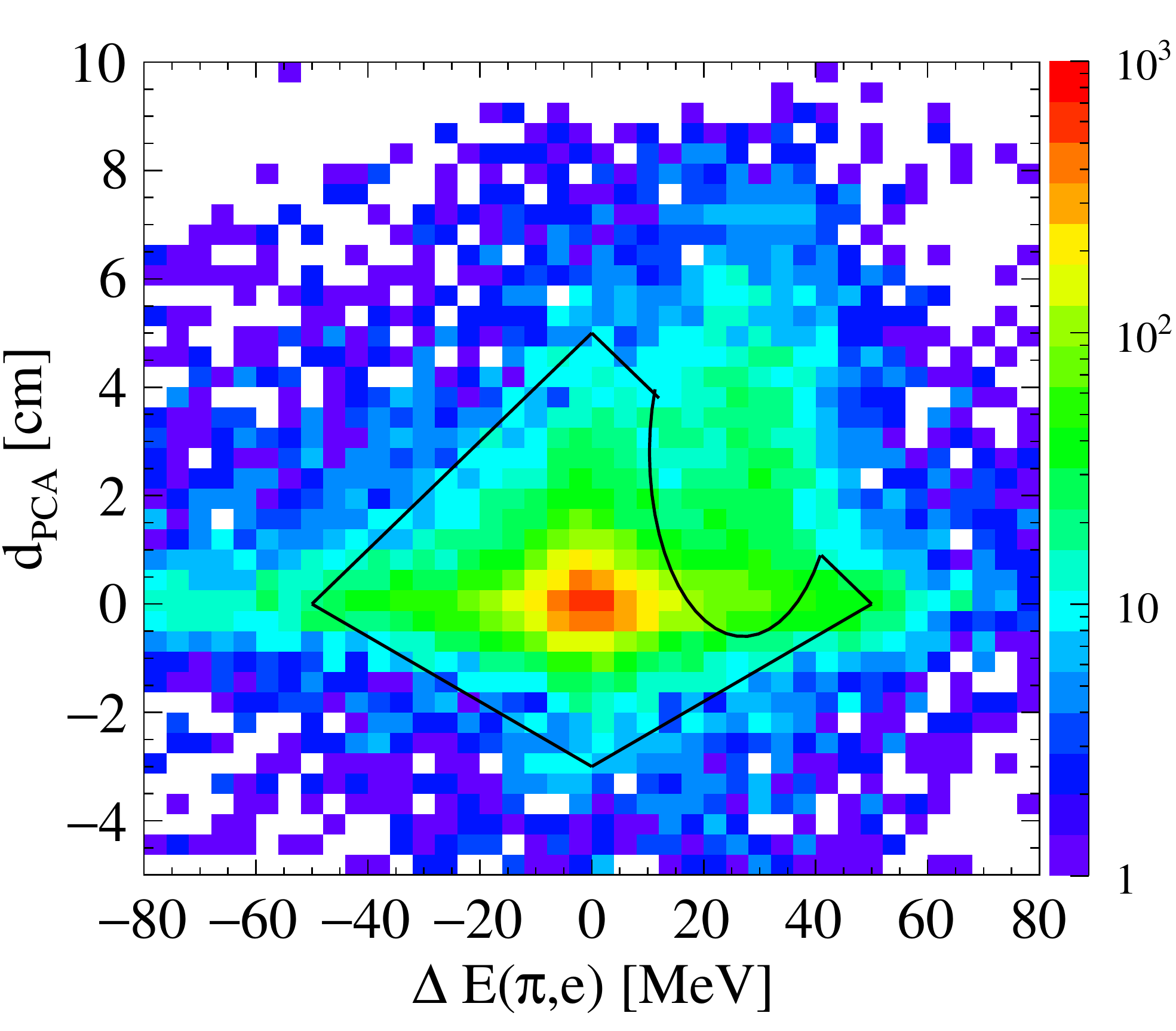}
    \caption{All data events}
  \end{subfigure}
  \caption{Difference between two tracks' distances of closest approach to the $\phi$ decay point vs\. difference between missing energy and momentum in the kaon decay. Events are accepted if they lie inside the region defined by a rhombus with exclusion of an elliptical region, marked with black solid line.}\label{fig:dpca_de_cut}
\end{figure}

To recognize such events, each of the two tracks is extrapolated backwards to its point of closest approach (PCA) to the average $\phi$ vertex. If their distances of closest approach to $\phi$ measured in the $xy$ plane of the detector are denoted as $d_{xy}(\pi)$ and $d_{xy}(e)$, the following variable:
\begin{equation}
  \label{eq:dpca}
  d_{PCA} = d_{xy}(e) - d_{xy}(\pi),
\end{equation}
is sensitive to tracks not originating at the same $\Ks$ decay vertex as in the case of muon track from $\Ks\to\pi^+\pi^-\to \pi\mu\nu$. For the tracks from $\Ks\to\pi e\nu$ originating close to the primary vertex, the $d_{PCA}$ distribution peaks around zero as shown in~\fref{fig:dpca_de_cut}. To enhance the background discrimination power of this variable, it was correlated with a difference between missing energy and missing momentum in a decay defined as:
\begin{equation}
  \label{eq:de_pi_e}
  \Delta E(\pi,e) = E_{miss}(\pi,e) - p_{miss} = E_{K_S}-E_{\pi}-E_{e} - |\vec{p}_{K_S}-\vec{p}_{\pi}-\vec{p}_{e}|,
\end{equation}
where the energies of tracks identified as pion and electron/positron are calculated using respective particle mass hypotheses~\cite{Ambrosino:2006si,daria_memo}. This value is sensitive to all three background components listed above as the electron mass hypothesis is wrong for one of the tracks in such events. For signal events, $\Delta E(\pi,e)$ is expected to be close to zero. Although the resolution of $\Delta E(\pi,e)$ in case of the considered process $\Ks\Kl\to \pi e \nu\;3\pi^0$ is limited due to imperfect knowledge of the decaying $\Ks$ momentum, the two dimensional distributions of $d_{PCA}$ relative to $\Delta E(\pi,e)$ presented in~\fref{fig:dpca_de_cut} exhibit the background components clustered in an off-center region. Events are selected if they lie within the marked rhombus and outside an ellipse enclosing the background-populated area, which is equivalent to the following conditions:
\begin{eqnarray*}
  d_{PCA}/\text{cm} &<& -0.1\cdot\Delta E(\pi,e)/\text{MeV}+5, \\
  d_{PCA}/\text{cm} &>&  0.06\cdot\Delta E(\pi,e)/\text{MeV}-3, \\
  d_{PCA}/\text{cm} &>& -0.06\cdot\Delta E(\pi,e)/\text{MeV}-3, \\
  d_{PCA}/\text{cm} &<& 0.1\cdot\Delta E(\pi,e)/\text{MeV}+5, \\
  \left(\frac{d_{PCA}-2.8\:\text{cm}}{3.4\:\text{cm}}\right)^2 &+& \left(\frac{\Delta E(\pi,e)-27\:\text{MeV}}{16.7\:\text{MeV}}\right)^2 > 1.
\end{eqnarray*}

\subsection{Determination of lepton charge and kaon decay times' difference}\label{sec:charge_and_dt}
Determination of the \Ts-asymmetric ratios of double decay rates defined in Equations~\ref{eq:r2def} and~\ref{eq:r4def} requires that the $\Ks\Kl\to\pi e\nu\;3\pi^0$ dataset is split into two subsamples depending on whether the lepton in the $\Ks$ decay was an electron or a positron. This charge identification is based on the curvature of the track identified as corresponding to a particle of electron mass. Sign of the curvature with respect to the known direction of the magnetic field in KLOE is used to determine the sign of the lepton charge in an event.

Moreover, the $R_2^{exp}$ and $R_4^{exp}$ ratios are functions of the difference between proper times of the two kaon decays. Therefore, for each of the studied events, the following difference must be precisely calculated:
\begin{equation}
  \label{eq:dt_definition}
  \Delta t = t'_{3\pi^{0}} - t'_{\pi e \nu},  
\end{equation}
where $t'_{\pi e \nu}$ and $t'_{3\pi^{0}}$ are times of decays identified by respective final states, expressed in the frames of reference of the decaying kaons. In case of the $\pi e\nu$ decay, decay time is obtained from the path travelled by the kaon between the $\phi$ decay point and a common vertex of the two tracks and the velocity calculated from K meson momentum. As transformation to the kaon rest frame is performed along the travelled path, its proper decay time reads:
\begin{eqnarray}
  \label{eq:ks_proper_time}
  t'_{\pi e\nu} &=& \frac{L}{c\beta}\frac{1}{\gamma} = \frac{Lm_{\kaon}}{c|\vec{p}_{\kaon}|},\\
  L &=& |\vec{\mathbf{v}}_{\phi} - \vec{\mathbf{v}}_{\pi e \nu}|,
\end{eqnarray}
where $\vec{\mathbf{v}}_{\phi}$ and $\vec{\mathbf{v}}_{\pi e \nu}$ denote position vectors of the respective vertices.

The time of the other kaon decay into $3\pi^0$ may be estimated in a similar manner using the reconstructed decay vertex position or, alternatively, using the time resulting directly from trilaterative reconstruction (see~\sref{sec:gps_kloe}). However, due to limited accuracy of this reconstruction based on times and positions recorded by the electromagnetic calorimeter (see~\fref{fig:resolutions_nofit}), neither of this methods has a time resolution comparable with the one of $t_{\pi e \nu}$ which is at the level of 1~$\tau_S$. As the estimation of proper $\Kl\to 3\pi^{0}$ decay time analogical to~\eref{eq:ks_proper_time} depends on both the decay vertex location (through travelled path $L_{\Kl}$) and kaon momentum, its resolution can be improved by means of a kinematic fit imposing additional constraints binding $L_{K_L}$ and $\vec{p}_{\Kl}$ as described in~\sref{sec:kinfit}.

\subsection{Rejection of background from $\Ks\to\pi^0\pi^0$}\label{sec:ks2pi0_rejection}
\begin{figure}[h!]
  \centering
  \begin{tikzpicture}
    \node[anchor=south west,inner sep=0] at (0,0) 
    {\includegraphics[width=0.45\textwidth]{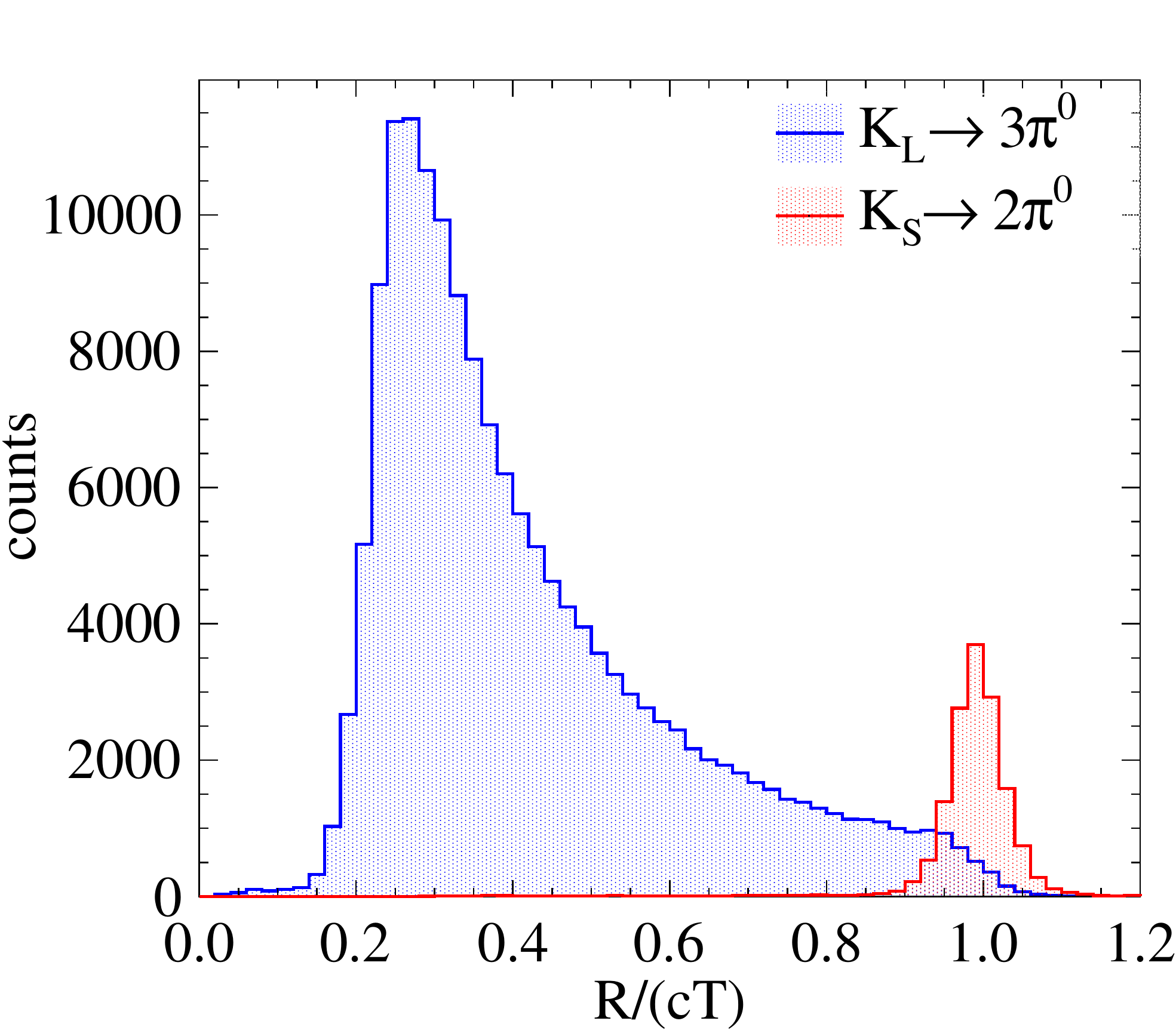}};
    \draw[black, thick, dashed] (5.22,0.8) -- (5.22,4.0);
  \end{tikzpicture}
  \hspace{1em}
    \begin{tikzpicture}
    \node[anchor=south west,inner sep=0] at (0,0) 
    {\includegraphics[width=0.45\textwidth]{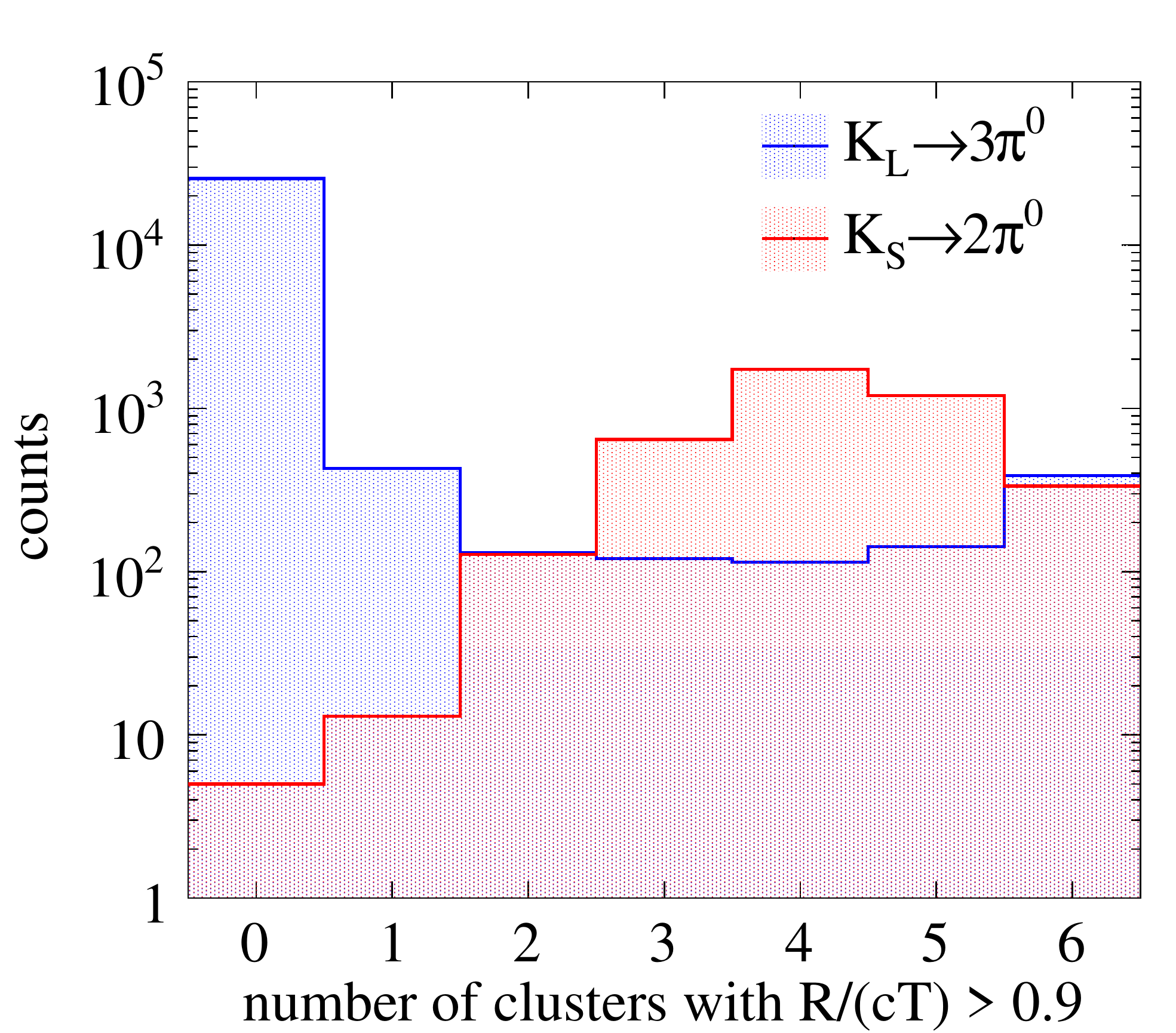}};
    \draw[black, thick, dashed] (2.65,0.8) -- (2.65,5.46);
  \end{tikzpicture}  
  \caption{Left: Ratio of distance travelled by a hypothetical prompt photon from the primary interaction vertex to an EMC cluster to the product of cluster recording time and velocity of light. Distributions are presented for MC-simulated events of $\Ks\to\pi^0\pi^0$ and $\Kl\to 3\pi^0$. Right: multiplicity of clusters with $R/(cT_{clu})$ larger than 0.9 (see the dashed line in the left panel). Dashed vertical lines denote selection cuts used in the analysis.}\label{fig:rtc} 
\end{figure}

After the selection discussed up to this point, as tested with KLOE MC simulations, about \SI{12}{\percent} of the chosen events are constituted by the $\Ks\to\pi^0\pi^0$ and $\Kl\to\pi e\nu$ processes where the $\Ks$ decay along with additional EMC clusters originating from cluster splitting or machine background was misidentified as an early $\Kl\to 3\pi^0$ decay. Such cases are characterized by a small difference between both kaon decay times as otherwise the semileptonic decay vertex would not be recorded in the fiducial volume around the $\phi$~decay point. To remove this background component, EMC clusters coming from prompt photons (i.e.\ photons which originate close to the primary interaction vertex) are identified by testing the value of $R/(cT_{clu})$~\cite{kloe_memo_146}, where $R$ is the distance between average $\phi$~decay location and position of an EMC cluster, $T_{clu}$ is the cluster recording time and $c$ is the velocity of light. This ratio is close to unity for prompt photons from $\Ks\to\pi^0\pi^0$ which differentiates them from those created in $\Kl$ decay occurring further in the detector volume as shown in~\fref{fig:rtc} (left).

In order to detect the $\Ks\to\pi^0\pi^0$ events, clusters from prompt photons for which $R/(cT_{clu})>0.9$ (compare~\fref{fig:rtc} (left)) were counted in each event. The resulting prompt photon multiplicities checked with MC simulations for $\Ks\to\pi^0\pi^0$ and $\Kl\to 3\pi^0$ decays are presented in the right panel of~\fref{fig:rtc}. Strong discrimination of the $\pi^0\pi^0$ background is obtained by rejecting events which contain two or more clusters for which $R/(cT_{clu})>0.9$.
%
% TODO: odnosnik do sekcji z wydajnosciami i do sekcji gdzie dyskutuje sie uzyty zakres delta t
%
The cost of this requirement is a significant reduction of efficiency for the $\Kl\to 3\pi^0$ signal events with an early $\Kl$ decay, corresponding to time differences between both kaon decays in the range (0;12)~$\tau_S$. This region, however, is not crucial for the \Ts~symmetry test considered in this work.

\subsection[Rejection of background from $\Ks\to\pi^+\pi^-(\to\pi\mu\nu)$\newline and \mbox{$\Ks\to\pi^+\pi^-(\gamma)$} decays]{Rejection of background from $\Ks\to\pi^+\pi^-(\to\pi\mu\nu)$ and \mbox{$\Ks\to\pi^+\pi^-(\gamma)$} decays}\label{sec:pimu_rejection}
\begin{figure}[h!]
  \captionsetup[subfigure]{justification=centering}
  \centering
  \begin{subfigure}{0.45\textwidth}
  \includegraphics[width=1.0\textwidth]{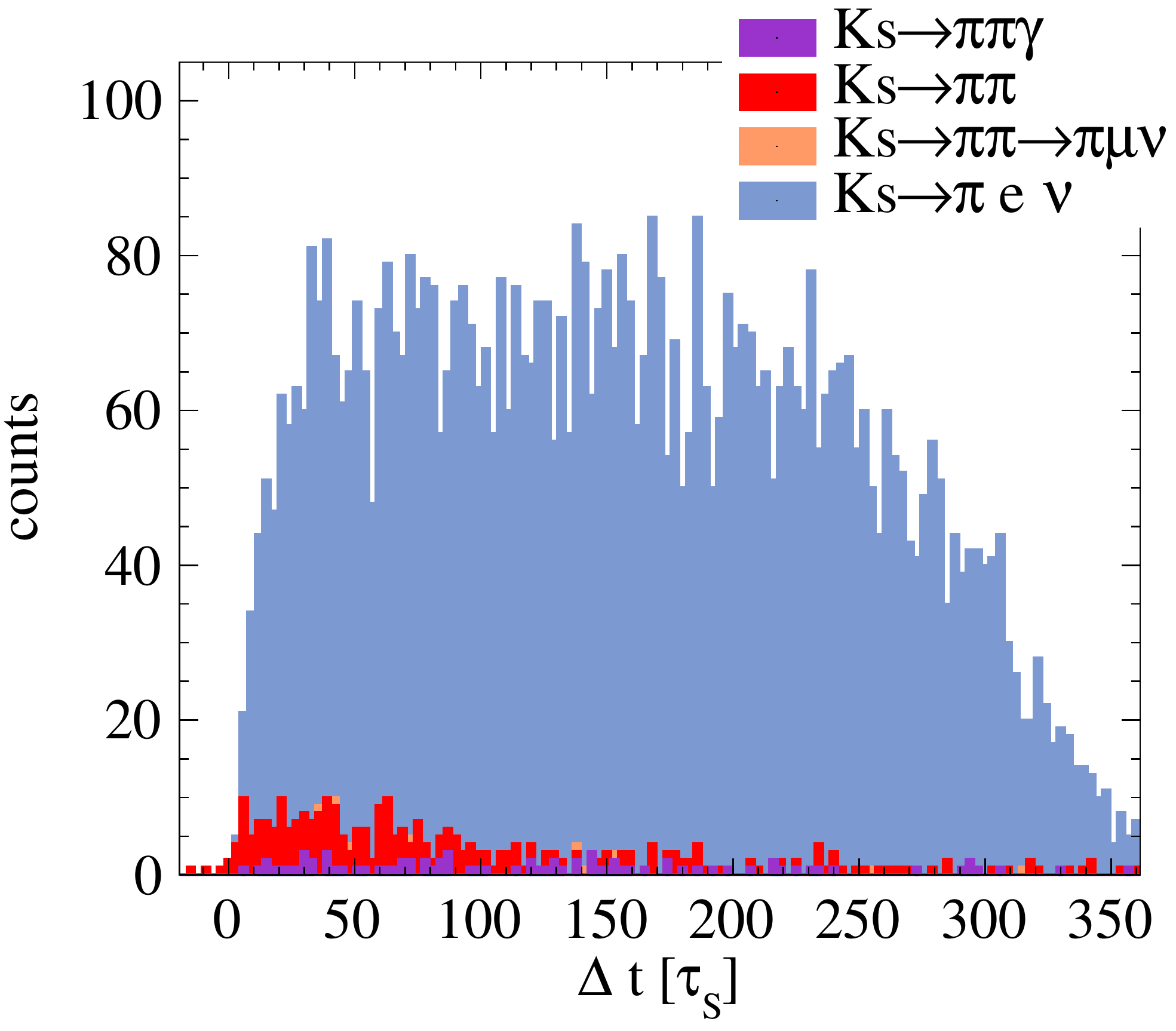}
  \caption{Events with a track identified as electron.}
  \end{subfigure}
  \begin{subfigure}{0.45\textwidth}
  \includegraphics[width=1.0\textwidth]{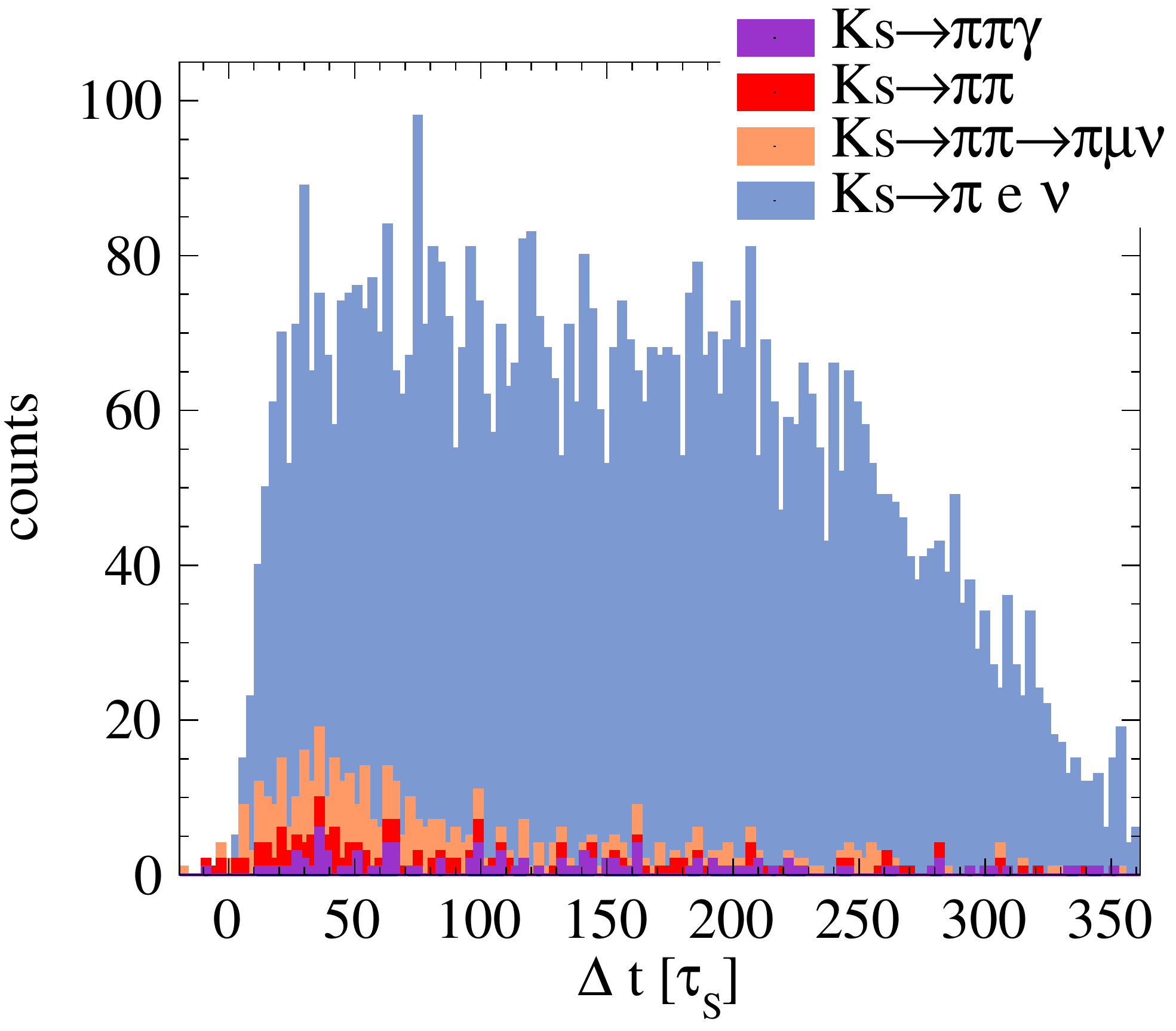}
  \caption{Events with a track identified as positron.}
  \end{subfigure}
  \caption{Stacked MC-based distributions of kaons' decay time difference for signal and background events surviving the selection up to (inclusive) the cut on $d_{PCA}$ and $\Delta E(\pi,e)$.}\label{fig:csps-dt-before}
\end{figure}

%
% Zaktualizować S/B
%
Although the cuts presented in the previous section significantly improve signal to background ratio in the selected event sample (from about 2.7 to about 11.3), almost \SI{10}{\percent} of the selected events was still constituted by the $\Ks\to\pi^+\pi^-(\gamma)$ and $\Ks\to\pi^+\pi^-\to\pi\mu\nu$ decays. The remaining background is especially problematic due to its presence over a large range of time differences between kaons' decays and strong charge asymmetry visible in the two MC-based distributions in~\fref{fig:csps-dt-before}, plotted separately for events with negative and positive charge of the particle identified as lepton in the $\Ks\to\pi e\nu$ decay.

In order to reduce the remaining background components, a set of classifiers based on the properties of tracks and clusters was prepared to distinguish electrons and positrons from pion and muons. Classification relied primarily on the fact that electromagnetic showers created in the KLOE calorimeter by interactions of electrons are considerably different than in case of charged pions and muons. While the former tend to release a large amount of energy at their first interaction in the calorimeter, muons and charged pions deposit energy rather uniformly along their path travelled through the material, with a possible greater deposit at a large depth due to the Bragg peak~\cite{graziani_anns}. To probe the structure of such electromagnetic showers, the EMC segmentation into layers was used (see~\sref{sec:emc}). As every calorimeter module in KLOE is composed of five layers located at increasing depths, energies deposited by a given particle in particular layers as well as the total number of layers with a non-zero deposited energy provide information specific to the type of interacting particle. Additionally, the total deposited energy of the whole EMC cluster and the particle momentum estimated using the associated DC track were also used for classification. 

Two classifiers were constructed using a neural network implementation from the Toolkit for Multivariate Analysis~\cite{Hocker:2007ht}. Each classifier acted on an ensemble constituted by a DC track and its associated EMC cluster and attempted to distinguish those originating from an electron/positron and from a pion ($e/\pi$ classifier) or a muon ($e/\mu$ classifier). Multilayer perceptrons (MLPs) with five inputs and one output were employed as classifiers. Details of input variables and implemented classifiers are presented in~\aref{appendix:clasifiers}.

\begin{figure}[h!]
  \centering
  \includegraphics[width=0.45\textwidth]{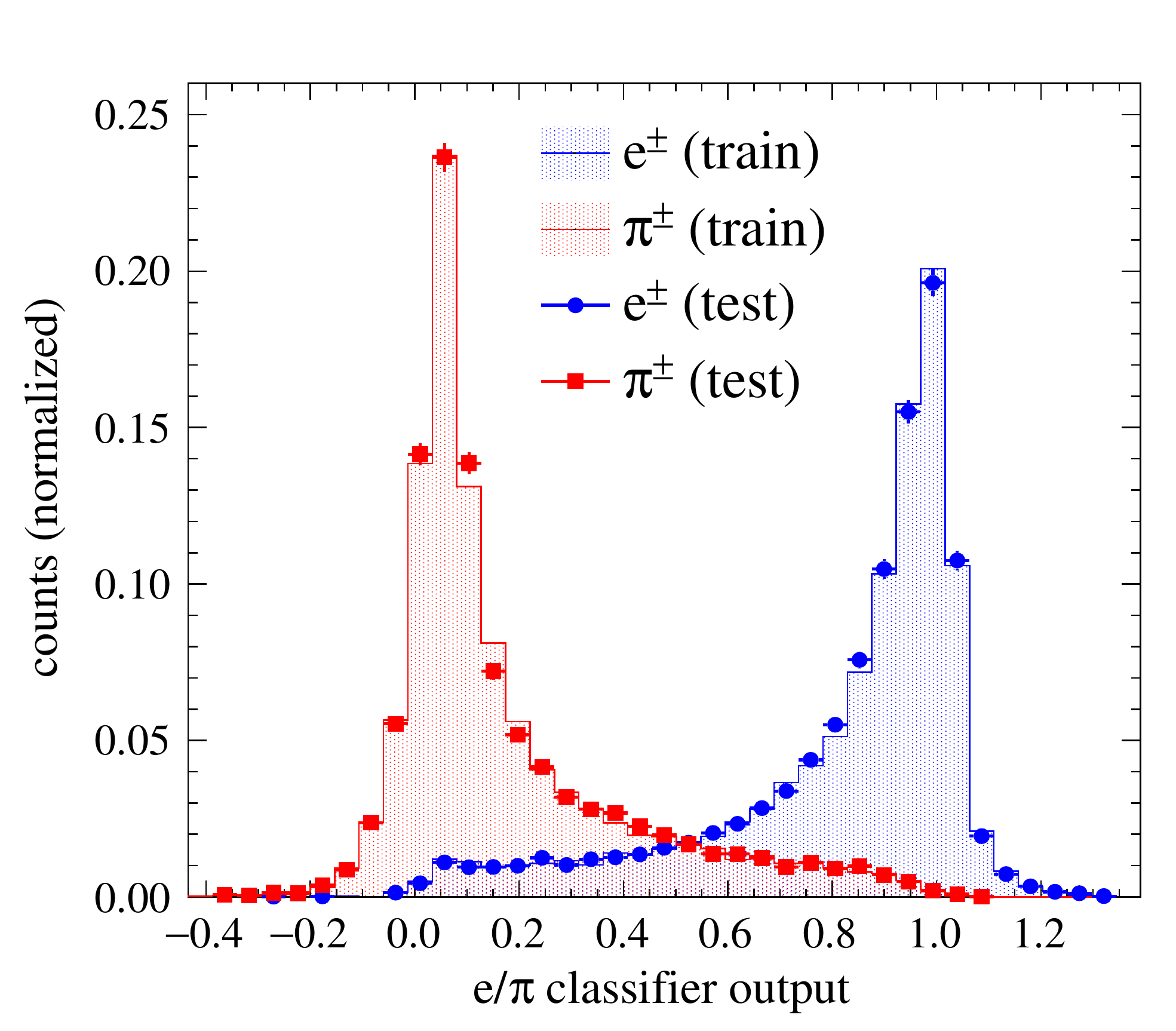}
  \hspace{1em}  
  \includegraphics[width=0.45\textwidth]{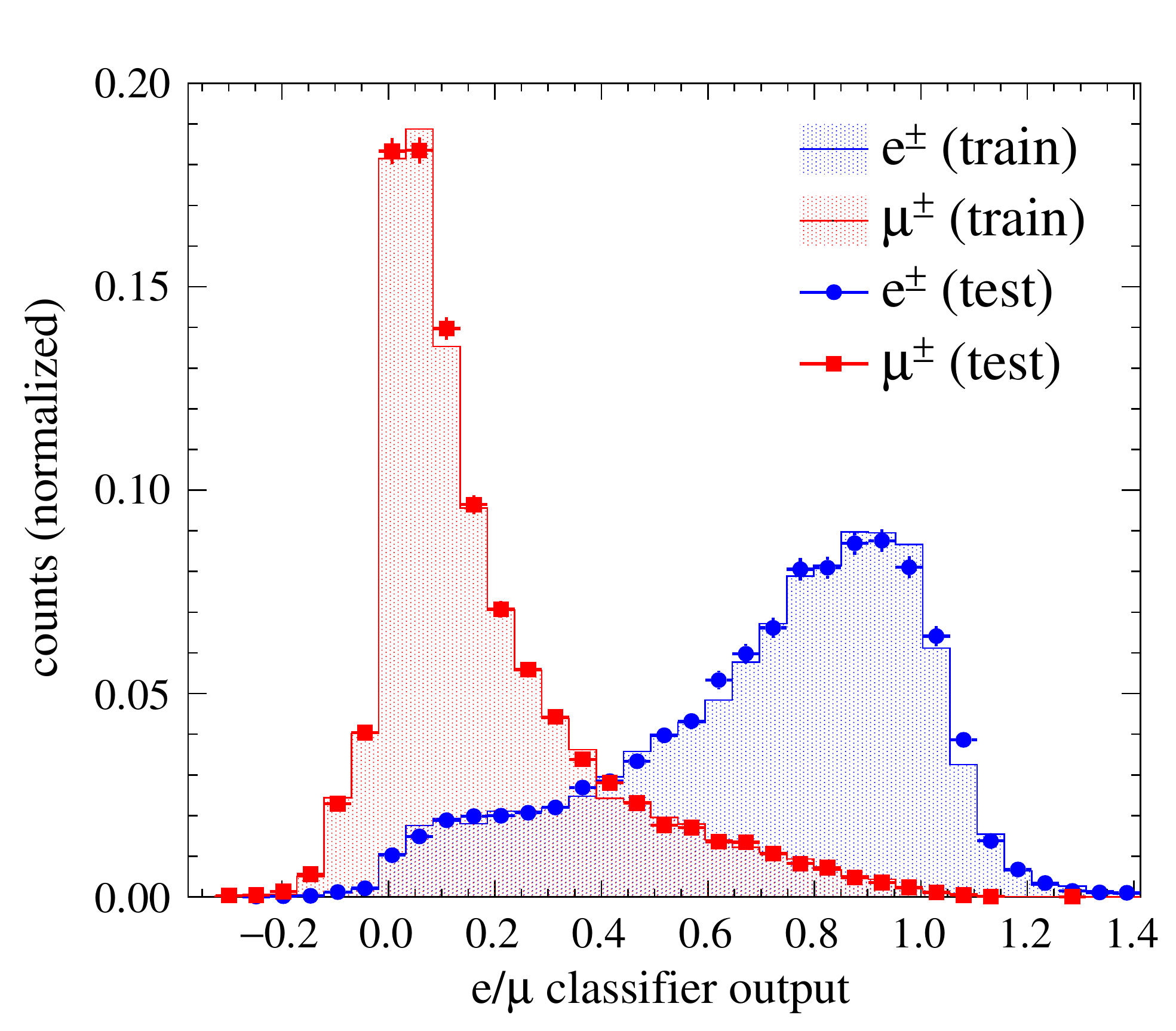}
  \caption{Output values of artificial neural networks used as electron/pion classifier (left) and as electron/muon classifier (right), obtained with test and training data samples. }\label{fig:mlp_outputs}
\end{figure}

\begin{figure}[h!]
  \centering
  \includegraphics[width=0.45\textwidth]{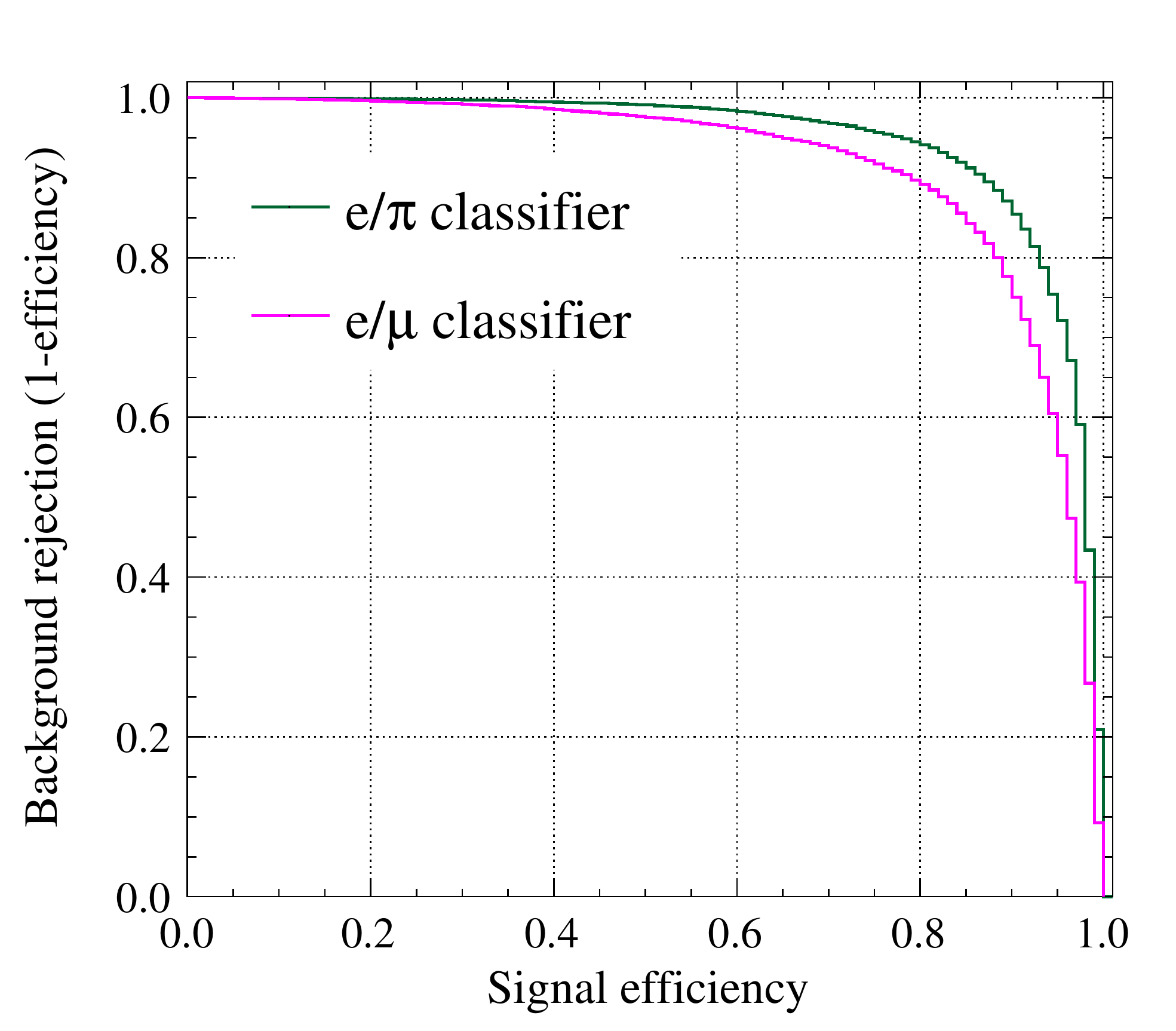}
  \caption{Comparison of background discrimination power and signal efficiency for the $e/\pi$ and $e/\mu$ classifiers.}\label{fig:mlp_rocs}
 \end{figure}

In order to avoid adjusting of the MLPs to artifacts specific to MC simulations, the classifiers were trained solely using data. As the properties of track and clusters under study do not depend on the origin of the particles, semileptonic decays of the long-lived neutral kaon were used to obtain training and test samples of electron/positron, muon and charged pion tracks. To this end, the event selection of $\Ks\Kl\to\pi^+\pi^-\;\pi e\nu$ described in~\sref{sec:t-analysis-2} was used directly to obtain a $\Kl\to\pi e \nu$ sample with a purity at the level of \SI{99}{\percent}. Subsequently, the selection cuts were adjusted to select semileptonic $\Kl$ decays with a muon, yielding the second training sample with a similar purity. Output values of the MLPs used for both classifiers, obtained with test and training samples, are shown in~\fref{fig:mlp_outputs}. The agreement between distributions resulting from training and test data shows that no MLP overtraining effects are present. Although the background discrimination power of the two classifiers, compared in~\fref{fig:mlp_rocs}, is slightly worse in case of electron/muon, both allow for rejection of a large part of the background related to pion and muon tracks while preserving more than \SI{90}{\percent} of signal events with electron/pion tracks.

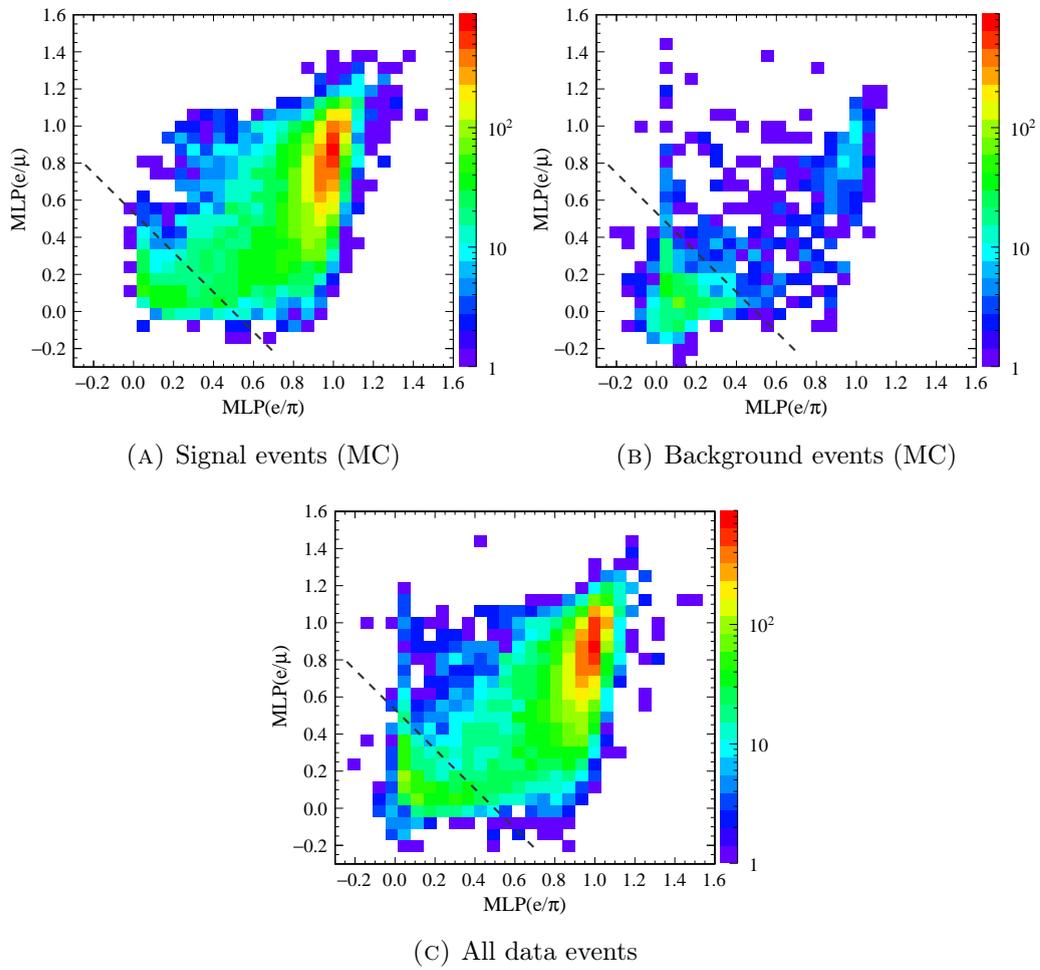
\begin{figure}[h!]
  \centering
\captionsetup[subfigure]{justification=centering}
  \begin{subfigure}{0.45\textwidth}
    \begin{tikzpicture}
      \node[anchor=south west,inner sep=0] at (0,0) 
      {\includegraphics[width=1.0\textwidth]{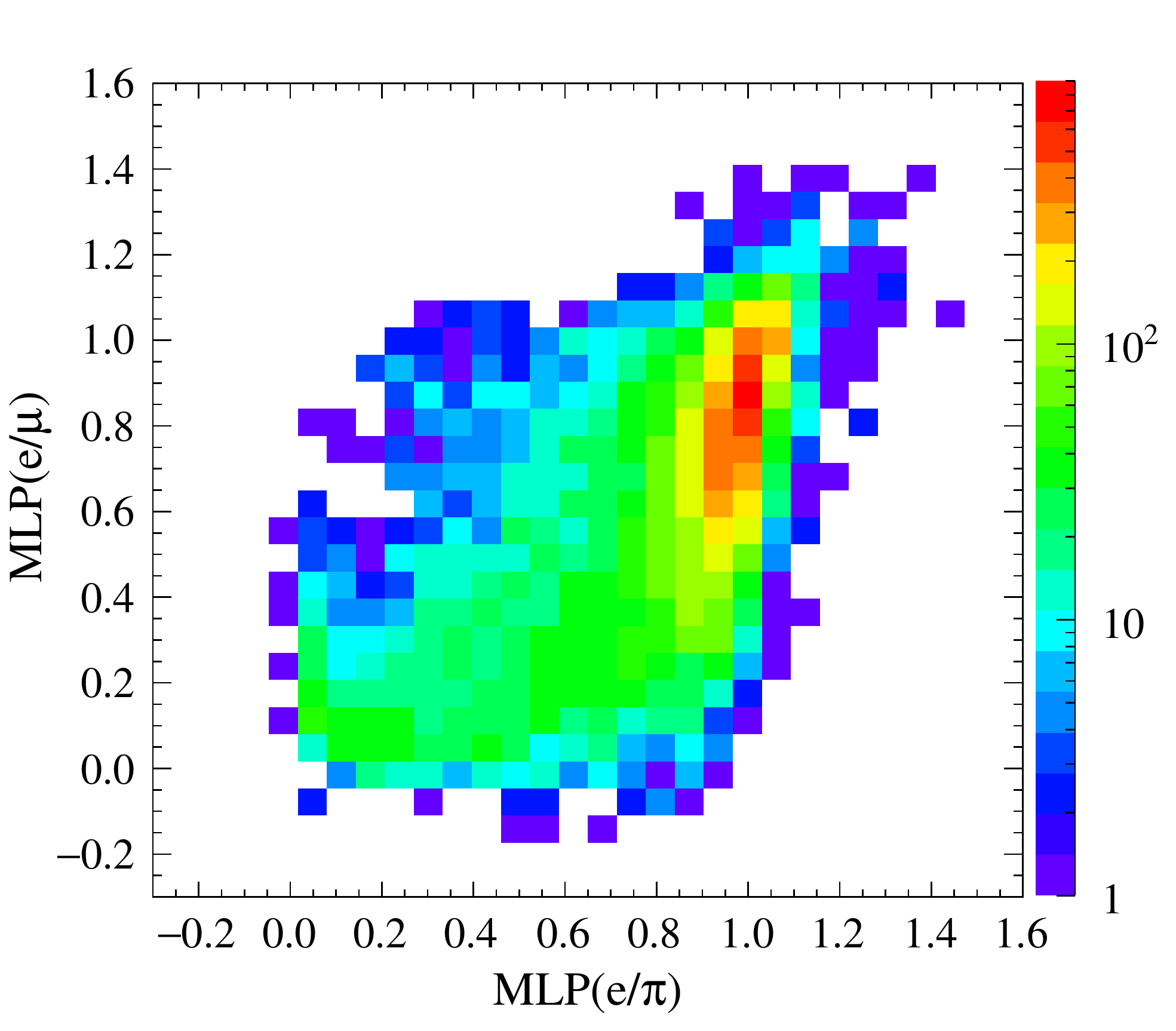}};
      \draw[black, thick, dashed] (3.5,1.0) -- (1.0,3.5);
    \end{tikzpicture}  
    \caption{Signal events (MC)}
  \end{subfigure}
  \begin{subfigure}{0.45\textwidth}
    \begin{tikzpicture}
      \node[anchor=south west,inner sep=0] at (0,0) 
      {\includegraphics[width=1.0\textwidth]{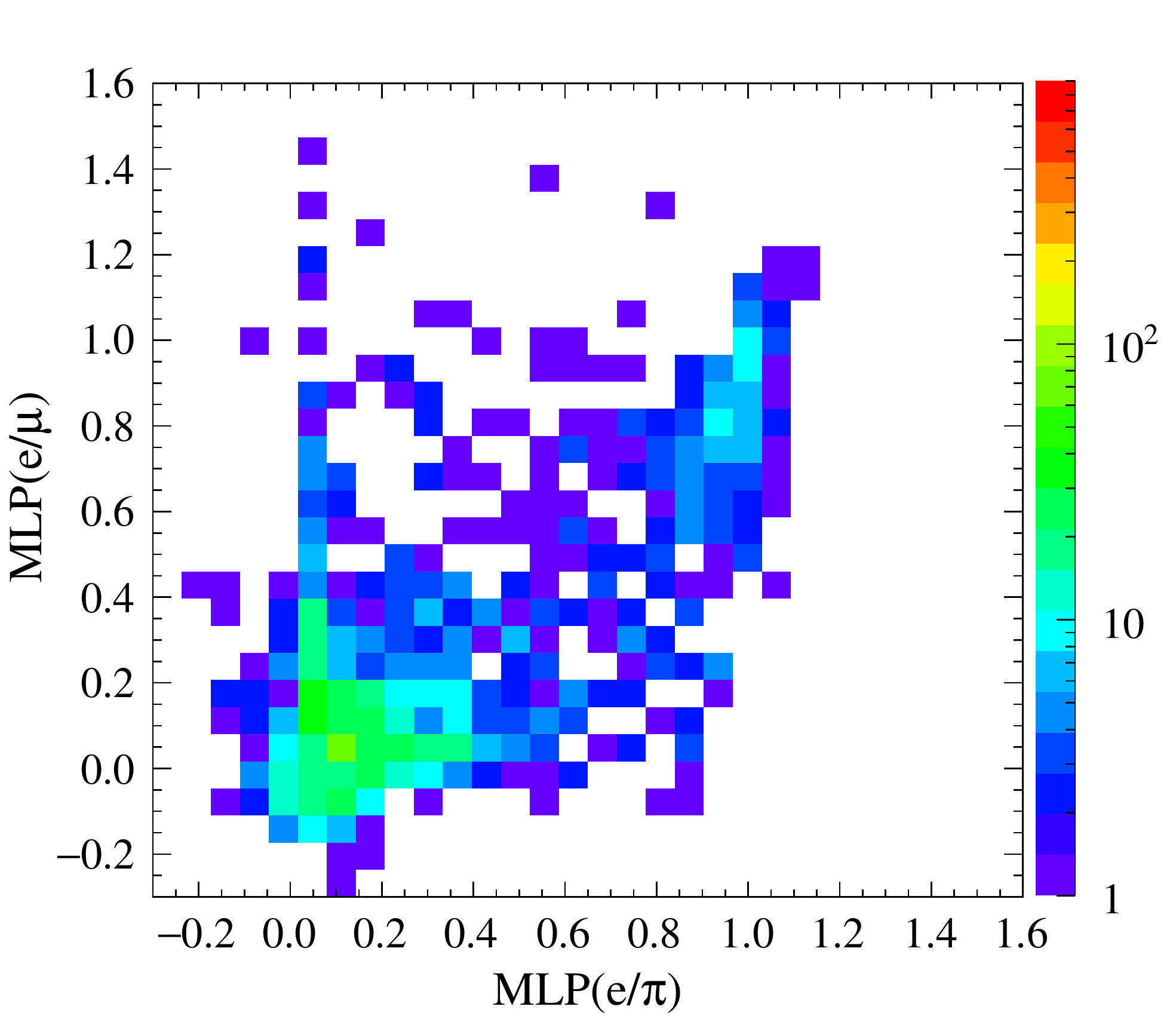}};
      \draw[black, thick, dashed] (3.5,1.0) -- (1.0,3.5);
    \end{tikzpicture}  
    \caption{Background events (MC)}
  \end{subfigure}
  \begin{subfigure}{0.45\textwidth}
    \begin{tikzpicture}
      \node[anchor=south west,inner sep=0] at (0,0) 
      {\includegraphics[width=1.0\textwidth]{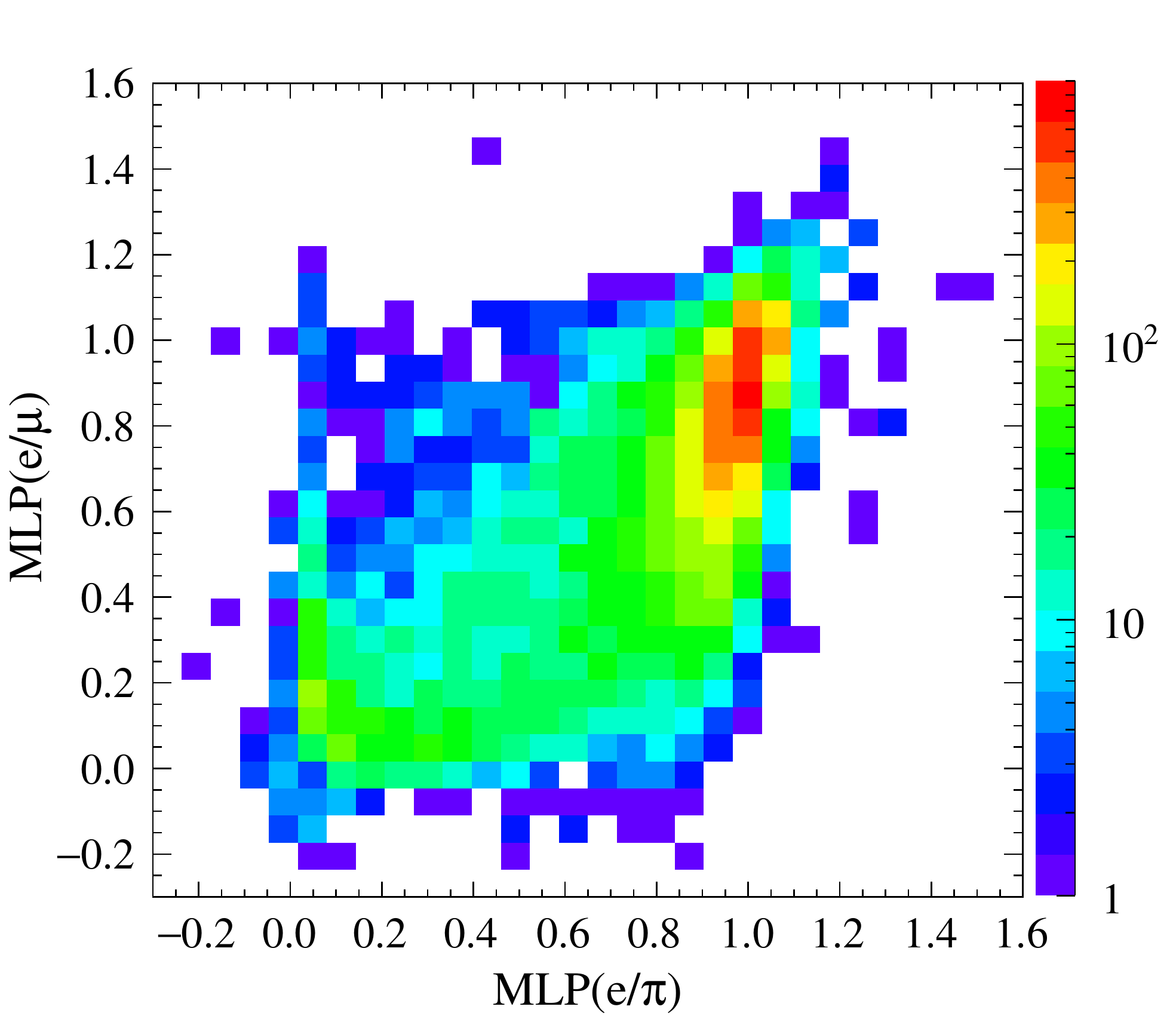}};
      \draw[black, thick, dashed] (3.5,1.0) -- (1.0,3.5);
    \end{tikzpicture}  
    \caption{All data events}
  \end{subfigure}
  \caption{Relative outputs of the $e/\pi$ and $e/\mu$ classifiers applied to the track identified as electron/positron in the selected event sample. Events are retained if they lie above the dashed line.}\label{fig:mlp_2d}
\end{figure}

 Discrimination of background was performed by subsequent application of the $e/\pi$ and $e/\mu$ classifiers to the ensemble constituted by DC track identified as electron or positron in the previous analysis steps and its associated EMC cluster. A two-dimensional distribution of both classifier output values displayed in~\fref{fig:mlp_2d} reveals good separation of signal and background events. Events are chosen for the final $\Ks\Kl\to\pi e\nu\;3\pi^0$ sample if they satisfy the following constraint on the sum of both classifiers' outputs:
 \begin{equation*}
   \text{MVA}(e,\pi) + \text{MVA}(e,\mu) > 0.5,
 \end{equation*}
corresponding to an anti-diagonal cut marked in~\fref{fig:mlp_2d}.
 
\begin{figure}[h!]
  \captionsetup[subfigure]{justification=centering}
  \centering
  \begin{subfigure}{0.45\textwidth}
  \includegraphics[width=1.0\textwidth]{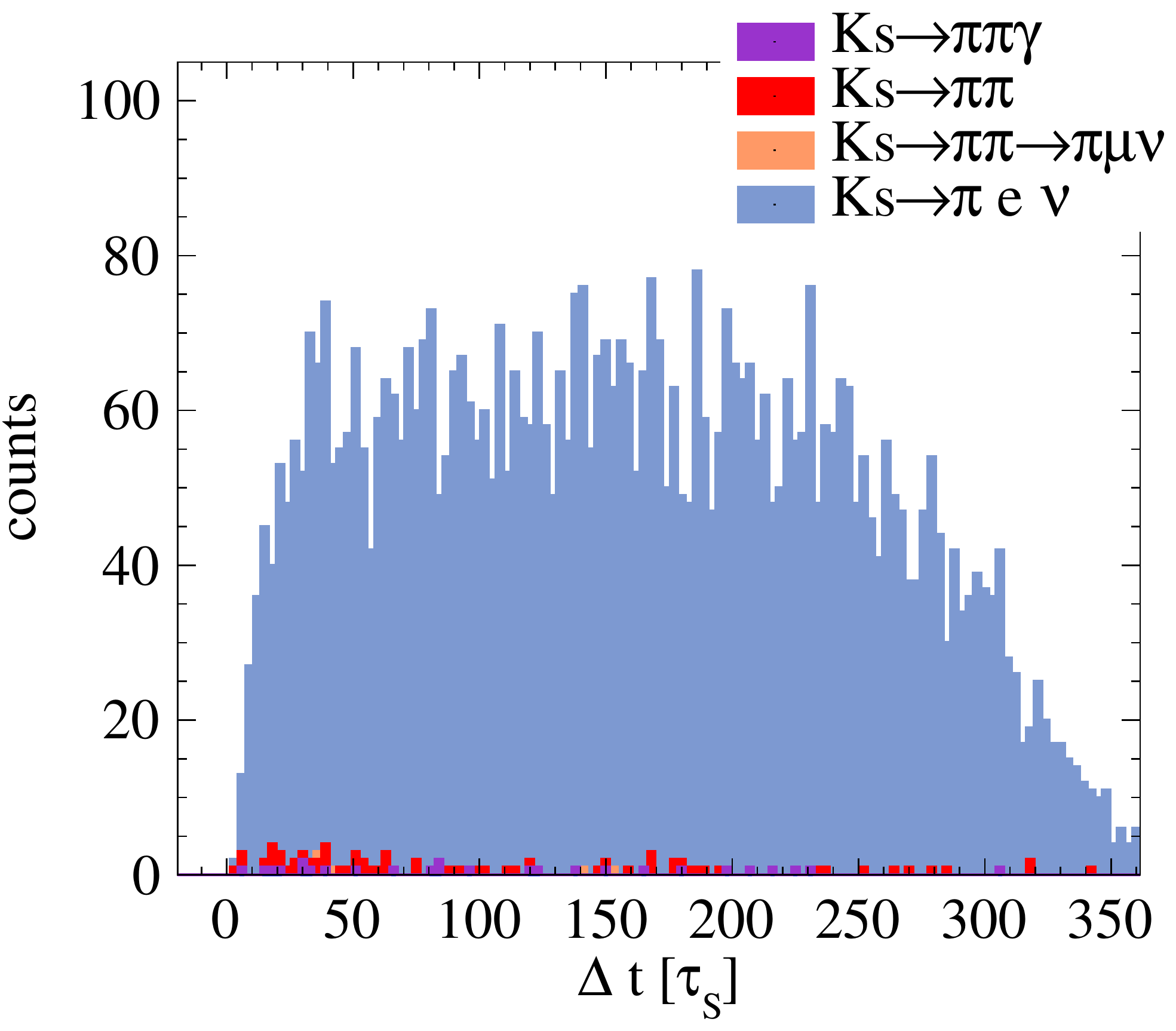}
  \caption{Events with a track identified as electron.}
  \end{subfigure}
  \begin{subfigure}{0.45\textwidth}
  \includegraphics[width=1.0\textwidth]{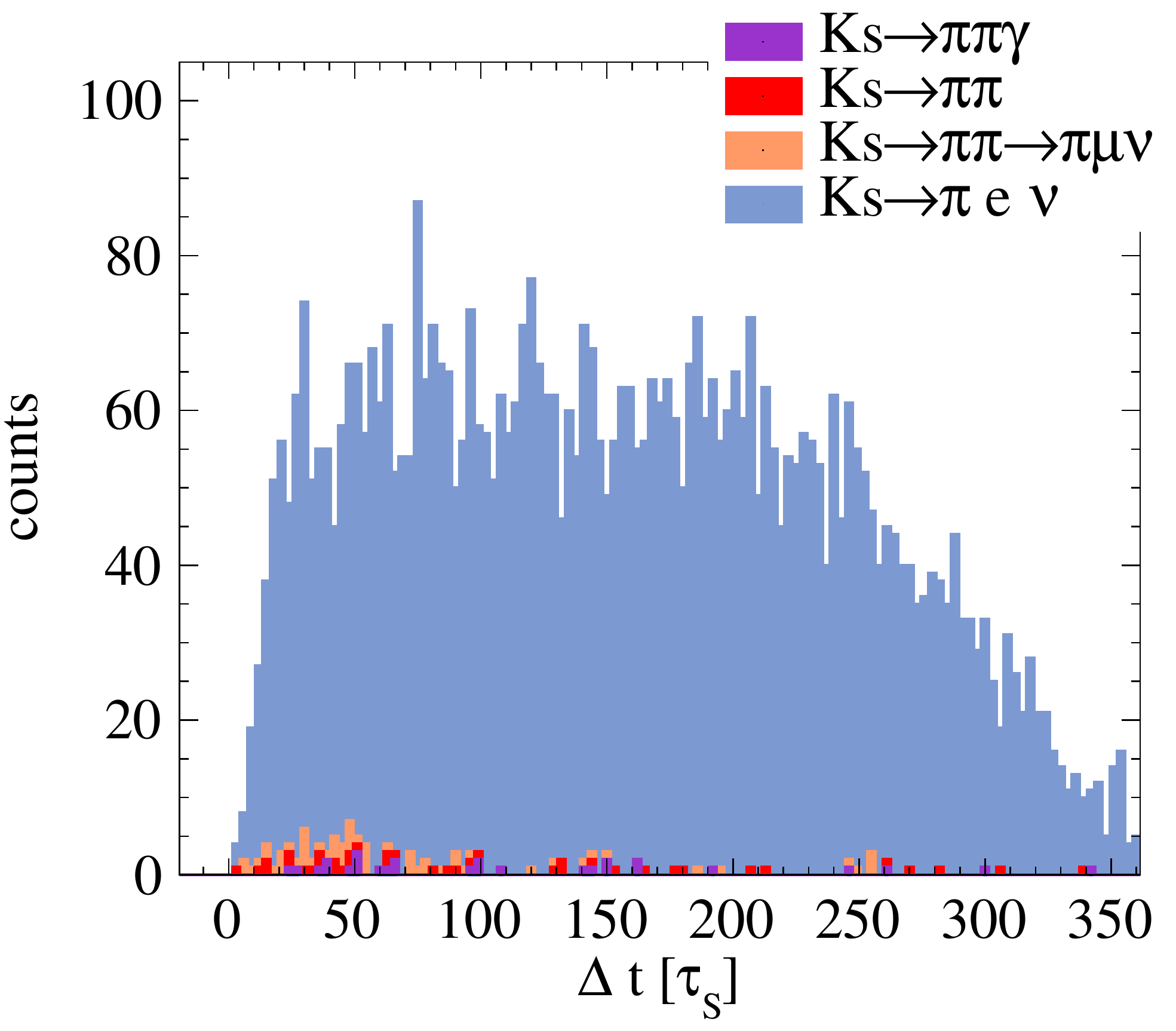}
  \caption{Events with a track identified as positron.}
  \end{subfigure}
  \caption{Stacked MC-based distributions of kaons' decay time difference for signal and background events after application of $e/\pi$ and $e/\mu$ ANN-based track classifiers.}\label{fig:csps-dt-after}
\end{figure}

After the track and cluster classification described above, the signal to background ratio of the surviving event sample is increased from about 11.3 to 33.5. The contribution of the charge-asymmetric background components $\Ks\to\pi^+\pi^-(\gamma)$ and $\Ks\to\pi^+\pi^-\to\pi\mu\nu$ to the $\Delta t$ distributions used further in this study is strongly reduced as shown in~\fref{fig:csps-dt-after}.

\subsection{Kinematic fit}\label{sec:kinfit}
A distinctive property of the trilaterative reconstruction of $\Kl\to 3\pi^{0}\to 6\gamma$ decay is its capability to determine the time of decay in addition to vertex position, based predominantly on the event timing. Estimation of decay time based on the distance between the kaon decay and creation point (alike the one performed for the other decay, see~\eref{eq:ks_proper_time}) may therefore be considered as an independent information. Constraining the event topology and timing to enforce equality of both time estimates thus allows for an reduction of the errors on quantities measured in the event, and consequently, for an improvement of the resolution of decay time itself. This aim is achieved with a kinematic fit, in which the following 39 experimentally measured quantities are varied:
\begin{itemize}
\item $x,y,z$, $t$ and $E_{dep}$ for each of the 6 EMC clusters identified as corresponding to the photons from $\Kl\to 3\pi^{0}\to 6\gamma$,
\item 3 Cartesian coordinates of the average $\phi$ decay point,
\item 3 momentum components for each of the two products of the $\Ks\to\pi e\nu$ decay,
\end{itemize}
in order to find a configuration which best satisfies a set of 3 constraints. First constraint is based on the aforementioned agreement between $\Kl$ decay time estimated using trilaterative decay reconstruction ($t_{3\pi^{0}}$) and using the path travelled by the kaon, i.e.:
\begin{equation}
  \label{eq:constraint_vts}
  C_1 = \frac{|\vec{\mathbf{v}}_{3\pi^{0}} - \vec{\mathbf{v}}_{\phi}|}{\upsilon_{K_L}} - t_{3\pi^{0}} = 0,
\end{equation}
where $\vec{\mathbf{v}}_{\phi}$ and $\vec{\mathbf{v}}_{3\pi^{0}}$ are position vectors of the respective vertices and $\upsilon_{K_L}$ is the velocity of the kaon obtained from its momentum.

The second constraint imposed in the kinematic fit requires that invariant mass of the decaying $\Kl$, reconstructed using the momenta of 6 photons, is in agreement with mass of a neutral K meson:
\begin{equation}
  \label{eq:constraint_m6g}
  C_2 = \text{M}(6\gamma) - m_{\kaon} = {\left( \sum_{i=1}^{6} \abs{\vec{p_{\gamma}}^{(i)}} \right)}^2 - \abs{\sum_{i=1}^{6} \vec{p_{\gamma}}^{(i)}}^{2} - m_{\kaon} = 0.
\end{equation}

Finally, the last constraint refers to the semileptonic decay and enforces minimization of the following variable based on the mass attributed to the track identified as electron:
\begin{equation}
  \label{eq:constraint_mtrk}
  C_3 = \text{M}_{trk}^2(e) = \left(E_{\Ks}-E_{\pi}-\abs{\vec{p}_{miss}(\pi,e)}\right)^2 - \abs{\vec{p}_e}^2 = 0,
\end{equation}
where $E_{\Ks}$ and $E_{\pi}$ are the energies of the decaying kaon and the produced charged pion, respectively, $\vec{p}_{miss}(\pi,e)$ is the momentum missing in the decay with an assumption that the products were a charged pion and an electron, and $\vec{p}_e$ is the momentum attributed to the DC track identified as $e^{\pm}$. This variable, expected to be close to zero for semileptonic events, is sensitive to both momentum resolution of the DC tracks corresponding to decay product and to accuracy of $\Ks$ momentum estimation coming from the $\Kl$ decay vertex reconstruction.

The fit was performed as an iterative procedure based on the constrained least squares method~\cite{optimization}. In each step the updated values of measured parameters were used to determine new estimates of the $\Kl\to 3\pi^0$ decay vertex and time with the trilaterative technique, and, subsequently, new estimates of both kaons' momenta were calculated using the procedure described in~\sref{sec:ks_from_kl}. Only then were the values of the constraints recalculated and a set of corrections $\mathbf{\Delta x}$ to the vector of measured quantities $\mathbf{x}$ was found which minimizes the following $\chi^2$-like function:
\begin{equation}
  \label{eq:general_chisq}
  \chi^2 = \mathbf{\Delta x}^T \mathbf{V}^{-1} \mathbf{\Delta x} + \mathbf{\lambda}^T(\mathbf{D\Delta x} + \mathbf{C}(\mathbf{x})),
\end{equation}
where $\mathbf{V}$ is a covariance matrix of the varied parameters, $\mathbf{\lambda}$ is a vector of Lagrange multipliers, $\mathbf{D}$ denotes a matrix of derivatives of the constraints $C_{1-3}$ over the varied parameters, and $\mathbf{C}$ stands for a vector of constraint values at point $\mathbf{x}$ in the parameter space.
%The exact recipe for the minimization procedure used in the kinematic fit is described in~\aref{appendix:kinfit}.

\begin{figure}[h!]
  \centering
  \includegraphics[width=0.45\textwidth]{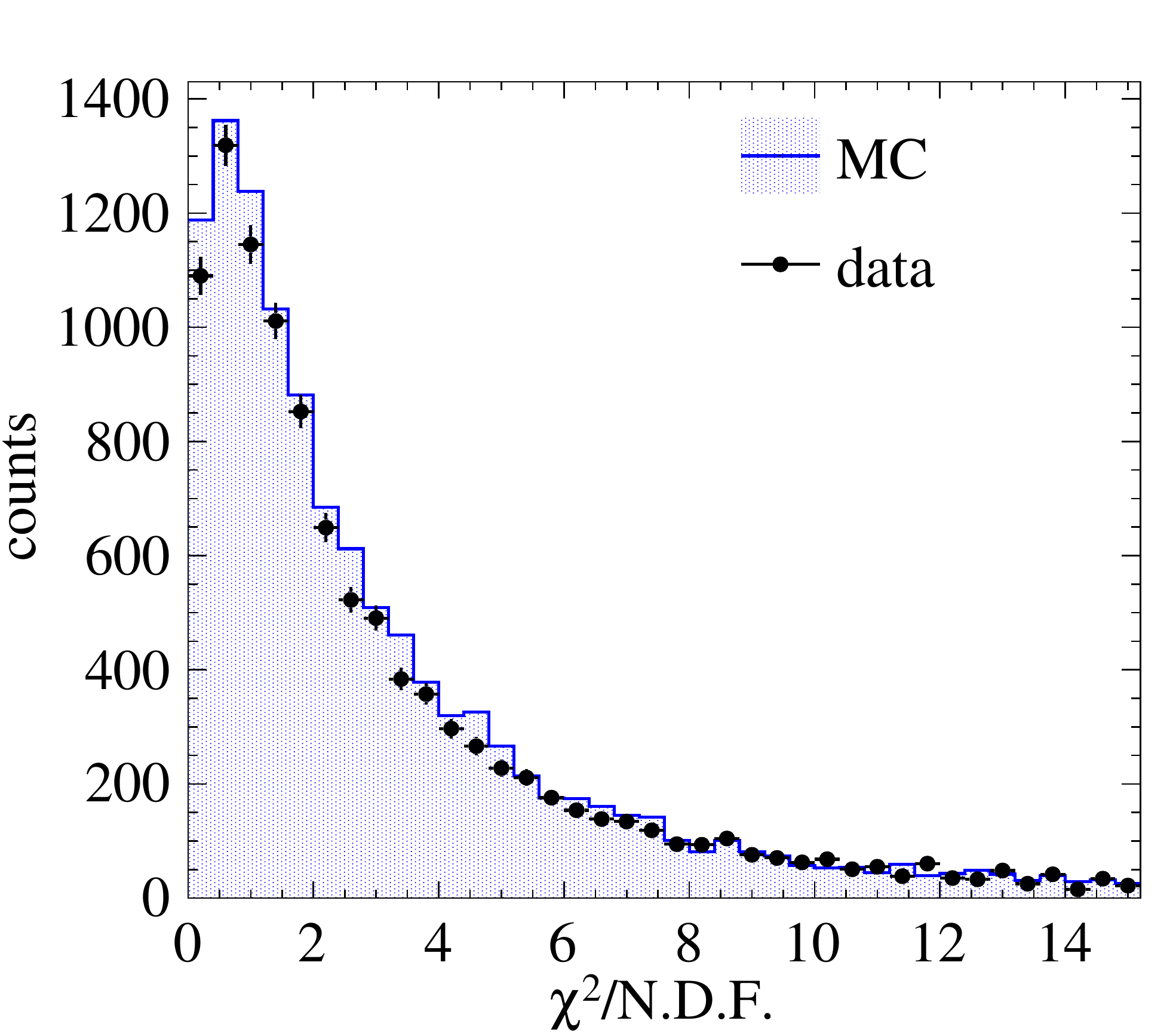}
  \caption{Distribution of $\chi^{2}$ normalized to the number of degrees of freedom resulting from the kinematic fit.}\label{fig:chisq}
\end{figure}

The kinematic fit was applied to all events surviving the selection steps described in the previous Sections. The reason for its introduction at this stage is that the first constraint~(\eref{eq:constraint_vts}), crucial for improvement of reconstruction of $\Kl$ travelled path length and decay time, relies on measurement of the 6 photons' interaction time in the KLOE EMC with a correct event start time, which is only determined after the time of flight analysis.

\begin{figure}[h!]
  \centering
  \includegraphics[width=0.45\textwidth]{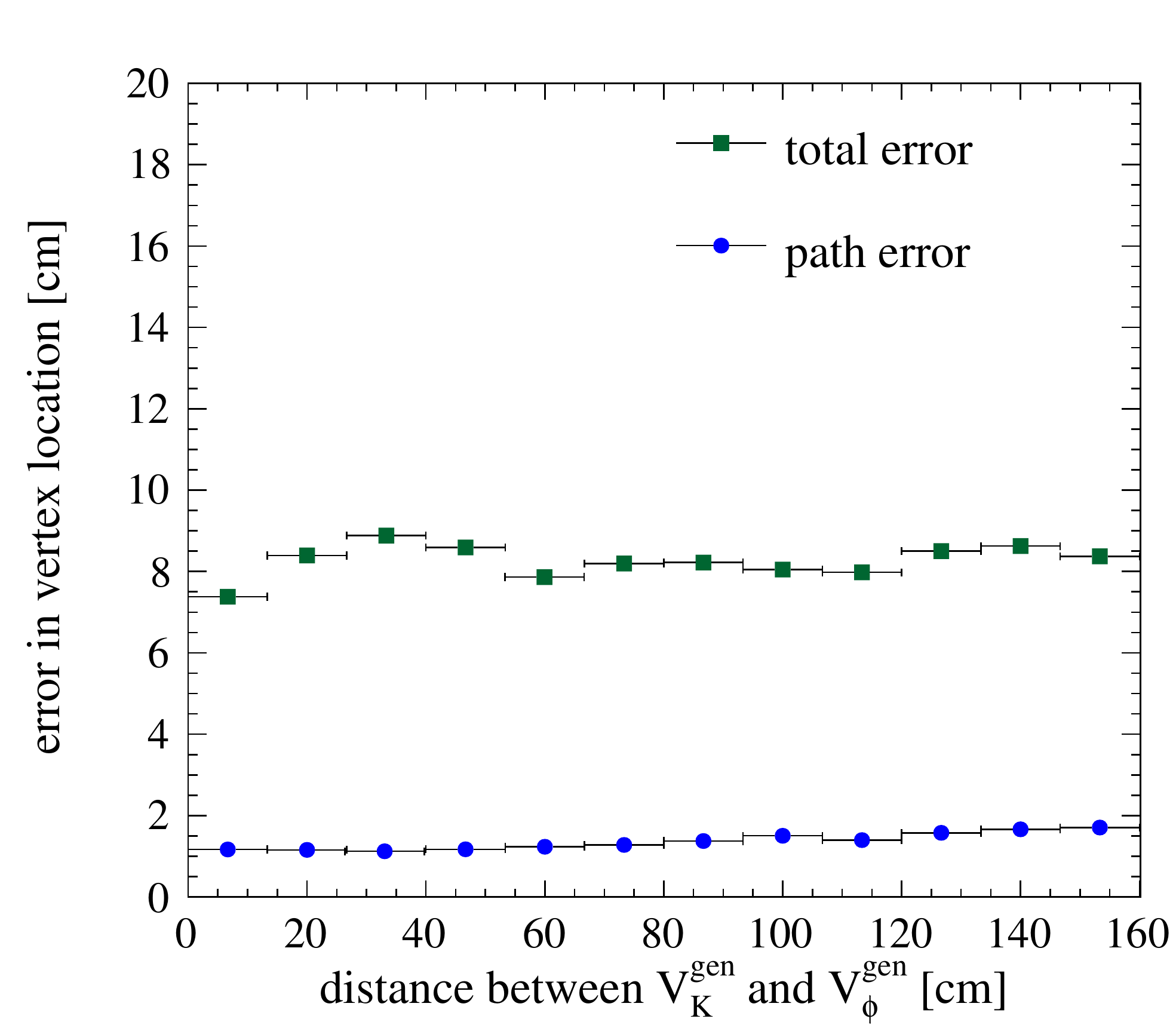}
  \hspace{1em}
  \includegraphics[width=0.45\textwidth]{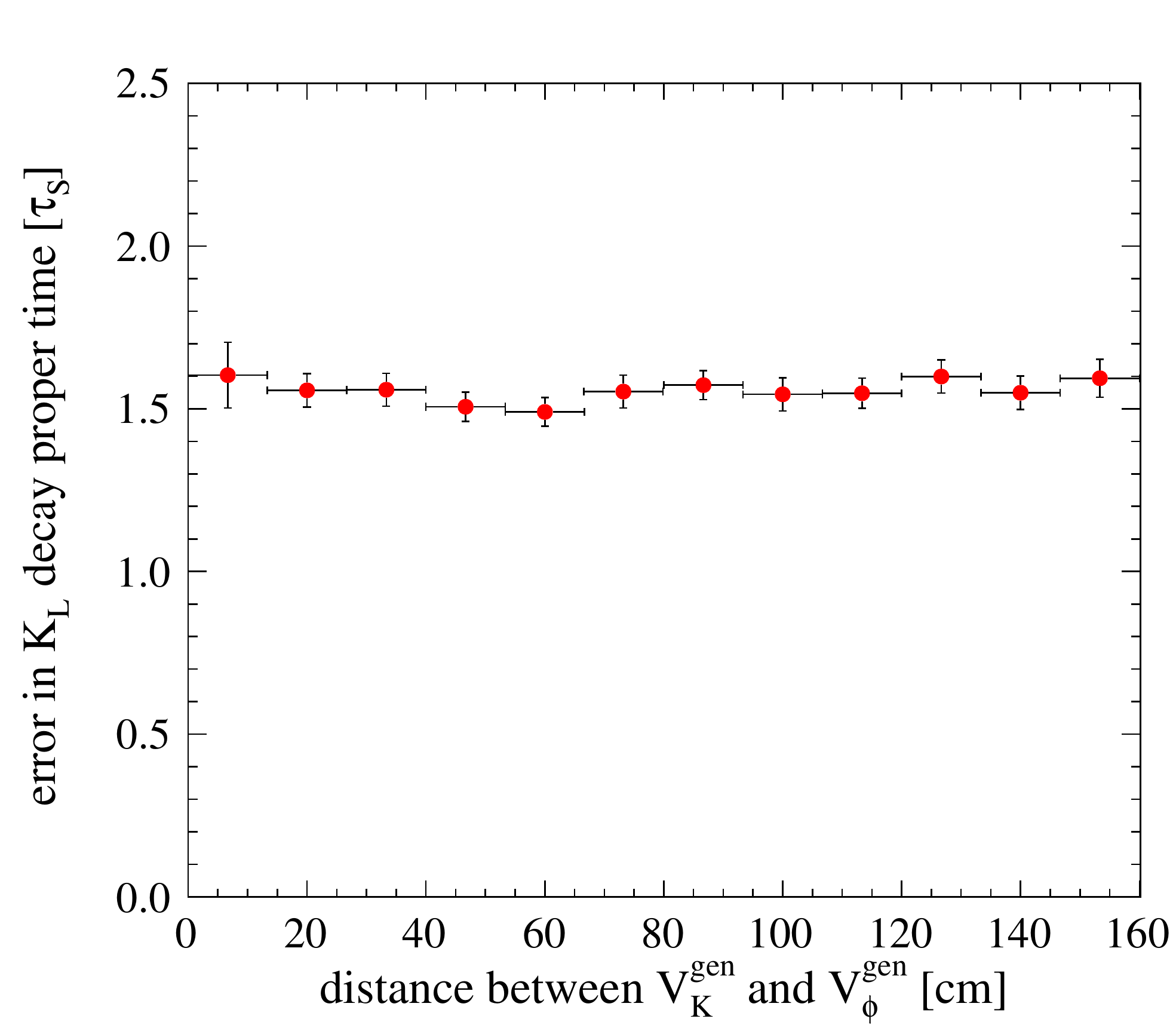}
  \caption{MC-based study of the resolution of reconstructed $\Kl\to 3\pi^0$ decay position (left) and proper time (right) obtained after the kinematic fit. Resolution is defined as $\sigma$ of an error on reconstructed w.r.t.\ MC-generated decay in subsequent ranges of decay distance from $\phi$.}\label{fig:resolution_fit}
\end{figure}

\fref{fig:chisq} presents a distribution of minimal $\chi^{2}$ values obtained with the procedure described above, normalized to the number of degrees of freedom in the fit. The agreement between data and MC-simulated events indicates similar performance of the fitting procedure in both cases. The resolution of $\Kl\to 3\pi^0$ decay point and time obtained after the kinematic fit was studied with a MC sample in a similar manner as for the reconstruction without the fit (see~\sref{sec:kl3pi0}). The resulting resolution of the spatial vertex location and proper decay time of the kaon is shown in the left and right panels of~\fref{fig:t2-kspipi}, respectively. The improvement of of spatial reconstruction, especially of the length of the path travelled by the kaon (compare~\fref{fig:resolutions_nofit}) leads to a resolution of the $\Kl$ decay proper time at the level of approximately~1.6~$\tau_S$ independently of the location of the decay in the KLOE detector~\cite{gajos-acta,Gajos:2015ija}. According to Monte Carlo-simulation studies, this resolution combined with the one of $\Ks\to \pi e \nu$ decay time reeults in the resolution of both decays' time difference $\Delta t$ at the level of 2.3~$\tau_S$. Consequently, in the further considerations of any distributions as functions of $\Delta t$, a bin width of 3~$\tau_S$ is used.
%
% TODO: wyrysowac rozdzielczosc delta t i uzasadnic dlaczego wybrane binowanie 3 tau_s
% 
\section{Selection and reconstruction of the $\Ks \Kl \to \pi^{+}\pi^{-} \pi^{\pm}e^{\mp}\nu$ process}\label{sec:t-analysis-2}

The process where an earlier kaon decays into two charged pions and the later one through a semileptonic channel requires a more straightforward event selection than $\Ks\Kl\to\pi e\nu\;3\pi^0$, mostly due to the fact that the combined branching ratio of the required decays amounts to about \SI{28}{\percent}, over 2000 times larger than in case of the other class of processes. Moreover, the $\Ks\to\pi^+\pi^-$ decay is well identified by tracks of both products recorded by the KLOE drift chamber, whose sum of momenta additionally provides a precise estimate of $\vec{p}_{\Ks}$.

\subsection{Selection of $\Ks$ decays into charged pions}\label{sec:kspipi}
Similarly as in the case of the first class of events (see~\sref{sec:t-analysis-1}), the selection of $\Ks \Kl \to \pi^{+}\pi^{-} \pi e\nu$ process is started at the level of the neutral kaon stream of KLOE data and corresponding Monte Carlo-simulated events. As a first criterion, presence of two tracks in the drift chamber is required which share a common vertex located within a cylindrical volume limited by $r_T<$15~cm and $|z|<$10~cm, where a majority of $\Ks$ decays are expected. In the process of interest, however, the decay vertex of an early semileptonic decay of the long-lived neutral kaon can satisfy the same requirement. To avoid confusing the vertices in signal events, for all 2-track vertices located in the fiducial volume for $\Ks$ decays the invariant mass of decaying kaon M$(\pi,\pi)$ is reconstructed using product tracks with an assumption that both correspond to charged pions. If more than one such vertex is present, the one is selected for which the calculated invariant mass is closest to the neutral kaon mass.

\begin{figure}[h!]
  \centering
  \begin{tikzpicture}
    \node[anchor=south west,inner sep=0] at (0,0) 
    {\includegraphics[width=0.45\textwidth]{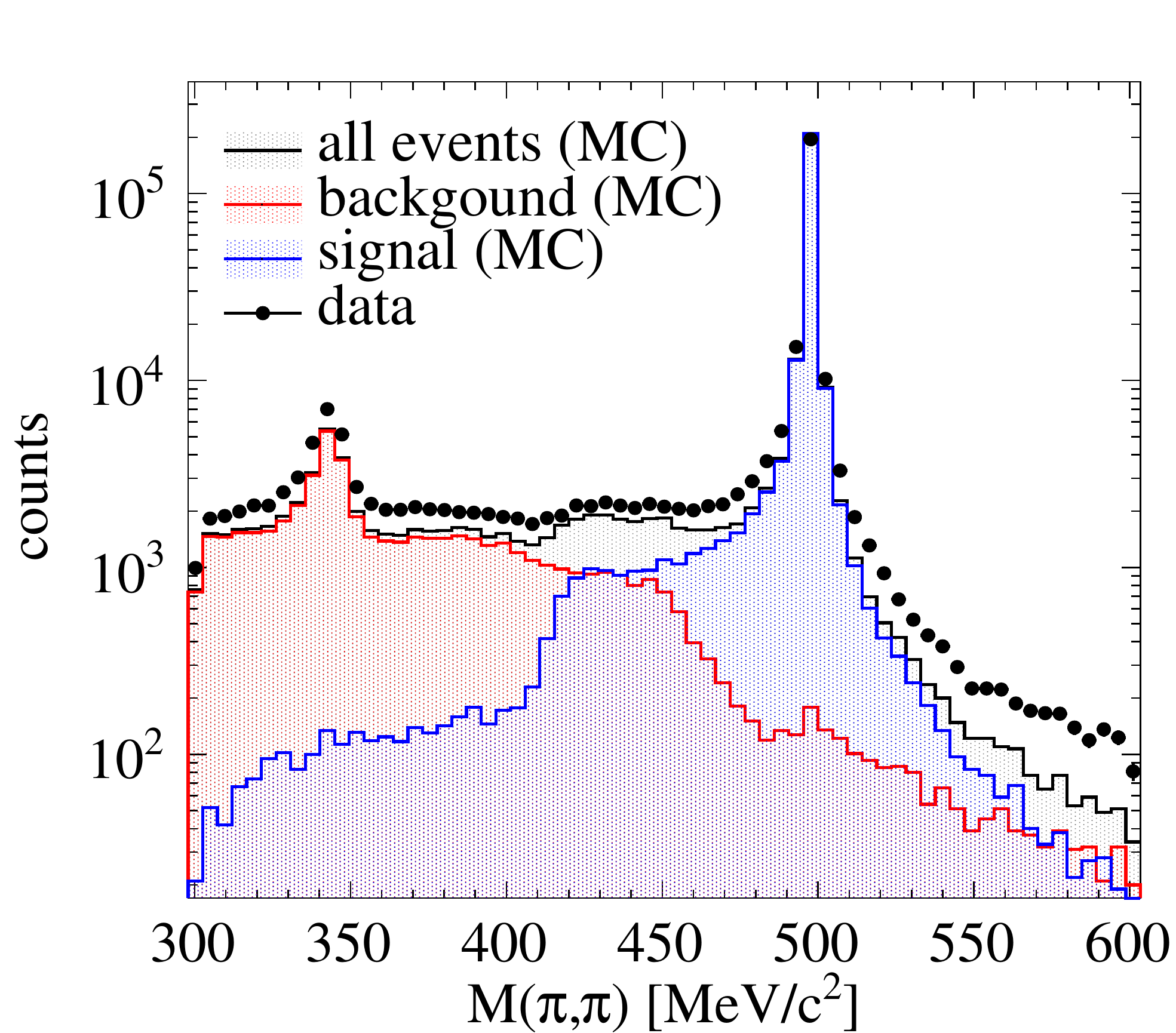}};
    \draw[black, thick, dashed] (4.47,0.8) -- (4.47,5.46);
    \draw[black, thick, dashed] (4.85,0.8) -- (4.85,5.46);
    \draw[ultra thick, black!70!white, <->] (4.44, 1.2) -- (4.87, 1.2);
  \end{tikzpicture}
  \hspace{1em}
  \begin{tikzpicture}
    \node[anchor=south west,inner sep=0] at (0,0) 
    {\includegraphics[width=0.45\textwidth]{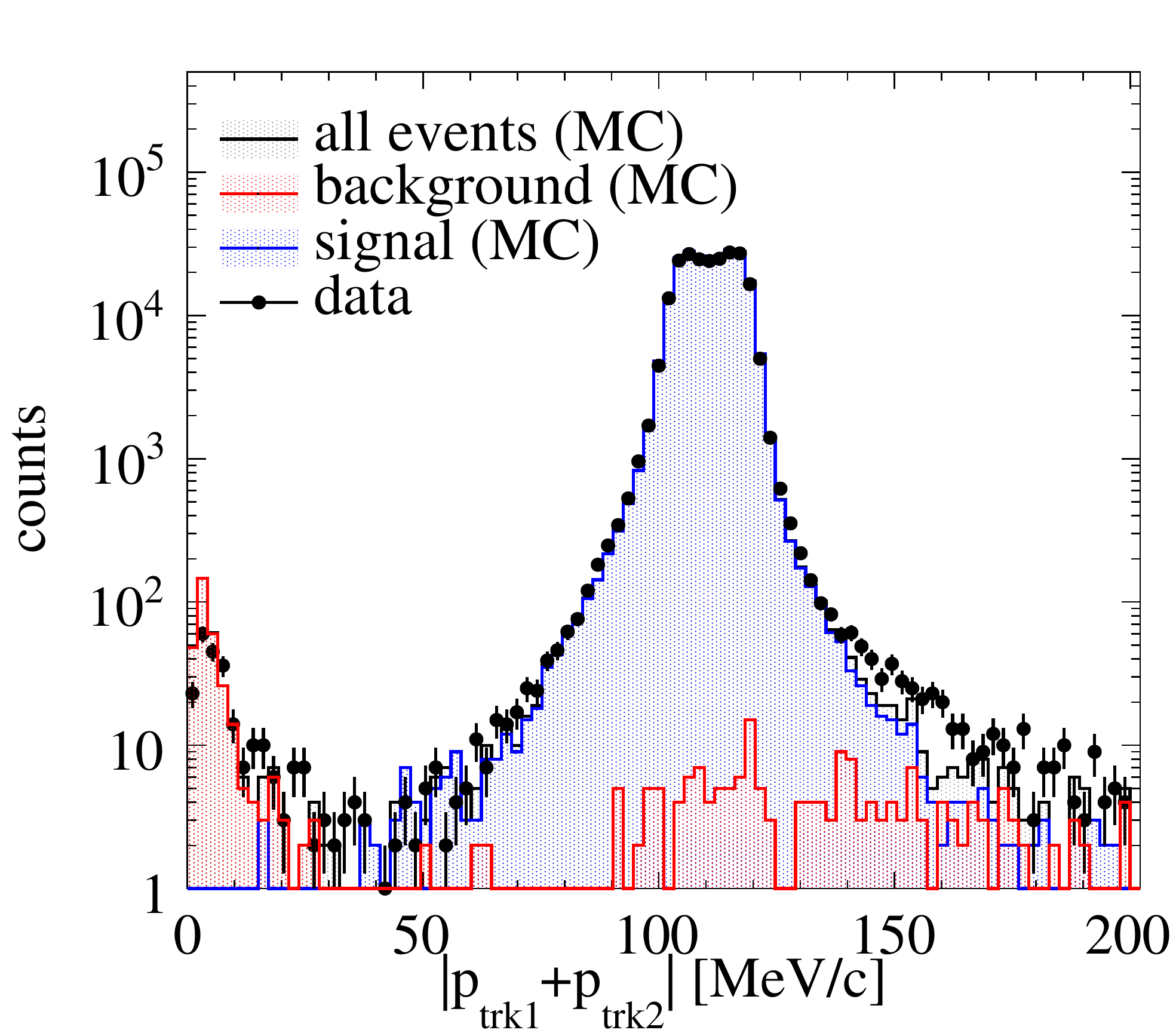}};
    \draw[black, thick, dashed] (3.4,0.8) -- (3.4,4.0);
    \draw[black, thick, dashed] (4.75,0.8) -- (4.75,5.46);
    \draw[ultra thick, black!70!white, <->] (3.42, 2.2) -- (4.73, 2.2);
  \end{tikzpicture}
  \caption{Left: invariant mass of a decaying kaon reconstructed with momenta corresponding to the two DC tracks assuming both were charged pions. Right: Sum of momenta corresponding to the two DC tracks after a cut on the $\pi,\pi$ invariant mass. Dashed lines denote values of the cuts used and gray arrows indicate the accepted regions.}\label{fig:t2-kspipi}
\end{figure}
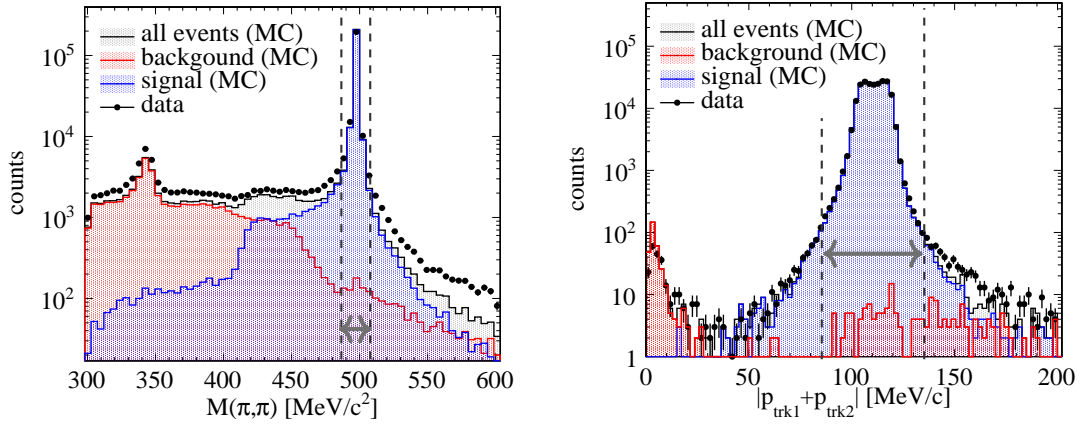

Distribution of the invariant masses for the selected 2-track decay vertices is shown in the left panel of~\fref{fig:t2-kspipi}. As evident from the MC-simulated spectra, for the $\Ks\to\pi^+\pi^-$ recorded tracks of the produced pions yield an accurate estimate of the decaying kaon mass and thus a highly pure set of these events is obtained with the following cut on the reconstructed invariant mass with $\pi^+\pi^-$ mass assumption:
\begin{equation}
  \label{eq:t2-invmass-cut}
  \abs{\text{M}(\pi,\pi) - m_{\kaon}} < 10\:\text{MeV}.
\end{equation}

Subsequently, the sum of momenta associated with the two DC tacks is restricted to be close to the nominal momentum modulus of a neutral kaon originating from a $\phi$ decay from the DA$\Phi$NE collider which amounts to about 110~MeV/c:
\begin{equation}
  \label{eq:t2-ptot-cut}
  \abs\Big{\abs{\vec{p}_{trk1} + \vec{p}_{trk2}} - 110\:\text{MeV/c}} < 25\:\text{MeV/c}.
\end{equation}
Although this cut only removes a small amount of background surviving the invariant mass cut as shown in the right panel of~\fref{fig:t2-kspipi}, its purpose is also to limit the further analysis to events with well reconstructed momenta of the pions.

\subsection{Selection of $\Kl\to \pi^{\pm}e^{\mp}\nu$ decays}\label{sec:klsemil}
Semileptonic decays of the long-lived neutral K meson may be identified by analyzing the times of flight of products particles from the decay point to their incidence on the calorimeter, in a similar manner as in case of $\Ks\to\pi e \nu$ (\sref{sec:t1_tof}). Preparation of a high-purity event sample is easier in this case due to considerably larger branching fraction for semileptonic decays of $\Kl$ (see~\tref{tab:kaon_properties}). A major difference, however, is that the kaon decay may occur anywhere inside the detector. In presence of background processes, this may require selecting a correct vertex among several decay vertices recorded by the drift chamber in a single event.

Previous analyses of $\Kl\to\pi e\nu$ at KLOE~\cite{kloe_memo_280,kloe_memo_322,kloe_memo_334,kloe_kl3pi0_br} utilized the accurate estimate of $\Kl$ momentum direction obtained from $\Ks\to\pi^+\pi^-$ and momentum conservation in the $\phi$~decay to search for a 2-track DC vertex closest to the $\Kl$ line of flight.
%
% TODO: dodac ref do sekcji z opisem kanalu kontrolnego z Ks->pi0pi0
%
Such approach, however, is not feasible in case of the analysis presented in this work as the channel which is later used as a control sample for estimation of selection efficiency of $\Kl\to\pi e \nu$ is a process where the latter is accompanied by a $\Ks\to\pi^0\pi^0$ decay. In such case, the accuracy of $\vec{p}_{\Ks}$ obtained from photons' momenta is considerably lower, resulting in poor angular resolution of $\Kl$ line of flight and a  degradation of selection efficiency. Therefore, in this analysis, an alternative approach was devised which does not rely on momentum estimation obtained form the other kaon in the same event.

\begin{figure}[h!]
  \centering
  \includegraphics[width=0.45\textwidth]{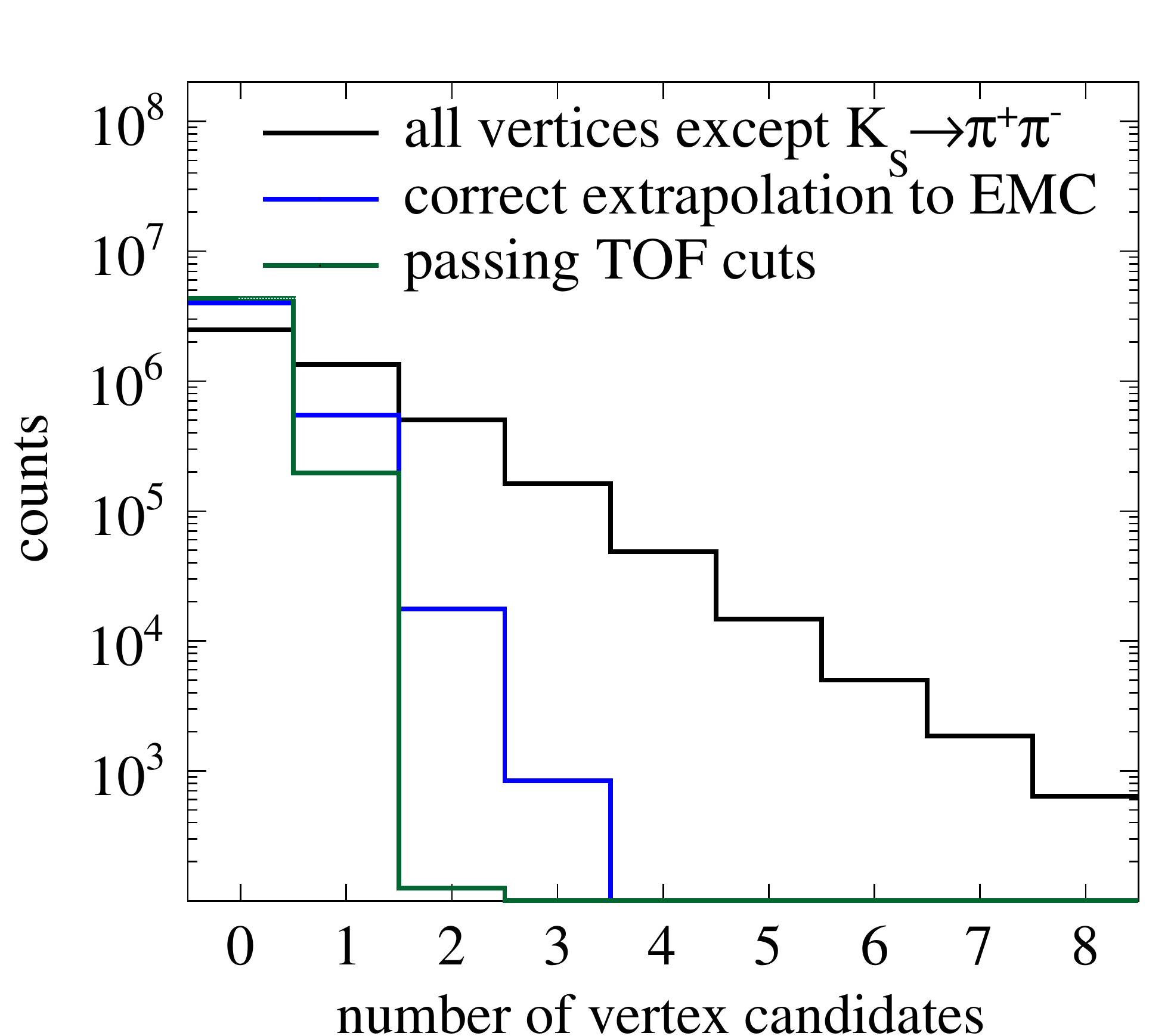}
  \caption{Multiplicity of $\Kl$ decay vertex candidates, ie.\ all vertices except the one identified as $\Ks\to\pi^+\pi^-$ and subsequent pion decays (black), vertices for which both associated tracks were correctly extrapolated to the EMC surface and associated to a cluster (blue) and vertices for which the particles associated to the tracks passed the time of flight cuts (green). Distributions were obtained with data.}
  \label{fig:klvtxcount}
\end{figure}

In order to find the decay vertex and tracks corresponding to an electron/positron and to a charged pion, all vertices constituted by two tracks in the drift chamber were considered as candidates with an exception of the previously identified vertex of $\Ks\to\pi^+\pi^-$. As the pions may decay in the DC, vertices connected to tracks associated to $\pi^+\pi^-$ from the $\Ks$ decay were also excluded. Multiplicity of vertex candidates is shown as a black histogram in~\fref{fig:klvtxcount}. In order to reduce the ambiguity of vertex choice,
for each of the candidates 
an extrapolation to the inner surface of the calorimeter was attempted
for its two associated DC tracks. If a successful extrapolation was followed by finding a corresponding EMC cluster for both tracks, the vertex was considered in further analysis and rejected otherwise. After this step, the amount of events with more than one vertex candidate was still at the level of several percent as shown by the blue histogram in~\fref{fig:klvtxcount}. Therefore, each of the surviving candidates was subject to a time of flight analysis.

The TOF analysis applied to the 2 tracks associated with each vertex candidate was similar to the one used for $\Ks\to\pi e\nu$ (see~\sref{sec:t1_tof}). Again, the difference between recorded time of flight (based on the EMC cluster registration time) and the time expected from particle path length along the extrapolated DC track and its velocity, was used to define the variables of the TOF analysis. In case of $\Kl$ decays, however, the expected time of flight included an additional component related to the flight of the kaon before its decay,~$\frac{L_K}{\upsilon_K}$:
\begin{equation}
  \label{eq:dtof_t2}
  \delta t({m_x}) = t_{cl} - \frac{L}{c\beta(m_x)}  - \frac{L_K}{\upsilon_K},
\end{equation}
where $L_K$ is the distance from the average $\phi$ decay point to the location of considered $\Kl$ decay vertex candidate, and $\upsilon_K$ is the kaon velocity obtained from its momentum.

In order to avoid any dependency of the TOF analysis on resolution of $\Ks$ momentum (with a view to its application on a $\Ks\Kl\to\pi^0\pi^0\;\pi e\nu$ control sample), the $\abs{\vec{p}_{\Kl}}$ value was calculated using momentum conservation in the $\phi\to\Ks\Kl$ decay as described in~\sref{sec:ks_from_kl} by assuming that the $\Kl$ momentum direction was parallel to the kaon path spanning between the $\phi$ vertex and current vertex candidate.

\begin{figure}[h!]
  \captionsetup[subfigure]{justification=centering}
  \centering
  \begin{subfigure}{0.45\textwidth}
    \includegraphics[width=1.0\textwidth]{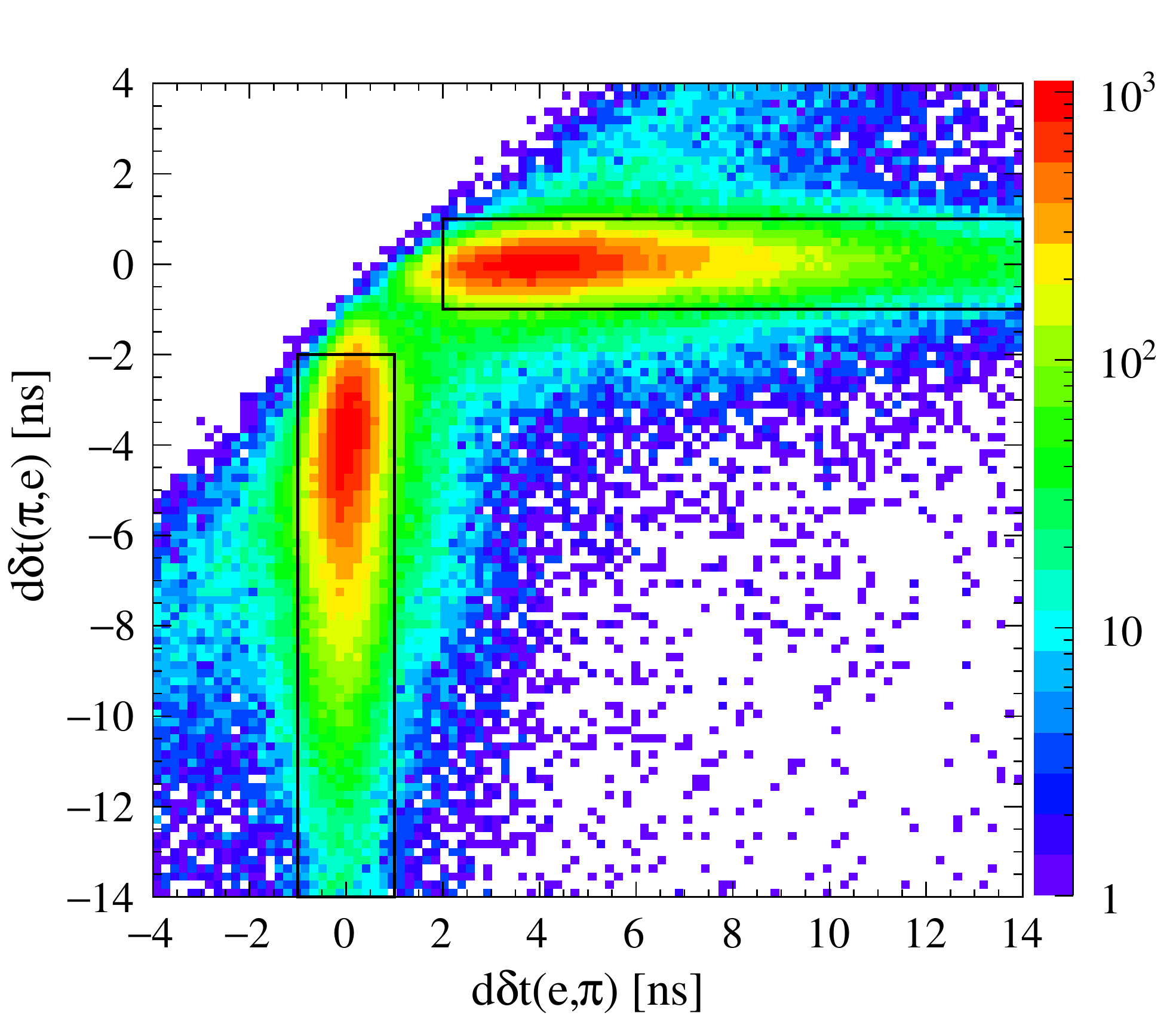}
    \caption{Signal events (MC)}
  \end{subfigure}
  \begin{subfigure}{0.45\textwidth}
    \includegraphics[width=1.0\textwidth]{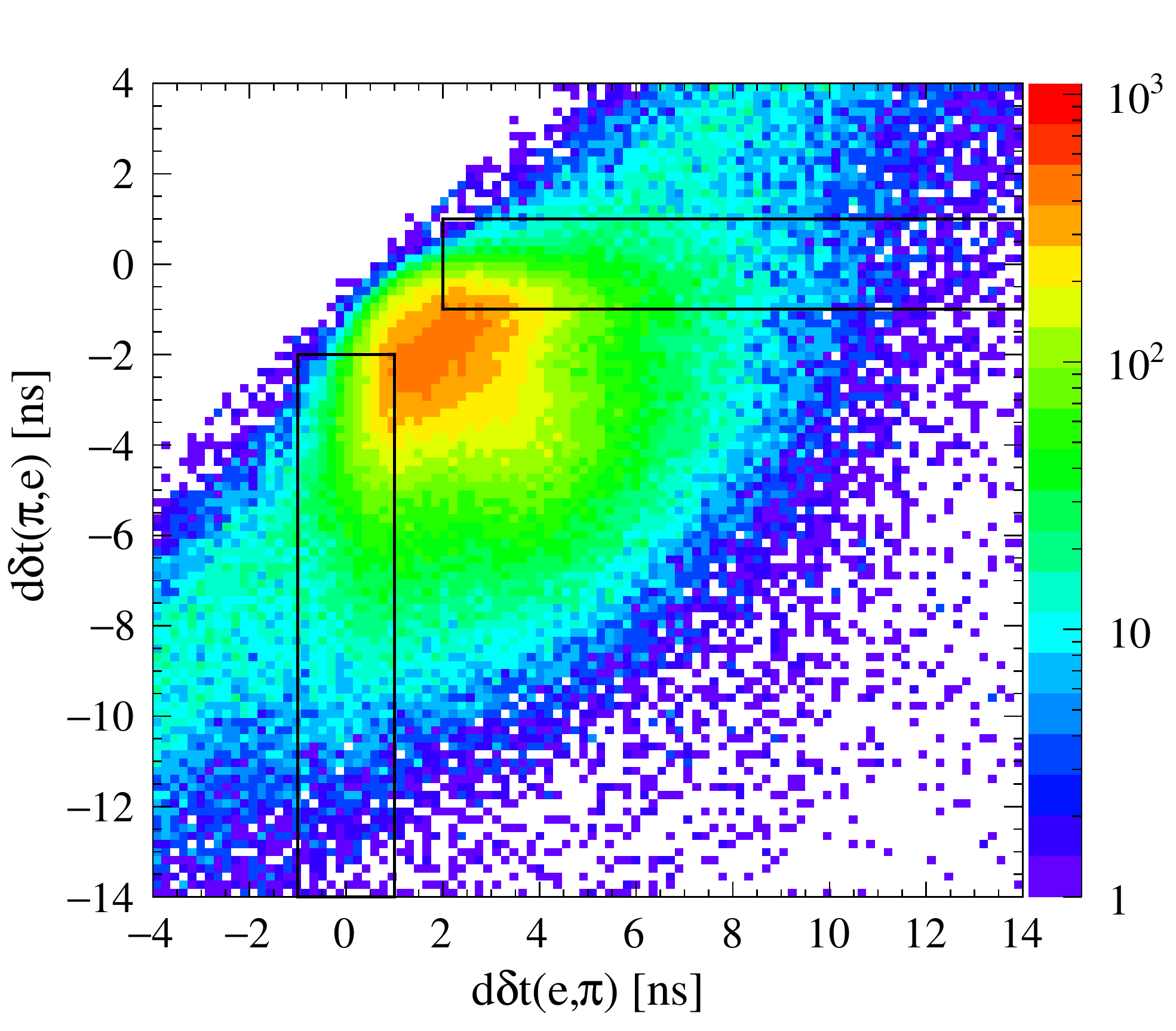}
    \caption{Background events (MC)}
  \end{subfigure}
  \begin{subfigure}{0.45\textwidth}
    \includegraphics[width=1.0\textwidth]{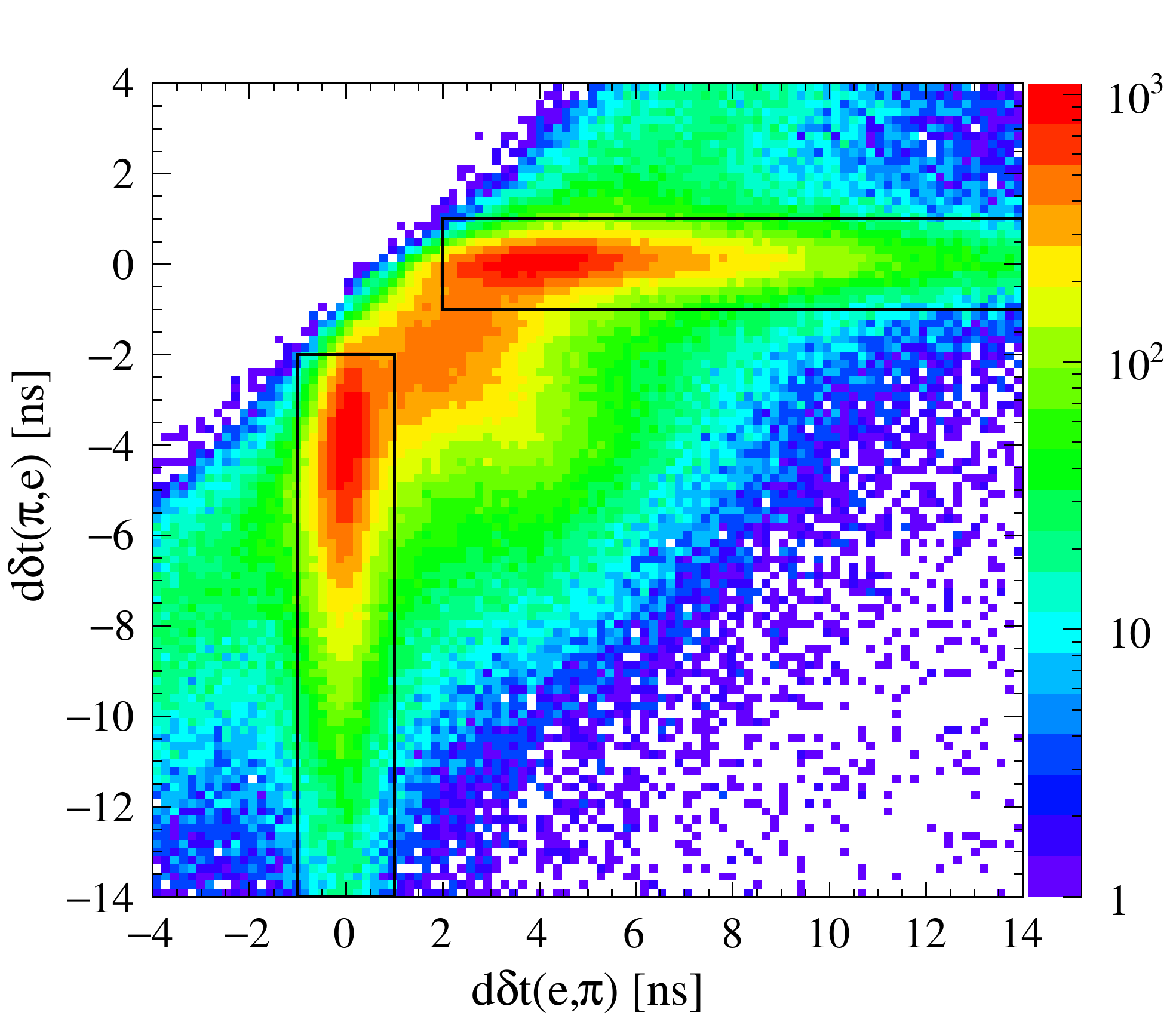}
    \caption{All data events}
  \end{subfigure}  
  \caption{Errors of two possible $\pi^{\pm}$ and $e^{\mp}$ mass hypotheses assignments to tracks. Solid lines denote accepted regions, each corresponding to one of the possible mass hypotheses assignment. The MC background plot includes other $\Kl$ decays as well as combinatorial background due to incorrect choice of a DC vertex.}\label{fig:t2-tof1}
\end{figure}

Similarly as in the case of $\Ks\to\pi e \nu$, further analysis requires that each of the charged particle tracks in the decay is identified as either an electron/positron or a pion. To this end, the relative distribution of $d\delta t_{x,y} = \delta t_1(m_x) - \delta t_2(m_y),$ for two possible assignments of particle mass hypotheses to tracks was used%
\footnote{Similarly as in the case of semileptonic $\Ks$ decays, this variable is constructed so as to cancel out the event start time $T_0$. Even though the $\Kl$ decay time is also cancelled in this expression, it is retained in the definition of $\delta t(m_x)$ as later in the analysis its absolute values are used.}.
The distributions obtained with simulated $\Kl\to\pi e \nu$ events and a correct choice of the DC vertex, as well as Monte-Carlo simulated background processes and wrong vertex choices for signal events (combinatorial background), are presented in~\fref{fig:t2-tof1} along with the total spectrum obtained with KLOE data. As visible in the upper left panel, for signal events and a correctly identified $\Kl\to\pi e\nu$ decay vertex and tracks, one of the $d\delta t$ values tends to be close to zero, indicating that the assumed hypothesis on particle masses attributed to tracks was correct. A DC vertex with 2 tracks is thus considered a valid candidate if its calculated $d\delta t(e,\pi)$ and $d\delta t(\pi,e)$ values lie within one of the marked rectangular regions. 

\begin{figure}[h!]
  \captionsetup[subfigure]{justification=centering}
  \centering
  \begin{subfigure}{0.45\textwidth}
    \includegraphics[width=1.0\textwidth]{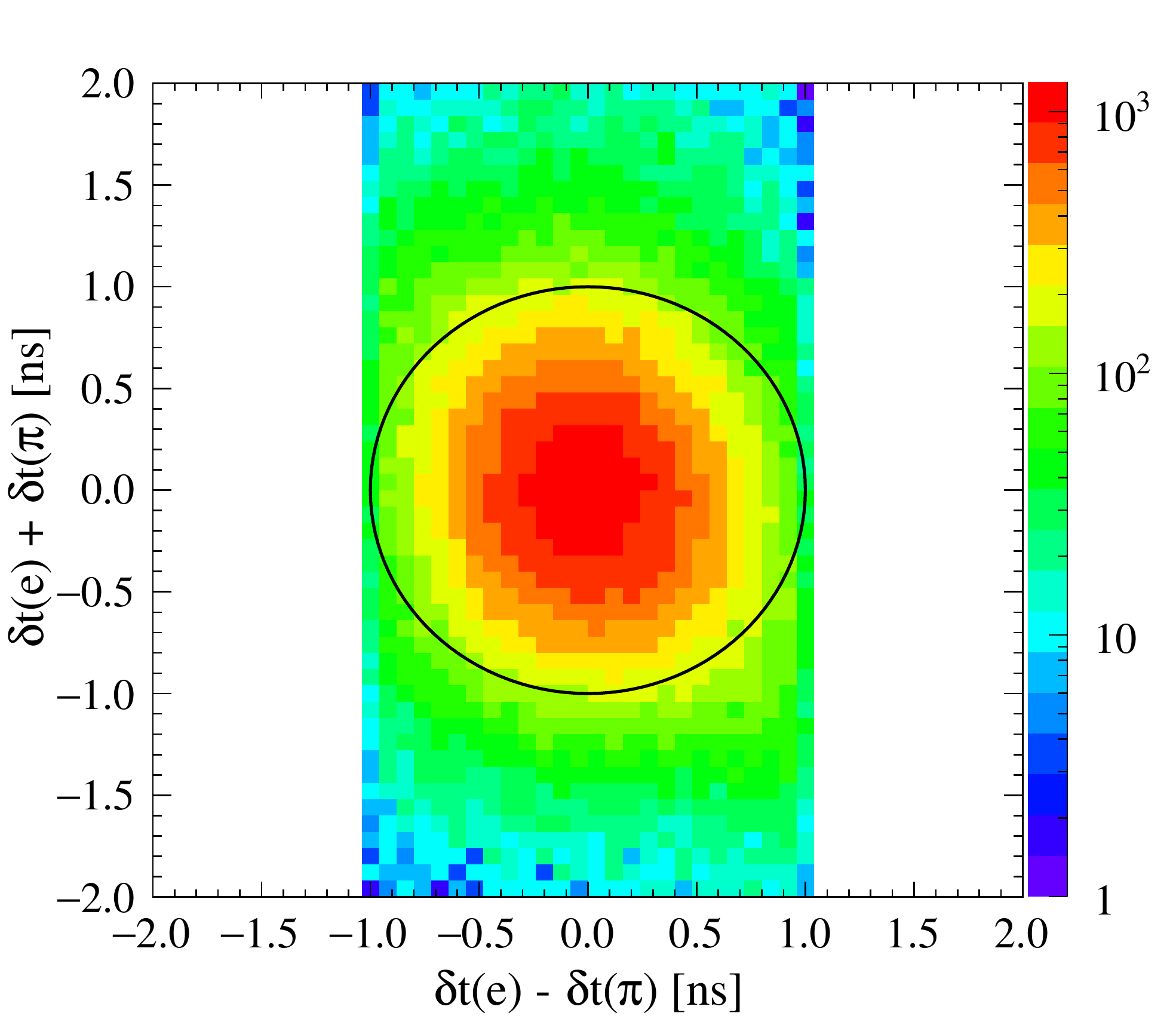}
    \caption{Signal events (MC)}
  \end{subfigure}
  \begin{subfigure}{0.45\textwidth}
    \includegraphics[width=1.0\textwidth]{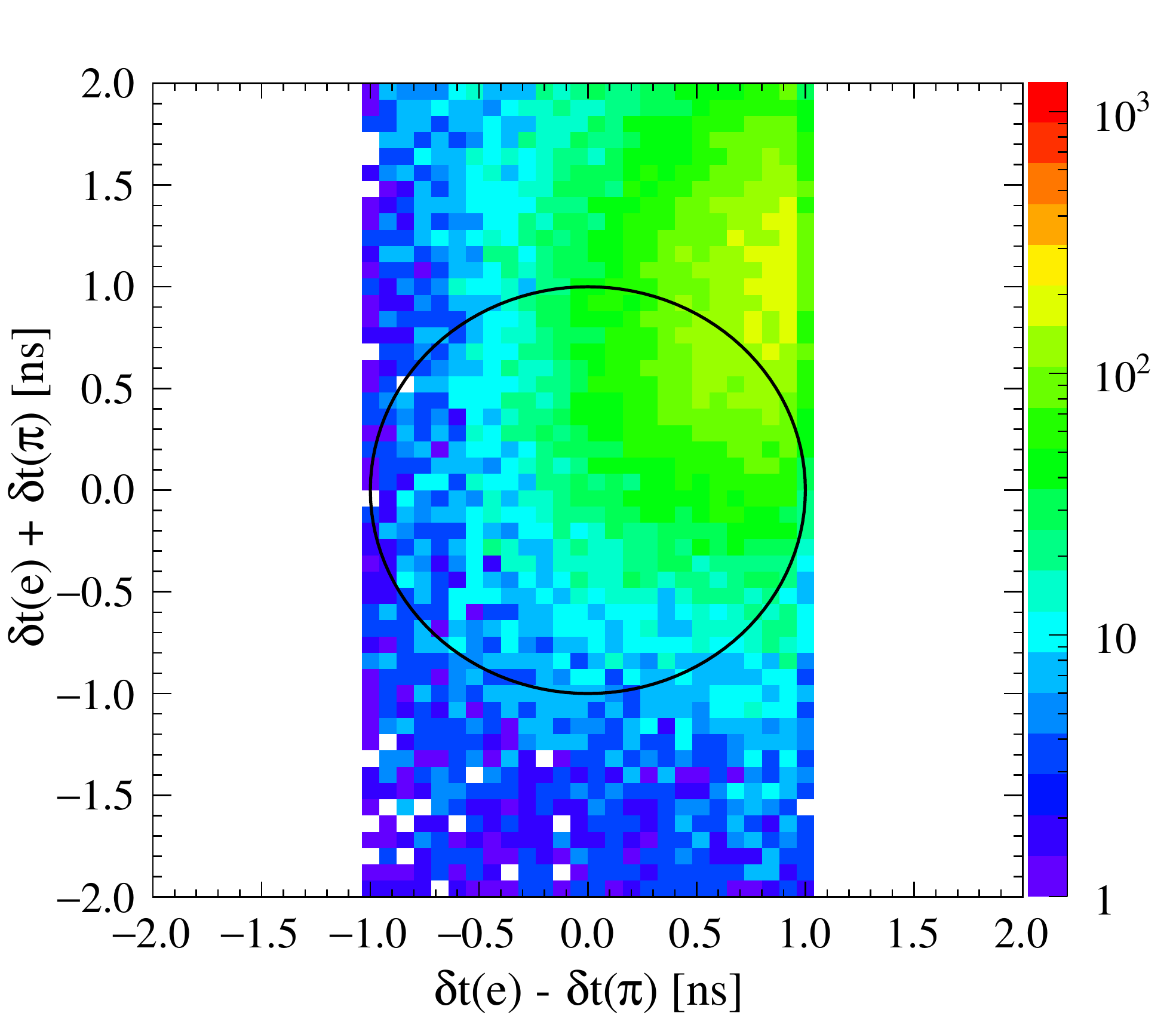}
    \caption{Background events (MC)}
  \end{subfigure}
  \begin{subfigure}{0.45\textwidth}
    \includegraphics[width=1.0\textwidth]{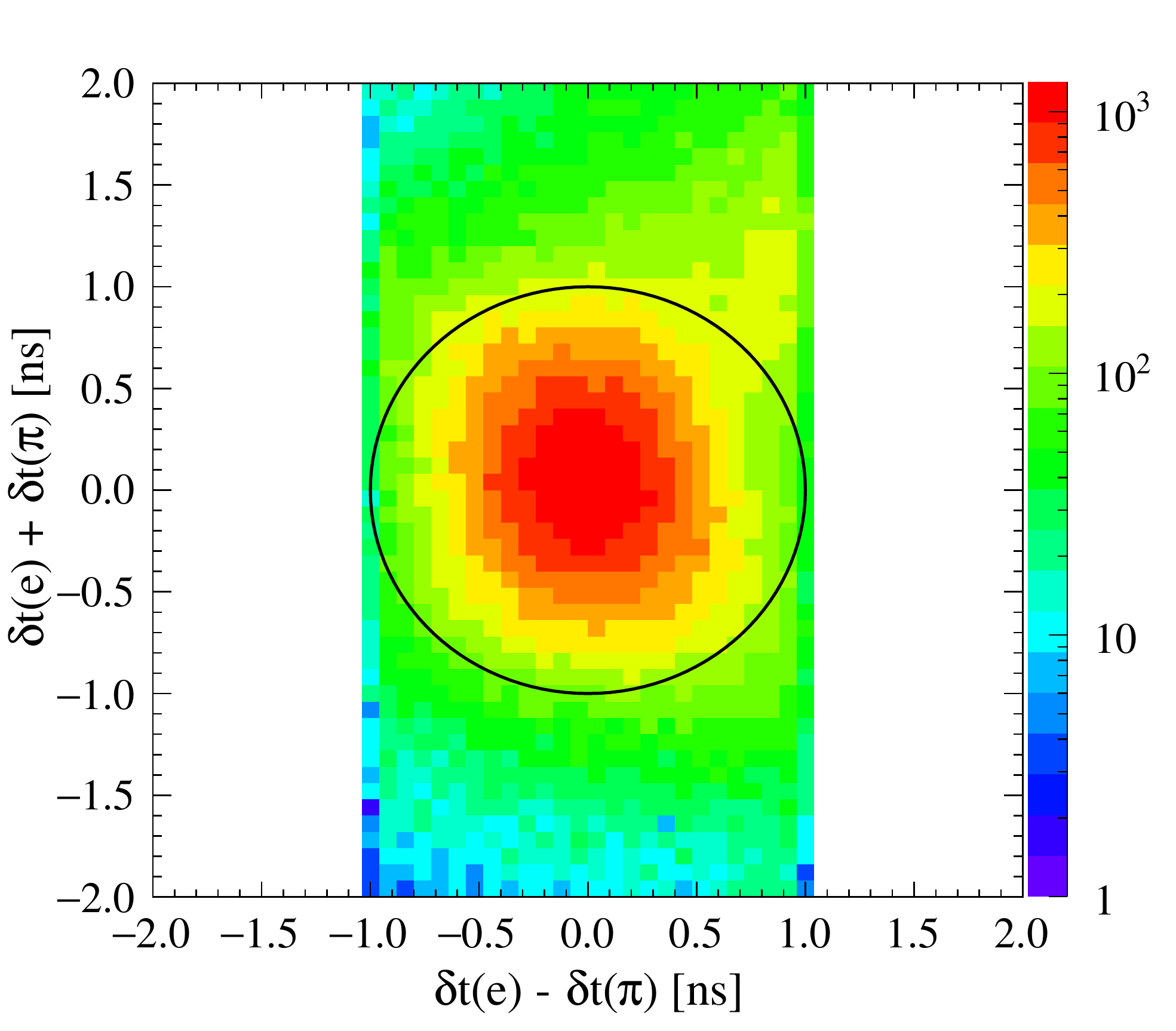}
    \caption{All data events}
  \end{subfigure}  
  \caption{Relative distribution of the sum and difference of times of flight for the decay products identified as electron/positron and pion. Limits of the populated region result form the previous TOF cut (compare~\fref{fig:t2-tof1}). The MC background plot includes other $\Kl$ decays as well as combinatorial background due to incorrect choice of a DC vertex. Vertex candidates which lie outside the circle are discarded.}\label{fig:t2-tof2}
\end{figure}

Once the tracks associated with a given vertex candidate had been identified as $e^{\pm}$ and $\pi^{\pm}$, the correct discrepancies between expected and recorded time of flight $\delta t$ can be calculated for each of these particles. A relative distribution of their sum and difference is useful for discrimination of the remaining background~\cite{kloe_memo_334}. Using the distributions presented in~\fref{fig:t2-tof2} vertex candidates were selected if their respective values were located in the central region limited by a circle marked with a solid line:
\begin{equation*}
  \left(\delta t(e)-\delta t(\pi)\right)^{2} +  \left(\delta t(e)+\delta t(\pi)\right)^{2} < (0.9\:\text{ns})^2.
\end{equation*}

The green histogram in~\fref{fig:klvtxcount} shows the multiplicity of $\Kl$ decay vertex candidates per event surviving the above selection. If none of the DC vertices present in a event passed the TOF-based criteria, the event was rejected. Events with more than one remaining candidate were discarded as well to avoid further ambiguity, resulting in a loss of signal at a \SI{0.1}{\percent} level only.

Once the $\Kl\to\pi e \nu$ decay vertex and tracks of its products were identified, proper times of decay were calculated for each of the kaons in the process using their travelled path lengths and velocities as in~\eref{eq:ks_proper_time} and track curvature with respect to KLOE magnetic field was used to determine the charge of lepton in the decay.

\subsection{Requirements on the mass attributed to product tracks}
\label{sec:trackmass}
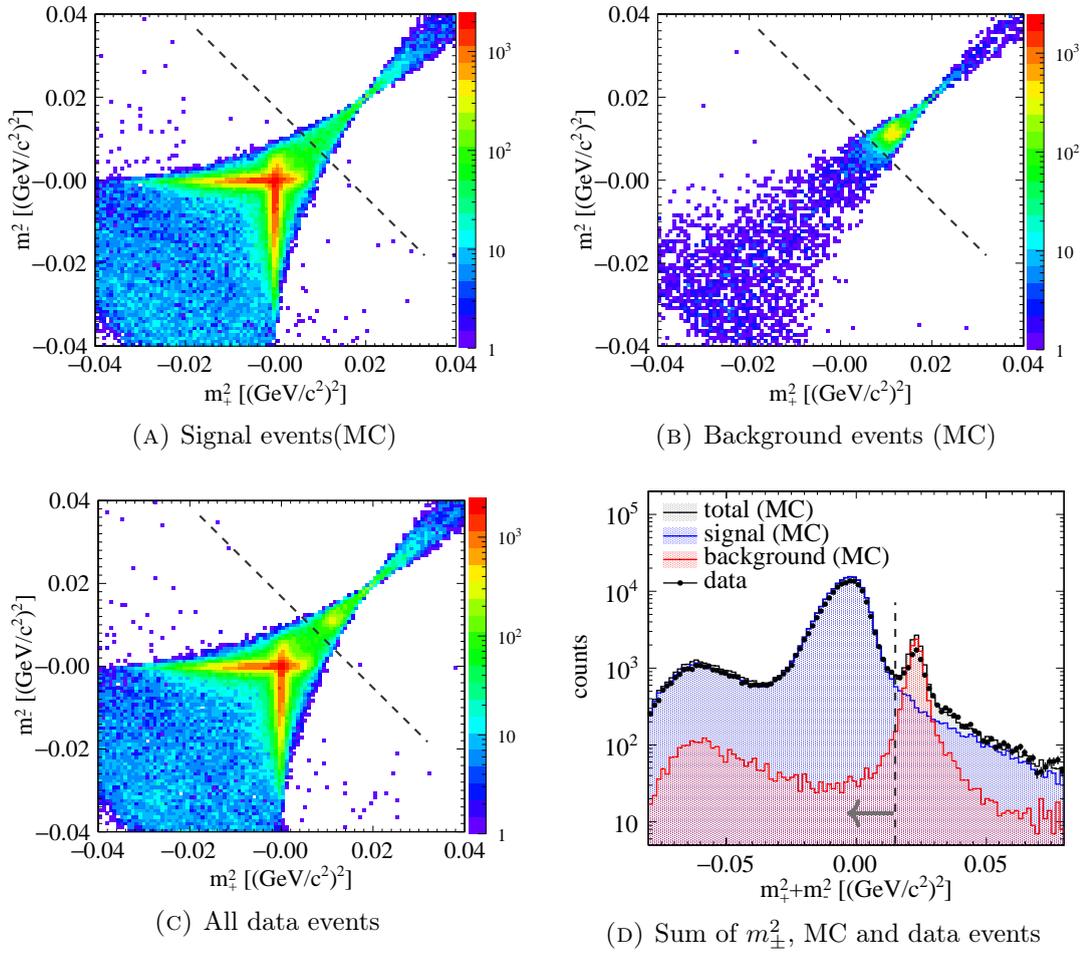
\begin{figure}[h!]
  \captionsetup[subfigure]{justification=centering}
  \centering
  \begin{subfigure}{0.45\textwidth}
  \begin{tikzpicture}
    \node[anchor=south west,inner sep=0] at (0,0) 
    {\includegraphics[width=1.0\textwidth]{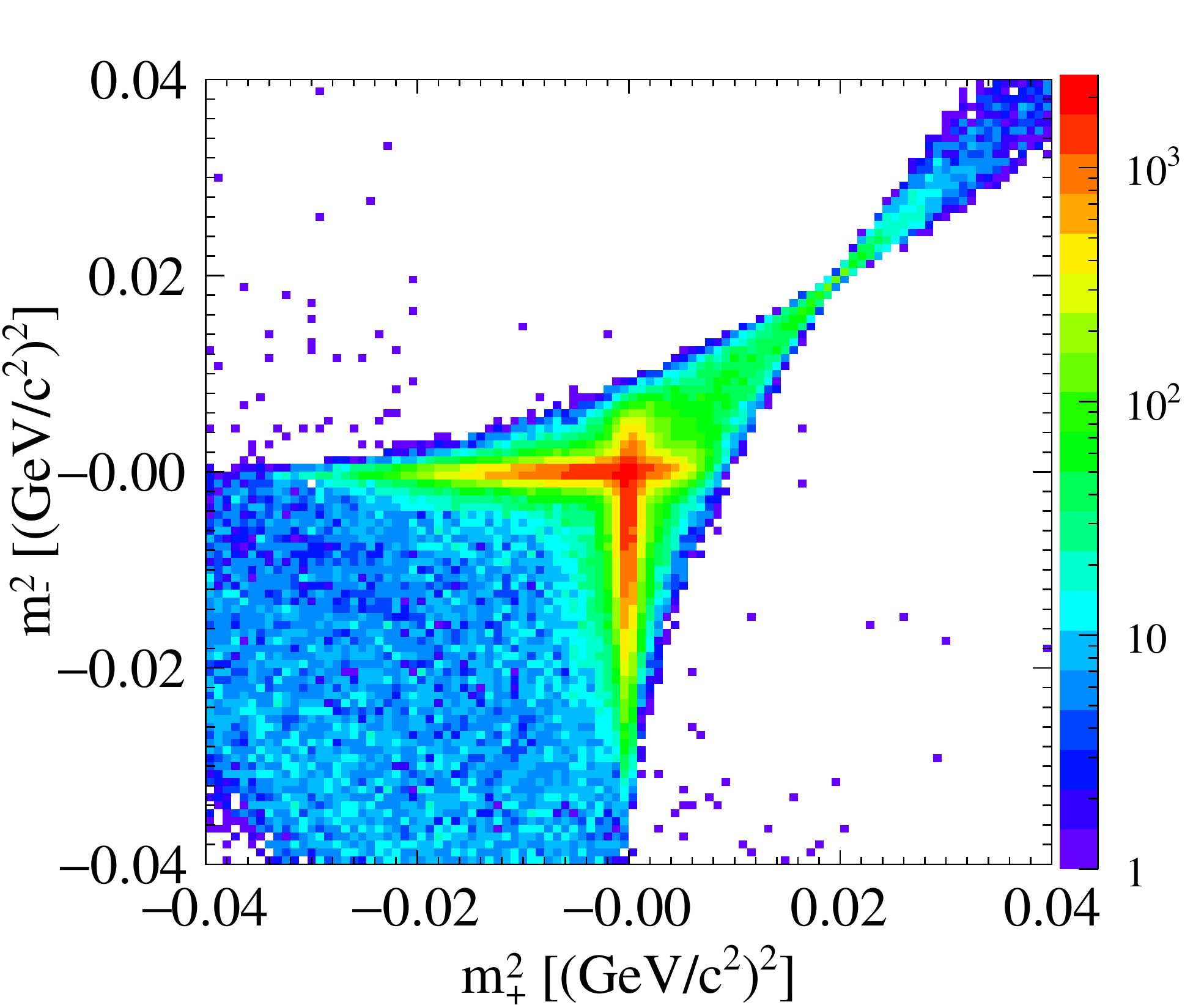}};
    \draw[black, thick, dashed] (2.50,5.0) -- (5.50,2.0);
  \end{tikzpicture}
    \caption{Signal events(MC)}
  \end{subfigure}
  \hspace{1em}
  \begin{subfigure}{0.45\textwidth}
  \begin{tikzpicture}
    \node[anchor=south west,inner sep=0] at (0,0) 
    {\includegraphics[width=1.0\textwidth]{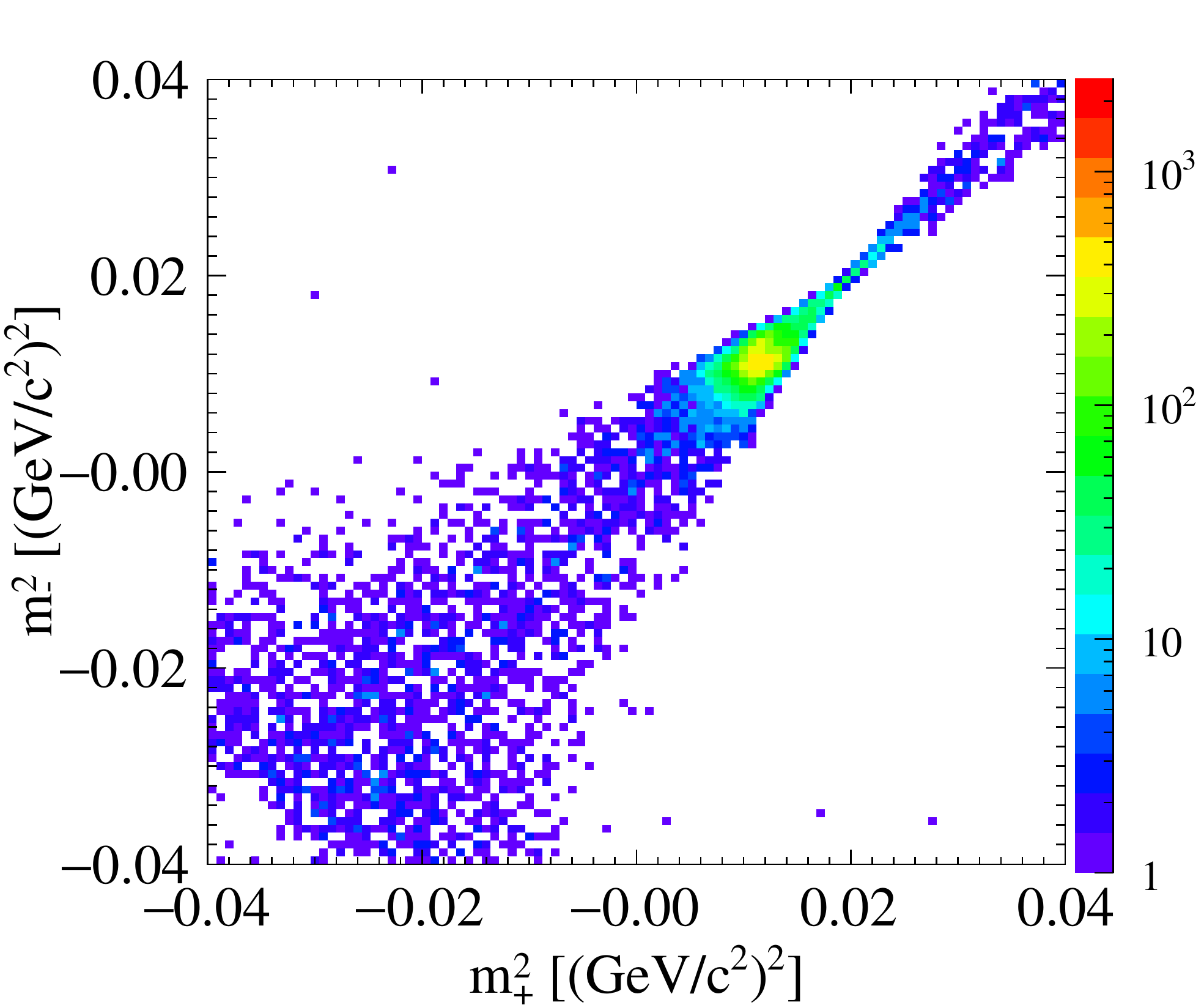}};
    \draw[black, thick, dashed] (2.50,5.0) -- (5.50,2.0);
  \end{tikzpicture}
    \caption{Background events (MC)}
  \end{subfigure}
  \begin{subfigure}{0.45\textwidth}
  \begin{tikzpicture}
    \node[anchor=south west,inner sep=0] at (0,0) 
    {\includegraphics[width=1.0\textwidth]{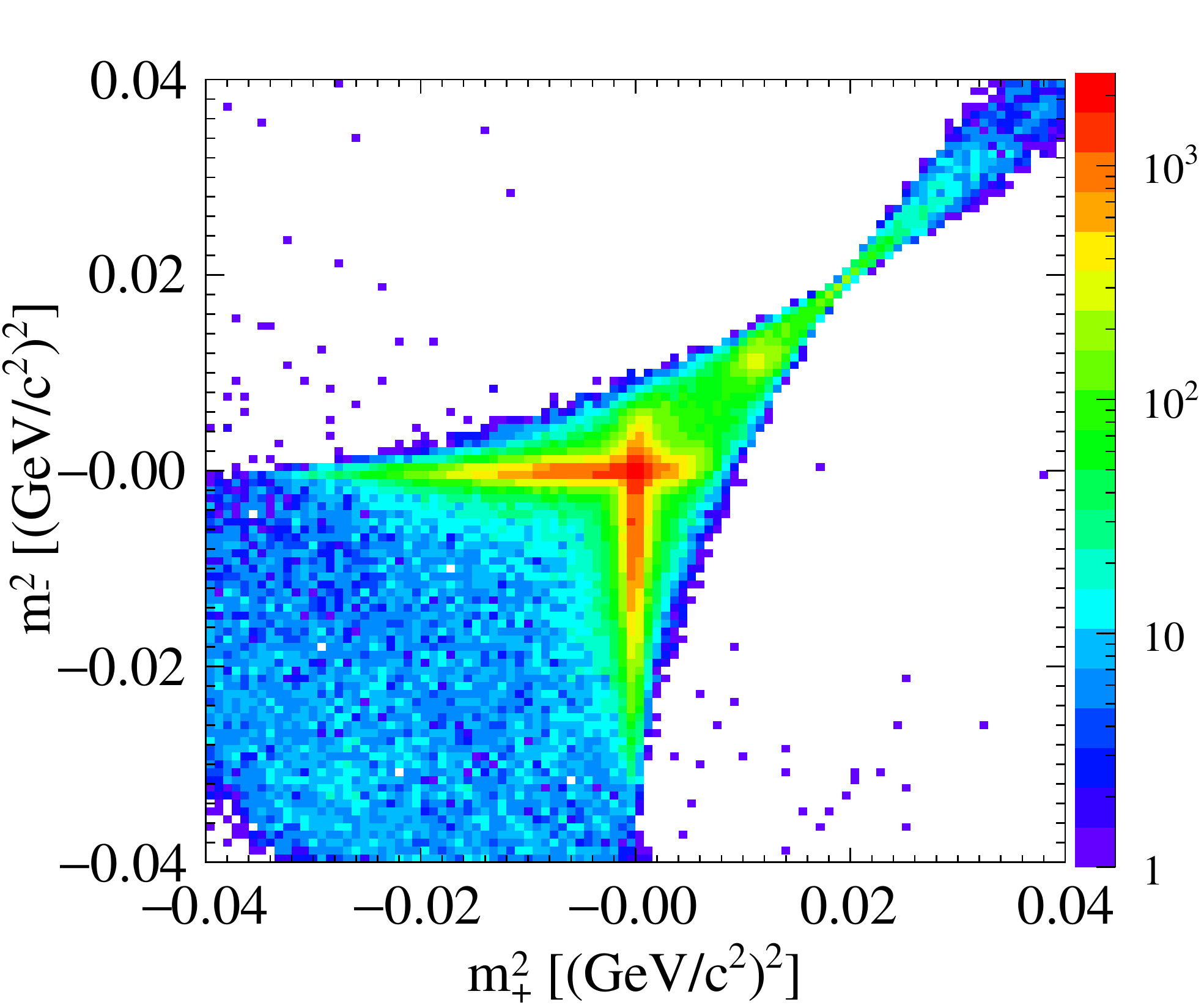}};
    \draw[black, thick, dashed] (2.50,5.0) -- (5.50,2.0);
  \end{tikzpicture}
    \caption{All data events}
  \end{subfigure}
  \hspace{0.8em}
  \begin{subfigure}{0.45\textwidth}
    \begin{tikzpicture}
      \node[anchor=south west,inner sep=0] at (0,0) 
      {\includegraphics[width=1.0\textwidth]{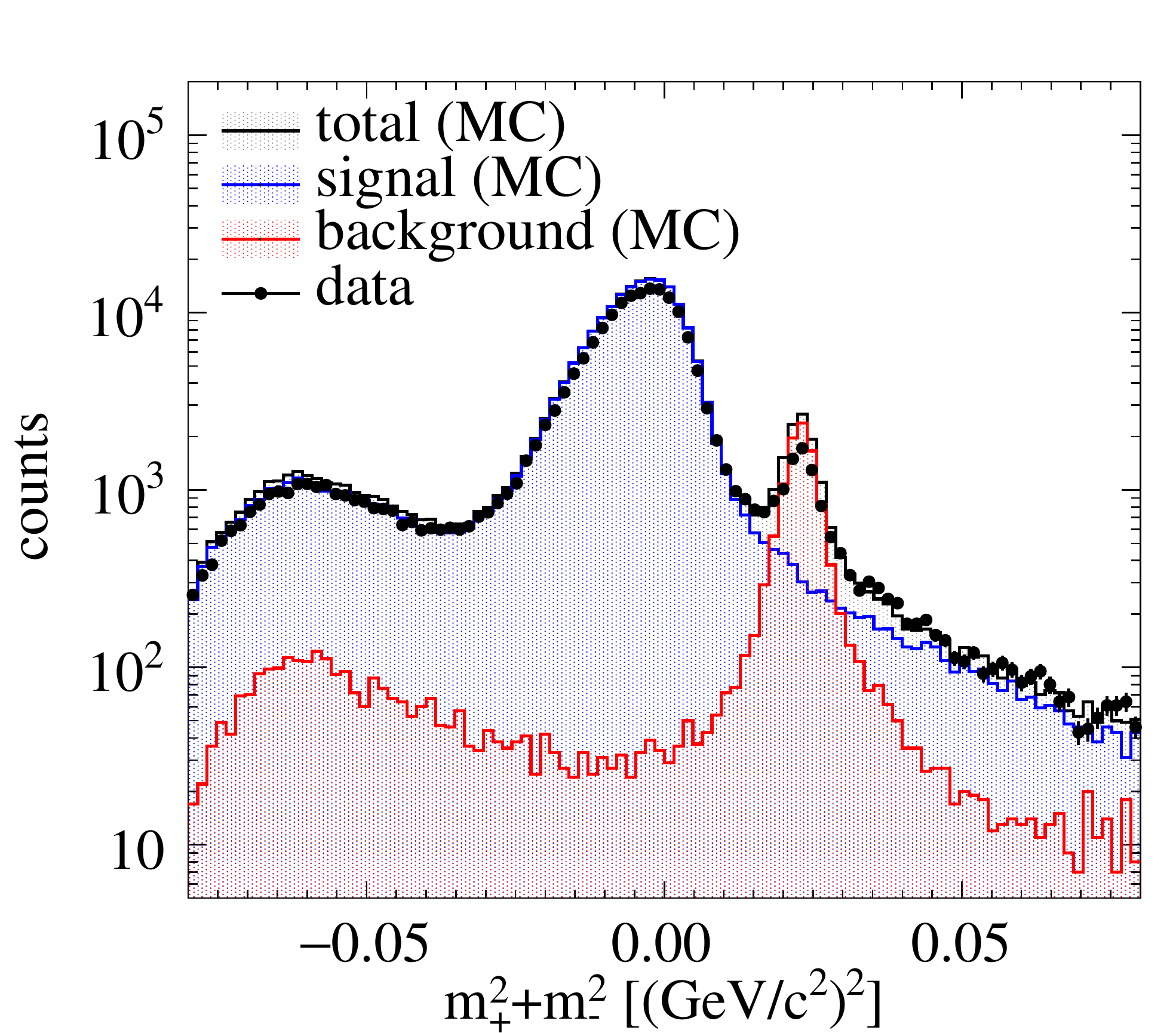}};
      \draw[black, thick, dashed] (4.34,0.8) -- (4.34,4.0);
      \draw[ultra thick, black!70!white, ->] (4.32, 1.2) -- (3.7, 1.2);
    \end{tikzpicture}
    \caption{Sum of $m^2_{\pm}$, MC and data events}\label{fig:t2-mmiss:sum}
  \end{subfigure}
  \caption{Distributions of the $m^2_{+}$ and $m^2_{-}$ variables corresponding to squared masses of particles attributed to the tracks with positive and negative charge, with assumption that the other track corresponded to a charged pion. Dashed line denotes the value of a cut imposed on the sum of $m^2_{+}$ and $m^2_{-}$ whose distribution is displayed in (D).}\label{fig:t2-mmiss}
\end{figure}

After the $\Kl$ decay vertex identification involving time of flight analysis presented in the previous Section, the signal to background ratio among the surviving events is about 13.9. The remaining background, mostly originating from $\Kl\to\pi\mu\nu$ decays, can be discriminated using quantities related to missing mass in the decay. To this end, the following variables were defined for each of the two tracks identified as $\Kl$ decay products (referred to by the particle charge):
\begin{equation}
  \label{eq:mplus_mminus}
  m^2_{\pm} = (E_K-E(\pi)_{\mp}-|\vec{p}_{miss}|)^2-\abs{\vec{p}_{\pm}}^2,
\end{equation}
where $E_K$ is the energy of the decaying kaon, $E(\pi)_{\mp}$ is the energy attributed to the track of negative (positive) charge with an assumption it was a charged pion, $\vec{p}_{\pm}$ is the momentum corresponding to positive (negative) charge track and $\vec{p}_{miss}$ is a difference between momentum of the kaon and its recorded products. As these quantities may be understood as squared masses related to each of the particle tracks in the decay, semileptonic decays with an electron or positron, whose mass is negligible compared to $\pi^{\pm}$, are characterized by $m^2_{+}$ and $m^2_{-}$ close to zero as opposed to decays with much heavier muons. Relative distributions of $m^2_{+}$ and $m^2_{-}$ displayed in~\fref{fig:t2-mmiss} reveal good separation of signal and background. To reject $\Kl\to\pi\mu\nu$, only events lying below the marked anti-diagonal line were accepted, which corresponds to the following cut on the $m^2_{+}+m^2_{-}$ sum presented in the last panel of~\fref{fig:t2-mmiss:sum}:
\begin{equation*}
m^2_{+} + m^2_{-} < 0.015\:(\text{GeV}/c^{2})^2.
\end{equation*}

The cut on $m^2_{+} + m^2_{-}$ is very powerful in terms of discrimination of background from decays other than $\Kl\to\pi e\nu$, increasing the total signal to background ratio in the selected sample of $\Ks\Kl\to\pi^+\pi^-\;\pi e\nu$ from about 14 to 64. It should be noted, however, that the same cut could not have been employed to the selection of semileptonic decays of $\Ks$ in the study of $\Ks\Kl\to\pi e\nu\;3\pi^0$ as the calculated mass values strongly depend on resolution of the input kaon momentum, which is significantly lower when obtained from the $\Kl\to 3\pi^0$ decay. After this stage of selection, the S/B ratio among the events selected as $\Ks\Kl\to\pi^+\pi^-\;\pi e\nu$ was almost twice as large as for the other class of processes, therefore no further purification of the event sample was attempted.

\section{Determination of event selection efficiencies}
\label{sec:efficiencies}

\subsection{Study of applied selection criteria performance with Monte Carlo simulations}
\label{sec:eff-mc}
Performance of each of the event selection cuts described in the previous Sections was
assessed using Monte Carlo-simulated events by evaluating the cuts' efficiency for signal events as well as signal to background ratio obtained after each cut. Results are presented in Tables~\ref{tab:eff_mc_t1} and~\ref{tab:eff_mc_t2} respectively for selection of the $\Ks\Kl\to\pi e\nu\;3\pi^0$ and $\Ks\Kl\to\pi^+\pi^-\;\pi e\nu$ processes.
In case of several cuts used to select semileptonic decays which rely on interaction of the decay products in the electromagnetic calorimeter, there is a discrepancy in the efficiency between $\pi^+e^-$ and $\pi^-e^+$. This effect is caused by different interactions allowed for positively and negatively charged pions in the EMC material, leading to a higher registration probability for $\pi^+$~\cite{Marin:1998zf}.

\begin{table}[h!]
  \centering
  \caption{Subsequent cuts and reconstruction requirements applied in the selection of $\Ks\Kl\to\pi e\nu\;3\pi^0$ events with the signal efficiencies of each step and signal to background ratio after that step. Signal is understood as events defined by both kaon decays, i.e.\ $\Ks\Kl\to\pi e\nu\;3\pi^0$. Efficiencies were obtained by applying the selection to Monte Carlo-simulated events. For the steps where efficiency differs significantly depending on the charge of lepton on the semileptonic decay, separate efficiency values were given.}\label{tab:eff_mc_t1}
  \begin{tabular}{cccc}
    \toprule
    Cut description & Section & \makecell{Cut\\efficiency [\%]} & \makecell{S/B\\after cut} \\ 
    \midrule
    \addlinespace[1ex]
    \makecell{2 DC tracks with common vertex\\
    within $r_T=<$15~cm and $|z|<$10~cm} & \ref{sec:preselection-t1} & 85.26 & 0.001 \\
    \addlinespace[1ex]
    \makecell{6 or more EMC clusters with E>20 MeV\\
    and not associated to DC tracks} & \ref{sec:preselection-t1} & 57.88 & 0.007 \\
    \addlinespace[1ex]
    \makecell{At least one 6-cluster set\\
    matching the $\Kl\to 3\pi^0\to 6\gamma$ hypothesis} & \ref{sec:kl3pi0} & 82.48 & 0.009 \\
    \addlinespace[1ex]
    \makecell{$   \ang{40} < \alpha_{CM} < \ang{170}$ and\\
    $\text{M}(\pi,\pi) < 490\:\text{MeV/c}^{2}$} & \ref{sec:ksemil} & 91.45 & 0.013\\
    \addlinespace[1ex]
    \makecell{Correct extrapolation of both\\
    DC tracks to EMC clusters} & \ref{sec:t1_tof} & 43.61 & 0.014\\
    \addlinespace[1ex]
    \makecell{TOF cuts on $|d\delta t_{\pi,\pi}| \in (1.5,\;10)\:\text{ns}$\\
    and $d\delta t_{e,\pi}$ vs $d\delta t_{\pi,e}$} & \ref{sec:t1_tof} & \makecell{89.34 ($\pi^+e^-\bar{\nu}$)\\86.33 ($\pi^-e^+\nu$)} & 0.178 \\
    \addlinespace[1ex]
    \makecell{TOF cut on $\delta t(\pi)$ vs $\delta t(e)$} & \ref{sec:t1-fine-selection} & \makecell{92.33 ($\pi^+e^-\bar{\nu}$)\\90.35 ($\pi^-e^+\nu$)} & 1.05 \\
    \addlinespace[1ex]
    \makecell{Rejection of $\Ks\to\pi^0\pi^0$ by identification\\
    of prompt photons with R/(cT)> 0.9} & \ref{sec:ks2pi0_rejection} & 95.78 & 2.7 \\
    \addlinespace[1ex]
    \makecell{Cut on $d_{PCA}$ vs $\Delta E(\pi,e)$} & \ref{sec:t1-fine-selection} & \makecell{85.93 ($\pi^+e^-\bar{\nu}$)\\86.82 ($\pi^-e^+\nu$)} & 11.3 \\
    \addlinespace[1ex]
    \makecell{$\Ks\to\pi^+\pi^-(\to\pi\mu\nu)$ rejection\\
    using $e/\pi$ and $e/\mu$ classifiers} & \ref{sec:pimu_rejection} & \makecell{94.28 ($\pi^+e^-\bar{\nu}$)\\96.31 ($\pi^-e^+\nu$)} & 33.5 \\
    \bottomrule
  \end{tabular}
\end{table}

As the observables of the \Ts~symmetry test, the $R_2^{exp}$ and $R_4^{exp}$ ratios of double decay rates will be considered as functions of the time difference between kaon decays in an event, the efficiencies used to correct the experimentally recorded decay rate spectra must be studies separately for intervals of possible $\Delta t$ values. Such $\Delta t$-dependent efficiencies were determined using MC as well as using KLOE data and independent control samples discussed in the next subsection.

\begin{table}[h!]
  \centering
  \caption{Subsequent cuts and reconstruction requirements applied in the selection of $\Ks\Kl\to\pi^+\pi^-\;\pi e\nu$ events with the signal efficiencies of each step and signal to background ratio after that step. Signal is understood as events defined by both kaon decays, i.e.\ $\Ks\Kl\to\pi^+\pi^-\;\pi e\nu$. Efficiencies were obtained by applying the selection to Monte Carlo-simulated events. For the steps where efficiency differs significantly depending on the charge of lepton on the semileptonic decay, separate efficiency values were given.}\label{tab:eff_mc_t2}
  \begin{tabular}{cccc}
    \toprule
    Cut description & Section & \makecell{Cut\\efficiency [\%]} & \makecell{S/B\\after cut} \\
    \midrule
    \makecell{2 DC tracks with common vertex\\
    within $r_T=<$15~cm and $|z|<$10~cm} & \ref{sec:kspipi} & 82.58 & 0.238 \\
    \addlinespace[1ex]
    $\abs{\text{M}(\pi,\pi) - m_{\kaon}} < 10\:\text{MeV}$ & \ref{sec:kspipi} & 86.63 & 0.276 \\
    \addlinespace[1ex]
    $\abs\Big{\abs{\vec{p}_{trk1} + \vec{p}_{trk2}} - 110\:\text{MeV/c}} < 25\:\text{MeV/c}$ & \ref{sec:kspipi} & 99.51 & 0.277 \\
    \addlinespace[1ex]
    \makecell{at least 1 set of 2 DC tracks with\\
    a common vertex excluding $\Ks\to\pi^+\pi^-$} & \ref{sec:klsemil} & 67.08 & 0.515 \\
    \addlinespace[1ex]
    \makecell{at least 1 DC vertex with correct\\
    extrapolation of 2 tracks to EMC clusters} & \ref{sec:klsemil} & \makecell{46.95 ($\pi^+e^-\bar{\nu}$)\\46.77 ($\pi^-e^+\nu$)} & 1.23 \\
    \addlinespace[1ex]
    \makecell{at least 1 DC vertex passing TOF cuts} & \ref{sec:klsemil} & \makecell{60.59 ($\pi^+e^-\bar{\nu}$)\\57.72 ($\pi^-e^+\nu$)} & 13.9 \\
    \addlinespace[1ex]
    \makecell{exactly 1 surviving $\Kl\to\pi e\nu$ vertex candidate} & \ref{sec:klsemil} & 99.89 & 14.1 \\
    \addlinespace[1ex]
    $m^2_{+} + m^2_{-} < 0.015\:(\text{GeV}/c^{2})^2$ & \ref{sec:trackmass} & 96.15 & 64.5 \\
    \bottomrule
  \end{tabular}
\end{table}

\subsection{Evaluation of selection efficiencies with KLOE data and control samples}
\label{sec:eff_data}
Total event selection efficiencies included in the final results were obtained using KLOE data
%
% TODO: add footnote about dpca exception
%
in order to avoid artifacts from possible inconsistencies between properties of MC-simulated and real events. As each of the two classes of processes studied in this analysis is constituted by two distinct neutral kaon decays, four independent control samples of events were selected from KLOE data:
\begin{itemize}
\item $\Ks\to\pi e \nu$ events where the associated $\Kl$ reached the electromagnetic calorimeter and interacted therein --- for selection efficiency of semileptonic $\Ks$ decays,
\item $\Ks\to\pi^+\pi^-$ events with a similar $\Kl$ interaction in the EMC --- for efficiency of selection of $\Ks\to\pi^+\pi^-$,
\item $\Ks\Kl\to\pi^+\pi^-\;3\pi^0$ events --- for the $\Kl\to 3\pi^0\to 6\gamma$ selection efficiency,
\item $\Ks\Kl\to\pi^0\pi^0\;\pi e\nu$ --- for selection efficiency of semileptonic decays of $\Kl$.
\end{itemize}

For each of the control samples, selection of the decay whose efficiency was under study was applied to $\phi\to\Ks\Kl$ events in the same form as selection of the main processes of interest, $\pi e\nu\;3\pi^0$ and $\pi^+\pi^-\;\pi e\nu$. Events with neutral K mesons were identified (tagged) by presence of a certain type of interaction of the kaon accompanying the one whose selection was tested. Number of the registered interactions tagging $\phi\to\Ks\Kl$ ($N_{tag}$) was multiplied by the branching ratio for the investigated kaon decay $\mathrm{K}\to\mathrm{X}$ to obtain the number of expected events:
\begin{equation}
  \label{eq:eff_n_expected_ks}
  N_{exp}^{\mathrm{K}\to\mathrm{X}} = N_{tag}\cdot \text{BR}(\mathrm{K}\to\mathrm{X}),
\end{equation}
for the case of efficiency for decays of $\Ks$, which is not limited by detector size. For estimation of the $\Kl$ decays' detection efficiency, the differential number of expected events was determined:
\begin{equation}
  \label{eq:eff_n_expected_kl}
  \frac{dN_{exp}^{\mathrm{K}\to\mathrm{X}}}{d t_{k}} = N_{tag}\cdot \text{BR}(\mathrm{K}\to\mathrm{X}) \frac{1}{\tau_L}e^{-t_K/\tau_{L}},
\end{equation}
where $t_K$ is the time of the long-lived kaon decay and $\tau_{L}$ denotes its mean lifetime.

In case of the study of early kaon decays, i.e.\ $\Ks\to\pi e\nu$ and $\Ks\to\pi^+\pi^-$, their selection efficiency was estimated as a single value independent of the kaon decay time:
\begin{equation}
  \label{eq:eff_scalar_efficiency}
  \varepsilon^{\mathrm{K}\to\mathrm{X}} = \frac{N^{\mathrm{K}\to\mathrm{X}}_{obs}}{N^{\mathrm{K}\to\mathrm{X}}_{exp}},
\end{equation}
where $N^{\mathrm{K}\to\mathrm{X}}_{obs}$ is the number of observed events, i.e.\ number of events which passed the selection under study. Uncertainty of the efficiency was evaluated using the binomial approach~\cite{behnke, statisctical_methods}:
\begin{equation}
  \label{eq:eff_binomial_error}
  \sigma\left({\varepsilon^{\mathrm{K}\to\mathrm{X}}}\right) = \frac{\sqrt{N^{\mathrm{K}\to\mathrm{X}}\left(1-\varepsilon^{\mathrm{K}\to\mathrm{X}}\right)}}{N^{\mathrm{K}\to\mathrm{X}}_{exp}}.
\end{equation}

For the two late kaon decays, $\Kl\to 3\pi^0$ and  $\Kl\to\pi e\nu$, selection efficiency may depend on the location of decay inside the detector and was therefore studied separately in subsequent intervals of the proper time of kaon decay $t'_{\Kl}$. To this end, an expected $\Kl$ decay time distribution was obtained by scaling a binned exponential decay $\exp\left(-t/\tau_L\right)$ so that its integral for $t\in[0;\infty)$ was equal to the expected number of events obtained as shown in~\eref{eq:eff_n_expected_kl}. For events surviving the selection whose efficiency was investigated, the observed $t'_{\Kl}$ spectrum was obtained with the same binning as the expected exponential decay. Finally, the efficiency was evaluated for each $t'_{\Kl}$ interval as a ratio of observed to expected counts in a given bin of the $t'_{\Kl}$ distributions. Uncertainty of efficiency in each bin was again calculated assuming binomial errors as in~\eref{eq:eff_binomial_error}. Efficiencies of selection of the semileptonic decays were determined separately for each of the two lepton charges.

Total efficiencies of selection of the two main classes of studied processes, $\pi e\nu\;3\pi^0$ and $\pi^+\pi^-\;\pi e\nu$, which are required for further analysis, must be expressed as functions of the difference of kaons' proper decay times. As the aforementioned efficiency estimation scheme results in independent values for each of the kaon decays, total efficiencies for double kaon decay events were calculated with the following approximations for the two processes of interest:
\begin{equation*}
%  \label{eq:eff_approximation}
  \begin{split}
    \Delta t & = t_{3\pi^{0}} - t_{\pi e \nu} \\
             & \approx\ t_{3\pi^{0}} \text{\ \ for\ \ }{t_{3\pi^{0}} \gg t_{\pi e \nu}},  \\  
   \Delta t & = t_{\pi e \nu} - t_{\pi^+\pi^-} \\
            & \approx\ t_{\pi e \nu} \text{\ \ for\ \ } {t_{\pi e \nu} \gg t_{\pi^+\pi^-}},
          \end{split}
\end{equation*}
valid for $\Delta t\gg 0$, which is sufficient to obtain a correct efficiency estimate for the asymptotic region of the double kaon decay rates. Consequently, the efficiency for large time differences was obtained from the time-dependent efficiency distributions for the late kaon decays, uniformly scaled with the single-value efficiencies for the early kaon decays.

\subsection{The $\Ks\to\pi e\nu$ and $\Ks\to\pi^+\pi^-$ control samples}
In the two control samples used for evaluation of efficiency for $\Ks\to\pi e\nu$ and $\Ks\to\pi^+\pi^-$, presence of neutral kaons in an event was identified by requiring that an interaction of the long-lived neutral meson was recorded in the EMC. As this happens for about \SI{60}{\percent} of $\phi\to\Ks\Kl$ decays, a clean sample of neutral kaon pairs is obtained this way.
Identification of $\Kl$ interactions in the calorimeter was based on the procedure described in References~\cite{daria_article, daria_memo, silarski_phd}.
Such events are characterized by a large energy deposit not related to any drift chamber track, therefore all clusters without track association and $E>100$~MeV were considered as $\Kl$ interaction candidates. Moreover, due to the p-wave distribution of polar angles $\theta$ of kaons' emission in a $\phi$ decay, given by:
\begin{equation}
  \label{eq:pwave}
  \frac{dN}{d\Omega} \propto \sin^2\theta,
\end{equation}
a majority of the $\Kl$ reaching the calorimeter are expected to interact in the barrel part. EMC clusters located in the end-caps are thus excluded to discriminate machine background. As in the analysis of $\Kl\to 3\pi^0$ the momentum direction of the long-lived kaon is not restricted (not to compromise selection efficiency), Monte Carlo simulations were used to confirm that the efficiency for selection of $\Ks$ decays does not exhibit a significant dependence on the momentum direction of the associated $\Kl$.

%
% MEMO TODO: dodać stosowny obrazek z rozkładem kątowym i liczby dowodzące, że to nie ma znaczenia
%
As the average velocity of a kaon produced in a $\phi$ decay at DA$\Phi$NE is well defined and amounts to about $\beta\approx 0.22$, the hypothetical $\Kl$ velocity was calculated for each candidate cluster as $\beta_{cl} = \frac{R_{cl}}{cT_{cl}}$ where $R_{cl}$ denotes distance between average $\phi$ decay point and the EMC cluster recorded at time $T_{cl}$ and $c$ is the speed of light (see~\sref{sec:ks2pi0_rejection} where a similar variable is used to identify clusters from prompt photons). The $\beta_{cl}$ value was transformed to the CM reference frame of $\phi$ and restricted to the region:
\begin{equation}
  \label{eq:betacl}
   0.2 < \beta^{CM(\phi)}_{cl} < 0.25,
\end{equation}
%
% MEMO TODO: dodac wykresy energii i bety
%
whose upper limit takes into account a possible shift due to incorrect $T_0$ in an event. To achieve a high sample purity, the selected events were finally limited to ones where the energy deposited in the EMC cluster was above 250~MeV. Tests with MC revealed that background contamination of the selected $\phi\to\Ks\Kl$ samples was about \SI{0.8}{\percent}. 

The vector spanning between average $\phi$ decay point at KLOE and location of the EMC cluster corresponding to $\Kl$ interaction provides a precise estimate of the kaon momentum direction due to large distance from detector center to the calorimeter. Momentum of the $\Ks$ meson, inferred using momentum conservation in the $\phi\to\Ks\Kl$ and $\vec{p}_{\Kl}$ direction as described in~\sref{sec:ks_from_kl} is therefore characterized by a considerably higher resolution compared to the one obtained with $\Kl\to 3\pi^{0}$. Consequently, as several steps of $\Ks\to\pi e \nu$ event selection rely on $\Ks$ momentum estimate obtained from the partner kaon, the total selection performance is prone to behave differently when applied to the control sample rather than to $\Kl\to 3\pi^{0}$ events. Therefore, the $\vec{p}_{\Kl}$ direction obtained using the respective EMC cluster location in case of the control sample was subjected to an artificial smearing with double Gaussian resolution model obtained with Monte Carlo-simulated $\Ks\Kl\to 3\pi^0\;\pi e\nu$ events. In case of the other control sample, $\Ks\to\pi^+\pi^-$, the event selection does not use the momentum estimate from $\Kl$ and therefore no smearing was introduced.
%
% MEMO TODO: Dokładny opis i wykresy do smearingu
%

\subsection{The $\Ks\Kl\to\pi^+\pi^-\;3\pi^0$ and $\Ks\Kl\to\pi^0\pi^0\;\pi e\nu$ control samples}
For estimation of the $\Kl\to 3\pi^0$ selection efficiency, these decays were tagged by a $\Ks$ decaying into charged pions. The applied selection of $\Ks\to\pi^+\pi^-$ was based on the already devised cuts discussed in~\sref{sec:kspipi}, which allow for preparation of a $\phi\to\Ks\Kl$ sample with a background contamination at a level below \SI{1}{\percent}. As the studied selection of $\Kl$ decays into neutral pions does not use any information related to the $\Ks$ decay, it was applied to the selected sample without further preparations.

In the control sample used to determine the selection efficiency of semileptonic decays of the long-lived neutral kaon, $\phi\to\Ks\Kl$ events were tagged by the $\Ks\to\pi^0\pi^0$ decay. Selection of neutral hadronic decays of Ks was performed following the approach described in Reference~\cite{memo_225}. Firstly, presence of at least 3 EMC clusters was required whose total energy was above 300~MeV. Each of the clusters matching these requirements was required to have a deposited energy in the range $(20;300)$~MeV and a polar angle above 21$^\circ$ to differentiate them from machine background. Subsequently, the clusters were required to match a prompt photon hypothesis, i.e.\ $0.9<\frac{R_{cl}}{cT_{cl}}<1.2$ (compare~\sref{sec:ks2pi0_rejection}). If 4 of 5 EMC clusters satisfying the aforementioned criteria were found in an event, with a total invariant mass of the photons within the range (390; 600)~MeV and at least one of the clusters recorded in the calorimeter barrel, the event was considered a $\Ks\to\pi^0\pi^0$ candidate. Additionally, a $\Ks\to\pi^+\pi^-$ veto was applied by negation of the selection conditions described in~\sref{sec:kspipi}.
According to MC-based studies, purity of the selected $\Ks\to\pi^0\pi^0$ event sample was at the level of \SI{98}{\percent}. Although the $\Ks$ momentum estimate available from photons' momenta in the $\Ks\to\pi^0\pi^0$ is considerably worse than in case of the charged pion channel, the selection of $\Kl\to\pi e\nu$ decays was prepared so as to be independent of accompanying kaon decay and its properties, and thus the tested cuts were applied directly to the control sample.

\subsection{MC-based correction to $\pi e\nu\;3\pi^0$ efficiency obtained with data}\label{sec:dpca_correction}
\begin{figure}[h!]
  \centering
  \includegraphics[width=0.7\textwidth]{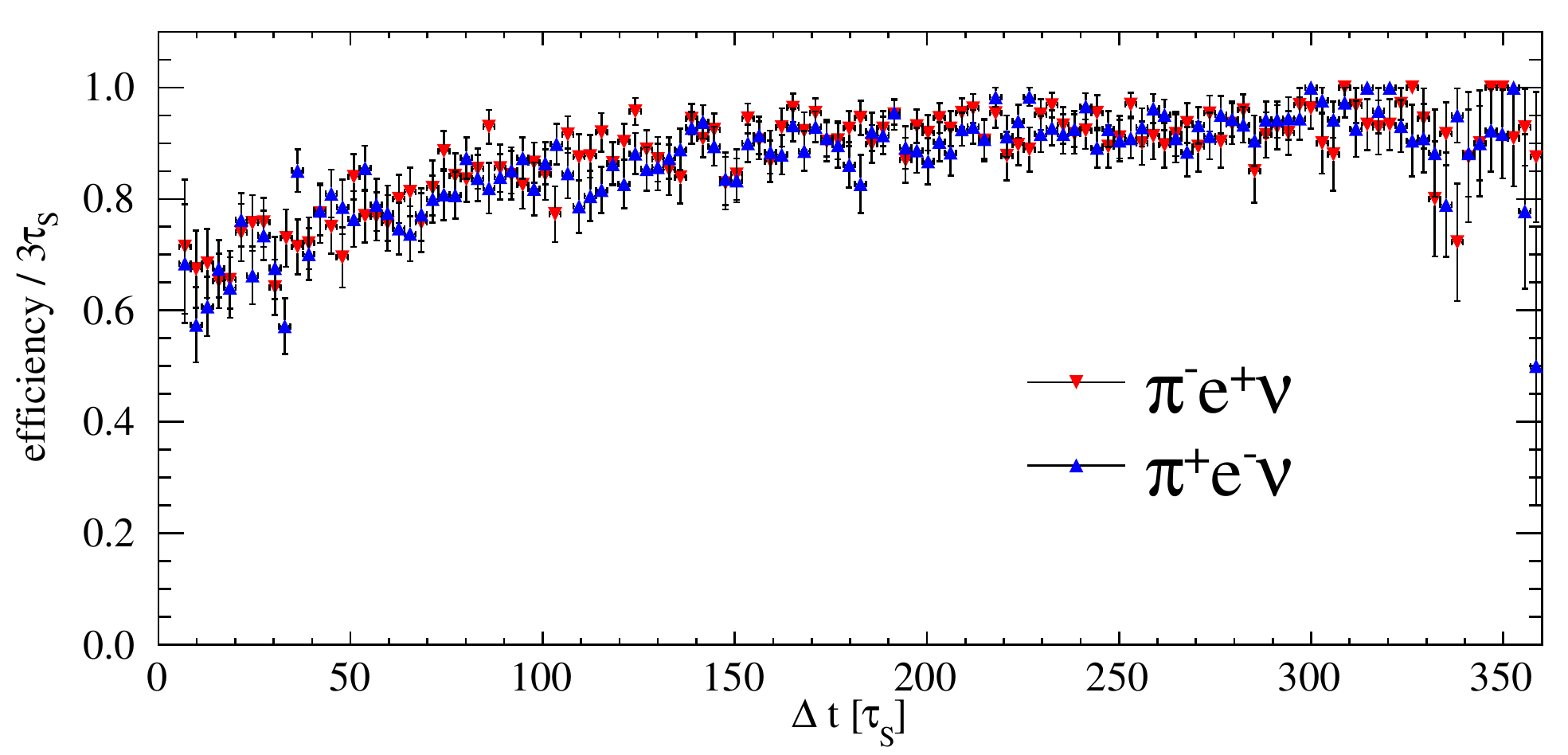}
  \caption{Shape of the $d_{PCA}$ vs. $\Delta E(\pi,e)$ cut efficiency as a function of kaons' decay time difference based on Monte Carlo-simulated $\pi e\nu\;3\pi^{0}$ events.}\label{fig:dpca_eff_mc}
\end{figure}

The scheme of determination of efficiencies based on independent control samples for each of the kaon decays relies on an assumption that the efficiency for $\Ks$ decay selection does not depend on the time of decay of $\Kl$ in the same event, so that it may be approximated by a single number uniformly scaling the $t_{\Kl}$-based efficiency for the associated $\Kl$ decay. It was tested with Monte Carlo simulations that this assumption is valid for all of the analysis steps except for the cut on $d_{PCA}$ vs. $\Delta E(\pi,e)$ described in~\sref{sec:t1-fine-selection}. In case of this cut, the $\Delta E(\pi,e)$ value is calculated using i.a.\ the total momentum of the decaying $\Ks$ meson. As the latter is inferred indirectly from the $\phi$ decay kinematics and $\Kl$ momentum direction, this variable is sensitive to the angular resolution of estimated $\Kl$ flight direction. Although the spatial resolution for $\Kl\to 3\pi^0$ is mostly constant (see~\fref{fig:t2-kspipi}, left), decays, the angular resolution of $\Kl$ direction varies strongly with decay distance from the primary interaction point, influencing $\vec{p}_{\Ks}$ resolution as visible in~\fref{fig:ks_angle_resolution}. Consequently, the shape of efficiency of the $d_{PCA}$ vs. $\Delta E(\pi,e)$ cut as a function of $\Delta t$ is not uniform as in~\fref{fig:dpca_eff_mc}. However, this selection step is powerful in terms of background rejection (see~\tref{tab:eff_mc_t1}) and thus cannot be removed from the analysis chain. Whereas it could be replaced by constraints imposed on other kinematical quantities sensitive to the $\Ks\to\pi^+\pi^-(\to\pi\mu\nu)$ background such as missing mass, the same effect is still expected as such variables also depend on accuracy of the $\Ks$ momentum estimation.

As shape of this efficiency cannot be reproduced using independent control samples for $\Ks\to\pi e\nu$ and $\Kl\to 3\pi^0$, the efficiency for selection of $\Ks$ semileptonic decays was estimated using control data sample and applying all selection steps except the requirement on $d_{PCA}$ vs. $\Delta E(\pi,e)$. Subsequently, a Monte Carlo simulation-based correction for the efficiency of this cut presented in~\fref{fig:dpca_eff_mc} was imposed on the data-derived efficiencies to obtain the total efficiencies used in the determination of the final result.

\subsection{Comparison of selection efficiencies from data and MC}
\label{sec:eff_comparison}
The event selection steps described in Sections~\ref{sec:t-analysis-1} and~\ref{sec:t-analysis-2} resulted in four sets of events, each characterized by a certain pair of time-ordered neutral kaon decays:
\begin{itemize}
\item $\Ks\Kl\to\pi^+ e^- \bar{\nu}\;3\pi^{0}$,
\item $\Ks\Kl\to\pi^- e^+ {\nu}\;3\pi^{0}$,
\item $\Ks\Kl\to\pi^+\pi^-\;\pi^+ e^- \bar{\nu}$,
\item $\Ks\Kl\to\pi^+\pi^-\;\pi^- e^+ {\nu}$.
\end{itemize}

\begin{figure}[h!]
  \centering
  \includegraphics[width=0.9\textwidth]{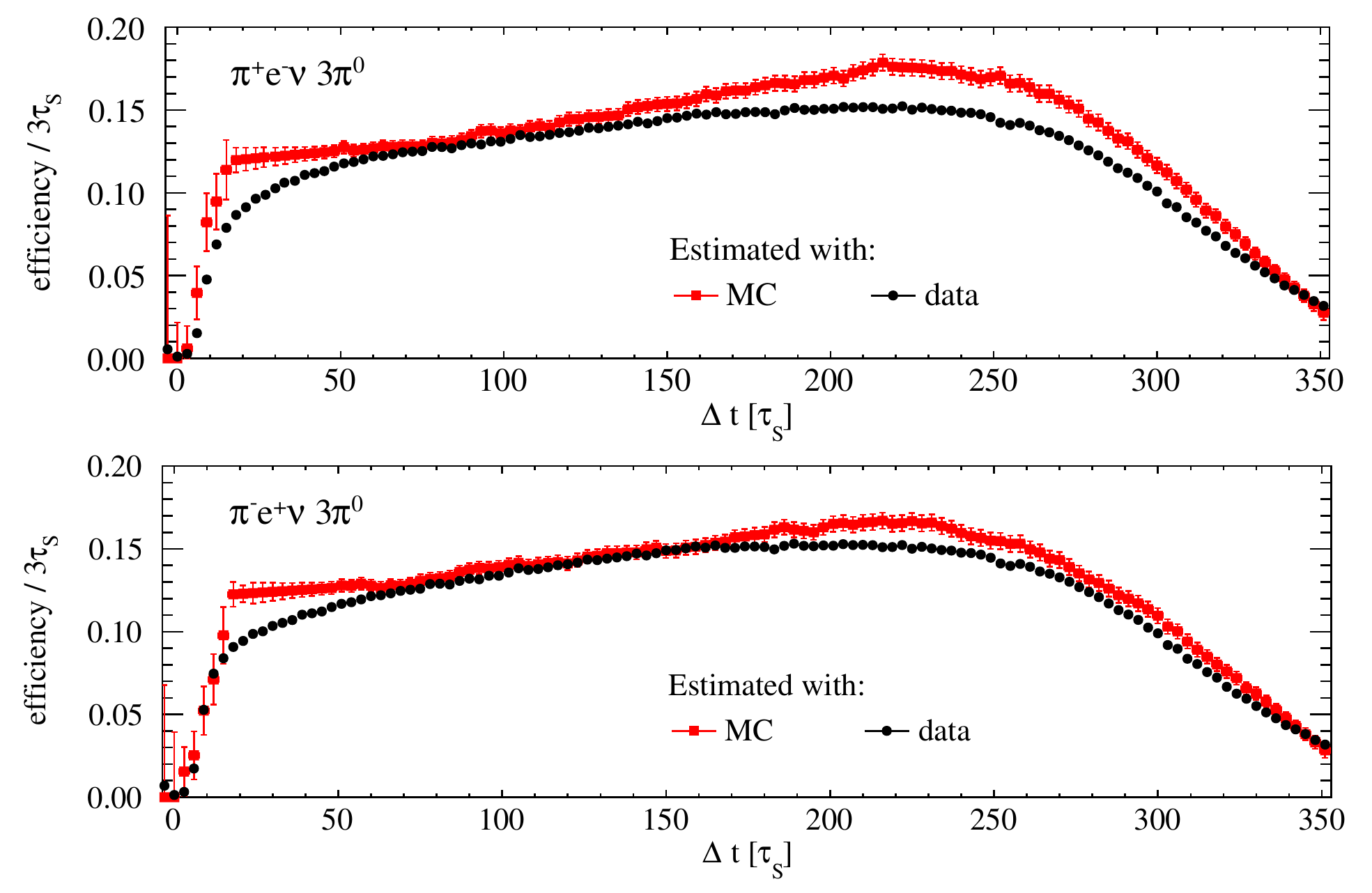}  
  \includegraphics[width=0.9\textwidth]{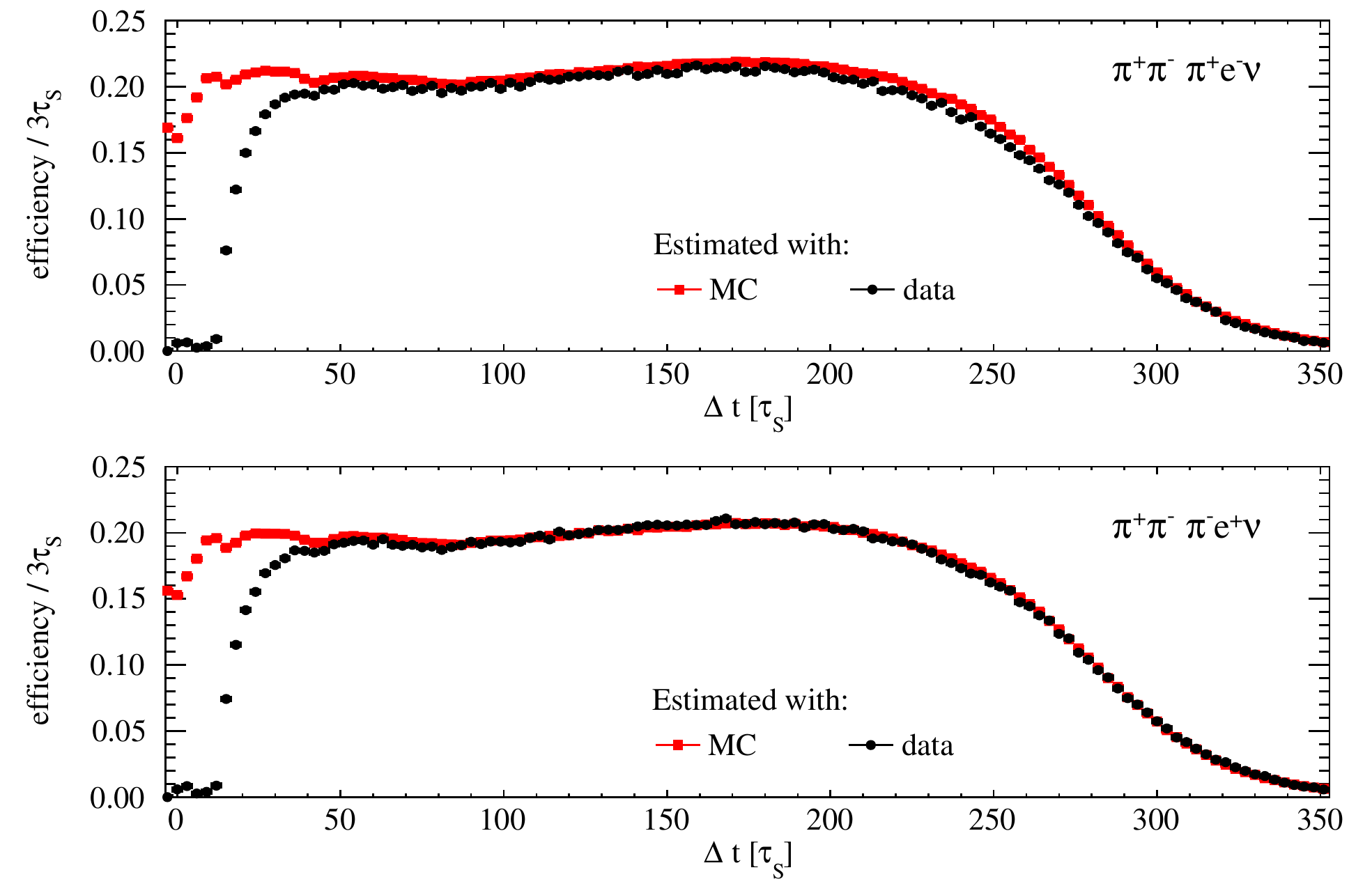}  
  \caption{Efficiencies of event selection as functions of the kaons' proper decay time difference for the four event subsamples used in the analysis. Efficiency obtained using data and an independent control sample for each kaon decay (black), used further in the determination of $R_2$ and $R_4$, is compared with efficiency obtained with generated Monte Carlo events (red). The discrepancy in the region of about $\Delta t < 50\ \tau_{S} $ visible for all subsamples is an artifact of the efficiency estimation method with control samples.}\label{fig:eff_comparison}
\end{figure}

Each of the four subsamples listed above is characterized by a separate $\Delta t$-dependent event selection efficiency. Such efficiencies obtained from data and control samples as described in the previous Section are presented as black points in~\fref{fig:eff_comparison}. Efficiencies obtained using MC-generated events are shown in red points for a comparison. In each of these efficiencies, a discrepancy between data and MC-derived distributions is visible for low time differences of approximately $\Delta t < 50 \tau_{S}$, caused by the nature of the employed method of efficiency estimation using independent control samples for $\Ks$ and $\Kl$ decays. However, as the final estimation of the \Ts-asymmetric observables is concentrated on the asymptotic $\Delta t \gg \tau_{S}$ region, the further considerations are limited to $\Delta t$ ranges above this discrepancy.

Whereas data and MC-derived efficiencies for the $\pi^+\pi^-\;\pi e\nu$ process and large decay time differences exhibit good agreement, the efficiency for $\pi e \nu\;3\pi^0$ is overestimated in the Monte Carlo simulations, especially in terms of the excess of recorded $\pi^+e^-\bar{\nu}$ events due to different interactions of $\pi^+$ and $\pi^-$ in the KLOE calorimeter, which is less pronounced in data. This discrepancy, demonstrated in~\fref{fig:eff_comparison} is the major reason for which the efficiencies used in the determination of the final $R_2(\Delta t)$ and $R_2(\Delta t)$ distributions are based on data and control samples rather than MC simulations.

\section{Determination of probability asymmetry distributions for the T test with KLOE}\label{sec:ratios}
\begin{figure}[h!]
  \centering
  \includegraphics[width=0.9\textwidth]{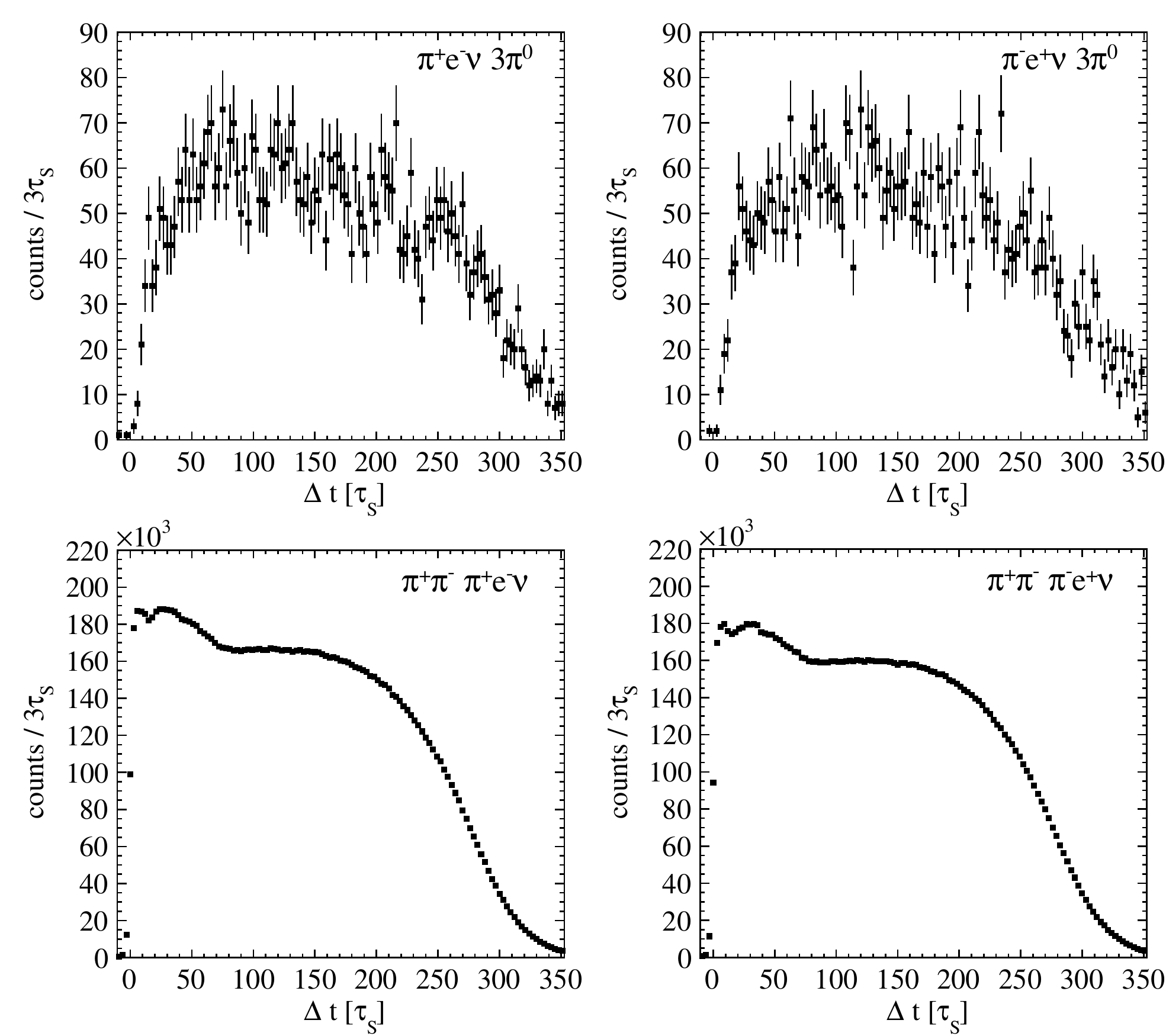}
  \caption{Double kaon decay rates as a function of the difference of kaon proper decay times for each of the studied class of events, as obtained after event selection and reconstruction. Presented distributions are not corrected for selection efficiencies. Errors on the number of counts in each bin of the histograms are Poissonian.}
  \label{fig:dt_plots}
\end{figure}

In order to evaluate the desired observables of the \Ts~symmetry test, i.e.\ the following ratios $\Delta t$-dependent double decay rates:
\begin{eqnarray}
  \label{eq:final_r2_and_r4}
  R_2(\Delta t) &= \frac{\mathrm{I}(\pi^+e^-\bar{\nu},3\pi^0;\Delta t)}{\mathrm{I}(\pi^+\pi^-,\pi^-e^+\nu;\Delta t)} \cdot \frac{1}{D},\\
  {R_4(\Delta t)} & = \frac{\mathrm{I}(\pi^-e^+\nu,3\pi^0;\Delta t)}{\mathrm{I}(\pi^+\pi^-,\pi^+e^-\bar{\nu};\Delta t)}  \cdot \frac{1}{D},
\end{eqnarray}
where $D$ is the constant factor discussed in~\sref{sec:d_determination}, each of the event samples listed above, along with their selection efficiencies, was divided into $\Delta t$ intervals (bins) of 3~$\tau_{S}$ width.
\fref{fig:dt_plots} presents the resulting double decay rate distributions for each of the four processes. The $R_2(\Delta t)$ and $R_4(\Delta t)$ ratios were calculated for each $i$-th bin of $\Delta t$ using the event counts in the respective distributions (referred to as $N(f_1,\:f_2;\Delta t_i)$ for final states $f_1$ and $f_2$), including selection efficiencies obtained with control data samples and the $D$ factor:
\begin{eqnarray}
  \label{eq:r2_and_r4_bin_value}
  R_{2}(\Delta t_i) &= \frac{N(\pi^+ e^-\bar{\nu},\:3\pi^0;\:\Delta t_i) / \varepsilon(\pi^+ e^-\bar{\nu},\:3\pi^0;\:\Delta t_i)}{N(\pi^+\pi^-,\:\pi^-e^+\nu;\:\Delta t_i) / \varepsilon(\pi^+\pi^-,\:\pi^-e^+\nu;\:\Delta t_i)}  \cdot \frac{1}{D},\\
  R_{4}(\Delta t_i) &= \frac{N(\pi^- e^+{\nu},\:3\pi^0;\:\Delta t_i) / \varepsilon(\pi^- e^+{\nu},\:3\pi^0;\:\Delta t_i)}{N(\pi^+\pi^-,\:\pi^+e^-\bar{\nu};\:\Delta t_i) / \varepsilon(\pi^+\pi^-,\:\pi^+ e^-\bar{\nu};\:\Delta t_i)}  \cdot \frac{1}{D}.
\end{eqnarray}

\begin{figure}[h!]
  \centering
  \includegraphics[width=0.9\textwidth]{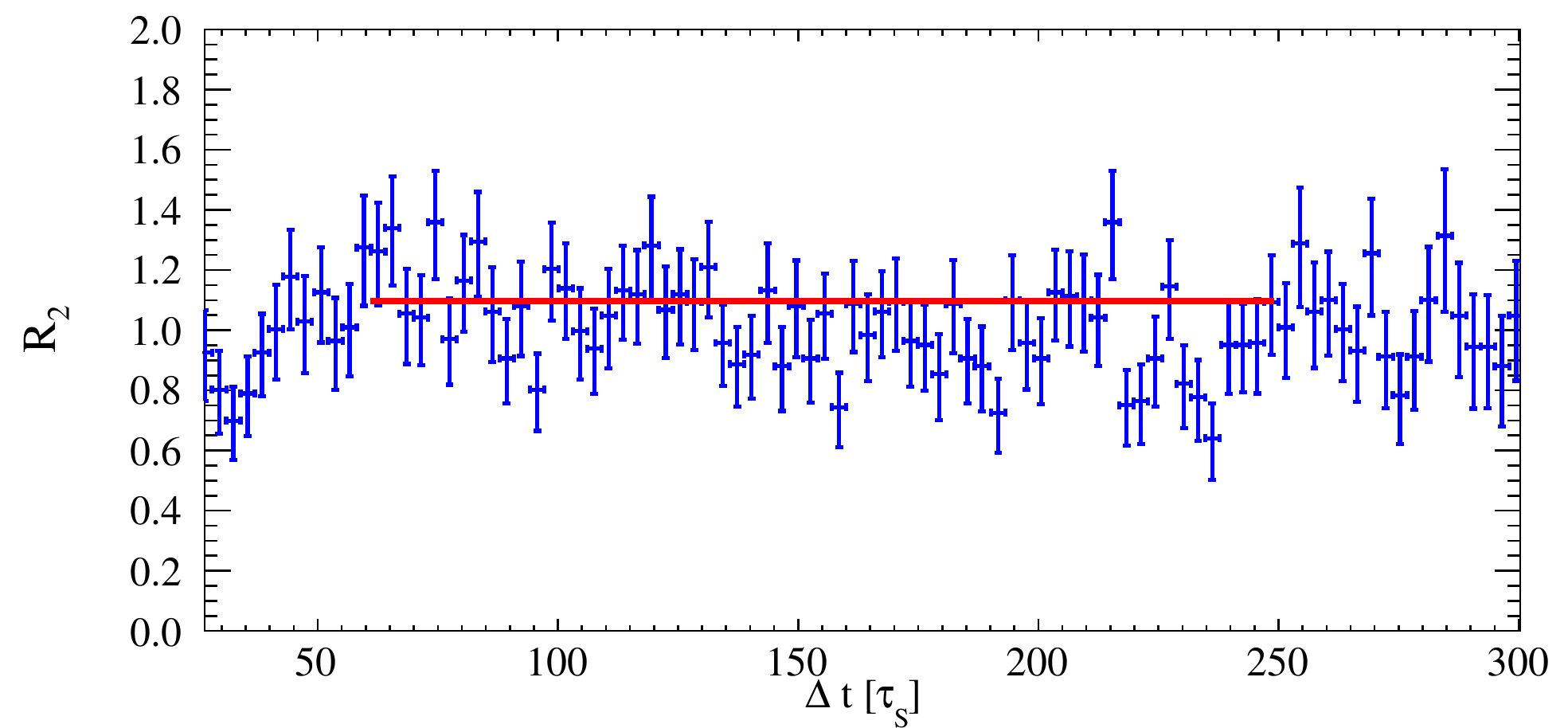}\\
  \includegraphics[width=0.9\textwidth]{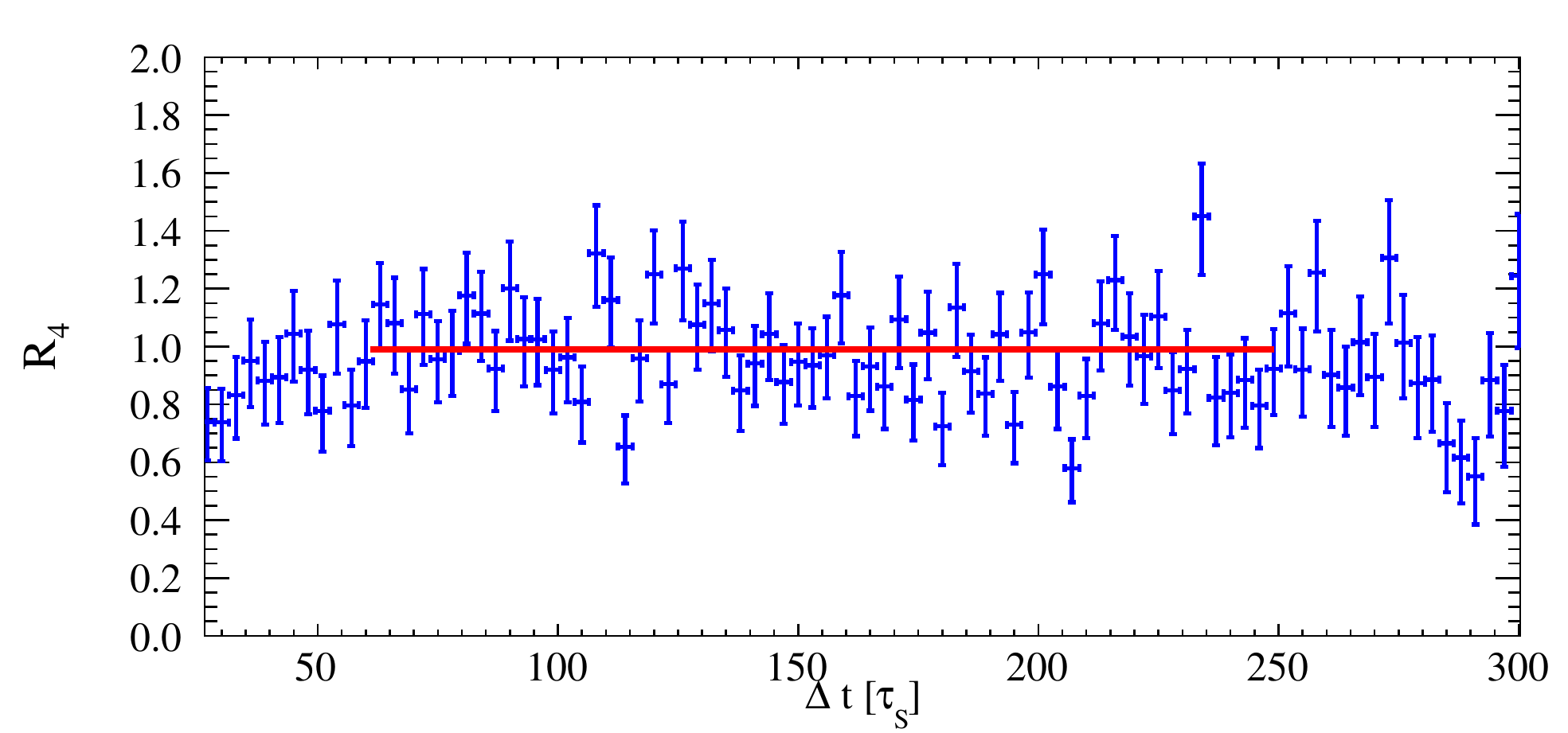}
  \caption{Two ratios of double decay rates sensitive to $\Ts$ symmetry violation obtained in the analysis of KLOE experiment data presented in this Thesis. The red line presents results of maximum likelihood fits of a constant level to the ratios in the $\Delta t \gg \tau_{S}$ region.}
  \label{fig:final_ratios}
\end{figure}

The \Ts-violation-sensitive ratios obtained in the manner described above are displayed in~\fref{fig:final_ratios}. Although for the test of symmetry under reversal in time pursued in this work only the region of kaon decay time differences significantly larger than $\Ks$ lifetime is relevant (see~\sref{sec:strategies}), the whole $\Delta t$ range available at KLOE is shown for completeness. Is should be noted, however, that in the region of small $\Delta t$ arising from comparable decay times of $\Ks$ and $\Kl$,
the assumed scheme of efficiency estimation with independent control samples for both kaons may not reproduce the true efficiency correctly. 
%the applied corrections for selection efficiencies do not reproduce the efficiency correctly
Moreover, a discrepancy between the efficiency based on data and Monte Carlo simulations revealed in~\fref{fig:eff_comparison} requires a further careful investigation and therefore for the determination of the asymptotic values of $R_2$ and $R_4$ in this work the region of $\Delta t$ below 60~$\tau_S$ was discarded.

%(compare) and thus this region may not be used to determine 
On the other hand, the region of $\Delta t$ above about 250~$\tau_S$ corresponds to $\Kl$ decays close to outer limits of the KLOE detector where limited acceptance reduces selection efficiencies leading to large statistical uncertainties on the event counts used to calculate the ratios.
%
% TODO: update based on the final efficiency distributions
%

In order to quantify the deviations from unity of the asymptotic levels of $R_2(\Delta t \gg \tau_S)$ and $R_4(\Delta t \gg \tau_S)$, constant values were fitted to the ratios in the following range, characterized by relatively high and stable event selection efficiencies and their agreement between control samples and MC simulations:
\begin{equation}
  \label{eq:fit_range}
  60\:\tau_S < \Delta t < 250\:\tau_S.
  %
  % TODO: update if necessary!!!
  % 
\end{equation}
Impact of the choice of upper and lower $\Delta t$ limit used for estimation of the ratios' level was incorporated into systematic uncertainty of the result. To properly account for the nature of uncertainties on the $R_2$ and $R_4$ points, the constant level calculation was performed with a dedicated binned maximum likelihood fit described in the next Section.

\subsection{Fit of the asymptotic level of double decay rates}
\label{sec:ml_fit}
Each point in the graphs of the $R_2(\Delta t)$ and $R_4(\Delta t)$ ratios shown in~\fref{fig:final_ratios} results from a division of two counts of events characterized by their $\Delta t$ value lying in a certain interval (compare~\eref{eq:r2_and_r4_bin_value}). This may be generally expressed for the $i$-th time difference interval as:
\begin{equation}
  \label{eq:fit_general_rpoint}
\forall_i:  R_i \equiv R(\Delta t_i) = \frac{N_i}{N_i'}\frac{\varepsilon_i'}{\varepsilon_i} \frac{1}{D},
\end{equation}
where $\varepsilon_i$ and $\varepsilon_i'$ are the selection efficiencies for the events counted as $N_i$ and $N_i'$ respectively.
Each of the event counts $N_i$ and $N_i'$  may be assumed to come from a Poissonian distribution with unknown mean. Although the values of $R$ depend also on the selection efficiencies and the $D$ factor, their uncertainties are negligible compared to the Poissonian errors on the event counts for the process involving the relatively rare semileptonic decays of $\Ks$. Thus, in order to incorporate the nature of the dominating uncertainty, a maximum likelihood fit to the $R_2$ and $R_4$ points was performed. 

If the initially unknown value of the ratio asymptotic level is denoted by $r$, it may be concluded from~\eref{eq:fit_general_rpoint} that the following is true for every point:
\begin{equation}
  \label{eq:fit_true_r}
  \forall_i: r = \frac{\lambda}{\lambda'} \frac{\varepsilon_i'}{\varepsilon_i} \frac{1}{D},
\end{equation}
where $\lambda_i$ and $\lambda_i'$ denote means of the assumed Poisson distributions from which the experimentally counted numbers $N_i$ and $N_i'$ were drawn. As rates of the processes whose count enter the denominator in~\eref{eq:fit_general_rpoint} are larger by a three orders of magnitude, it may be approximated that:
\begin{equation}
  \label{eq:fit_approx}
  \lambda_i'{\approx} N_i' \text{\ \ for\ \ } {N_i'\gg\sigma(N_i')},  
\end{equation}
from which it follows that:
\begin{equation}
  \label{eq:fit_n_mean}
  \forall_i:\lambda_i = r N_i'D\frac{\varepsilon_i}{\varepsilon_i'}.
\end{equation}
Consequently, the likelihood of recording $N_i$ counts of the $\pi e\nu\;3\pi^0$ process in all the considered $\Delta t_i$ bins with an assumed level of $r$ reads:
\begin{equation}
  \label{eq:fit_likelihood}
  \mathcal{L}(r) = \prod_{i \text{ in fit limits}} p\left( N_i,\: r N_i'D\frac{\varepsilon_i}{\varepsilon_i'} \right),
\end{equation}
where $p(n,\:\lambda)$ is the Poissonian probability of observing $n$ counts with the distribution mean of $\lambda$.
In the fit, the $-\log\left(\mathcal{L}(r)\right)$ function was scanned to find an optimal value of $r$. Uncertainties were determined as shifts of $r$ with respect to the optimum which change the log likelihood value by $\frac{1}{2}$~\cite{behnke}. As no significant asymmetry between lower and upper error was observed, a symmetric uncertainty is quoted for the following results.

Constant levels of $R_2(\Delta t)$ and $R_4(\Delta t)$ obtained with the described fit procedure are marked with red solid lines in~\fref{fig:final_ratios}. The obtained values amount to:
\begin{eqnarray}
  \label{eq:fit_results_statonly}
  R_2 &= 1.020 \pm 0.017_{stat},\\
  R_4 &= 0.990 \pm 0.017_{stat}.
\end{eqnarray}

\section{Study of systematic uncertainties}\label{sec:systematics}
The systematic uncertainty of the asymptotic levels of the $R_2$ and $R_4$ ratios determined in the fit results from several factors which must be taken into account.
Firstly, the requirements imposed on the studied data and control samples at the stage of event selection may introduce a bias into the final event samples, affecting the results. The impact of the applied selection criteria on $R_2$ and $R_4$ was studied separately for each analysis step by varying the limit value of a single requirement with the remainder of event selection fixed in the same form as used to determine the result. In order to search for significant dependence of the result on a certain cut, each of the cut values was varied by several multiples of the resolution ($\sigma$) of a variable on which a cut was imposed. Systematic uncertainty introduced by a given selection requirement was estimated separately for $R_2$ and $R_4$ as an average shift of the result observed as a consequence of cut variation by $\pm1\:\sigma$. \tref{tab:systematics}~presents the resulting uncertainties attributed to particular analysis steps. Selection requirements, for which the results did not exhibit a significant dependence on the cut variation, have been neglected in the study. However, for certain analysis steps considerable systematic shifts of $R_2$ and $R_4$ levels have been  observed. These steps comprise the fine selection of semileptonic $\Ks$ decays, which requires most stringent background discrimination.
In case of the two cuts for which systematic effects are largest, i.e.\ the constraint imposed on the $\delta t(\pi)$ and $\delta t(e)$ time of flight variables as well as the cut on the relative distribution of $d_{PCA}$ and $\Delta E(\pi,e)$ (both described in~\sref{sec:t1-fine-selection}), the regions of the relevant distributions where signal and background events are discriminated are characterized by a large excess of background over signal events, making the applied cuts very sensitive.
Conversely, only two of the requirements used in the less complicated selection of the $\Ks\Kl\to\pi^+\pi^-\;\pi e\nu$ class of processes give statistically significant contributions and none of them is dominant in the total systematic uncertainty.

\begin{table}[h!]
  \centering
  \caption{Systematic error contributions of particular steps of the data analysis and determination of $R_2$ and $R_2$ asymptotic levels. Only significant effects are listed.}\label{tab:systematics}
  \begin{tabular}{lcc}
    \toprule
    Step of the analysis & \multicolumn{2}{l}{Contribution to systematic error} \\ 
    {} & $R_2$  & $R_4$ \\
    \midrule
%    \addlinespace[1ex]
    \multicolumn{3}{l}{\bf Selection of \boldmath$\Ks\Kl\to\pi e\nu\; 3 \pi^0$} \\
%    \midrule
    Requirement of $<40^o\alpha_{CM}<170^o$ & 0.0073 & 0.0105\\
    Requirement of $\text{M}(\pi,\pi) < 490 \text{MeV/c}^2$ & 0.0013 & 0.0009 \\
    TOF requirement $|d\delta t_{\pi,\pi}| \in (1.5,\;10)\:\text{ns}$ & 0.0026 & 0.0045 \\
    TOF requirements on $d\delta t_{e,\pi}$ vs $d\delta t_{\pi,e}$ & 0.0039 & 0.0035 \\
    Requirements on $\delta t(\pi)$ and $\delta t(e)$ & 0.0124 & 0.0129 \\
    Requirements on $d_{PCA}$ vs $\Delta E(\pi,e)$ & 0.0145 & 0.0120 \\
    $e/\pi$ and $e/\mu$ track and cluster classification & 0.0081 &  0.0099\\
    \addlinespace[1ex]
    % \midrule
    \multicolumn{3}{l}{\bf Selection of \boldmath$\Ks\Kl\to \pi^+\pi^-\;\pi e\nu$} \\
%    \midrule
    Requirement on $\abs{\vec{p}_{trk1} + \vec{p}_{trk2}}$ & 0.0021 & 0.0022 \\
    TOF requirements on $d\delta t_{e,\pi}$ vs $d\delta t_{\pi,e}$ & 0.0010 & 0.0012 \\
    \addlinespace[1ex]
    % \midrule
    \multicolumn{3}{l}{\bf Other effects} \\
%    \midrule
    Lower limit of $R_2$ and $R_4$ fit $\Delta t$ range & 0.0102 & 0.0041 \\
    Upper limit of $R_2$ and $R_4$ fit $\Delta t$ range & 0.0151  & 0.0040 \\
    $\Delta t$ bin width in $R_2$ and $R_4$ determination & 0.0259 & 0.0300 \\
    \midrule
    \textbf{Total systematic error} & 0.0345  & 0.0386  \\
    \bottomrule
  \end{tabular}
\end{table}
    % Requirement of $<40^o\alpha_{CM}<170^o$ & 0.00728 & 0.01045\\
    % Requirement of $\text{M}(\pi,\pi) < 490 \text{MeV/c}^2$ & 0.00133 & 0.00088 \\
    % TOF requirement $|d\delta t_{\pi,\pi}| \in (1.5,\;10)\:\text{ns}$ & 0.0026 & 0.0045 \\
    % TOF requirements on $d\delta t_{e,\pi}$ vs $d\delta t_{\pi,e}$ & 0.00389 & 0.00345 \\
    % Requirements on $\delta t(\pi)$ and $\delta t(e)$ & 0.01235 & 0.01288 \\
    % Requirements on $d_{PCA}$ vs $\Delta E(\pi,e)$ & 0.01449 & 0.01196 \\
    % $e/\pi$ and $e/\mu$ track and cluster classification & 0.00811 &  0.00995\\
    % \addlinespace[1ex]
    % % \midrule
    % \multicolumn{3}{l}{Selection of $\Ks\Kl\to \pi^+\pi^-\;\pi e\nu$} \\
    % \midrule
    % Requirement of $\abs\Big{\abs{\vec{p}_{trk1} + \vec{p}_{trk2}} - 110\:\text{MeV/c}} < 25\:\text{MeV/c}$ & 0.00208 & 0.00216 \\
    % TOF requirements on $d\delta t_{e,\pi}$ vs $d\delta t_{\pi,e}$ & 0.00101 & 0.00121 \\

Another source of systematic uncertainty is caused by the deviation from linearity visible in the distributions of $R_2(\Delta t)$ and $R_4(\Delta t)$. This effect, especially pronounced in case of $R_2(\Delta t)$ (see~\fref{fig:final_ratios}), may result from the presence of residual background from $\Ks\to\pi^+\pi^-(\to \pi \mu \nu)$ in the event samples entering the numerators of the ratios (compare~\fref{fig:csps-dt-after}), from imperfect reproduction of event selection efficiencies using control data samples or from inconsistencies between data and Monte Carlo simulations used to determine the signal efficiency of the $d_{PCA}$ and $\Delta E(\pi,e)$ cut. As a result, the fit of a constant level of the asymptotic region
of the ratios exhibits a dependence on the chosen fit limits as well as on the used width of the $\Delta t$ intervals.
To account for these effects, the fit range was varied around the limits given in~\eref{eq:fit_range} by several multiples of the default 3~$\tau_{S}$ bin width and largest deviations of the fit results were adopted as an estimate of systematic uncertainty. The width of the $\Delta t$ intervals was scanned from 1~$\tau_{S}$ to 10~$\tau_{S}$ and its contribution to systematic error was determined in a similar manner. As shown in ~\tref{tab:systematics}, effects of the fit results dependence on the fit region and bin width account for a large part of the total systematic uncertainty. 

\section{Discussion of the result}
\label{sec:kloe-discussion}
The values of observables of the studied \Ts~symmetry test, constituted by asymptotic levels of ratios of double kaon decay rates (defined in~\eref{eq:final_r2_and_r4}) and determined using the dataset of $1.7\:\text{fb}^{-1}$ collected by the KLOE experiment in the years 2004--5, amount to:
\begin{eqnarray}
  \label{eq:fit_results_stat_syst}
  R_2 &= 1.020 \pm 0.017_{stat} \pm 0.035_{syst},\\
  R_4 &= 0.990 \pm 0.017_{stat} \pm 0.039_{syst}.
\end{eqnarray}

As expected on the basis of size of the available data sample, the achieved precision is not sufficient to probe the violation of the symmetry under reversal in time, as the expected deviations of the above observables from unity in presence of \Ts~noninvariance (see~\eref{eq:theor_deviations}) are at the level of about $6.4\times10^{-3}$~\cite{pdg2016}. It should be emphasized, however,
that the objective of this work was to
demonstrate the feasibility of performing a direct \Ts~symmetry test in the conditions of the KLOE and KLOE-2 experiments and to
devise data analysis and reconstruction methods allowing
to perform a statistically significant test with the data to be collected by KLOE-2.
In the analysis presented herein, no major obstacles have been encountered, which proves that the studies whose concept was presented in~\cref{chapter:test_kloe} may be realized using the KLOE/KLOE-2 detector.

However, the presently obtained results are dominated by systematic uncertainty, twice exceeding the statistical error specific to the KLOE 2004--5 data sample. Therefore, consideration of a statistically significant test of the symmetry under reversal in time with a larger dataset must be started with a careful investigation of the systematic effects observed in the analysis presented in this work. Improvements are necessary concerning the selection of semileptonic decays of the short-lived neutral kaon where two of the applied event selection criteria, despite high power to purify the event sample, are likely to introduce a bias to the final $\Ks\Kl\to \pi e \nu \; 3\pi^0$ event samples. The second effect which must be mitigated is the apparent negative slope visible especially in the $R_2(\Delta t)$ distribution.

A number of enhancements to the presented scheme of KLOE data analysis is hence worth considering for the future measurement with a larger data sample. Firstly, the remaining background contamination in the $\Ks\Kl\to\pi e\nu\;3\pi^0$ originating from the $\Ks\to\pi^+\pi^-$ and subsequent $\pi\to\mu\nu$ decays (compare~\fref{fig:csps-dt-after}), should be addressed with more stringent event selection or additional dedicated selection cuts. Presence of this residual background, asymmetric both in its $\Delta t$ distribution and identified lepton charge, strongly affects the obtained observables $R_{2(4)}(\Delta t)$, causing a noticeable increase of these ratios' values in the lower part of the stable-efficiency $\Delta t$ region used to fit the asymptotic levels. This effect, herein incorporated into systematic uncertainty of the results, should be removed in the future studies by a further purification of the $\Ks\Kl\to\pi e\nu\;3\pi^0$ event sample. The latter can be attempted in several manners, with a most viable effect by improving performance of the track and cluster electron/pion and electron/muon classifiers. Their realization in this work (see~\sref{sec:pimu_rejection}) is relatively simple and may be improved e.g.\ by considering separate classifiers for different momentum and incidence angles of the particles on the calorimeter surface~\cite{graziani_anns} as well for the lepton charge subsamples.

Another desirable enhancement of the presented analysis would be replacing the selection cut based in the values of $d_{PCA}$ and $\Delta E(\pi,e)$ described in~\sref{sec:t1-fine-selection} by a selection step whose efficiency does not exhibit such a strong dependence on the resolution of $\Kl$ momentum direction, correlated with the decay time difference in an event. In its present form, signal efficiency of this cut could not be properly reproduced using data and control samples and required insertion of MC-based efficiency into determination of the results (see~\sref{sec:dpca_correction}). A possible alternative to this selection cut may come from the aforementioned improved $e/\pi$ and $e/\mu$ classifiers which are capable of providing large background discrimination power without relying on an estimate of the $\Ks$ momentum as using only properties of its decay product DC tracks and associated EMC clusters instead.

%
% TODO: napisac, cos ze mozna szukac lamania T w rejonie interferencyjnym ale ze "beyond the scope of this work"
%

%%% Local Variables:
%%% TeX-master: "../main"
%%% End: 
\chapter{Feasibility study of discrete symmetry violation searches with the J-PET detector}
\label{chapter:analysis_jpet}

The J-PET detector aims at recording a large sample of ortho-positronium decays into three photons in order to measure the expectation values of several angular correlation operators odd under the \Ts, \CPs~and \CPTs~symmetries (see~\tref{tab:jpet_operators}). Each of these operator involves spin of the decaying \ops/, which, however, cannot be measured directly in the experiment.
While previous studies of \ops/$\to 3\gamma$ decays used strong external magnetic field~\cite{cp_positronium} or~limited the momenta of positrons creating \ops/ atoms to a hemisphere and incorporated the further positronium polarization into systematic uncertainty~\cite{cpt_positronium}, a different approach was proposed by the J-PET experiment.
Alike the latter of aforementioned experiments, J-PET will estimate spin directions of ortho-positronium atoms using momentum of the original positrons. However, the latter may be reconstructed separately for each recorded event of \ops/ decay using the procedure presented in~\sref{sec:ops_polarization}.

The key requirements to implement this measurement scheme comprise:
\begin{itemize}
\item the possibility to identify \ops/$\to 3\gamma$ events among recorded background,
\item reconstructing the decay points inside the \ops/ production medium,
\end{itemize}
where achievable resolution of such reconstruction will dictate the level of precision at which the positronium spin polarization is controlled.

\section{Resolution tests with Monte Carlo Simulations}
\label{sec:jpet_mc_tests}
First tests of the the ortho-positronium decay reconstruction were performed with dedicated Monte Carlo simulations
of \ops/$\to 3\gamma$ decays recorded by a detection setup similar to J-PET. 
Prior to positronium annihilations, the simulation involved emission of positrons from a $^{22}$Na $\beta^+$ source and their thermalization in a porous medium with a possibility to form an \ops/ atom. The detector implemented in the simulation was built of plastic scintillator strips with the same properties as J-PET (compare~\sref{sec:jpet}) but had an idealized geometrical acceptance obtained with four concentric cylindrical layers of detection modules mounted without gaps between them. Further details of the simulated setup can be found in Reference~\cite{gajos_gps}.

Like in the J-PET detector, photon interaction point in a scintillator was resolved in the simulation up to the strip location, resulting in a precision of about 0.5$^{\circ}$. The assumed resolutions of longitudinal ($z$) $\gamma$ interaction position and its time were varied in order to explore the sensitivity of the reconstruction to uncertainty of each input parameter. For a number of $10^{5}$ of simulated and fully recorded \ops/$\to 3\gamma$ events, their simulated origins were compared to the decay points reconstructed with the technique presented in~\cref{chapter:gps}. For each of the Cartesian coordinates, the $\sigma$ value of the resulting distributions of simulated-reconstructed discrepancies was used as an estimate of spatial resolution.

\begin{figure}[h!]
  \centering
  \includegraphics[width=0.45\textwidth]{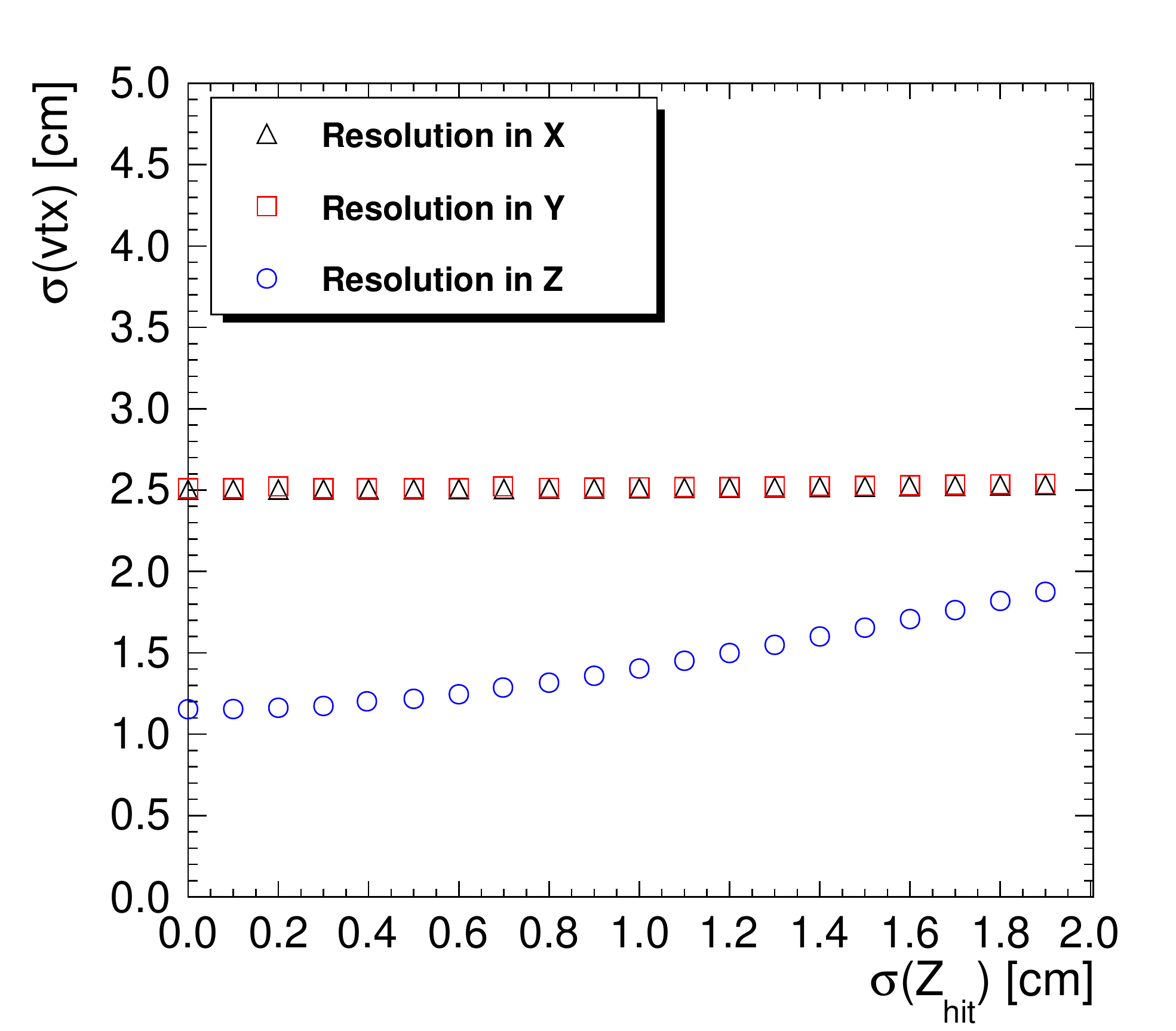}
  \hspace{1em}
  \includegraphics[width=0.45\textwidth]{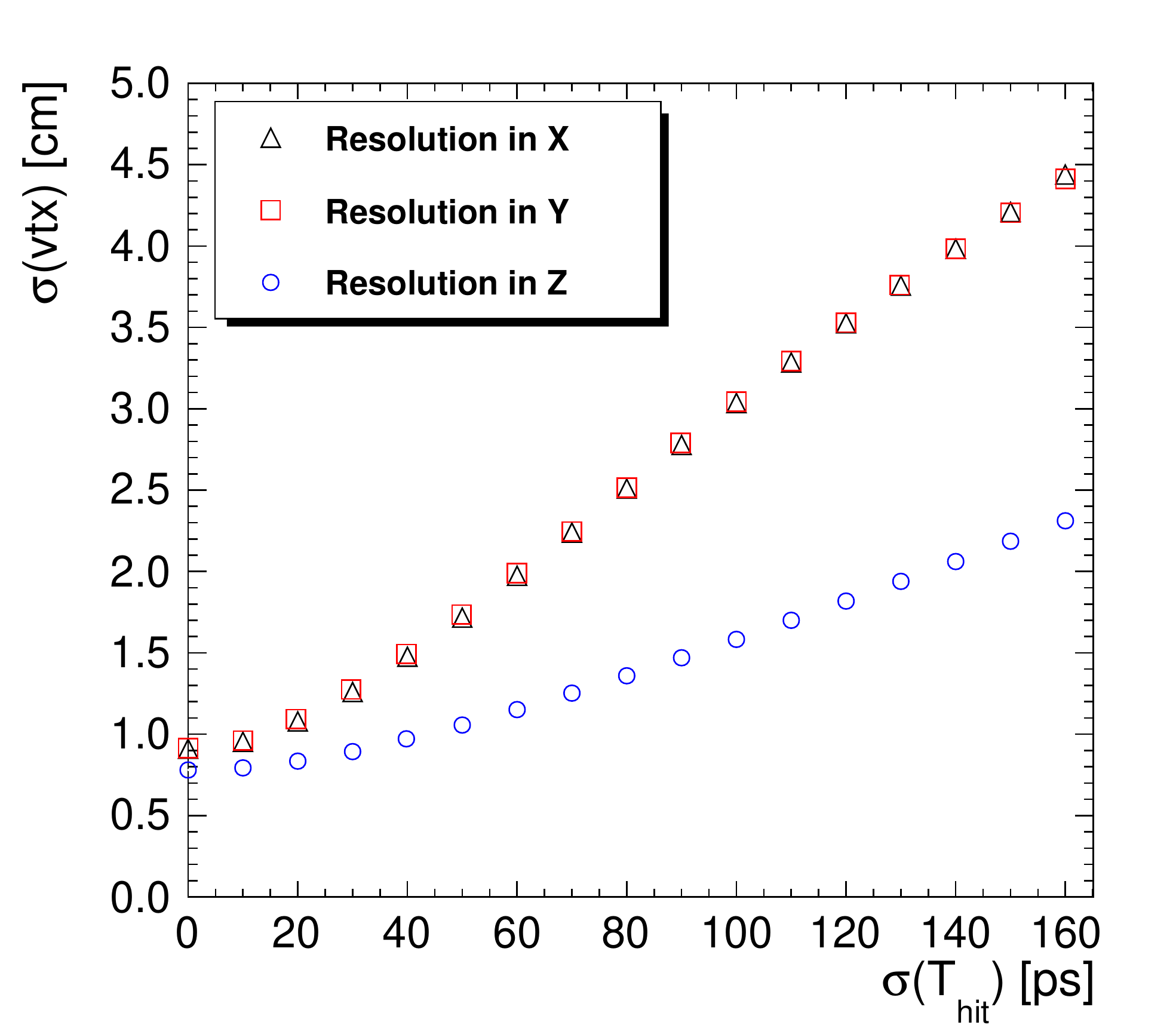}
  \caption{Spatial resolution of Cartesian coordinates of the \ops/$\to 3\gamma$ decay point as a function of the detector resolution of $\gamma$ interactions in the scintillator strips, in terms of the interaction longitudinal position (left) and interaction time (right). Figures adapted from~\cite{gajos_gps}.}\label{fig:jpet_mc_res}
\end{figure}

\fref{fig:jpet_mc_res} presents the obtained distributions of the resolution depending on the accuracy of $\gamma$ interaction properties recorded by the detector. As visible in the right panel, the tested reconstruction method exhibits largest sensitivity to temporal resolution.
At the same time, the impact of $z$ coordinate determination for photon interactions in scintillator strips is only visible in the decay point resolution along the $z$ axis, distinguished by the cylindrical geometry of the detector as well as a different resolution than in the transverse plane.
The pronounced sensitivity of reconstruction to time of recorded photon interactions with respect to their $z$-coordinate resolutions is caused by the fact that while $\sigma(t_{\gamma})$ and $\sigma(z_{\gamma})$ are connected approximately by a factor equal to the effective velocity of light in a scintillator strip ($\upsilon_{eff}\approx 12\:\frac{\text{cm}}{\text{ns}}$), the errors on time determination propagate to the final result multiplied by the velocity of light in vacuum (see~\aref{appendix:jpet_solution}), larger by a factor of~2.5.
%
% TODO: dopisać o kątowej rozdzielczości i jej wpływie na polaryzację
%

The demonstrated sensitivity of this reconstruction technique to errors of photon recording times may, however, be turned into an advantage when this method is employed for background discrimination.
The positron source assumed in the MC simulations had an activity 10~MBq, of the same order as sources planned for J-PET measurements. This allowed to test the capability of the trilaterative reconstruction to reject background originating from accidental photon coincidences. As estimated in a separate study~\cite{kowalski_scatter_fraction}
the rate of such coincidences expected with the aforementioned source activity and a time window of 5~ns would be about 1/5 of the total rate. Monte Carlo simulations performed for such conditions yielded an accidental coincidence rate of 2800/s~\cite{gajos_gps}. Since in the accidental $3\gamma$ sets at least one of the quanta does not originate in the same decay as the other ones, event slight inconsistencies of the photons' creation times lead to the reconstructed annihilation point located either in a non-physical region or at a considerable distance from the \ops/ creation medium. Therefore, a requirement of the reconstructed point distance from the volume of possible positronium decays allowed to reduce the rate of accidental triple coincidences by a factor of \SI{89}{\percent} in the MC-based studies~\cite{gajos_gps}.

As discussed in~\sref{sec:ops_polarization}, the polarization along a given axis for positrons from a $\beta^+$ decay whose direction of momentum is known with an angular uncertainty described by a cone around this axis and an opening angle $\alpha$ is given by the relation:
\begin{equation}
  \label{eq:jpet_positron_polarization}
  P_{e+} = \frac{\upsilon_{e+}}{c}\frac{1+\cos(\alpha)}{2},
\end{equation}
where $\upsilon_{e+}$ is the average positron velocity and $c$ denotes the velocity of light~\cite{Coleman}. The level of average polarization of the \ops/ atoms inferred from the incident positrons hence depends on the angular resolution of $e^+$ momentum if a setup like the one presented in~\fref{fig:ops_spin_determination} is used. The tests with MC-simulated $3\gamma$ events originating in the wall of the cylindrical chamber have shown that, after inclusion of a kinematical fit restricting the reconstructed decay points in the transverse radius cylindrical coordinate to lie in the chamber material, the angular uncertainty of the positron motion direction can be reduced to about~\SI{15}{\degree}. It follows from~\eref{eq:jpet_positron_polarization} that
such event-wise estimation of the $e^+$ spin direction
allows to retain $\frac{1}{2}(1+\cos(15^{\circ}))\approx$\SI{98}{\percent} of the polarization resulting from parity violation in the $\beta$ decay~\cite{gajos_gps}.

\section{Analysis of first J-PET measurement with $3 \gamma$ annihilation medium}\label{sec:jpet_firs_data}

\subsection{The test experimental setup}
\label{sec:jpet_test_setup}
\begin{figure}[h!]
  \centering
\begin{tikzpicture}[
  scale=0.5,
  >={Stealth[inset=0pt,length=6pt,angle'=50,round]}
  ]
  % dummy rect
  \draw[white] (-5.5,-6) rectangle (9,6);
  
  % detector
  \draw[fill=black!50!white] (-5,4.1) rectangle (5,4.3) node[right, yshift=-7] {\large Layer 1};
  \draw[fill=black!50!white] (-5,4.6) rectangle (5,4.8) node[right] {\large Layer 2};
  \draw[fill=black!50!white] (-5,5.6) rectangle (5,5.8) node[right] {\large Layer 3};

  \draw[fill=black!50!white] (-5,-4.3) rectangle (5,-4.1);
  \draw[fill=black!50!white] (-5,-4.8) rectangle (5,-4.6);
  \draw[fill=black!50!white] (-5,-5.8) rectangle (5,-5.6);
  
  \draw[line width=3pt, black!50!white] (-5,-0.65) -- (-5,0.65) -- (5,0.65) -- (5,0.2) -- (5.4, 0.2);
  \draw[line width=3pt, black!50!white] (-5,-0.65) -- (5,-0.65) -- (5,-0.2) -- (5.4, -0.2);
  \node[] at (6.8, 0) {\large $\rightarrow$ vacuum};
  \node[] at (7.2, -0.7) {\large system};
  
  \draw[line width=3pt, black] (0, -0.65) -- (0, -0.15);
  \draw[line width=3pt, black] (0, 0.65) -- (0, 0.15);
  \draw[line width=1pt, black] (0, 0.15) -- (0, -0.15);

  \draw[thick,<->] (2,-0.65) -- (2,0.65) node[midway, right] {\large 14~cm};

  \draw[black] (0.1, 0.3) -- (1.5, 2.0) -- (2.0, 2.0);
  \node[right] at (2.0, 2.0) {\large $\beta^+$ source holder};

  % max and min paths
  % \draw[darkgray, line width=1pt, dashed, ->] (-4.9,-0.65) -- (-4.9,-4.1);
  % \draw[darkgray, line width=1pt, dashed, ->] (-4.9,0.65) -- (4.9,-5.6);
  
\end{tikzpicture}
\hspace{1ex}
\includegraphics[width=0.40\textwidth]{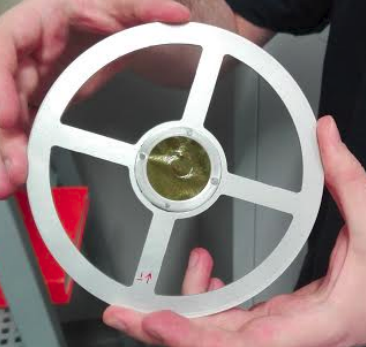}
  \caption{Left: cross section view scheme of the setup used in the test measurement with a 3$\gamma$ event production chamber. Right: setup of the $^{22}$Na $\beta^+$ source mounted in the center of the $3\gamma$ production chamber. The source was enclosed between two layers of kapton foil spanned on an aluminum frame mounted to the inner walls of the chamber. The chamber and source were constructed at the Maria Curie-Sk\l{}odowska University by a subgroup of the J-PET collaboration.}\label{fig:jpet-source}
\end{figure}
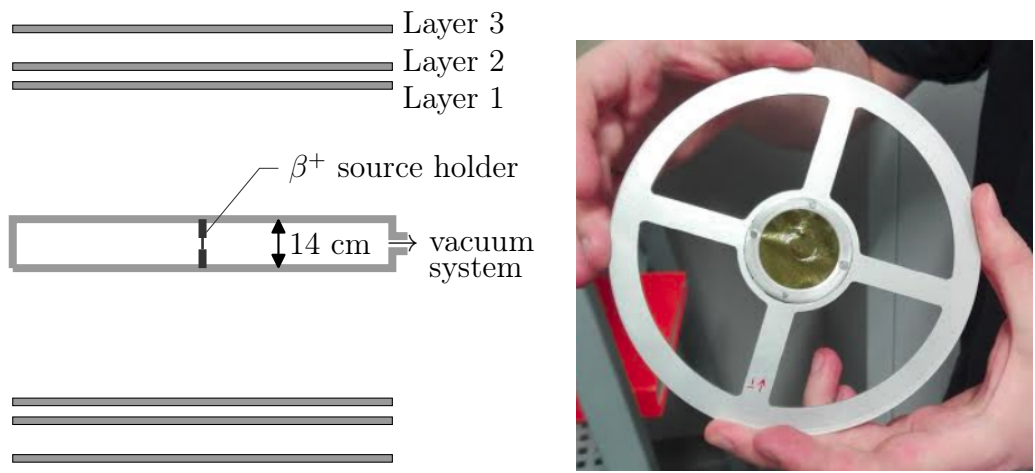

The first measurement with the J-PET detector implementing the scheme discussed in~\sref{sec:ops_polarization} was carried out with an aluminum vacuum chamber depicted in~\fref{fig:jpet-source} (left). The chamber was shaped as a cylinder with a diameter of about 14~cm. In the center of the chamber a $\beta^+$ source was placed in the form of $^{22}$Na enclosed by two layers of kapton foil spanned by a metal frame as presented in ~\fref{fig:jpet-source} (right). A vacuum system connected to the chamber ensured pressure at a level of about 8~mbar inside its volume.

As the porous aerogel material intended as a positronium creation medium was in preparation at the time of the test measurement, the chamber walls contained no additional material besides aluminum.
Although the probability of ortho-positronium formation in aluminum is negligibly low, direct electron-positron annihilation into three photons is possible and may be used to test the procedures for identification and reconstruction of such events regardless of the origin of the 3$\gamma$ state.

As the yield of 3$\gamma$ events from direct annihilation is smaller than the expected rate from decays of ortho-positronium produced in aerogel by a factor of about 400, a demonstration of the possibility to identify such events in the data of such test measurement would guarantee the feasibility of handling the \ops/$\to 3\gamma$ decays in the future experiments with J-PET\@.

\subsection{Data reconstruction and preselection}
\label{sec:jpet_preselection}
The test measurement was performed for two days, resulting in a total of 5.5~TB of collected raw data. Reduction of such a high data flux, resulting from the triggerless mode of J-PET data acquisition~\cite{greg_daq}, requires stringent discrimination of background in order to allow for effective analysis of the J-PET data. The analysis of the test data was performed using the J-PET Analysis Framework software~\cite{Krzemien:2015hkb}. At the first stages of data reconstruction, single times recorded at certain voltage thresholds applied to the PMT electric signals were assembled into representations of these signals as presented in~\fref{fig:jpet_signal}. For each signal, the time over threshold (TOT) value was calculated (see~\eref{eq:tot_i}) using information on all available thresholds. Subsequently, pairs of signals (referred to as \textit{hits} in the further considerations) coming from the same $\gamma$ interaction were identified. Signals were paired if they originated from distinct sides of the same detection module (scintillator strip) and their arrival times (estimated using time at the leading edge on a threshold lowest in terms of absolute voltage) were separated by no more than 5~ns. The last stage of early reconstruction comprised assembling event candidates as sets of hits contained within a time window of 20~ns. Although such a time window is broad with respect to possible time differences in a physical event, its purpose was a reduction of data volume without limiting further fine selection of event candidates by more strict timing requirements. Only event candidates with at least 3 hits within 20~ns were accepted for further analysis which allowed for reduction of the raw data by about \SI{98.8}{\percent}.

After preselection of event candidates described above, a total sum of times over threshold for both signals comprising each hit was calculated (see~\eref{eq:tot_def}) as a measure of the deposited energy of a corresponding $\gamma$ quantum. However, such estimation of the deposited energy through size of the electric signals is sensitive to effects of properties of the detector such as gain of the photomultipliers, which may vary between modules. Therefore, a simple calibration of the TOT values was performed. 

\subsection{Calibration of time over threshold (TOT) values}
\label{sec:jpet_tot_calib}
The TOT calibration was performed using a sample of direct $e^+e^-\to 2\gamma$ annihilation events for which the initial energy of the photons interacting in J-PET scintillators is well defined (511~keV). Such events were selected using criteria described in Reference~\cite{monika_2gamma_imaging}. For each of the 192 detection modules of J-PET, a separate distribution of the TOT$_{\gamma}$ values was obtained using hits from the selected two-photon annihilation events. In each spectrum, mode TOT$_{\gamma}$ value was identified as a weighted average of the centers of the most populated bin in the histogram and its neighboring bins. Thus obtained mode corresponds to the most probable deposited energy in a Compton spectrum for 511~keV photons, which amounts to about 280~keV.
In order to equalize the TOT responses of particular detection modules,
the following relation between TOT and deposited energy was used:
\begin{equation}
  \label{eq:tot_edep}
  E_{dep}(TOT_{\gamma}) \text{[keV]} = \exp\left( \left( TOT\text{[ps]} + 1.1483\times 10^{5} \right) / 23144  \right),
\end{equation}
which had been determined experimentally using J-PET data and an approach similar as discussed in Reference~\cite{tot_edep}. For each detection module $k$, a correction factor was calculated as a ratio of the most probable deposited energy to the time over threshold mode TOT$_{M}$ obtained for that module and transformed to deposited energy:
\begin{equation}
  \label{eq:tot_corr}
  \alpha_{k} = \frac{280\ \text{keV}}{E_{dep}(TOT^{(k)}_{M})}.
\end{equation}

Calibration factors obtained in this manner for each of the 192 scintillators of the detector were applied in the analysis of the test measurement with 3$\gamma$ production chamber. For every recorded hit, its calculated TOT value was transformed to deposited energy, multiplied by the correction factor relevant to the detection module in which it was recorded, and transformed back to time over threshold.
\begin{figure}[h!]
  \centering
  \begin{tikzpicture}
    \node[anchor=south west,inner sep=0] at (0,0) 
    {\includegraphics[width=0.45\textwidth]{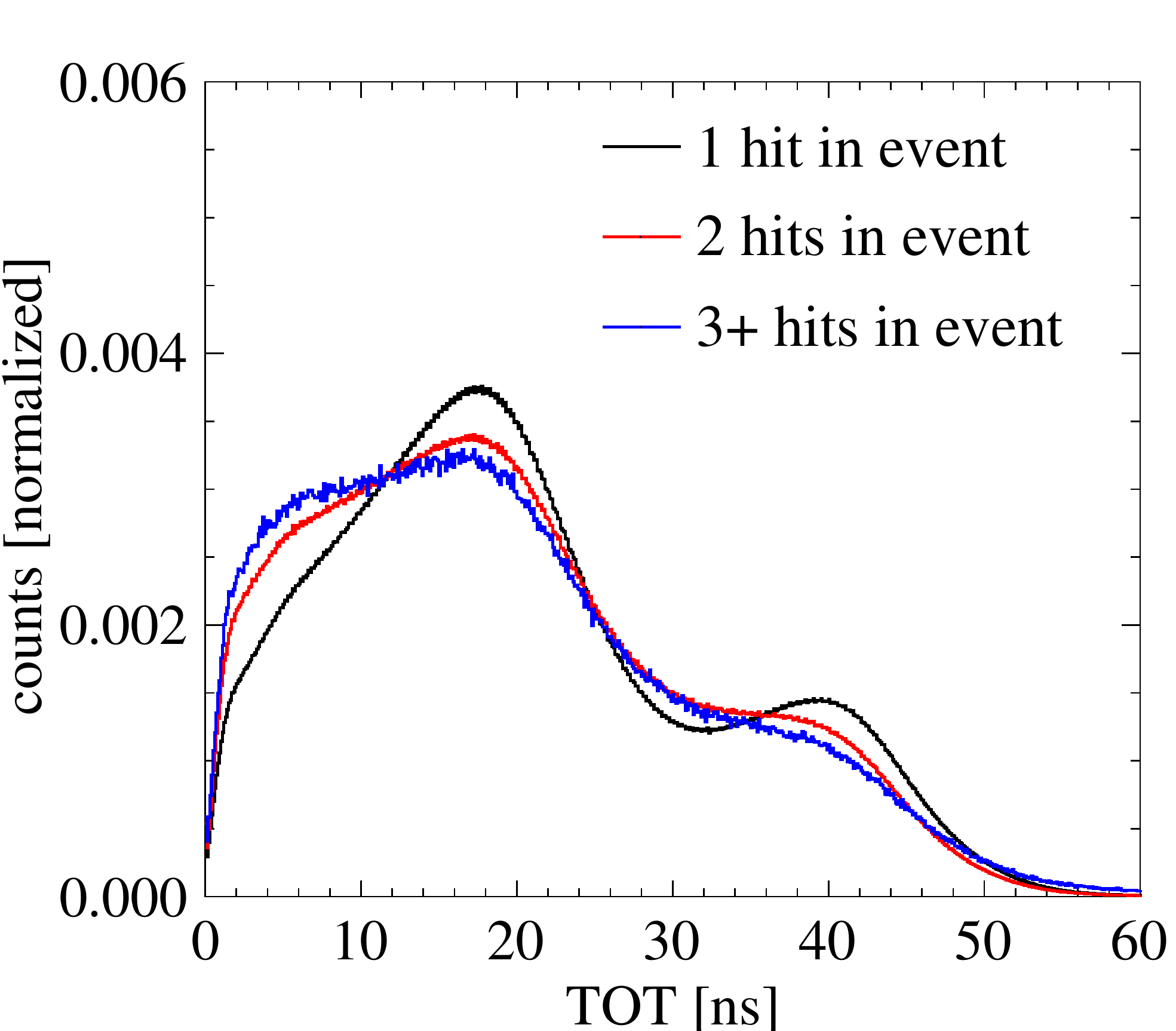}};
%    \draw[black, thick, dashed] (2.50,5.0) -- (5.50,2.0);
  \end{tikzpicture}
  \hspace{1em}
    \begin{tikzpicture}
    \node[anchor=south west,inner sep=0] at (0,0) 
    {\includegraphics[width=0.45\textwidth]{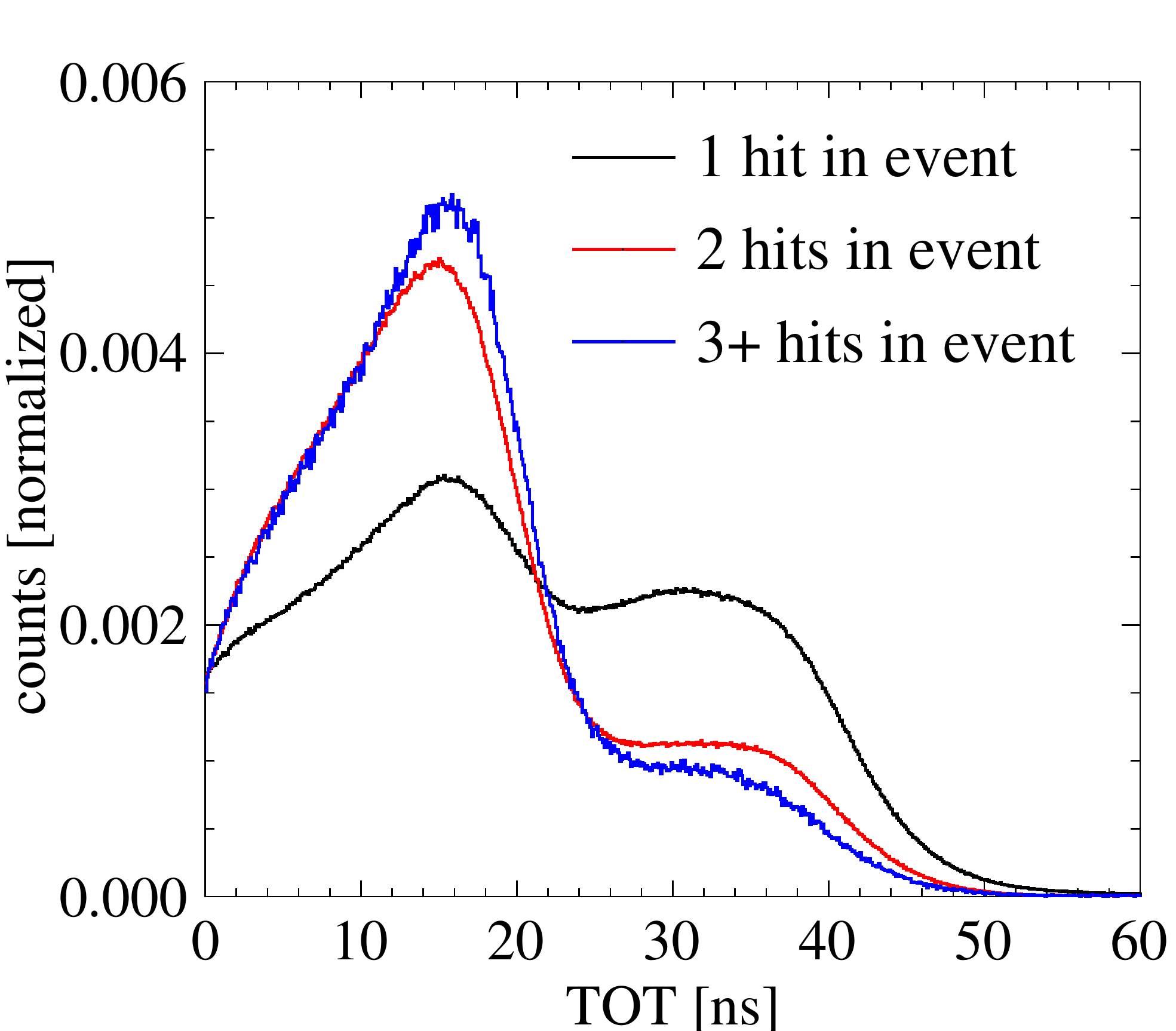}};
    \draw[black, thick, dashed] (1.25, 0.8) -- (1.25,5);
    \draw[black, thick, dashed] (2.98, 0.8) -- (2.98,5);
    \draw[black!60!white, ultra thick, <->] (1.25, 1.3) -- (2.98, 1.3);
  \end{tikzpicture}
  \caption{Distributions of time over threshold (TOT) values calculated for $\gamma$ interactions in all J-PET detection modules, before (left) and after calibration (right). Different colors denote TOT values of $\gamma$ hits observed in groups of 1, 2 and more within a 2.5~ns event time window. In the right panel the region of TOT used to select annihilation photon candidates is marked with dashed lines and a gray arrow.}\label{fig:tot_calib}
\end{figure}
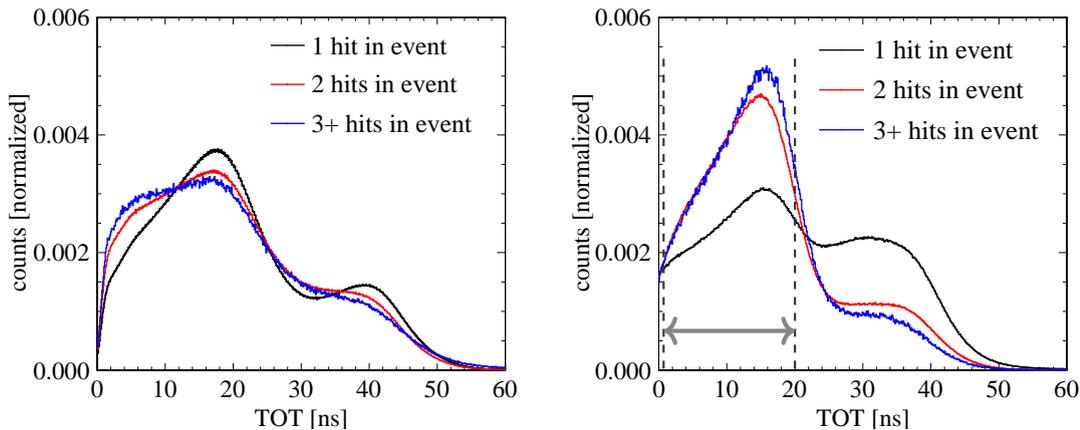

\fref{fig:tot_calib} shows a comparison of the total TOT spectra from the whole detector before and after calibration. The distributions are presented separately for three classes of events, where an event is a set of hits contained within a narrow time window of 2.5~ns. In~cases where only one $\gamma$ interaction was recorded in a time window, the probability that the observed hit corresponds to~a high-energy deexcitation photon (1275~keV for the $^{22}$Na isotope used in the measurement) is almost even with the chance of observing a photon from annihilation. Conversely, for an increasing number of hits required in coincidence, the influence of high-energy photons on the TOT spectrum is decreased. Although slopes corresponding to Compton edges are visible in the TOT distributions both before and after calibration, its application results in a better separation of the structures from Compton spectra for photons from annihilation and deexcitation as well as in sharper Compton edges. This, in turn, allows for imposing requirements on the TOT value for identification of annihilation photon candidates as discussed in the next Section.

\subsection{3$\gamma$ event selection}\label{sec:3g_selection}
The preselection of data only required presence of at least three $\gamma$ interactions (hits) in a broad time window of 20~ns. Calibrated TOT values allow for restricting the further considerations to hits whose deposited energy, measured as TOT, could correspond to a~photon from 3$\gamma$ annihilation. Therefore, in further analysis only the hits whose TOT satisfied the following criterion:
\begin{equation*}
 0.5\ \text{ns} < TOT_{\gamma} < 20\ \text{ns},
\end{equation*}
(where the lower limit is introduced to reject poorly reconstructed hits) were considered as annihilation photon candidates. Similarly, location of the measured TOT$_{\gamma}$ value in the rightmost part of the TOT spectra presented in~\fref{fig:tot_calib} may be used to identify candidates for interactions of high-energy deexcitation quanta.
%
% TODO: dopisać, że prompty nie analizowane bo nie mają znaczenia bez o-Ps
%

\begin{figure}[h!]
  \centering
  \includegraphics[width=0.45\textwidth]{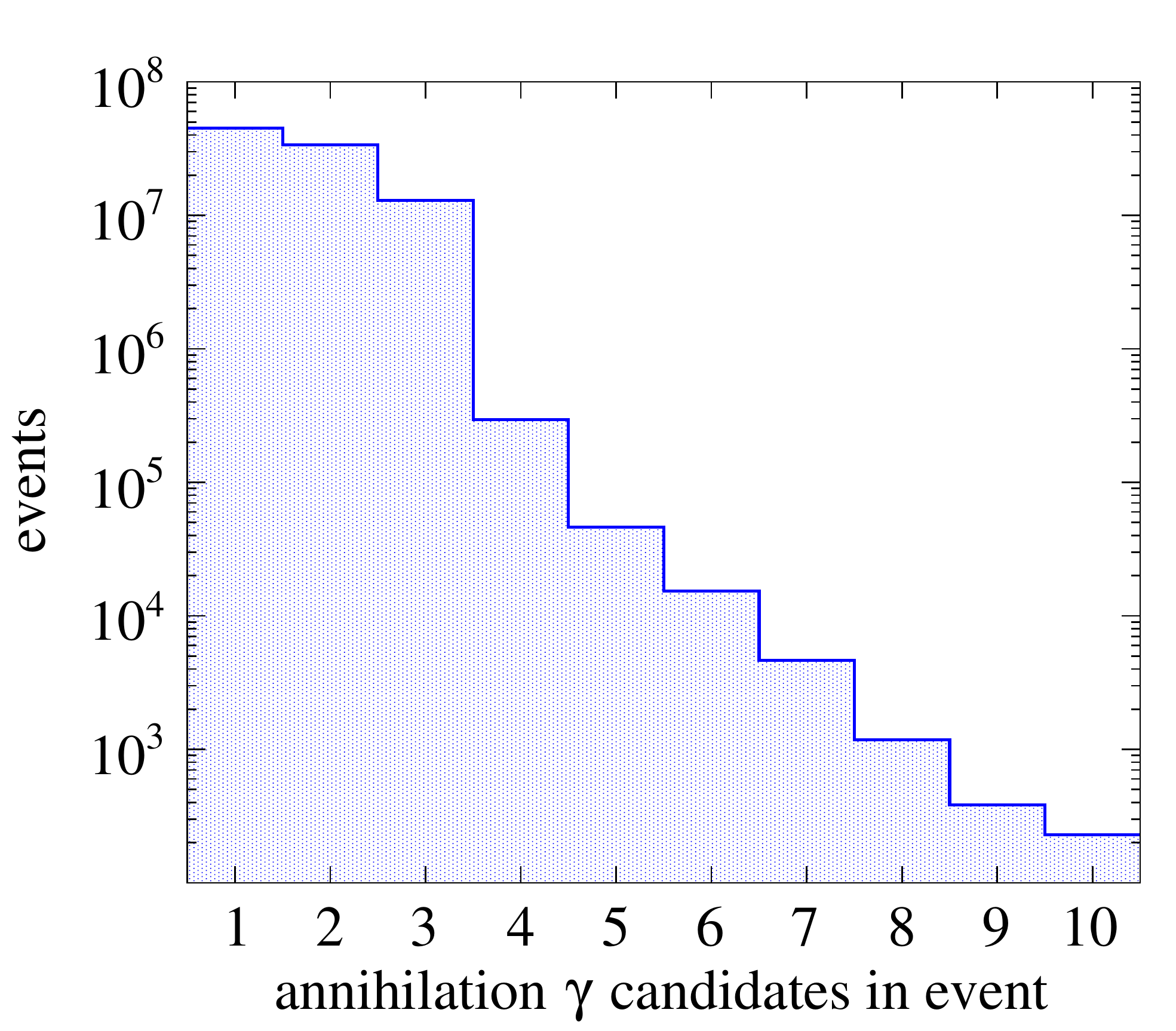}
  \caption{Multiplicity of annihilation photon candidates (identified using TOT) in a 2.5~ns time window.}\label{fig:hit_multiplicity}
\end{figure}

In order to accommodate all possible $3\gamma$ event topologies and account for time resolution effects without allowing an excess of accidental coincidences, a time window width of 2.5~ns was chosen for further studies. A set of hits contained within such a time window will be later on referred to as an \textit{event}. Multiplicity of identified annihilation photon candidates observed in a single event is presented in~\fref{fig:hit_multiplicity}. Although further selection of hits corresponding to a single $3\gamma$ annihilation
is possible
in case 4 and more interactions were recorded in~coincidence,
the analysis presented in this work was restricted to events with exactly three annihilation photon candidates to avoid combinatorial issues.

One of the major sources of background in the search for 3$\gamma$ annihilations in J-PET comes from secondary interactions of primary photons Compton-scattered in the scintillators. Registration of such scattered quanta is most probable in the detection modules directly neighboring the ones where the primary photon interacted, therefore events were rejected if they contained a pair of modules whose azimuthal coordinates were closer than \SI{8}{\degree}. Such criterion removes possible scatterings in modules next to each other in a single detector layer as well in pairs neighboring between the layers.

\begin{figure}[h!]
  \centering
  \begin{tikzpicture}
    \node[anchor=south west,inner sep=0] at (0,0) 
    {\includegraphics[width=0.45\textwidth]{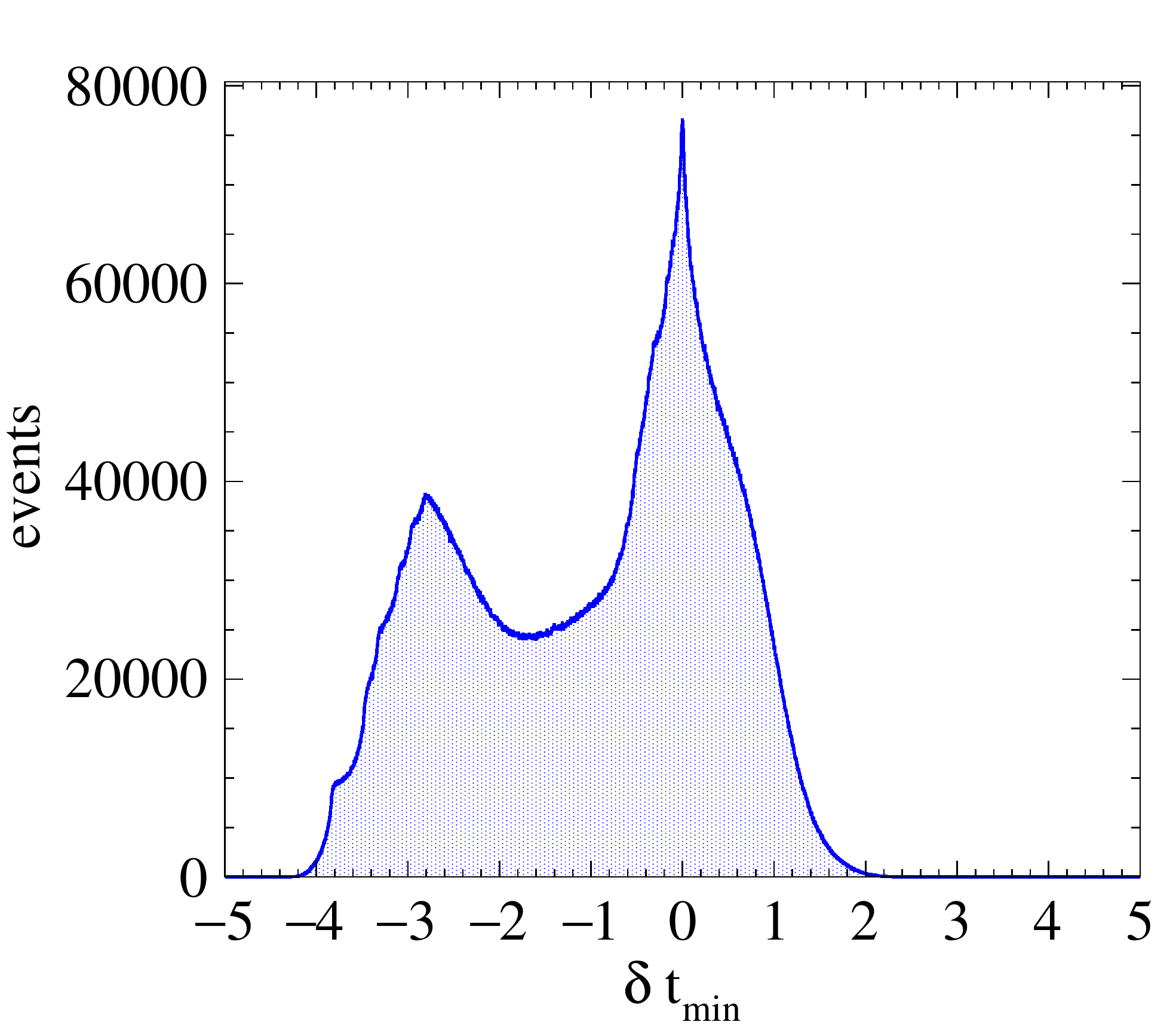}};
    \draw[black, thick, dashed] (3.0,0.9) -- (3.0,5.3);
    \node[] at (4.7, 0.25)  {[ns]};
    \draw[ultra thick, black!50!white, <->] (1.3, 3.5) -- (3.0, 3.5);
  \end{tikzpicture}    
  \caption{Distribution of the discrepancy between hits' recording times and their spatial separation divided by velocity of light, for the 2 out of 3 hits in an event which give a value closest to zero. Events with $\delta_{min} < 1.8$~ns are considered in further analysis as marked with the dashed line and gray arrow.}
  \label{fig:dvt}
\end{figure}
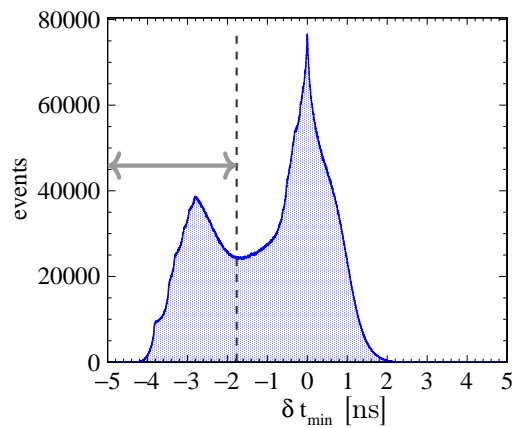

The second selection criterion targeting scattered photons was based on testing a hypothesis that the time difference between recorded hits corresponds to a time of flight of a photon between the two reconstructed $\gamma$ interaction positions. For all possible choices of 2 out of 3 hits in an event, the following variable was calculated:
\begin{equation}
  \label{eq:dvt}
  \delta t_{ij} = \abs{t_i - t_j} - \frac{1}{c}\abs{\vec{r}_i-\vec{r}_j},
\end{equation}
where $t_i$ and $\vec{r}_i$ denote respectively the recording time and position vector  of $i$-th photon interaction in an event and $c$ is the velocity of light.
A value of $\delta t_{ij}$ close to zero corresponds to a pair of hits created by subsequent Compton scatterings of the same photon in different detection modules.
Distribution of $\delta t_{min}$, defined as the smallest (in terms of absolute value) of three possible $\delta t_{12}$, $\delta t_{13}$ and $\delta t_{23}$ values for each event is displayed in~\fref{fig:dvt}.
In order to avoid possible contamination of the 3$\gamma$ event sample with secondary scatterings, a three-hit event was rejected if its $\delta t_{min}$ was greater than 1.8~ns.

\begin{figure}[h!]
  \centering
  \begin{tikzpicture}
    \node[anchor=south west,inner sep=0] at (0,0) 
    {\includegraphics[width=0.45\textwidth]{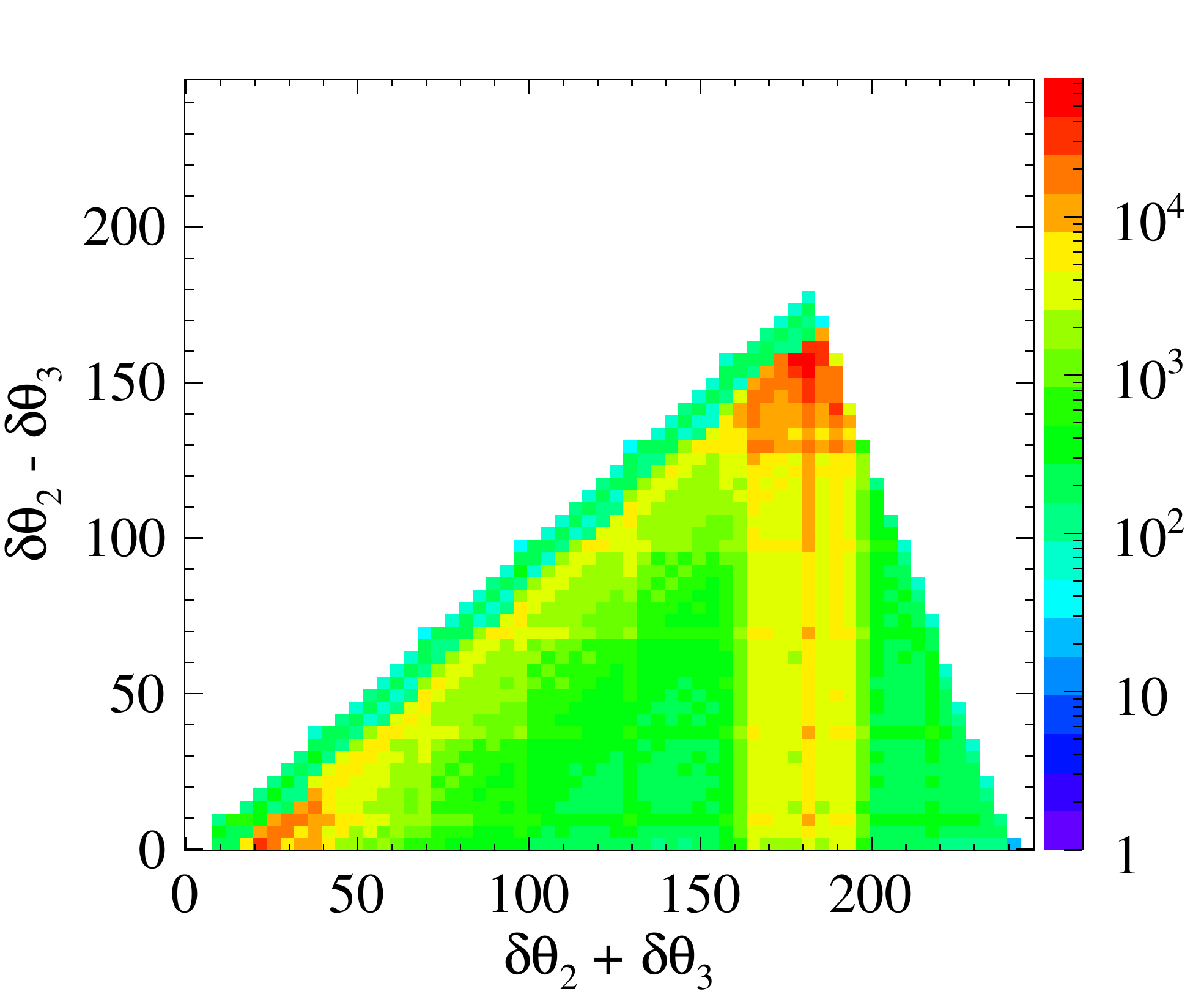}};
    \draw[black, thick, dashed] (5.0,0.9) -- (5.0,5.3);
    \draw[ultra thick, black!50!white, <->] (5.0, 3.5) -- (5.8, 3.5);
  \end{tikzpicture}  
  \caption{Relative distribution of the sum and difference of the two smaller azimuthal angles between three detection modules where hits were recorded in an event ($\delta\theta_1 > \delta\theta_2 > \delta\theta_3$). Events located to the right of the dashed line are selected as 3$\gamma$ annihilation candidates.}\label{fig:jpet_angles}
\end{figure}

The last criterion imposed on the three-hit events to identify annihilations into three photons was based on the angular topology of the event. Preliminary studies with Monte-Carlo simulations had proven that 3$\gamma$ events can be distinguished from accidental coincidences of a 2$\gamma$ events with a third photon by the relative angles between the three photons' momenta~\cite{kowalski_scatter_fraction}.
Since the origin of the photons was not yet reconstructed at that stage, differences between azimuthal angles of the particular detection modules were used to calculate estimates of such relative angles, defined as:
\begin{equation}
  \label{eq:jpet_delta_theta}
  \delta \theta_{ij} = \abs{\theta_i - \theta_j}, \qquad ij=12,23,31,
\end{equation}
where $\theta_i$ is the azimuthal angle of the location of a detection module where $i$-th hit in an event was recorded. If such values of the three relative angles are labeled according to their descending order so that $\delta \theta_1 > \delta \theta_2 > \delta \theta_3$, the two smallest angles constitute useful variables in the 3$\gamma$ and 2$\gamma$ event discrimination. \fref{fig:jpet_angles} presents the relative distribution of the difference and sum of the two smallest relative $\theta$ angles. With such a choice of variables on the axes, events from two-photon annihilations, which contain two $\gamma$ quanta with opposite momenta, are clustered in a vertical band around $\delta\theta_2 + \delta \theta_3 \approx 180^{\circ}$. The 3$\gamma$ annihilations, on the other hand, are characterized by a large value of the sum of smallest angles and thus only events with:
\begin{equation*}
  \delta \theta_2 + \delta \theta_3 > 205^{\circ},
\end{equation*}
were considered as 3$\gamma$ candidates and subjected to the trilaterative reconstruction of the annihilation point as described in~\sref{sec:jpet_3g_imaging}. However, before the reconstruction results are presented, effects of imaging with two-photon annihilations using the same measurement will be briefly discussed as a useful benchmark for the study of 3$\gamma$ events.

\subsection{Benchmark imaging of the 3$\gamma$ production chamber with 2$\gamma$ annihilations}
\label{sec:2g_imaging}
In order to understand the limits of available reconstruction of points of annihilation taking place in the wall of the 3$\gamma$ production chamber, classical tomography with $e^+e^-$ annihilations to two photons may be used as a benchmark. Due to the probability of such annihilations larger by over two orders of magnitude than in case of direct annihilation to 3$\gamma$ as well as simple back-to-back configuration of the photons' momenta, a pure sample of such events is selected more easily at the same time yielding larger statistics for the reproduction of the chamber tomographic image.

\begin{figure}[h!]
  \centering
%  \captionsetup[sub]{margin=1ex}
  \captionsetup[subfigure]{justification=centering}
  \begin{subfigure}{0.45\textwidth}
  \includegraphics[width=1.0\textwidth]{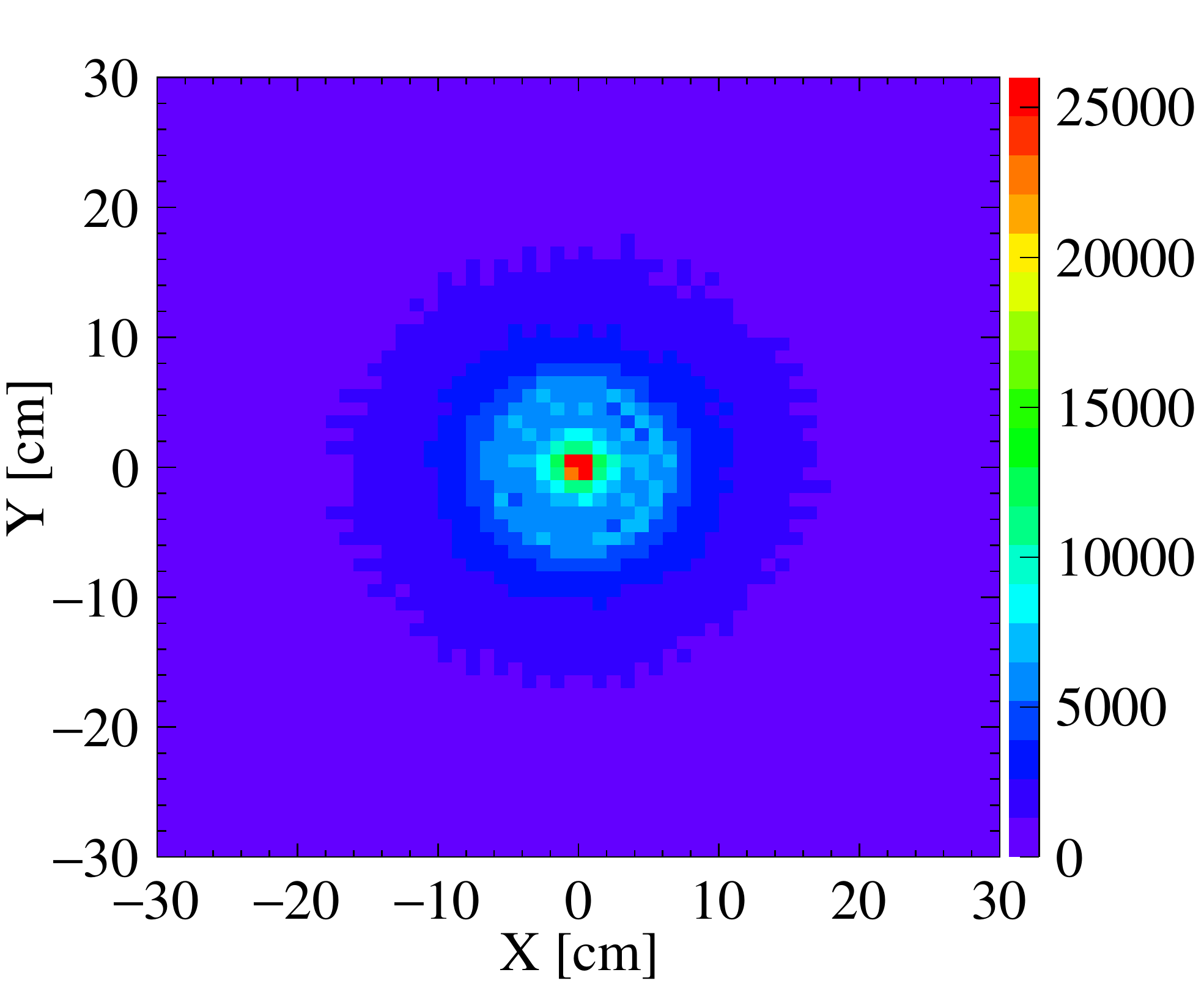}
  \caption{Transverse view image,\\ all $z$ values projected}\label{fig:2g_image_a}
  \end{subfigure}
  \hspace{1em}
  \begin{subfigure}{0.45\textwidth}
  \includegraphics[width=1.0\textwidth]{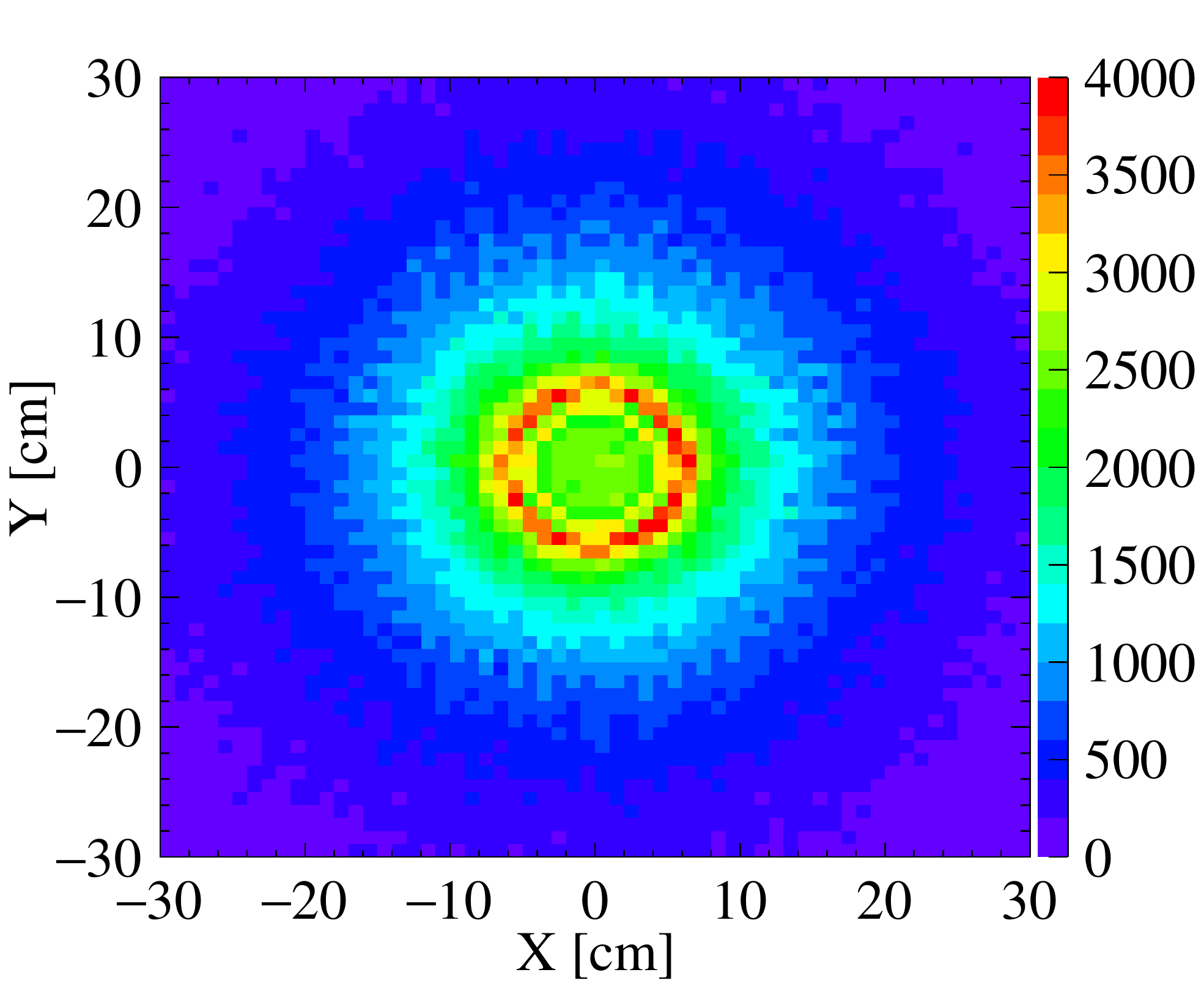}
  \caption{Transverse view image,\\ $\abs{z} > 4$~cm}\label{fig:2g_image_c}
\end{subfigure}
  \begin{subfigure}{0.45\textwidth}
  \includegraphics[width=1.0\textwidth]{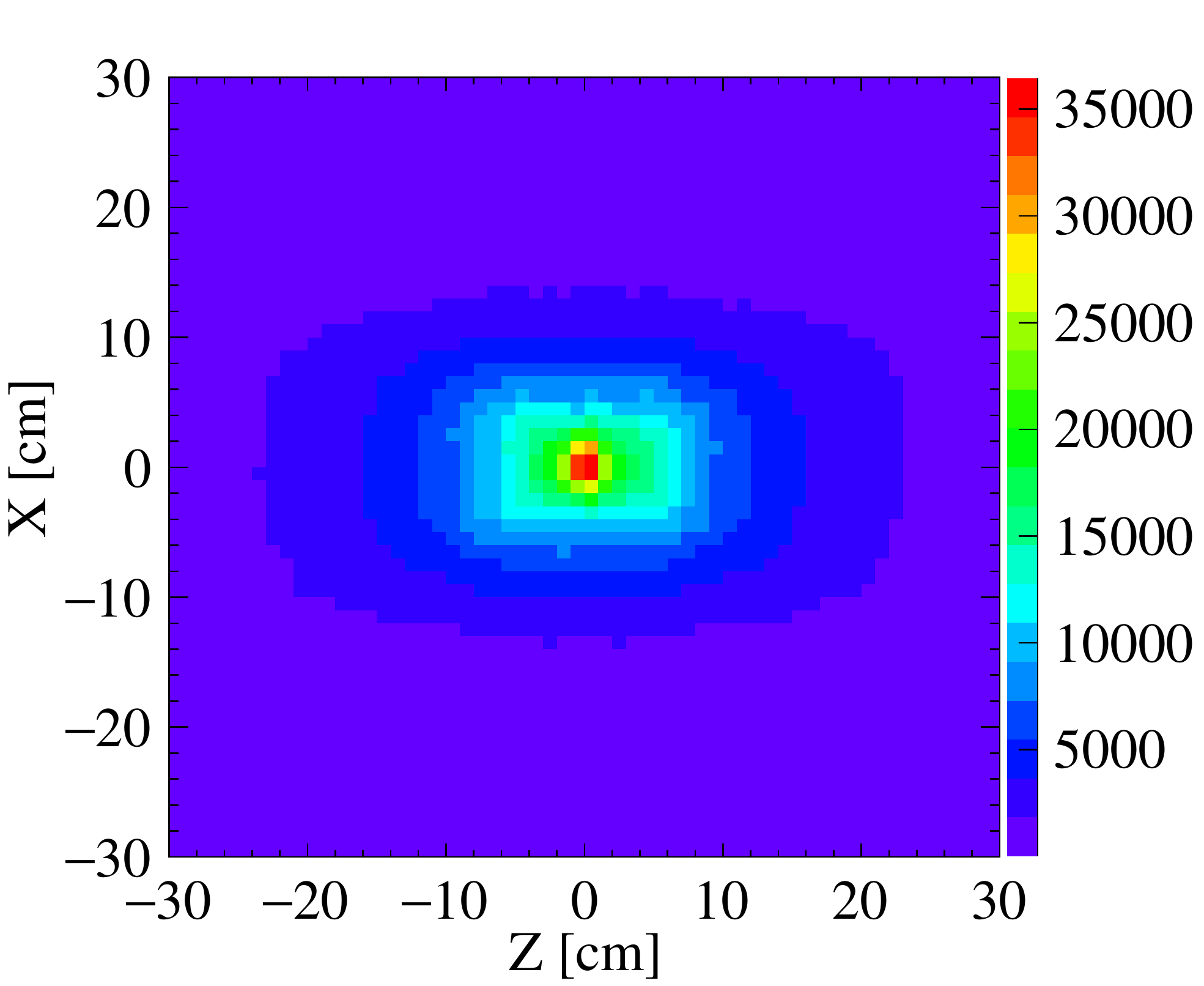}
  \caption{Side view image}\label{fig:2g_image_b}
\end{subfigure}
  \caption{Tomographic images of the annihilation target cylinder and $\beta^+$ source obtained using $e^+e^-\to 2\gamma$ annihilations used as a benchmark for the 3$\gamma$ annihilation reconstruction studies.}\label{fig:2g_image}
\end{figure}

About \SI{15}{\percent} of the data taken in the test run were analyzed using the procedures devised for tomographic imaging with J-PET. Identified two-photon annihilation events were used to reconstruct lines of response
\footnote{In PET tomography, a line of response is a line connecting two places of photon interactions recorded in the detector, on which the 2$\gamma$ annihilation point must be located.}
(LORs) and difference between each photon time of flight allowed to find the annihilation point along a LOR. Detailed description of the applied imaging procedure may be found in Reference~\cite{monika_2gamma_imaging}. Thus obtained tomographic images of the aluminum chamber as well as the $\beta^+$ source setup used in the measurement are presented in~\fref{fig:2g_image}. This benchmark images lead to two observations useful for further analysis of for 3$\gamma$ reconstruction performance. Firstly, it is visible that the largest amount of $e^+e^-$ annihilations take place directly in elements of the setup supporting the $^{22}$Na source (see~\fref{fig:jpet-source}, right), located at $z=0$. Therefore, a clear observation of the reconstruction of annihilations originating in the chamber walls requires exclusion of its central part where the source was mounted. \fref{fig:2g_image_c} shows the transverse plane location of $2\gamma$ annihilation points whose longitudinal coordinate satisfied:
\begin{equation*}
  \abs{z} > 4\ \text{cm},
\end{equation*}
in which case annihilations taking place in the chamber material are clearly visible.
A similar requirement will be used for the demonstration of $3\gamma$ annihilations' reconstruction discussed in the next Section.

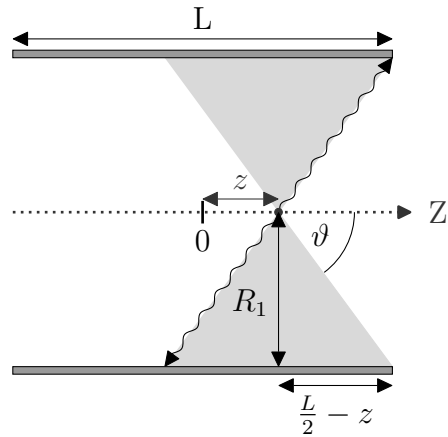
\begin{figure}[h!]
  \centering
\begin{tikzpicture}[
  scale=0.5,
  >={Stealth[inset=0pt,length=6pt,angle'=50,round]}
  ]
  % dummy rect
  \draw[white] (-5.5,-6) rectangle (9,6);

  % cones
  \draw[fill, gray!30!white] (-1,4.1) -- (5, 4.1) -- (2.0, 0.0);
  \draw[fill, gray!30!white] (-1,-4.1) -- (5, -4.1) -- (2.0, 0.0);

  % photons
  \draw[->,decorate,decoration={snake, amplitude=0.4mm}] (2.0, 0.0) -- (5.0, 4.1);
  \draw[->,decorate,decoration={snake, amplitude=0.4mm}] (2.0, 0.0) -- (-1.0, -4.1);
  
  % detector
  \draw[<->] (-5,4.6) -- (5,4.6) node[midway, above] {\large L};
  \draw[fill=black!50!white] (-5,4.1) rectangle (5,4.3);% node[right] {\large Layer 1};

  \draw[fill=black!50!white] (-5,-4.3) rectangle (5,-4.1);
  
  \draw[line width=1pt,dotted, black, ->] (-5,0) -- (5.5,0) node[xshift=10] {\large Z};
  \draw[black, <->] (0,0.35) -- (2.0,0.35) node[midway, above] {\large $z$};
  \draw[line width=1pt] (0.0, -0.3) -- (0.0, 0.3) node[below, yshift=-7] {\large 0};

  % annihilation
  \draw[fill, black] (2.0, 0.0) circle (0.1);
  \node[] at (3.1,-0.6) {\large $\vartheta$};
  \draw[] (4.0, 0.0) arc(0:-53:2.0);
  
  % radius
  \draw[<->] (2.0, 0.0) -- (2.0, -4.1) node[midway, left, yshift=-5] {\large $R_1$};

  \draw[<->] (2.0, -4.6) -- (5.0, -4.6) node[midway, below, yshift=0] {\large $\frac{L}{2} - z$};
  
\end{tikzpicture}
\caption{Schematic presentation of the geometrical acceptance limits on reconstruction of 2$\gamma$ annihilations along the $z$ axis of the detector. Shaded region denotes the volume in which emitted $e^+e^-$ pairs may be detected.}\label{fig:stereo}
\end{figure}

Second conclusion following from the benchmark $2\gamma$ images is the size of the effective sensitive volume inside the detector along the $z$ coordinate. Although the aluminum chamber spanned the whole length of the latter, \fref{fig:2g_image_b} reveals that only in the region of about~$|z| < 8$~cm the annihilations in the chamber material are efficiently reproduced. This is caused by two geometrical effects:
\begin{itemize}
\item for larger $z$ values, a constant element of the chamber wall surface corresponds to a smaller solid angle of positrons' emission from a point-like source; a decrease of $e^+e^-$ annihilation rate is thus expected with increasing distance from the source along $z$,
\item the geometrical acceptance of the detector barrel is reduced for annihilations taking place closer to its edges.
\end{itemize}
The first argument, concerning rate of positron interactions in a section of the chamber wall surface, is directly relevant also in case of the 3$\gamma$ annihilations. The second factor, in case of two photon annihilations, results from the fact that after registration of one $\gamma$ quantum, the other one with opposite momentum must also reach a detector module.
\fref{fig:stereo} schematically depicts this geometrical acceptance limitations for a simplified case of annihilations originating on the detector $z$ axis.
Effectively, only $\gamma\gamma$ pairs emitted in a volume limited by two cones coaxial with the detector with an opening angle $\vartheta = \text{arctan}\frac{R_1}{L/2-z}$ (where $L$ is the scintillator length and $R_1$ is the radius of the innermost scintillator layer) can be registered.
The stereo angle corresponding to this volume is given by:
\begin{equation}
  \label{eq:stereo_angle}
  \Omega(z) = 2\int_{0}^{2\pi}d\phi \int_{\vartheta(z)}^{\pi/2} \sin\theta d\theta = 4\pi \cos\vartheta(z) = 4\pi\frac{\frac{L}{2}-z}{\sqrt{R_1^2+\left(\frac{L}{2}-z\right)^2}}.
\end{equation}
The above relation approximately describes the $z$-dependent decrease in the detection efficiency for recording 2$\gamma$ annihilations. Such considerations can also be used to approximate the drop in efficiency for 3$\gamma$ events if a decay plane is considered instead of the line constituted by two photons' opposite momenta. A conclusion which follows from the above benchmark is therefore that good performance of the reconstruction of three photon annihilations should be expected in approximately the same range as in case of the 2$\gamma$ benchmark image, i.e.\ for about $\abs{z} < 8$~cm.

\subsection{Reconstruction of 3$\gamma$ annihilation points}
\label{sec:jpet_3g_imaging}
\begin{figure}[h!]
  \centering
%  \captionsetup[sub]{margin=1ex}
  \captionsetup[subfigure]{justification=centering}
  \begin{subfigure}{0.45\textwidth}
  \includegraphics[width=1.0\textwidth]{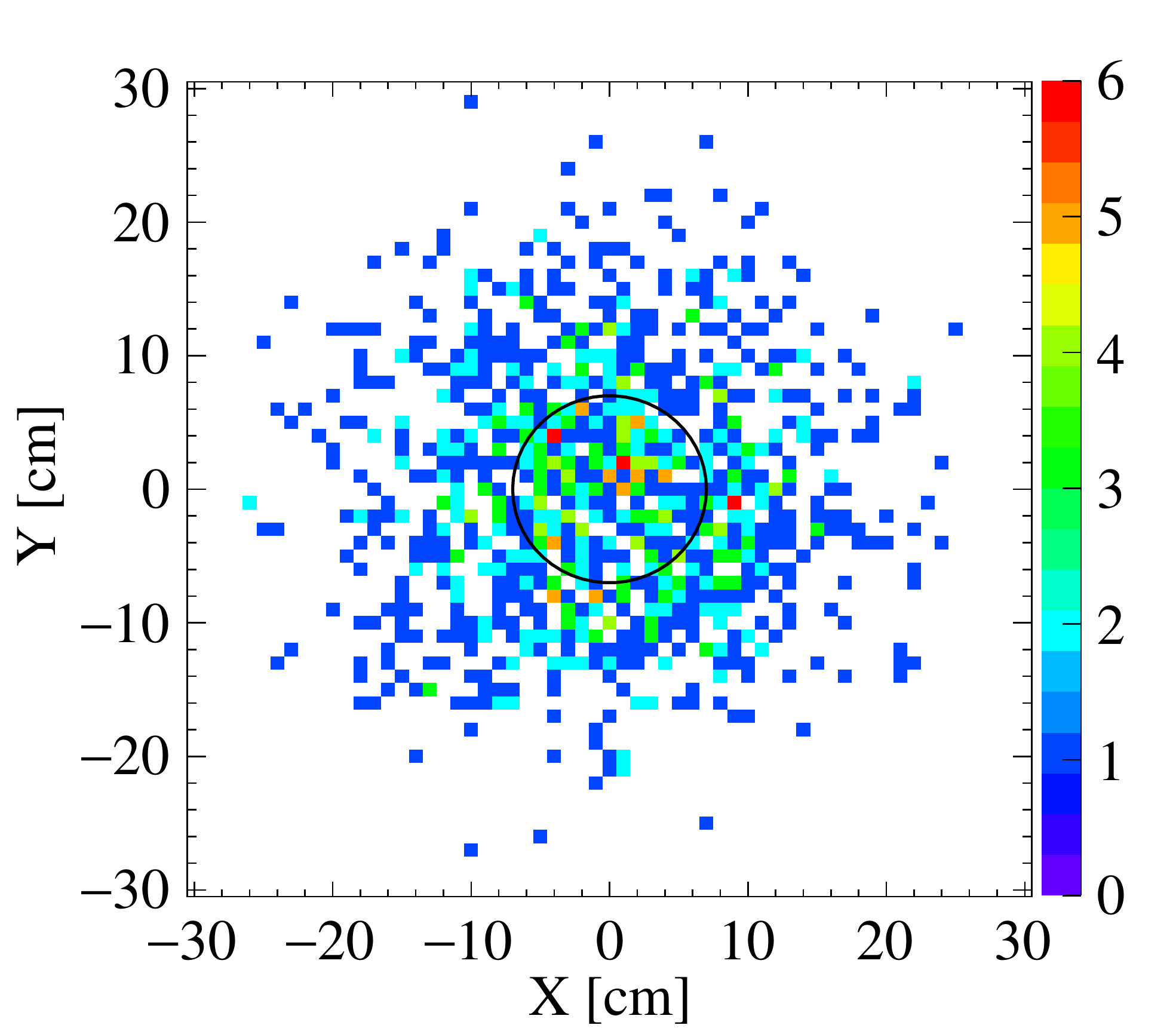}
  \caption{Transverse view image,\\ all $z$ values projected}\label{fig:3g_image_a}
  \end{subfigure}
  \hspace{1em}
  \begin{subfigure}{0.45\textwidth}
  \includegraphics[width=1.0\textwidth]{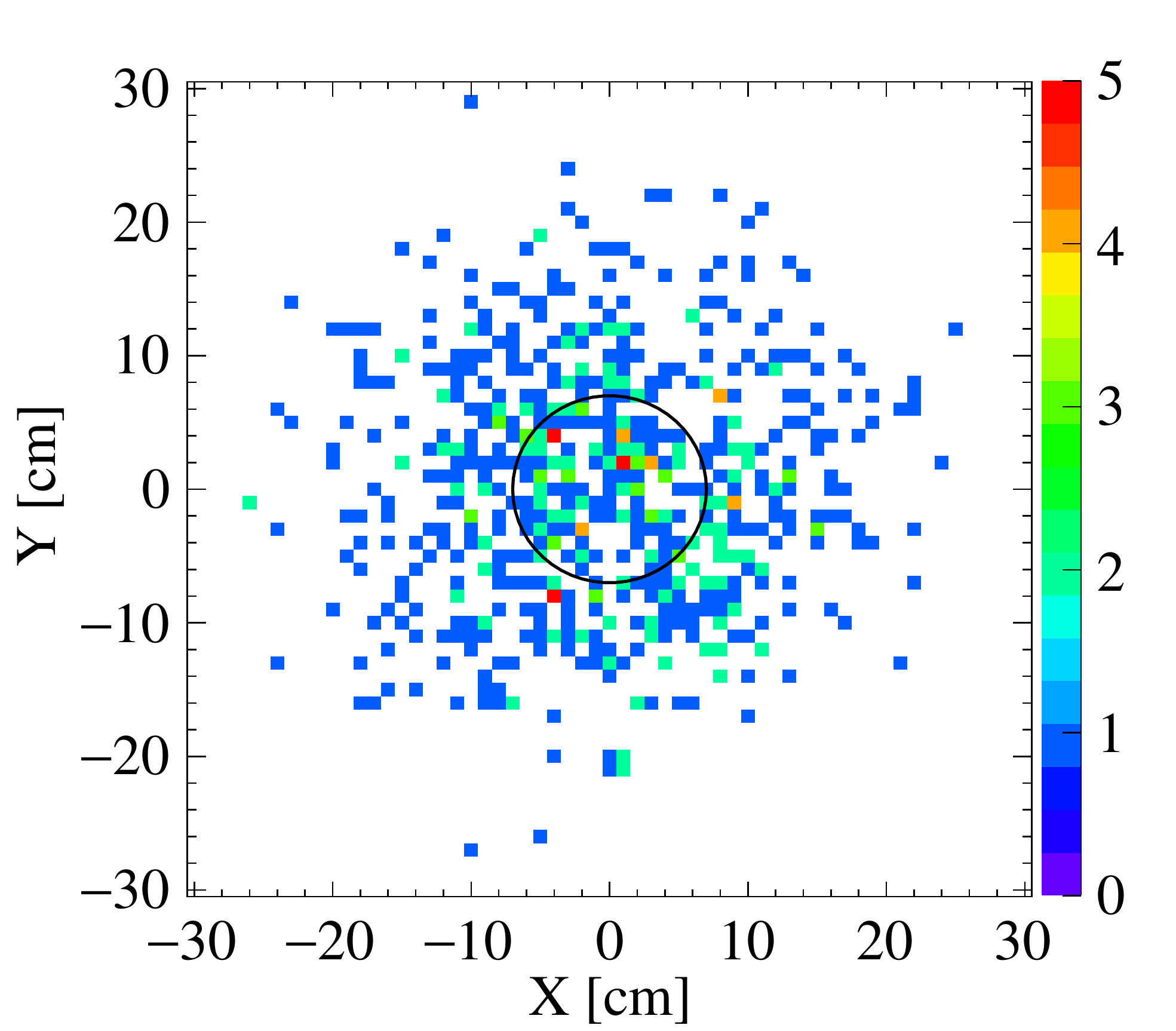}
  \caption{Transverse view image,\\ $\abs{z} > 4$~cm}\label{fig:3g_image_c}
\end{subfigure}
  \begin{subfigure}{0.45\textwidth}
  \includegraphics[width=1.0\textwidth]{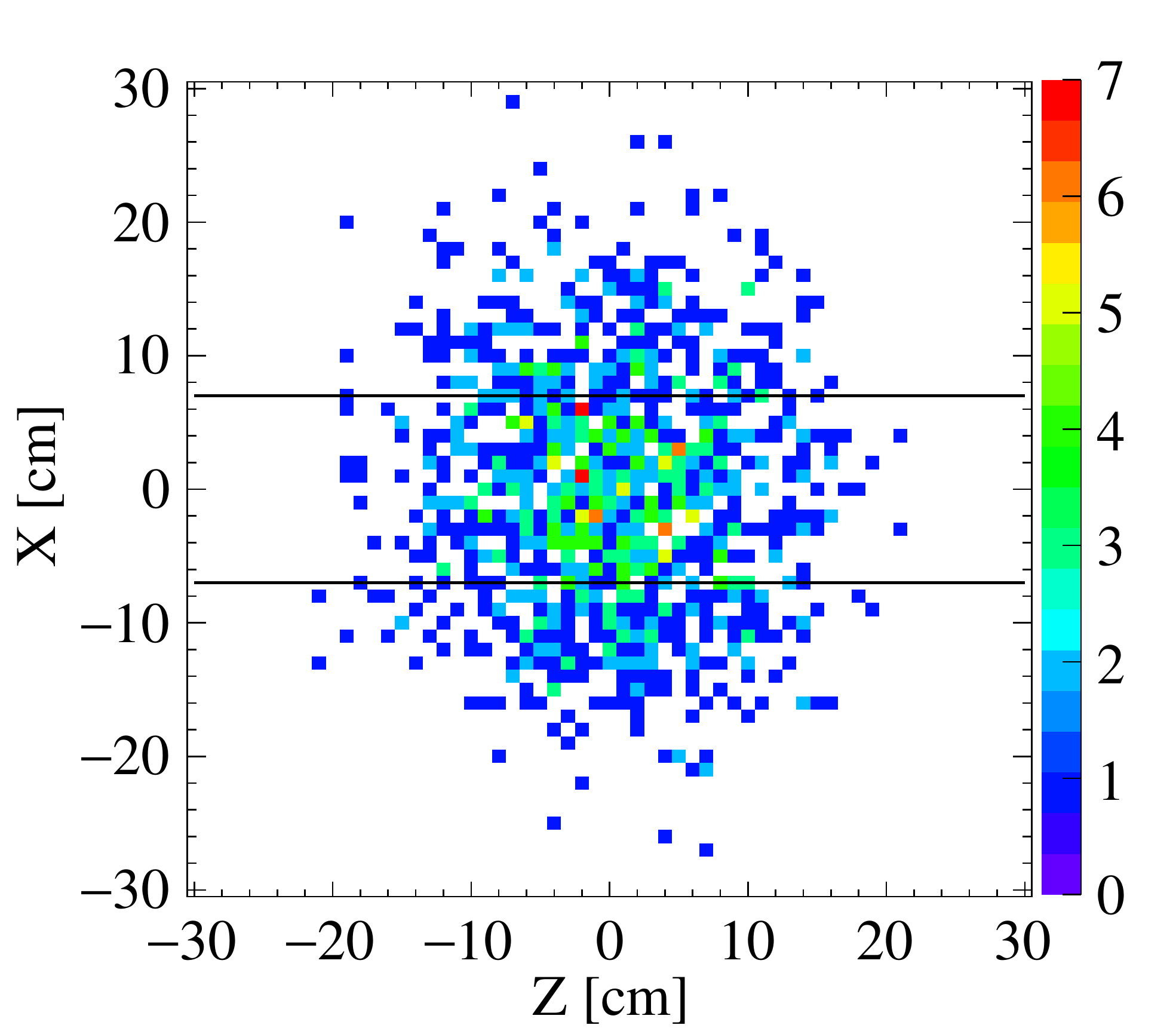}
  \caption{Side view image}\label{fig:3g_image_b}
\end{subfigure}
\caption{Distribution of the reconstructed origin points of 3$\gamma$ annihilation events identified in the test measurement data. Black lines indicate limits of the 3$\gamma$ production chamber used used in the measurement.}\label{fig:3g_image}
\end{figure}

The selection of 3$\gamma$ events described in~\sref{sec:3g_selection} applied to the test measurement data yielded 1164 event candidates. For each candidate, the annihilation point was reconstructed using the trilateration-based method presented in~\sref{sec:gps_jpet}. Distributions of the resulting annihilation points are presented in~\fref{fig:3g_image}.
The transverse view images are shown separately for the complete projection along detector $z$ axis as well as for the region of $\abs{z}>4$~cm, i.e.\ excluding the volume were most annihilations originate in the support of the $\beta^+$ source rather than in the chamber walls as demonstrated with the 2$\gamma$ images.

The statistical limitations of the data sample obtained with the 2-day measurement in absence of an ortho-positronium production medium do not allow for a quantitative estimation of the reconstruction resolution. However, the reconstructed points are grouped in proximity of the geometrical limits of the 3$\gamma$ production chamber, indicated with black lines in~\fref{fig:3g_image}. Moreover, in case of the longitudinal image (\fref{fig:3g_image_b}), the populated region along detector $z$ axis is similar to the $z$ range of efficient 2$\gamma$ imaging (compare~\fref{fig:2g_image_b}), as expected from the considerations presented in the previous Section.

\subsection{Discussion of the results and perspectives}
\label{sec:jpet_discussion_perspectives}

Distributions of reconstructed annihilation points obtained with the test run, in conjunction with the validation of the trilaterative reconstruction method using Monte Carlo simulations show good prospects for future experiments with the \mbox{J-PET} detector with a view to testing the fundamental discrete symmetries.
The present result is limited in terms of statistics, purity of the event sample used and detector resolution. Each of these effects, however, is expected to be improved in the planned measurements.
Once an ortho-positronium production medium is included in the setup, the yield of 3$\gamma$ annihilations will increase by about two orders of magnitude, allowing for easier selection of a pure 3$\gamma$ event sample. The latter will be further aided by the fact that genuine \ops/$\to 3\gamma$ events, as opposed to direct $e^+e^-$ direct annihilations used in this test study, will involve a characteristic time interval between an associated deexcitation photon (corresponding approximately to the \ops/ formation) and recording of the annihilation photons. This property may be used for more sophisticated identification of the ortho-positronium annihilations.

The angular resolution of reconstructed $\gamma$ annihilations points, crucial for control of polarization in the experiment, may be improved by means of a kinematic fit requiring the radial cylindrical coordinate of the reconstruction results to be contained within the annihilation medium as demonstrated using Monte-Carlo simulated events~\cite{gajos_gps}. Moreover, the procedures for calibration and reconstruction of times in the \mbox{J-PET} detector are being improved on at the time of writing of this Thesis, with a view to improving the time resolution which has a deciding impact on the performance of trilateration-based 3$\gamma$ event reconstruction. Advanced signal reconstruction techniques have also been prepared and tested using a \mbox{J-PET} detector prototype~\cite{lech_compressive, neha_synchronized} and their inclusion in the future experiments would provide a further resolution enhancement.

%%% Local Variables:
%%% TeX-master: "../main"
%%% End:

\chapter{Summary and perspectives}
\label{chapter:conclusions}

The aim of this Thesis was to  pursue the investigations of fundamental discrete symmetries with two new measurements. Firstly, a direct test of the symmetry under time reversal using transitions between flavour and CP-definite states of quantum-entangled neutral K mesons was performed using data collected by the KLOE detector, with a view to realization of a statistically sensitive test with the KLOE-2 experiment.
Secondly, a fesibility study of planned searches of discrete symmetry violations in purely leptonic systems, i.e.\ in the decays of ortho-positronium atoms into photons, manifested as non-vanishing angular correlations of final state photons' momenta, was carried out using the J-PET experimental setup.

As both of these objectives relied on reconstruction of the location and time of neutral particle decays into photons, a novel reconstruction procedure based on trilateration was devised and applied to $\Kl\to 3\pi^0\to 6\gamma$ decays recorded by the KLOE detector and to \ops/$\to 3\gamma$ in J-PET.

It was demonstrated that the trilaterative reconstruction of $\Kl$ decays into neutral pions provides a resolution of $\Kl$ proper decay time at a constant level of about $1.6\:\tau_{S}$, only two times less precise than in case of using drift chamber vertices for reconstruction of kaon decays into charged particles. The possibility to reconstruct $\Kl\to 3\pi^0$ decays at KLOE allowed for extraction of samples of $\Ks\Kl\to \pi e \nu\; 3\pi^0$ and $\Ks\Kl\to\pi^+\pi^-\;\pi e \nu$ from 1.7~fb$^{-1}$ of the $e^+e^-\to\phi\to\Ks\Kl$ data collected by the KLOE experiment in 2004--2005. The extracted event sets, split into subsamples by charge of leptons in the semileptonic kaon decay, were used to construct two ratios of double kaon decay rates expressed as functions of the kaons' proper decay time difference $\Delta t$. A quantitative measure of the level of \Ts~noninvariance was determined as the constant level of these ratios in the $\Delta t \gg \tau_{S}$ with respect to unity. The asymptotic level of the \Ts-violation sensitive ratios was estimated by means of a dedicated maximum likelihood fit. The obtained values:
\begin{eqnarray}
  \label{eq:result_in_conclusions}
  R_2 &= 1.020 \pm 0.017_{stat} \pm 0.035_{syst},\\
  R_4 &= 0.990 \pm 0.017_{stat} \pm 0.039_{syst},
\end{eqnarray}
are in agreement with unity within the achieved precision, insufficient for sensitivity to \Ts-violating effects.
% as expected based on the statistics of KLOE dataset used.
However, the resulting uncertainties and performance of the devised analysis show perspectives for reaching the required sensitivity level with the data of the KLOE-2 experiment (which aims at collecting at least 5~fb$^{-1}$ of $\phi$ decay data) provided that the presently dominant systematic errors are mitigated. The results obtained with KLOE data are still affected by certain background contamination of the extracted data samples and imperfections of the used efficiency estimation method. These problems have been identified and possible solutions to be considered in the KLOE-2 data analysis have been proposed.

In addition to the planned test of the symmetry under time reversal at KLOE-2, the prepared data analysis steps may be used in a direct test of the \CPTs~symmetry with neutral kaons at KLOE and KLOE-2 using a similar principle~\cite{theory:bernabeu-cpt}. Such a test has never been performed to date and work is in progress with a view to establishing the first result of \CPTs~test in transitions of neutral mesons with the KLOE data~\cite{KLOE-2:2017lyj,Gajos:2017tlb}.

In the second study comprised in this Thesis, the feasibility of the searches for non-zero average angular correlations in ortho-positronium decays into three photons was demonstrated using a test measurement featuring a cylindrical aluminum chamber for production of direct $e^+e^-\to 3\gamma$ annihilations. A simple selection of 3$\gamma$ event candidates was proposed and about 1100 candidates were identified in the test data. A dedicated reconstruction method was prepared for such decays, validated using Monte Carlo simulations and applied to the candidate events. Additionally, tomographic images of the annihilation chamber using 2$\gamma$ annihilations were obtained as a~benchmark. Although the limited statistics of the test measurement only allowed for qualitative conclusions on the performance of the 3$\gamma$ event identification and reconstruction, distribution of the obtained annihilation points around the location of the annihilation medium shows good prospects for future measurements of \ops/$\to 3\gamma$ events whose yield is expected at a level higher than in the test measurement by two orders of magnitude. Additionally, as the sensitivity of the trilateration-based reconstruction to timing resolution of the detector was demonstrated, a large improvement of the reconstruction performance should come from enhancements of the J-PET detector calibration and data reconstruction procedures, all of which are being elaborated on at the time of writing of this Thesis.

%%% Local Variables:
%%% TeX-master: "../main"
%%% End: 

%----------------------------------------------------------------------------------------
%	THESIS CONTENT - APPENDICES
%----------------------------------------------------------------------------------------
\addtocontents{toc}{\vspace{1em}} % Add a gap in the Contents, for aesthetics

%\appendix % Cue to tell LaTeX that the following 'chapters' are Appendices

% Include the appendices of the thesis as separate files from the Appendices folder
% Uncomment the lines as you write the Appendices
\begin{appendices}
\chapter{Least-squares solution of the trilateration reconstruction with 5~or~6 reference points}\label{appendix:numerical_6_equations}

Although only 4 reference points (i.e.~4 equations in the system) are required to obtain an analytical solution from~\ref{eq:gps_6_eqns}, information on additional available reference points allows for minimizing the uncertainties in the reconstruction. To this end, the overdetermined equation system can be solved numerically e.g.\ with an iterative least-squares minimization procedure presented below for the case of 6 reference points.

The equation system defining the reconstruction problem is of the following form (a square root has been taken sidewise in order for the right-hand-sides to be linear in the unknown~$t$):
\begin{equation}
  \label{eq:numerical_6_equations}
  \left((x-X_i)^2 + (y-Y_i)^2 + (z-Z_i)^2 \right)^{\frac{1}{2}} = c(T_i-t), \qquad i=1,\ldots,6,
\end{equation}
where capitals denote known values, $c$ is the velocity of light and subscripts indicate the number of reference point.

For each of these equations, characterized by subscript $i$, let us define the left-hand-side and right-hand-side values as:
\begin{eqnarray}
  \label{eq:numerical_lhs_rhs}
  LHS_i(x,y,z) &=& \left( (x-X_i)^2 + (y-Y_i)^2 + (z-Z_i)^2) \right)^{\frac{1}{2}} \equiv R_i(x,y,z),   \label{eq:numerical_lhs} \\
  RHS_i(t) &=& c(T_i-t).   \label{eq:numerical_rhs}
\end{eqnarray}

The optimal solution $(x,y,z,t)$ should minimize the following residual error of the equations:
\begin{equation}
  \label{eq:numerical_residual}
  \sum_{i=1}^{6} \left(LHS_i(x,y,z) - RHS(t)\right)^2.
\end{equation}

In the iterative procedure, the optimal solution is sought by starting from an initial guess $\mathbf{x}^{(0)} = (x^{(0)},y^{(0)},z^{(0)},t^{(0)})^T$ (which can be obtained e.g.\ from an analytical solution using any 4 out of 6 equations) and solving a linearized system for a set of corrections \mbox{$\mathbf{\Delta x} = (\Delta x, \Delta y\, \Delta z, \Delta t)^T$} so that after $k$-th iteration step:
\begin{equation}
  \label{eq:numerical_corrections}
  \mathbf{x}^{(k+1)} = \mathbf{x}^{(k)} + \mathbf{\Delta x}^{(k)}.
\end{equation}

Let $\mathbf{\vec{x}}^{(k)}$ denote the spatial part of solution vector so that:
\begin{eqnarray}
  \label{eq:numerical_spatial_part}
  \mathbf{\vec{x}}^{(k)} &=& (x^{(k)},y^{(k)},z^{(k)})^T, \\
  \mathbf{\Delta \vec{x}}^{(k)}  &=& (\Delta x^{(k)} , \Delta y^{(k)} , \Delta z^{(k)} )^T, \\
  \mathbf{x}^{(k)} &=& (\mathbf{\vec{x}}^{(k)}, ct^{(k)})^T, \quad \mathbf{\Delta x}^{(k)}  = (\mathbf{\Delta \vec{x}}^{(k)} , c\Delta t^{(k)})^T,
\end{eqnarray}
then linearized system allowing to calculate the corrections for step $k+1$ can be obtained by approximating the left-hand-side expressions from~\eref{eq:numerical_lhs} up to the first term of a Taylor expansion around the solution of step $k$:
\begin{equation}
  \label{eq:numerical_taylorexp}
  LHS_i(\mathbf{\vec{x}}^{(k+1)}) \approx R_i(\mathbf{\vec{x}}^{(k)}) + \grad R_i(\mathbf{\vec{x}}^{(k)})\mathbf{\Delta \vec{x}}^{(k)} .
\end{equation}
As the right-hand-side of the original equations is:
\begin{equation}
  \label{eq:numerical_rhs_nextstep}
  RHS_i(t^{(k+1)}) = c\left(T_i - t^{(k+1)}\right) = cT_i - ct^{(k)} - c\Delta t^{(k)} ,
\end{equation}
using relations~\ref{eq:numerical_taylorexp} and~\ref{eq:numerical_rhs_nextstep}, the system of Equations~\ref{eq:numerical_6_equations} can be expressed in the matrix form:
\begin{equation}
  \label{eq:numerical_axb}
  \mathbf{A}^{(k)}\mathbf{\Delta x}^{(k)}  = \mathbf{b}^{(k)},
\end{equation}
where the $\mathbf{A}$ matrix is defined as:
\begin{equation}
  \label{eq:numerical_A}
  \mathbf{A}^{(k)} =
  \begin{bmatrix}
    \frac{x^{(k)}-X_1}{R_1(\mathbf{\vec{x}}^{(k)})} & \frac{y^{(k)}-Y_1}{R_1(\mathbf{\vec{x}}^{(k)})} & \frac{z^{(k)}-Z_1}{R_1(\mathbf{\vec{x}}^{(k)})}& c \\
    \vdots & \ddots & & \vdots \\
    \vdots & & \ddots & \vdots \\
    \frac{x^{(k)}-X_6}{R_6(\mathbf{\vec{x}}^{(k)})} & \frac{y^{(k)}-Y_6}{R_6(\mathbf{\vec{x}}^{(k)})} & \frac{z^{(k)}-Z_6}{R_6(\mathbf{\vec{x}}^{(k)})} & c 
  \end{bmatrix},
\end{equation}
and:
\begin{equation}
  \label{eq:numerical_b}
  \mathbf{b}^{(k)} =
  \begin{bmatrix}
    cT_1 - ct^{(k)} - R_1(\mathbf{\vec{x}}^{(k)}) \\
    \vdots\\
    cT_6 - ct^{(k)} - R_6(\mathbf{\vec{x}}^{(k)})
  \end{bmatrix}.
\end{equation}

For each step of the iteration, the vector of corrections can be obtained by solving Equation~\ref{eq:numerical_axb} using the pseudo-inverse of the $\mathbf{A}$ matrix~\cite{optimization}:
\begin{equation}
  \label{eq:numerical_solution}
  \mathbf{\Delta x}^{(k)}  = \left( (\mathbf{A}^{(k)})^T\mathbf{A}^{(k)} \right)^{-1} (\mathbf{A}^{(k)})^{T} \mathbf{b}^{(k)},
\end{equation}
after which the solution estimate is updated according to~\eref{eq:numerical_corrections}. Such iteration can be repeated until convergence is indicated by insignificant values of subsequent correction vectors $\mathbf{\Delta x}^{(k)}$.

%%%Local Variables:
%%% TeX-master: "../main"
%%% End:
\chapter{Analytical solution of the planar trilateration problem for \ops/$\to 3\gamma$ reconstruction}\label{appendix:jpet_solution}

The problem of reconstructing an \ops/$\to 3\gamma$ decay point in the decay plane is defined by the following system of equations:
\begin{eqnarray}
  \label{eq:sol_jpet_system}
  (X'_1-x')^2 + (Y'_1-y')^2 = c^2(T_1-t)^2, \\
  (X'_2-x')^2 + (Y'_2-y')^2 = c^2(T_2-t)^2, \\
  (X'_3-x')^2 + (Y'_3-y')^2 = c^2(T_3-t)^2,
\end{eqnarray}
where $x'$, $y'$ and $t$ are unknowns, measured parameters of reference points are denoted with capital letters and $c$ is the velocity of light.

A maximum of two linearly independent equations linear in the unknowns can be obtained by sidewise subtraction of pairs of the above equations. The resulting underdetermined linear system takes the form:
\begin{equation}
  \label{eq:sol_jpet_linear}
  \begin{bmatrix}
    A_{11} & A_{12} & A_{13} \\
    A_{21} & A_{22} & A_{23} 
  \end{bmatrix}
  \begin{bmatrix}
    x' \\
    y' \\
    t
  \end{bmatrix}
  =
  \begin{bmatrix}
    d_1 \\
    d_2
  \end{bmatrix},
\end{equation}
where the \textbf{A} matrix elements are:
\begin{eqnarray*}
  \label{eq:sol_jpet_elements}
  A_{11} =& 2(X'_1-X'_2), \qquad A_{21} =& 2(X'_2-X'_3), \\
  A_{12} =& 2(Y'_1-Y'_2), \qquad A_{22} =& 2(Y'_2-Y'_3), \\
  A_{13} =& 2c^2(T_2-T_1), \qquad A_{23} =&  2c^2(T_3-T_2),
\end{eqnarray*}
and the \textbf{d} vector contains:
\begin{eqnarray*}
  \label{eq:sol_jpet_belements}
  d_1 &=& X_1'^2-X_2'^2 + Y_1'^2 - Y_2'^2 - c^2T_1^2 + c^2T_2^2, \\
  d_2 &=& X_2'^2-X_3'^2 + Y_2'^2 - Y_3'^2 - c^2T_2^2 + c^2T_3^2.
\end{eqnarray*}

\eref{eq:sol_jpet_linear} can be solved e.g.\ for $x'$ and $y'$ parametrized by $t$:
\begin{eqnarray}
  x'(t) &= a_xt + b_x, \label{eq:sol_jpet_x} \\
  y'(t) &= a_yt + b_y, \label{eq:sol_jpet_y}
\end{eqnarray}
with:
\begin{eqnarray*}
  a_x =& \frac{A_{13}A_{22}-A_{12}A_{23}}{A_{12}A_{21}-A_{11}A_{22}}, \qquad b_x = \frac{A_{12}d_{2}-A_{22}d_{1}}{A_{12}A_{21}-A_{11}A_{22}}, \\
  a_y =& \frac{A_{11}A_{23}-A_{13}A_{21}}{A_{12}A_{21}-A_{11}A_{22}}, \qquad b_y = \frac{A_{21}d_{1}-A_{11}d_{2}}{A_{12}A_{21}-A_{11}A_{22}}.        
\end{eqnarray*}
Insertion of relations~\ref{eq:sol_jpet_x} and~\ref{eq:sol_jpet_y} into one of the original equations, e.g.~\eref{eq:sol_jpet_system} yields the following quadratic equation for $t$:
\begin{equation}
  \begin{split}
    \left[ a_x^2 + a_y^2 - c^2 \right] t^2 &+ 2\left[ a_x(b_x-X'_1) + a_y(b_y-Y'_1)  + c^2T_1 \right] t \\
    &+ (b_x-X'_1)^2 + (b_y-Y'_1)^2 - c^2T_1^2  = 0.
  \end{split}
\end{equation}
The above equation may have up to two distinct solutions, which, after insertion back into Equations~\ref{eq:sol_jpet_x} and~\ref{eq:sol_jpet_y}, result in a maximum of two sets of decay location and time in the decay plane:
\begin{equation}
  \label{eq:sol_jpet_solutions}
  (x',y',t)^{(j)}, \quad j=1,2.
\end{equation}

%%% Local Variables:
%%% TeX-master: "../main"
%%% End:
%\input{./Appendices/AppendixC}
\chapter{Electron/pion and electron/muon classifiers based on artificial neural networks}\label{appendix:clasifiers}

The rejection of background components where the DC track identified as electron or positron in fact corresponds to either a poorly recorded charged pion or to a muon from a $\pi^{\pm}\to\mu^{\pm}\bar{\nu}_{\mu}$ decay, presented in~\sref{sec:pimu_rejection}, is based on artificial neural networks. Other machine learning solution were considered as well, such as Boosted Decision Trees and the $k$-Nearest Neighbors algorithm (all in their implementation from the TMVA software package~\cite{Hocker:2007ht}). However, as the classification problem considered in this case is rather uncomplicated, all of these methods exhibited comparable performance, with the output from neural networks allowing for a combination of the two ($e$/$\pi$ and $e/\mu$) classifiers' results most convenient to design a background rejection cut.

The employed ANN-s were simple feed-forward networks with each with five inputs, a single 10-neuron hidden layer and a single output. The input variables were not subject to any preparatory transformations. The two first input variables used are based on the different longitudinal structure of electromagnetic showers from $e^{\pm}$ and from $\pi^{\pm}$/$\mu^{\pm}$ in the KLOE calorimeter. First variable is defined as normalized difference of energies deposited in the first and second calorimeter layers with non-zero recorded energy:
\[ \Delta E_{12} = \frac{E_{dep}^{1st\:layer} - E_{dep}^{2nd\:layer}}{E_{dep}^{1st\:layer} + E_{dep}^{2nd\:layer}}. \]  

\begin{figure}[h!]
  \centering
  \includegraphics[width=1.0\textwidth]{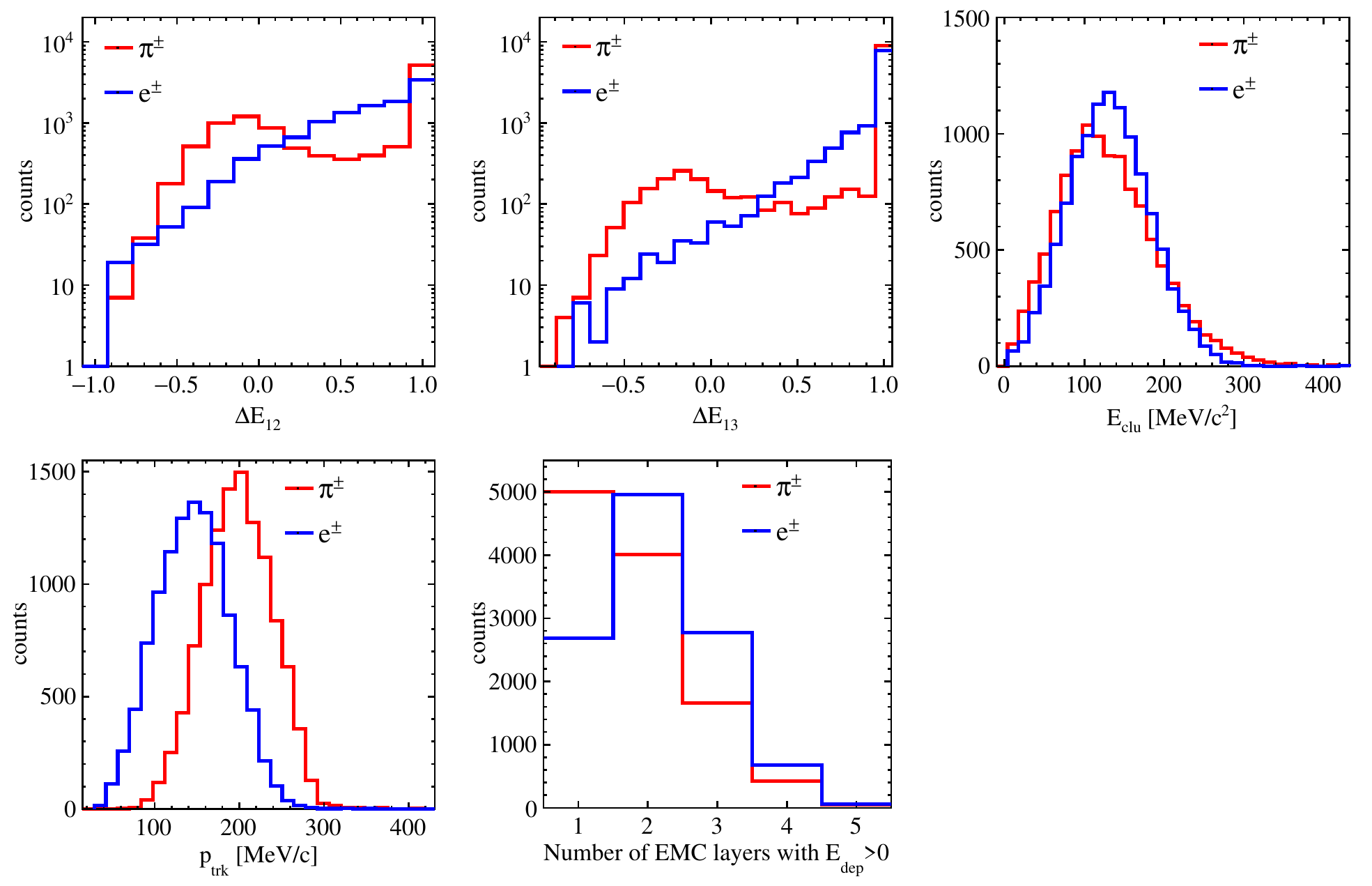}
  \caption{Distributions of five variables characterizing the EMC cluster identified as coming from a lepton and its associated DC track, used as input for the $e$/$\pi$ classifier. The distributions were obtained with a $\Kl\to\pi e\nu$ data sample used for training of the neural network.}\label{fig:mva_inputs_epi}
\end{figure}

\begin{figure}[h!]
  \centering
  \includegraphics[width=1.0\textwidth]{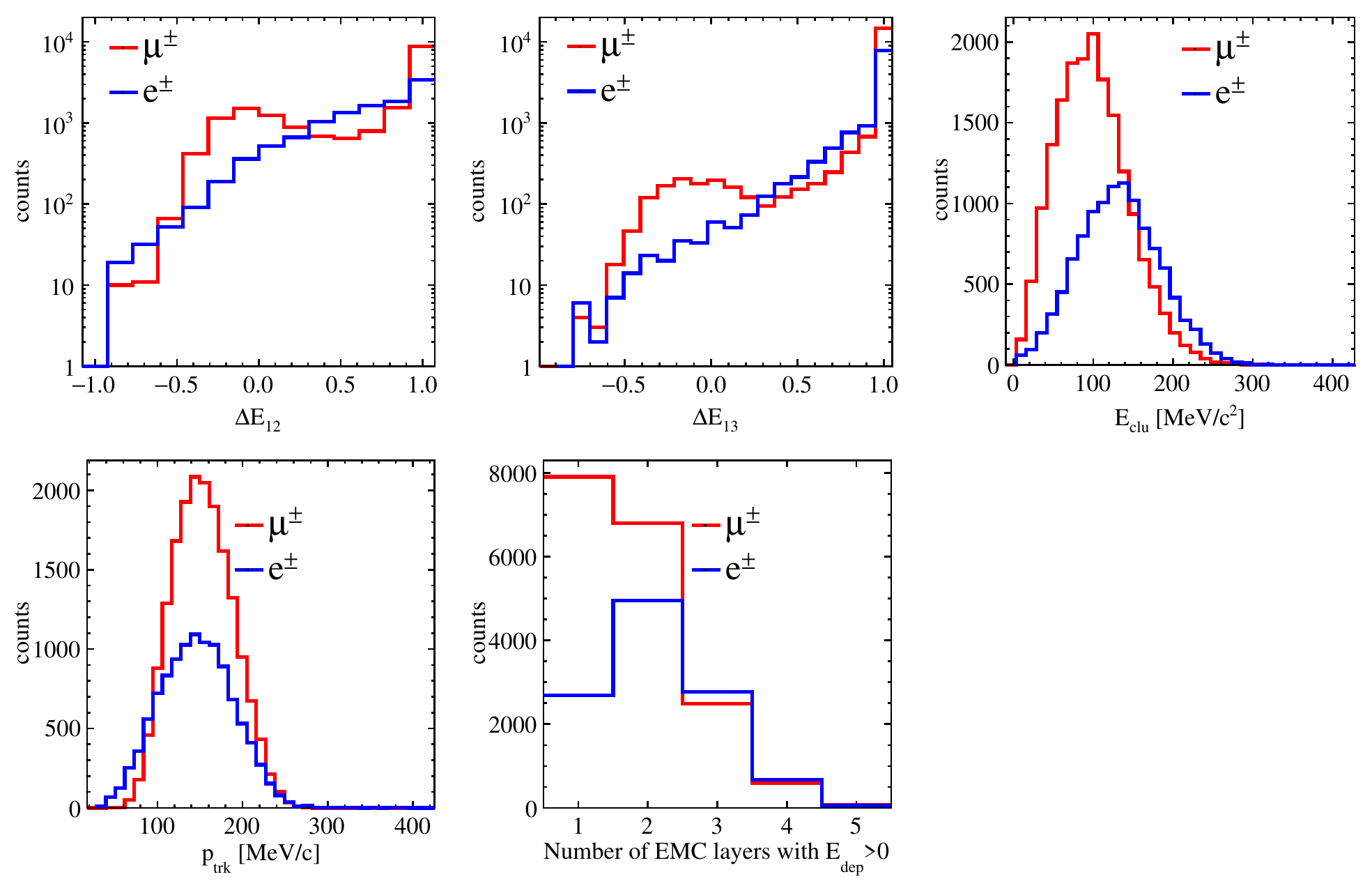}
  \caption{Distributions of five variables characterizing the EMC cluster identified as coming from a lepton and its associated DC track, used as input for the $e$/$\mu$ classifier. The distributions were obtained with a $\Kl\to\pi\mu\nu$ data sample used for training of the neural network.}\label{fig:mva_inputs_emu}
\end{figure}

Similarly, the second input variable is a relative difference between energies deposited in the first and third active EMC layers:
\[ \Delta E_{13} = \frac{E_{dep}^{1st\:layer} - E_{dep}^{3rd\:layer}}{E_{dep}^{1st\:layer} + E_{dep}^{3rd\:layer}}. \]  
As visible in Figures~\ref{fig:mva_inputs_epi} and~\ref{fig:mva_inputs_emu}, distributions of $\Delta E_{12}$ and $\Delta E_{13}$ are closer to unity for electrons/positrons, in accordance with the large deposition of energy at the first interactions expected from these particles. In a large number of events, however, the energy deposited in one or both of the subsequent layers is zero e.g.\ due to limited efficiency of the calorimeter for small energy deposits, which results in the values of normalized differences equal to unity. As this effect occurs for all considered types of particles alike, discrimination in such cases must be based on other input. Therefore, another input to the classifiers is constituted by the number of EMC layers which recorded a non-zero energy deposit. As shown in the last panel of Figures~\ref{fig:mva_inputs_epi} and~\ref{fig:mva_inputs_emu}, most electrons activate at least two layers of the calorimeter whereas both $\pi^{\pm}$ and $\mu^{\pm}$ tend to often deposit a registerable energy amount in a single layer only. Finally, the last two variables used as input of the classifiers are the particle momentum estimated with the DC track and total energy deposited in the EMC cluster associated to that track. As both these quantities are measured independently, their relative values allow for a good discrimination between electrons/positrons and heavier particles. Distributions of all five input variables compared between electrons/positrons and charged pions and muons are presented in Figures~\ref{fig:mva_inputs_epi} and~\ref{fig:mva_inputs_emu}, respectively.

%%%Local Variables:
%%% TeX-master: "../main"
%%% End:  
\end{appendices}

\addtocontents{toc}{\vspace{1em}} % Add a gap in the Contents, for aesthetics

%\backmatter

\cleardoublepage % Start a new page

%----------------------------------------------------------------------------------------%
%	ACKNOWLEDGEMENTS
%----------------------------------------------------------------------------------------
%\fancyhead[LO,RE]{\emph{Acknowledgements}} %
\pagestyle{empty}
\acknowledgements{%\addtocontents{toc}{\vspace{1em}} %Add a gap in the Contents, for aesthetics
\vspace{10em}

\setlength{\parskip}{1em}

During the years of my doctoral studies I had the luck to meet a number of great people, without whom this Thesis would never have been created. I would like to thank every one of them for their support, inspiration and friendship.

%
% Promotorzy
%

I would like to express my deepest gratitude to prof. Pawe\l{} Moskal for the opportunity to work in his research group, for his supervision over the preparation of this Thesis, and --- above all --- for his contagious passion for science.

The second person without whom this work would not be possible is dr Eryk Czerwi\'nski. I~am greatly indebted to Eryk for his guidance in all my research for the past six years and, perhaps even more importantly, for introducing me to virtually all aspects of academic life.
%
% Antonio
%

I would like to extend my special thanks to prof. Antonio Di Domenico, for giving me the possibility to work on the fascinating subject of direct symmetry tests with neutral kaons at KLOE and KLOE-2, and for his careful supervision on my analysis.

%
% Kamys i Jarczyk
%

I am grateful to prof.\ Bogus\l{}aw Kamys for the opportunity to work in the Department of Nuclear Physics of the Jagiellonian University and to prof.\ Lucjan Jarczyk for all his comments on my work, always motivating me to improve my research and presentation skills.

% 
% Koledzy z KLOE w Krakowie: Michal, Wojtek, Daria, Krzysiek
%

The time of my work in the Krak\'ow subgroup of KLOE was exceptional thanks my Colleagues Daria Kisielewska, Krzysztof Kacprzak, dr Wojciech Krzemie\'n and dr Micha\l{} Silarski. Thank you for the great and inspiring atmosphere and countless help you gave me during these years. 

%
% KLOE advice
%

I would like to thank prof.\ Filippo Ceradini, dr~Erika De~Lucia, dr~Antonio De~Santis, dr~Paolo Gauzzi and dr~Enrico Graziani for sharing their expertise in working with the KLOE data and for their suggestions which helped me overcome several dead ends in my work. I am also indebted to prof. Wojciech Wiślicki for his helpful advice on statistics.
%
% KLOE Colleagues
%

Moreover, my frequent visits in the Laboratories of Frascati would not be the same without the great people I met there and their hospitality, especially dr Elena Perez del~Rio, dr~Marcin Ber\l{}owski, dr~Paolo Fermani, dr~Gianfranco Morello and all my Colleagues from KLOE.

No less do I owe to my Colleagues from J-PET with whom I have worked on the second part of this Thesis. I would like to especially thank dr Magdalena Skurzok, Monika Pawlik-Niedźwiecka, dr Grzegorz Korcyl, Szymon Niedźwiecki and dr Sushil Sharma for providing me with building blocks for the calibration and analysis of J-PET data based on their great efforts.

Great thanks also to my officemate Krzysztof Nowakowski, for motivating me when it was time to write, and for saving my sanity with conversations about everything but science when it was time to take a break.

%
% Spradzić czy "to owe big" jest formalne
% Krzysiek?

%
% Rodzina
%

As so many others things in my life, this Thesis would never come to life without the continuous support of my wife Kasia, whom I would like to thank for her patience and unwavering belief that I will finish writing some day. I also owe great thanks to my son Andrzej, for bringing a completely new quality to my life in my last PhD student years, and for his unstoppable will to help me in writing of this text, even if by typing randomly on daddy's keyboard.

\begin{otherlanguage}{polish}
Na koniec dziękuję Wam, Mamo i Tato, za wiarę we mnie i za całą pomoc w dotarciu do tego momentu.
\end{otherlanguage}

\vspace{3cm}

This work was supported by the Polish National Science Centre through\\Projects No.~2014/14/E/ST2/00262 and 2016/21/N/ST2/01727.
%
% Może o grancie?
% 

%
% TODO: spraedzic czy gauzzi i Graziani to doktorzy
%

%%% Local Variables:
%%% TeX-master: "main"
%%% End: 
}

%----------------------------------------------------------------------------------------
%	BIBLIOGRAPHY
%---------------------------------------------------------------------------------------- 
\cleardoublepage
\pagestyle{fancy}
%\fancyhead[LO,RE]{\emph{\leftmark}} % Set the left side page header to chapter title
\addtotoc{\bf Bibliography}
\label{Bibliography}
%\lhead{\emph{Bibliography}} % Change the page header to say "Bibliography"

\setstretch{1.1} % Line spacing of 1.3
 \bibliographystyle{utphys}

\bibliography{bibtex/theory,bibtex/misc,bibtex/kloe,bibtex/jpet} % The references (bibliography) information are stored in the file named "Bibliography.bib"

\providecommand{\href}[2]{#2}\begingroup\raggedright\begin{thebibliography}{100}

\bibitem{bernabeu_colloquium}
J.~Bernabeu and F.~Martinez-Vidal, ``{Colloquium: Time-reversal violation with
  quantum-entangled B mesons},''
  \href{http://dx.doi.org/10.1103/RevModPhys.87.165}{{\em Rev. Mod. Phys.}
  {\bfseries 87} (2015) 165},
\href{http://arxiv.org/abs/1410.1742}{{\ttfamily arXiv:1410.1742 [hep-ph]}}.
%%CITATION = ARXIV:1410.1742;%%.

\bibitem{sozzi}
M.~S. Sozzi, ``{Discrete symmetries and CP violation: from experiment to
  theory},''
\newblock Oxford Graduate Texts. Oxford Univ. Press, New York, NY, 2008.

\bibitem{wigner1931}
E.~Wigner, ``{Group Theory and its Application to the Quantum Mechanics of
  Atomic Spectra},''. New York: Academic Press, 1959.

\bibitem{parity_violation}
C.~S. Wu, E.~Ambler, R.~W. Hayward, D.~D. Hoppes, and R.~P. Hudson,
  ``Experimental test of parity conservation in beta decay,''
  \href{http://dx.doi.org/10.1103/PhysRev.105.1413}{{\em Phys. Rev.} {\bfseries
  105} (Feb, 1957)  1413--1415}.

\bibitem{c_violation}
P.~C. Macq, K.~M. Crowe, and R.~P. Haddock, ``{Helicity of the Electron and
  Positron in Muon Decay},''
\href{http://dx.doi.org/10.1103/PhysRev.112.2061}{{\em Phys. Rev.} {\bfseries
  112} (1958) 2061--2071}.
%%CITATION = PHRVA,112,2061;%%.

\bibitem{t_violation_babar}
J.~P. Lees {\em et~al.} [{BaBar} Collaboration], ``{Observation of Time
  Reversal Violation in the $B^0$ Meson System},''
  \href{http://dx.doi.org/10.1103/PhysRevLett.109.211801}{{\em Phys. Rev.
  Lett.} {\bfseries 109} (2012) 211801},
\href{http://arxiv.org/abs/1207.5832}{{\ttfamily arXiv:1207.5832 [hep-ex]}}.
%%CITATION = ARXIV:1207.5832;%%.

\bibitem{cp_violation}
J.~H. Christenson, J.~W. Cronin, V.~L. Fitch, and R.~Turlay, ``Evidence for the
  $2\ensuremath{\pi}$ decay of the $k_{2}^{0}$ meson,''
  \href{http://dx.doi.org/10.1103/PhysRevLett.13.138}{{\em Phys. Rev. Lett.}
  {\bfseries 13} (Jul, 1964)  138--140}.

\bibitem{pdg2016}
C.~Patrignani {\em et~al.} [{Particle Data Group} Collaboration], ``{Review of
  Particle Physics},''
\href{http://dx.doi.org/10.1088/1674-1137/40/10/100001}{{\em Chin. Phys.}
  {\bfseries C40} no.~10, (2016) 100001}.
%%CITATION = CHPHD,C40,100001;%%.

\bibitem{cpt_positronium}
P.~A. Vetter and S.~J. Freedman, ``Search for $cpt$-odd decays of
  positronium,'' \href{http://dx.doi.org/10.1103/PhysRevLett.91.263401}{{\em
  Phys. Rev. Lett.} {\bfseries 91} (Dec, 2003)  263401}.

\bibitem{cp_positronium}
T.~Yamazaki, T.~Namba, S.~Asai, and T.~Kobayashi, ``Search for $cp$ violation
  in positronium decay,''
  \href{http://dx.doi.org/10.1103/PhysRevLett.104.083401}{{\em Phys. Rev.
  Lett.} {\bfseries 104} (Feb, 2010)  083401}.

\bibitem{theory-babar}
J.~P. Lees {\em et~al.} [{BaBar} Collaboration], ``{Observation of Time
  Reversal Violation in the $B^0$ Meson System},''
  \href{http://dx.doi.org/10.1103/PhysRevLett.109.211801}{{\em Phys. Rev.
  Lett.} {\bfseries 109} (2012) 211801},
\href{http://arxiv.org/abs/1207.5832}{{\ttfamily arXiv:1207.5832 [hep-ex]}}.
%%CITATION = ARXIV:1207.5832;%%.

\bibitem{theory:bernabeu-t}
J.~Bernabeu, A.~Di~Domenico, and P.~Villanueva-Perez, ``{Direct test of
  time-reversal symmetry in the entangled neutral kaon system at a
  $\phi$-factory},''
  \href{http://dx.doi.org/10.1016/j.nuclphysb.2012.11.009}{{\em Nucl. Phys.}
  {\bfseries B868} (2013) 102--119},
\href{http://arxiv.org/abs/1208.0773}{{\ttfamily arXiv:1208.0773 [hep-ph]}}.
%%CITATION = ARXIV:1208.0773;%%.

\bibitem{moskal_potential}
P.~Moskal,\ldots, A. Gajos {\em et~al.}, ``{Potential of the J-PET detector for
  studies of discrete symmetries in decays of positronium atom - a purely
  leptonic system},'' \href{http://dx.doi.org/10.5506/APhysPolB.47.509}{{\em
  Acta Phys. Polon.} {\bfseries B47} (2016) 509},
\href{http://arxiv.org/abs/1602.05226}{{\ttfamily arXiv:1602.05226 [nucl-ex]}}.
%%CITATION = ARXIV:1602.05226;%%.

\bibitem{sachs}
R.~Sachs, ``The physics of time reversal,''. University of Chicago Press, 1987.

\bibitem{wolfenstein_summary}
L.~Wolfenstein, ``{The search for direct evidence for time reversal
  violation},''
\href{http://dx.doi.org/10.1142/S0218301399000343}{{\em Int. J. Mod. Phys.}
  {\bfseries E8} (1999) 501--511}.
%%CITATION = IMPAE,E8,501;%%.

\bibitem{nedm}
J.~M. Pendlebury {\em et~al.}, ``{Revised experimental upper limit on the
  electric dipole moment of the neutron},''
  \href{http://dx.doi.org/10.1103/PhysRevD.92.092003}{{\em Phys. Rev.}
  {\bfseries D92} no.~9, (2015) 092003},
\href{http://arxiv.org/abs/1509.04411}{{\ttfamily arXiv:1509.04411 [hep-ex]}}.
%%CITATION = ARXIV:1509.04411;%%.

\bibitem{eedm}
J.~Baron {\em et~al.} [{ACME} Collaboration], ``{Order of Magnitude Smaller
  Limit on the Electric Dipole Moment of the Electron},''
  \href{http://dx.doi.org/10.1126/science.1248213}{{\em Science} {\bfseries
  343} (2014) 269--272},
\href{http://arxiv.org/abs/1310.7534}{{\ttfamily arXiv:1310.7534
  [physics.atom-ph]}}.
%%CITATION = ARXIV:1310.7534;%%.

\bibitem{operatory_kek}
M.~Abe {\em et~al.}, ``{Search for T-violating transverse muon polarization in
  the $K^+\to\pi^0\mu^+\nu$ decay},''
\href{http://dx.doi.org/10.1103/PhysRevD.73.072005}{{\em Phys. Rev.} {\bfseries
  D73} (2006) 072005}.
%%CITATION = PHRVA,D73,072005;%%.

\bibitem{bodek_li}
R.~Huber, J.~Lang, S.~Navert, J.~Sromicki, K.~Bodek, S.~Kistryn, J.~Zejma,
  O.~Naviliat-Cuncic, E.~Stephan, and W.~Haeberli, ``Search for time-reversal
  violation in the $\ensuremath{\beta}$ decay of polarized
  $^{8}\mathrm{L}\mathrm{i}$ nuclei,''
  \href{http://dx.doi.org/10.1103/PhysRevLett.90.202301}{{\em Phys. Rev. Lett.}
  {\bfseries 90} (May, 2003)  202301}.

\bibitem{bodek_free_n}
A.~Kozela, G.~Ban, A.~Bia\l{}ek, K.~Bodek, P.~Gorel, K.~Kirch, S.~Kistryn,
  O.~Naviliat-Cuncic, N.~Severijns, E.~Stephan, and J.~Zejma [{nTRV
  Collaboration} Collaboration], ``Measurement of the transverse polarization
  of electrons emitted in free neutron decay,''
  \href{http://dx.doi.org/10.1103/PhysRevC.85.045501}{{\em Phys. Rev. C}
  {\bfseries 85} (Apr, 2012)  045501}.

\bibitem{Kleinknecht:1994ns}
K.~Kleinknecht, ``{Results on rare decays of neutral kaons from the NA31
  experiment at CERN},''
{\em Frascati Physics Series} {\bfseries 3} (1994) 377--398.
%%CITATION = 00309,3,377;%%.

\bibitem{Winhart:2012bv}
A.~Winhart [{NA48/2 and NA62} Collaboration], ``{Kaon physics at CERN: Recent
  results from the NA48/2 and NA62 experiments},''
\href{http://dx.doi.org/10.1051/epjconf/20123701010}{{\em EPJ Web Conf.}
  {\bfseries 37} (2012) 01010}.
%%CITATION = 00776,37,01010;%%.

\bibitem{Wanke:2003vp}
R.~Wanke, ``{New results in kaon physics from NA48 and KTeV},'' {\em eConf}
  {\bfseries C030626} (2003) FRAT07,
\href{http://arxiv.org/abs/hep-ex/0309078}{{\ttfamily arXiv:hep-ex/0309078
  [hep-ex]}}.
%%CITATION = HEP-EX/0309078;%%.

\bibitem{kloe_results}
F.~Bossi, E.~De~Lucia, J.~Lee-Franzini, S.~Miscetti, and M.~Palutan [{KLOE}
  Collaboration], ``{Precision Kaon and Hadron Physics with KLOE},''
  \href{http://dx.doi.org/10.1393/ncr/i2008-10037-9}{{\em Riv. Nuovo Cim.}
  {\bfseries 31} (2008) 531--623},
\href{http://arxiv.org/abs/0811.1929}{{\ttfamily arXiv:0811.1929 [hep-ex]}}.
%%CITATION = ARXIV:0811.1929;%%.

\bibitem{interf_handbook}
INFN, ``{Handbook on neutral kaon interferometry at a $\Phi$-factory},''
  vol.~43 of {\em Frascati physics series}.
\newblock INFN, Frascati, Italy, 2007.
\newblock
\url{http://www-spires.fnal.gov/spires/find/books/www?cl=QC793.5.M427H25::2007}.
\newblock
%%CITATION = 00309,43,;%%.

\bibitem{book_cp_violation}
J.~P.~S. Gustavo Castelo~Branco, Luís~Lavoura, ``Cp violation,''
\newblock Monographs on Physics. Oxford Univ. Press, New York, NY, 1999.

\bibitem{fidecaro_pedagogical}
M.~Fidecaro and H.-J. Gerber, ``{The Fundamental symmetries in the neutral kaon
  system: A Pedagogical choice},''
  \href{http://dx.doi.org/10.1088/0034-4885/69/10/C01}{{\em Rept. Prog. Phys.}
  {\bfseries 69} (2006) 1713--1770},
  \href{http://arxiv.org/abs/hep-ph/0603075}{{\ttfamily arXiv:hep-ph/0603075
  [hep-ph]}}.
[Erratum: Rept. Prog. Phys.69,2841(2006)].
%%CITATION = HEP-PH/0603075;%%.

\bibitem{Kabir1970}
P.~K. Kabir, ``{What is not invariant under time reversal?},''
\href{http://dx.doi.org/10.1103/PhysRevD.2.540}{{\em Phys. Rev.} {\bfseries D2}
  (1970) 540--542}.
%%CITATION = PHRVA,D2,540;%%.

\bibitem{cplear}
A.~Angelopoulos {\em et~al.} [{CPLEAR} Collaboration], ``{First direct
  observation of time reversal noninvariance in the neutral kaon system},''
\href{http://dx.doi.org/10.1016/S0370-2693(98)01356-2}{{\em Phys. Lett.}
  {\bfseries B444} (1998) 43--51}.
%%CITATION = PHLTA,B444,43;%%.

\bibitem{wolfenstein_other_paper}
L.~Wolfenstein, ``{Violation of time reversal invariance in K0 decays},''
\href{http://dx.doi.org/10.1103/PhysRevLett.83.911}{{\em Phys. Rev. Lett.}
  {\bfseries 83} (1999) 911--912}.
%%CITATION = PRLTA,83,911;%%.

\bibitem{babar_zero_result}
J.~P. Lees {\em et~al.} [{BaBar} Collaboration], ``{Search for $CP$ Violation
  in $B^0$-$\bar{B}^0$ Mixing using Partial Reconstruction of $B^0 \to
  D^{*-}X\ell^+ \nu_\ell$ and a Kaon Tag},''
  \href{http://dx.doi.org/10.1103/PhysRevLett.111.159901,
  10.1103/PhysRevLett.111.101802}{{\em Phys. Rev. Lett.} {\bfseries 111}
  no.~10, (2013) 101802}, \href{http://arxiv.org/abs/1305.1575}{{\ttfamily
  arXiv:1305.1575 [hep-ex]}}.
[Addendum: Phys. Rev. Lett.111,no.15,159901(2013)].
%%CITATION = ARXIV:1305.1575;%%.

\bibitem{Ellis:1999xh}
J.~R. Ellis and N.~Mavromatos, ``{Comments on CP, T and CPT violation in
  neutral kaon decays},''
  \href{http://dx.doi.org/10.1016/S0370-1573(99)00058-7}{{\em Phys.Rept.}
  {\bfseries 320} (1999) 341--354},
\href{http://arxiv.org/abs/hep-ph/9903386}{{\ttfamily arXiv:hep-ph/9903386
  [hep-ph]}}.
%%CITATION = HEP-PH/9903386;%%.

\bibitem{Gerber:2004hc}
H.~Gerber, ``{Evidence for time-reversal violation?},''
\href{http://dx.doi.org/10.1140/epjc/s2004-01813-6}{{\em Eur.Phys.J.}
  {\bfseries C35} (2004) 195--196}.
%%CITATION = EPHJA,C35,195;%%.

\bibitem{banuls_first_bmesons}
M.~C. Banuls and J.~Bernabeu, ``{CP, T and CPT versus temporal asymmetries for
  entangled states of the B(d) system},''
  \href{http://dx.doi.org/10.1016/S0370-2693(99)01043-6}{{\em Phys. Lett.}
  {\bfseries B464} (1999) 117--122},
\href{http://arxiv.org/abs/hep-ph/9908353}{{\ttfamily arXiv:hep-ph/9908353
  [hep-ph]}}.
%%CITATION = HEP-PH/9908353;%%.

\bibitem{babar_theory}
J.~Bernabeu, F.~Martinez-Vidal, and P.~Villanueva-Perez, ``{Time Reversal
  Violation from the entangled $\mathrm{B}^0-\overline{\mathrm{B}}^0$
  system},'' \href{http://dx.doi.org/10.1007/JHEP08(2012)064}{{\em JHEP}
  {\bfseries 08} (2012) 064},
\href{http://arxiv.org/abs/1203.0171}{{\ttfamily arXiv:1203.0171 [hep-ph]}}.
%%CITATION = ARXIV:1203.0171;%%.

\bibitem{Einstein:1935rr}
A.~Einstein, B.~Podolsky, and N.~Rosen, ``{Can quantum mechanical description
  of physical reality be considered complete?},''
\href{http://dx.doi.org/10.1103/PhysRev.47.777}{{\em Phys. Rev.} {\bfseries 47}
  (1935) 777--780}.
%%CITATION = PHRVA,47,777;%%.

\bibitem{Babusci:2013tr}
D.~Babusci {\em et~al.} [{KLOE} Collaboration], ``{A new limit on the CP
  violating decay $K_S \to 3\pi^0$ with the KLOE experiment},''
  \href{http://dx.doi.org/10.1016/j.physletb.2013.05.008}{{\em Phys. Lett.}
  {\bfseries B723} (2013) 54--60},
\href{http://arxiv.org/abs/1301.7623}{{\ttfamily arXiv:1301.7623 [hep-ex]}}.
%%CITATION = ARXIV:1301.7623;%%.

\bibitem{kloe_kl3pi0_br}
F.~Ambrosino {\em et~al.} [{KLOE} Collaboration], ``{Measurements of the
  absolute branching ratios for the dominant K(L) decays, the K(L) lifetime,
  and V(us) with the KLOE detector},''
  \href{http://dx.doi.org/10.1016/j.physletb.2005.10.018}{{\em Phys. Lett.}
  {\bfseries B632} (2006) 43--50},
\href{http://arxiv.org/abs/hep-ex/0508027}{{\ttfamily arXiv:hep-ex/0508027
  [hep-ex]}}.
%%CITATION = HEP-EX/0508027;%%.

\bibitem{kloe_kl_lifetime}
F.~Ambrosino {\em et~al.} [{KLOE} Collaboration], ``{Measurement of the K(L)
  meson lifetime with the KLOE detector},''
  \href{http://dx.doi.org/10.1016/j.physletb.2005.08.022}{{\em Phys. Lett.}
  {\bfseries B626} (2005) 15--23},
\href{http://arxiv.org/abs/hep-ex/0507088}{{\ttfamily arXiv:hep-ex/0507088
  [hep-ex]}}.
%%CITATION = HEP-EX/0507088;%%.

\bibitem{kloe_kspipi_br}
F.~Ambrosino {\em et~al.} [{KLOE} Collaboration], ``{Precise measurement of
  $\Gamma(K(s) \to \pi^+ \pi^- (\gamma)) / \Gamma(K(s) \to \pi^0 \pi^0$) with
  the KLOE detector at DAFNE},''
  \href{http://dx.doi.org/10.1140/epjc/s10052-006-0021-9}{{\em Eur. Phys. J.}
  {\bfseries C48} (2006) 767--780},
\href{http://arxiv.org/abs/hep-ex/0601025}{{\ttfamily arXiv:hep-ex/0601025
  [hep-ex]}}.
%%CITATION = HEP-EX/0601025;%%.

\bibitem{kloe_ks_lifetime}
F.~Ambrosino {\em et~al.} [{KLOE} Collaboration], ``{Precision Measurement of
  $K_S$ Meson Lifetime with the KLOE detector},''
  \href{http://dx.doi.org/10.1140/epjc/s10052-011-1604-7}{{\em Eur. Phys. J.}
  {\bfseries C71} (2011) 1604},
\href{http://arxiv.org/abs/1011.2668}{{\ttfamily arXiv:1011.2668 [hep-ex]}}.
%%CITATION = ARXIV:1011.2668;%%.

\bibitem{Arbic:1988pv}
B.~K. Arbic, S.~Hatamian, M.~Skalsey, J.~Van~House, and W.~Zheng, ``{Angular
  Correlation Test of {CPT} in Polarized Positronium},''
\href{http://dx.doi.org/10.1103/PhysRevA.37.3189}{{\em Phys. Rev.} {\bfseries
  A37} (1988) 3189--3194}.
%%CITATION = PHRVA,A37,3189;%%.

\bibitem{Skalsey:1991vt}
M.~Skalsey and J.~Van~House, ``{First test of CP invariance in the decay of
  positronium},''
\href{http://dx.doi.org/10.1103/PhysRevLett.67.1993}{{\em Phys. Rev. Lett.}
  {\bfseries 67} (1991) 1993--1996}.
%%CITATION = PRLTA,67,1993;%%.

\bibitem{Bernreuther:1988tt}
W.~Bernreuther, U.~Low, J.~P. Ma, and O.~Nachtmann, ``{How to Test {CP}, $T$
  and {CPT} Invariance in the Three Photon Decay of Polarized $s$ Wave Triplet
  Positronium},''
\href{http://dx.doi.org/10.1007/BF01412589}{{\em Z. Phys.} {\bfseries C41}
  (1988) 143}.
%%CITATION = ZEPYA,C41,143;%%.

\bibitem{Harpen:2003zz}
M.~D. Harpen, ``{Positronium: Review of symmetry, conserved quantities and
  decay for the radiological physicist},''
\href{http://dx.doi.org/10.1118/1.1630494}{{\em Med. Phys.} {\bfseries 31}
  (2004) 57--61}.
%%CITATION = MPHYA,31,57;%%.

\bibitem{PhysRevLett.72.1632}
A.~H. Al-Ramadhan and D.~W. Gidley, ``New precision measurement of the decay
  rate of singlet positronium,''
  \href{http://dx.doi.org/10.1103/PhysRevLett.72.1632}{{\em Phys. Rev. Lett.}
  {\bfseries 72} (Mar, 1994)  1632--1635}.

\bibitem{PhysRevLett.90.203402}
R.~S. Vallery, P.~W. Zitzewitz, and D.~W. Gidley, ``Resolution of the
  orthopositronium-lifetime puzzle,''
  \href{http://dx.doi.org/10.1103/PhysRevLett.90.203402}{{\em Phys. Rev. Lett.}
  {\bfseries 90} (May, 2003)  203402}.

\bibitem{JINNOUCHI2003117}
O.~Jinnouchi, S.~Asai, and T.~Kobayashi, ``Precision measurement of
  orthopositronium decay rate using sio2 powder,''
  \href{http://dx.doi.org/https://doi.org/10.1016/j.physletb.2003.08.018}{{\em
  Phys. Lett.} {\bfseries B572} no.~3, (2003) 117--126}.

\bibitem{daria_epjc}
D.~Kamińska, A.~Gajos, {\em et~al.}, ``{A feasibility study of
  ortho-positronium decays measurement with the J-PET scanner based on plastic
  scintillators},''
  \href{http://dx.doi.org/10.1140/epjc/s10052-016-4294-3}{{\em Eur. Phys. J.}
  {\bfseries C76} no.~8, (2016) 445},
\href{http://arxiv.org/abs/1607.08588}{{\ttfamily arXiv:1607.08588
  [physics.ins-det]}}.
%%CITATION = ARXIV:1607.08588;%%.

\bibitem{Vignola:1996mt}
G.~Vignola {\em et~al.}, ``{DAPHNE, the first Phi-factory},''
{\em Conf. Proc.} {\bfseries C960610} (1996) 22--26.
%%CITATION = CONFP,C960610,22;%%.

\bibitem{CurceanuPetrascu:2007zz}
C.~Curceanu-Petrascu {\em et~al.}, ``{Precision measurements of kaonic atoms at
  DAFNE and future perspectives},''
\href{http://dx.doi.org/10.1140/epja/i2006-10197-2}{{\em Eur. Phys. J.}
  {\bfseries A31} (2007) 537--539}.
%%CITATION = EPHJA,A31,537;%%.

\bibitem{Lucherini:2003fk}
V.~Lucherini {\em et~al.}, ``{The DEAR kaon monitor at DAPHNE},''
\href{http://dx.doi.org/10.1016/S0168-9002(02)01753-9}{{\em Nucl. Instrum.
  Meth.} {\bfseries A496} (2003) 315--324}.
%%CITATION = NUIMA,A496,315;%%.

\bibitem{Agnello:2007rk}
M.~Agnello {\em et~al.}, ``{DAFNE monitored by FINUDA},''
\href{http://dx.doi.org/10.1016/j.nima.2006.09.113}{{\em Nucl. Instrum. Meth.}
  {\bfseries A570} (2007) 205--215}.
%%CITATION = NUIMA,A570,205;%%.

\bibitem{LeeFranzini:2007hj}
J.~Lee-Franzini and P.~Franzini, ``{A Flavor of KLOE},'' {\em Acta Phys.
  Polon.} {\bfseries B38} (2007) 2703--2730,
\href{http://arxiv.org/abs/hep-ex/0702016}{{\ttfamily arXiv:hep-ex/0702016
  [hep-ex]}}.
%%CITATION = HEP-EX/0702016;%%.

\bibitem{Adinolfi:2002me}
M.~Adinolfi {\em et~al.}, ``{The QCAL tile calorimeter of KLOE},''
\href{http://dx.doi.org/10.1016/S0168-9002(01)01929-5}{{\em Nucl. Instrum.
  Meth.} {\bfseries A483} (2002) 649--659}.
%%CITATION = NUIMA,A483,649;%%.

\bibitem{Adinolfi:2002uk}
M.~Adinolfi, F.~Ambrosino, A.~Andryakov, A.~Antonelli, M.~Antonelli, {\em
  et~al.}, ``{The tracking detector of the KLOE experiment},''
\href{http://dx.doi.org/10.1016/S0168-9002(02)00514-4}{{\em Nucl.Instrum.Meth.}
  {\bfseries A488} (2002) 51--73}.
%%CITATION = NUIMA,A488,51;%%.

\bibitem{data_handling}
F.~Ambrosino {\em et~al.}, ``{Data handling, reconstruction, and simulation for
  the KLOE experiment},''
  \href{http://dx.doi.org/10.1016/j.nima.2004.06.155}{{\em Nucl. Instrum.
  Meth.} {\bfseries A534} (2004) 403--433},
\href{http://arxiv.org/abs/physics/0404100}{{\ttfamily arXiv:physics/0404100
  [physics]}}.
%%CITATION = PHYSICS/0404100;%%.

\bibitem{Adinolfi:2002zx}
M.~Adinolfi, F.~Ambrosino, A.~Antonelli, M.~Antonelli, F.~Anulli, {\em et~al.},
  ``{The KLOE electromagnetic calorimeter},''
\href{http://dx.doi.org/10.1016/S0168-9002(01)01502-9}{{\em Nucl.Instrum.Meth.}
  {\bfseries A482} (2002) 364--386}.
%%CITATION = NUIMA,A482,364;%%.

\bibitem{kloe_web}
{KLOE Experiment webpage}
\newblock \url{http://www.lnf.infn.it/kloe/index2.html}. Accessed: 2017-11-30.

\bibitem{zdebik_mgr}
J.~Zdebik, ``{Merging and splitting of lusters in the electromagnetic
  calorimeter of the KLOE detector},'' Master thesis, Jagiellonian University,
  Krakow, Poland (2008).

\bibitem{difalco_phd}
S.~D. Falco, ``{Study of neutral decays of $K^0_L$ mesons with the KLOE
  detector},'' PhD thesis, Universita Degli Studi di Pisa, Pisa, Italy (2001).

\bibitem{memo_225}
F.~Ambrosino {\em et~al.}, ``The event classification procedures.'' KLOE Memo
  {\bf 228} (2000).

\bibitem{kloe_inner_tracker}
A.~Balla {\em et~al.}, ``{Status of the Cylindical-GEM project for the KLOE-2
  Inner Tracker},'' \href{http://dx.doi.org/10.1016/j.nima.2010.06.315}{{\em
  Nucl. Instrum. Meth.} {\bfseries A628} (2011) 194--198},
\href{http://arxiv.org/abs/physics.ins-det/1003.3770}{{\ttfamily
  arXiv:physics.ins-det/1003.3770 [physics.ins-det]}}.
%%CITATION = ARXIV:1003.3770;%%.

\bibitem{Cordelli:2013mka}
M.~Cordelli {\em et~al.}, ``{CCALT: A Crystal CALorimeter with Timing for the
  KLOE-2 upgrade},''
\href{http://dx.doi.org/10.1016/j.nima.2012.11.074}{{\em Nucl. Instrum. Meth.}
  {\bfseries A718} (2013) 81--82}.
%%CITATION = NUIMA,A718,81;%%.

\bibitem{Cordelli:2009xb}
M.~Cordelli, G.~Corradi, F.~Happacher, M.~Martini, S.~Miscetti, C.~Paglia,
  A.~Saputi, I.~Sarra, and D.~Tagnani, ``{QCALT: A Tile calorimeter for KLOE-2
  experiment},'' \href{http://dx.doi.org/10.1016/j.nima.2009.09.106}{{\em Nucl.
  Instrum. Meth.} {\bfseries A617} (2010) 105--106},
\href{http://arxiv.org/abs/0906.1133}{{\ttfamily arXiv:0906.1133
  [physics.ins-det]}}.
%%CITATION = ARXIV:0906.1133;%%.

\bibitem{Zobov:2007xw}
M.~Zobov, ``{DAFNE status and upgrade plans},''
  \href{http://dx.doi.org/10.1134/S1547477108070042}{{\em Phys. Part. Nucl.
  Lett.} {\bfseries 5} (2008) 560--565},
\href{http://arxiv.org/abs/0709.3696}{{\ttfamily arXiv:0709.3696
  [physics.acc-ph]}}.
%%CITATION = ARXIV:0709.3696;%%.

\bibitem{kloe2_general}
G.~Amelino-Camelia {\em et~al.}, ``{Physics with the KLOE-2 experiment at the
  upgraded DA$\phi$NE},''
  \href{http://dx.doi.org/10.1140/epjc/s10052-010-1351-1}{{\em Eur. Phys. J.}
  {\bfseries C68} (2010) 619--681},
\href{http://arxiv.org/abs/1003.3868}{{\ttfamily arXiv:1003.3868 [hep-ex]}}.
%%CITATION = ARXIV:1003.3868;%%.

\bibitem{moskal_patent}
P.~Moskal, ``Strip device and the method for the determination of the place and
  response time of the gamma quanta and the application of the device for the
  positron emission thomography.'' Patent number US2012112079, EP2454612,
  JP2012533734.

\bibitem{gajos_gps}
A.~Gajos {\em et~al.}, ``{Trilateration-based reconstruction of
  ortho-positronium decays into three photons with the J-PET detector},''
  \href{http://dx.doi.org/10.1016/j.nima.2016.02.069}{{\em Nucl. Instrum.
  Meth.} {\bfseries A819} (2016) 54--59},
\href{http://arxiv.org/abs/1602.07528}{{\ttfamily arXiv:1602.07528
  [physics.ins-det]}}.
%%CITATION = ARXIV:1602.07528;%%.

\bibitem{anna_scints}
A.~Wieczorek,\ldots, A. Gajos {\em et~al.}, ``Novel scintillating material
  2-(4-styrylphenyl)benzoxazole for the fully digital and mri compatible j-pet
  tomograph based on plastic scintillators,''
  \href{http://dx.doi.org/10.1371/journal.pone.0186728}{{\em PLOS ONE}
  {\bfseries 12} no.~11, (2017) 1--16}.

\bibitem{jpet_single_module}
P.~Moskal,\ldots, A. Gajos {\em et~al.}, ``{Test of a single module of the
  J-PET scanner based on plastic scintillators},''
  \href{http://dx.doi.org/10.1016/j.nima.2014.07.052}{{\em Nucl. Instrum.
  Meth.} {\bfseries A764} (2014) 317--321},
\href{http://arxiv.org/abs/1407.7395}{{\ttfamily arXiv:1407.7395
  [physics.ins-det]}}.
%%CITATION = ARXIV:1407.7395;%%.

\bibitem{Griffiths}
D.~Griffiths, ``{Introduction to elementary particles},''. Wiley-VCH, 2nd~ed.,
  2010.

\bibitem{marek_fee}
M.~Pałka,\ldots, A. Gajos {\em et~al.}, ``{Multichannel FPGA based MVT system
  for high precision time (20 ps RMS) and charge measurement},''
  \href{http://dx.doi.org/10.1088/1748-0221/12/08/P08001}{{\em JINST}
  {\bfseries 12} no.~08, (2017) P08001},
\href{http://arxiv.org/abs/1707.03565}{{\ttfamily arXiv:1707.03565
  [physics.ins-det]}}.
%%CITATION = ARXIV:1707.03565;%%.

\bibitem{greg_daq}
G.~Korcyl,\ldots, A. Gajos {\em et~al.}, ``{Sampling FEE and Trigger-less DAQ
  for the J-PET Scanner},''
  \href{http://dx.doi.org/10.5506/APhysPolB.47.491}{{\em Acta Phys. Polon.}
  {\bfseries B47} (2016) 491},
\href{http://arxiv.org/abs/1602.05251}{{\ttfamily arXiv:1602.05251
  [physics.ins-det]}}.
%%CITATION = ARXIV:1602.05251;%%.

\bibitem{lech_compressive}
L.~Raczyński,\ldots, A. Gajos {\em et~al.}, ``{Compressive Sensing of Signals
  Generated in Plastic Scintillators in a Novel J-PET Instrument},''
  \href{http://dx.doi.org/10.1016/j.nima.2015.03.032}{{\em Nucl. Instrum.
  Meth.} {\bfseries A786} (2015) 105--112},
\href{http://arxiv.org/abs/1503.05188}{{\ttfamily arXiv:1503.05188
  [physics.ins-det]}}.
%%CITATION = ARXIV:1503.05188;%%.

\bibitem{neha_synchronized}
P.~Moskal,\ldots, A. Gajos {\em et~al.}, ``{A novel method for the
  line-of-response and time-of-flight reconstruction in TOF-PET detectors based
  on a library of synchronized model signals},''
  \href{http://dx.doi.org/10.1016/j.nima.2014.12.005}{{\em Nucl. Instrum.
  Meth.} {\bfseries A775} (2015) 54--62},
\href{http://arxiv.org/abs/1412.6963}{{\ttfamily arXiv:1412.6963
  [physics.ins-det]}}.
%%CITATION = ARXIV:1412.6963;%%.

\bibitem{jpet_time_calibration}
M.~Skurzok,\ldots, A. Gajos {\em et~al.}, ``{Time calibration of the J-PET
  detector},'' \href{http://dx.doi.org/10.12693/APhysPolA.132.1641}{{\em Acta
  Phys. Polon.} {\bfseries A132} no.~5, (2017) 1641--1645},
\href{http://arxiv.org/abs/1710.05598}{{\ttfamily arXiv:1710.05598
  [physics.ins-det]}}.
%%CITATION = ARXIV:1710.05598;%%.

\bibitem{Jasinska:2016qsf}
B.~Jasińska,\ldots, A. Gajos {\em et~al.}, ``{Determination of the $3\gamma$
  fraction from positron annihilation in mesoporous materials for symmetry
  violation experiment with J-PET scanner},''
  \href{http://dx.doi.org/10.5506/APhysPolB.47.453}{{\em Acta Phys. Polon.}
  {\bfseries B47} (2016) 453},
\href{http://arxiv.org/abs/1602.05376}{{\ttfamily arXiv:1602.05376 [nucl-ex]}}.
%%CITATION = ARXIV:1602.05376;%%.

\bibitem{PhysRevLett.43.1281}
P.~W. Zitzewitz, J.~C. Van~House, A.~Rich, and D.~W. Gidley, ``Spin
  polarization of low-energy positron beams,''
  \href{http://dx.doi.org/10.1103/PhysRevLett.43.1281}{{\em Phys. Rev. Lett.}
  {\bfseries 43} (Oct, 1979)  1281--1284}.

\bibitem{Coleman}
P.~Coleman, ``{Positron beams and their applications},''. World Scientific,
  2000.

\bibitem{Langley2005TheMO}
R.~B. Langley, ``The mathematics of gps,'' (2005).

\bibitem{Carter1999PrinciplesOG}
C.~Carter and {OSBORNE ASSOCIATES}, ``Principles of gps,'' (1999).

\bibitem{gajos_mgr}
A.~Gajos, ``{A novel algorithm for the
  $K\to\pi^0\pi^0\to\gamma\gamma\gamma\gamma$},'' Master thesis, Jagiellonian
  University, Krakow, Poland (2013).

\bibitem{Kleusberg2003}
A.~Kleusberg, ``Analytical gps navigation solution,'',
  \href{http://dx.doi.org/10.1007/978-3-662-05296-9_7}{93--96}.
\newblock Springer Berlin Heidelberg, 2003.

\bibitem{imaging_patent}
A.~Gajos, E.~Czerwi\'nski, D.~Kami\'nska, and P.~Moskal, ``Multi-tracer
  morphometric image reconstruction method and system using fast analytical
  algorithm for calculating time and position of positron-electron annihilation
  into three gamma quanta.'' International patent application number
  PCT/PL2015/050038.

\bibitem{daria_memo}
D.~Kisielewska {\em et~al.}, ``Measurement of the charge asymmetry for the
  $k_s\to\pi e \nu$ with the kloe detector.'' KLOE Memo (2017).

\bibitem{daria_article}
D.~Kisielewska,\ldots, A. Gajos {\em et~al.} [{KLOE-2} Collaboration],
  ``Measurement of the lepton charge asymmetry for short-lived kaon with the
  kloe detector,'' in preparation (2018).

\bibitem{Ambrosino:2006si}
F.~Ambrosino {\em et~al.} [{KLOE} Collaboration], ``{Study of the branching
  ratio and charge asymmetry for the decay $K(s) \to \pi e \nu$ with the KLOE
  detector},'' \href{http://dx.doi.org/10.1016/j.physletb.2006.03.047}{{\em
  Phys. Lett.} {\bfseries B636} (2006) 173--182},
\href{http://arxiv.org/abs/hep-ex/0601026}{{\ttfamily arXiv:hep-ex/0601026
  [hep-ex]}}.
%%CITATION = HEP-EX/0601026;%%.

\bibitem{kloe_memo_146}
S.~Sinibaldi and T.~Spadaro, ``Classification algorithm of calorimeter
  clusters.'' KLOE Memo {\bf 146} (1998).

\bibitem{graziani_anns}
A.~Anastasi,\ldots, A. Gajos {\em et~al.} [{KLOE-2} Collaboration], ``{Search
  for dark Higgsstrahlung in $e^{+}e^{-} \to \mu^{+}\mu^-$ and missing energy
  events with the KLOE experiment},''
  \href{http://dx.doi.org/10.1016/j.physletb.2015.06.015}{{\em Phys. Lett.}
  {\bfseries B747} (2015) 365--372},
\href{http://arxiv.org/abs/1501.06795}{{\ttfamily arXiv:1501.06795 [hep-ex]}}.
%%CITATION = ARXIV:1501.06795;%%.

\bibitem{Hocker:2007ht}
A.~Hoecker, P.~Speckmayer, J.~Stelzer, J.~Therhaag, E.~von Toerne, and H.~Voss,
  ``{TMVA: Toolkit for Multivariate Data Analysis},'' {\em PoS} {\bfseries
  ACAT} (2007) 040,
\href{http://arxiv.org/abs/physics/0703039}{{\ttfamily arXiv:physics/0703039}}.
%%CITATION = PHYSICS/0703039;%%.

\bibitem{optimization}
P.~Gill and W.~Murray, eds., ``Numerical methods for constrained
  optimization,''. Academic Press London, 1974.

\bibitem{gajos-acta}
A.~Gajos [{KLOE-2} Collaboration], ``{Study of the $K_{\rm S}K_{\rm L}\to \pi
  \ell \nu 3\pi ^0$ Process for Time Reversal Symmetry Test at KLOE-2},''
  \href{http://dx.doi.org/10.5506/APhysPolB.46.13}{{\em Acta Phys. Polon.}
  {\bfseries B46} no.~1, (2015) 13--18},
\href{http://arxiv.org/abs/1501.04801}{{\ttfamily arXiv:1501.04801 [hep-ex]}}.
%%CITATION = ARXIV:1501.04801;%%.

\bibitem{Gajos:2015ija}
A.~Gajos [{KLOE-2} Collaboration], ``{A direct test of T symmetry in the
  neutral K meson system at KLOE-2},''
\href{http://dx.doi.org/10.1088/1742-6596/631/1/012018}{{\em J. Phys. Conf.
  Ser.} {\bfseries 631} no.~1, (2015) 012018}.
%%CITATION = 00462,631,012018;%%.

\bibitem{kloe_memo_280}
F.~Scuri, ``Study of the possibility to detect radiative decays of the $k^0_l$
  at kloe.'' KLOE Memo {\bf 280} (2002).

\bibitem{kloe_memo_322}
M.~Antonelli, M.~Dreucci, and C.~Gatti, ``Measurement of ke3 semileptonic form
  factor slopes in the decay $k_l\to\pi^{\pm}e^{\mp}\nu$ at kloe.'' KLOE Memo
  {\bf 322} (2005).

\bibitem{kloe_memo_334}
M.~Antonelli, M.~Dreucci, and C.~Gatti, ``Measurement of
  br($k^0_{e3\gamma}$)/br($k^0_{e3(\gamma)}$) for inner bremmstrahlung and
  direct emission contribution in semileptonic decay
  $k_l\to\pi^{\pm}e^{\mp}\nu(\gamma)$.'' KLOE Memo {\bf 334} (2007).

\bibitem{Marin:1998zf}
A.~Marin {\em et~al.}, ``{Detection of charged pions and protons in the
  segmented electromagnetic calorimeter TAPS},''
\href{http://dx.doi.org/10.1016/S0168-9002(98)00683-4}{{\em Nucl. Instrum.
  Meth.} {\bfseries A417} (1998) 137--149}.
%%CITATION = NUIMA,A417,137;%%.

\bibitem{behnke}
O.~Behnke, K.~Kr\"oninger, G.~Schott, and T.~Sch\"orner-Sadenius, ``Data
  analysis in high energy physics,''. WILEY-VCH Verlag GmbH \& Co. KGaA, 2013.

\bibitem{statisctical_methods}
F.~James, ``Statistical methods in experimental physics,''. World Scientific
  Publishing, 2nd~ed., 2006.

\bibitem{silarski_phd}
M.~Silarski, ``{Search for the CP symmetry violation in the decays of Ks mesons
  using the KLOE detector},'' PhD thesis, Jagiellonian University, Krak\'ow,
  Poland (2012).
\newblock \href{http://arxiv.org/abs/1302.4427}{{\ttfamily arXiv:1302.4427
  [hep-ex]}}.

\bibitem{kowalski_scatter_fraction}
P.~Kowalski,\ldots, A. Gajos {\em et~al.}, ``{Scatter fraction of the J-PET
  tomography scanner},'' \href{http://dx.doi.org/10.5506/APhysPolB.47.549}{{\em
  Acta Phys. Polon.} {\bfseries B47} (2016) 549},
\href{http://arxiv.org/abs/1602.05402}{{\ttfamily arXiv:1602.05402
  [physics.med-ph]}}.
%%CITATION = ARXIV:1602.05402;%%.

\bibitem{Krzemien:2015hkb}
W.~Krzemie\'n, A.~Gajos, {\em et~al.}, ``{Analysis framework for the J-PET
  scanner},'' \href{http://dx.doi.org/10.12693/APhysPolA.127.1491}{{\em Acta
  Phys. Polon.} {\bfseries A127} (2015) 1491--1494},
\href{http://arxiv.org/abs/1503.00465}{{\ttfamily arXiv:1503.00465
  [physics.ins-det]}}.
%%CITATION = ARXIV:1503.00465;%%.

\bibitem{monika_2gamma_imaging}
S.~Pawlik-Nied\'zwiecka,\ldots, A. Gajos {\em et~al.}, ``Preliminary studies of
  j-pet detector spatial resolution,''
  \href{http://dx.doi.org/10.12693/APhysPolA.132.1645}{{\em Acta Phys. Polon.}
  {\bfseries A132} no.~5, (2017) 1645}.

\bibitem{tot_edep}
A.~M. Grant and C.~S. Levin, ``A new dual threshold time-over-threshold circuit
  for fast timing in pet,''
  \href{http://dx.doi.org/10.1088/0031-9155/59/13/3421}{{\em Physics in
  Medicine \& Biology} {\bfseries 59} no.~13, (2014) 3421}.

\bibitem{theory:bernabeu-cpt}
J.~Bernabeu, A.~Di~Domenico, and P.~Villanueva-Perez, ``{Probing CPT in
  transitions with entangled neutral kaons},''
  \href{http://dx.doi.org/10.1007/JHEP10(2015)139}{{\em JHEP} {\bfseries 10}
  (2015) 139},
\href{http://arxiv.org/abs/1509.02000}{{\ttfamily arXiv:1509.02000 [hep-ph]}}.
%%CITATION = ARXIV:1509.02000;%%.

\bibitem{KLOE-2:2017lyj}
A.~Gajos [{KLOE-2} Collaboration], ``{Discrete symmetries and QM studies with
  entangled neutral kaons at KLOE-2},''
\href{http://dx.doi.org/10.1088/1742-6596/873/1/012025}{{\em J. Phys. Conf.
  Ser.} {\bfseries 873} no.~1, (2017) 012025}.
%%CITATION = 00462,873,012025;%%.

\bibitem{Gajos:2017tlb}
A.~Gajos [{KLOE-2} Collaboration], ``{Tests of discrete symmetries and quantum
  coherence with neutral kaons at the KLOE-2 experiment},''
  \href{http://dx.doi.org/10.5506/APhysPolB.48.1975}{{\em Acta Phys. Polon.}
  {\bfseries B48} (2017) 1975},
\href{http://arxiv.org/abs/1710.08197}{{\ttfamily arXiv:1710.08197 [hep-ex]}}.
%%CITATION = ARXIV:1710.08197;%%.

\end{thebibliography}\endgroup

% \begin{thebibliography}{99}
% \input{./Other/References.tex}
% \end{thebibliography}

\end{document}